\documentclass[a4paper,twoside]{report}

\usepackage{amsmath,amsfonts,amssymb,latexsym,color,hyperref,comment,tikz,amsthm,authblk}

\newcommand{\notes}[1]{\textcolor{blue}{\textsf{[TODO: #1]}}}

\newcommand{\rev}[1]{{#1}}
\newcommand{\added}[1]{{#1}}

\theoremstyle{plain}
  \newtheorem{theorem}{Theorem}[chapter]
  \newtheorem{lemma}{Lemma}[chapter]
\theoremstyle{definition}
  \newtheorem{definition}{Definition}[chapter]
  \newtheorem{algorithm}{Algorithm}[chapter]
  \newtheorem{example}{Example}[chapter]
    \newtheorem{openprob}{Open Problem}
\theoremstyle{remark}
  \newtheorem{remark}{Remark}[chapter]

\newcommand{\NDhat}{\widehat{\mathrm{ND}}}
\newcommand{\PDhat}{\widehat{\mathrm{PD}}}

\newcommand{\Ntilpos}{\widetilde{T}_{\mathrm{pos}}}

\newcommand{\Nneg}{T_{\mathrm{neg}}}

\newcommand{\comp}{\mathsf{c}}
\newcommand{\EE}{{\mathbb{E}}}
\newcommand{\II}{\boldsymbol{1}}
\newcommand{\PP}{{\mathbb{P}}}
\newcommand{\K}{\mathcal{K}}
\newcommand{\Khat}{\hat{\mathcal{K}}}
\renewcommand{\L}{\mathcal{L}}
\newcommand{\zero}{\mathtt{0}}
\newcommand{\one}{\mathtt{1}}
\newcommand{\zo}{\{\zero,\one\}}
\newcommand{\question}{\mathtt{?}}
\newcommand{\zoq}{\{\zero, \one, \question \}}
\newcommand{\YY}{{\mathcal{Y}}}
\newcommand{\ee}{\mathrm{e}}
\newcommand{\asym}{\sim} 
\newcommand{\Perr}{\PP(\mathrm{err})} 
\newcommand{\Psuc}{\PP(\mathrm{suc})}
\newcommand{\Perrd}{\mathbb{P}_d(\mathrm{err}) }
\newcommand{\dmax}{d}
\newcommand{\PerrL}{\mathbb{P}_{L}(\mathrm{error}) }

\newcommand{\olR}{\overline{R}}

\newcommand{\SSS}{\mathrm{SSS}}
\newcommand{\COMP}{\mathrm{COMP}}
\newcommand{\DD}{\mathrm{DD}}

\newcommand{\Bern}{\mathrm{Bern}}
\newcommand{\NCC}{\mathrm{NCC}}
\newcommand{\Mn}{\operatorname{Multinomial}}
\newcommand{\mmid}{\mathbin{\|}}

\renewcommand{\vec}[1]{\mathbf{#1}}
\newcommand{\mat}[1]{\mathsf{#1}}

\newcommand*\circled[1]{%
   \tikz[baseline=(C.base)]\node[draw,circle,inner sep=0.5pt](C) {#1};\!
}

\newcommand{\qq}{q}
\newcommand{\defn}[1]{\emph{#1}}
\newcommand{\etal}{\emph{et al.}}
\newcommand{\yspace}{$\phantom{xxx}$}

\newcommand{\nondefman}{\;\includegraphics[height=0.8cm]{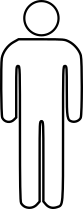}\;}
\newcommand{\defman}{\;\includegraphics[height=0.8cm]{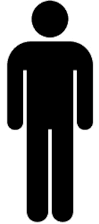}\;}
\newcommand{\quest}{\includegraphics[height=0.6cm]{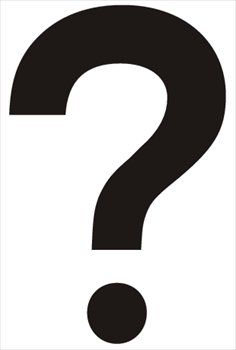}}
\newcommand{\bfone}{\circled{$\one$}}
\newcommand{\vc}[1]{{\mathbf{#1}}}
\renewcommand{\emptyset}{\varnothing}
\DeclareMathOperator*{\argmax}{arg\,max}
\newcommand{\bpnohat}[3]{#1_{#3 \rightarrow t}^{(#2)}}
\newcommand{\bphat}[3]{\widehat{#1}_{#3 \rightarrow i}^{(#2)}}

\DeclareUnicodeCharacter{202C}{} 

\usepackage{mwe}

\title{Group Testing: \\ An Information Theory Perspective \\ \bigskip \Large Second Edition}

\author[1]{Matthew Aldridge}
\author[2]{Oliver Johnson}
\author[3]{Jonathan Scarlett}

\affil[1]{University of Leeds, \texttt{m.aldridge@leeds.ac.uk}
}
\affil[2]{University of Bristol, \texttt{O.Johnson@bristol.ac.uk}
}
\affil[3]{National University of Singapore, \texttt{scarlett@comp.nus.edu.sg}}

\begin{document}

\maketitle

\tableofcontents

\renewcommand{\arraystretch}{1.25} 


%

\begin{abstract}
The group testing problem concerns discovering a small number of defective items within a large population by performing tests on pools of items. A test is positive if the pool contains at least one defective, and negative if it contains no defectives. This is a sparse inference problem with a combinatorial flavour, with applications in medical testing, biology, telecommunications, information technology, data science, and more.

In this monograph, we survey recent developments in the group testing problem from an information-theoretic perspective. We cover several related developments: efficient algorithms with practical storage and computation requirements, achievability bounds for optimal decoding methods, and algorithm-independent converse bounds.  We assess the theoretical guarantees not only in terms of scaling laws, but also in terms of the constant factors, leading to the notion of the {\em rate} of group testing, indicating the amount of information learned per test.  \added{For the noiseless setting, we present a series of results leading to optimal rates, which in turn imply optimality and suboptimality results of various algorithms depending on the sparsity regime.  We also survey analogous developments in noisy settings. }

In addition, we survey results concerning a number of variations on the standard group testing problem, including approximate recovery criteria, adaptive algorithms with a limited number of stages, sublinear-time algorithms, and settings with additional prior information, among others.

\end{abstract}

\chapter*{Preface to the second edition}
\markboth{\sffamily\slshape Preface to the second edition}{\sffamily\slshape Preface to the second edition}
\addcontentsline{toc}{chapter}{Preface to the second edition}

\added{
The first edition of `\emph{Group Testing: An Information Theory Perspective}' was published in late 2019. This publication turned out to arrive at a similar time to an important breakthrough work on the optimal rates of noiseless group testing, by Coja-Oghlan, Gebhard, Hahn-Klimroth and Loick \cite{coja2020optimal}, which we were only able to give a very brief mention at the end of the first edition.  The late 2019 timing also happened to coincide with the worldwide spread of the Covid-19 disease -- an extremely unfortunate event, but one that led to a surge of interest in group testing as a potential way to screen people for the disease while making efficient use of limited testing resources.  Progress has continued to be made in both the mathematics and applications of group testing, and the time seems right to update this monograph to cover the recent and ongoing developments.

This second edition follows the same structure as the first edition, but we have added material to cover a number of new developments that arose following the original publication.  Perhaps most notably, as mentioned above, the optimal rates of noiseless group testing were established by Coja-Oghlan \etal~\cite{coja2020optimal}, and were shown to be attainable with polynomial-time decoding (Section \ref{sec:optimal_gt}). The understanding of noisy group testing has also significantly improved, albeit with more gaps still remaining compared to the noiseless setting (Section \ref{sec:high_k_ach} as well as Section \ref{sec:near_const_noisy} and Section \ref{sec:lin_regime_noisy}). Several of the topics surveyed in Chapter \ref{ch:other-topics} also saw advances, including near-optimal sublinear-time algorithms (Section \ref{sec:sublinear}) and new rates in linear sparsity regimes (Section \ref{sec:linear}).  Moreover, notable new topics added to Chapter \ref{ch:other-topics} include `all-or-nothing' thresholds (Section \ref{sec:allnothing}), community-aware group testing (Section \ref{sec:community}), and computational aspects of group testing (Section \ref{sec:computation}).  We have also added a more detailed treatment of two group testing models with non-binary outcomes (Section \ref{sec:other_models}).

While the second edition remains focused on theory and algorithms rather than applications, we discuss Covid-19 testing in Section \ref{sec:applications}, and occasionally further highlight this application throughout the monograph. For deeper coverage of group testing as applied to Covid, including accounts of its use during the pandemic, we direct the reader to the survey paper `\emph{Pooled testing and its applications in the Covid-19 pandemic}' 
by Aldridge and Ellis \cite{aldridge2022pooled}.

This second edition, like its predecessor, concludes in Chapter \ref{ch:conclusion} with a list of open problems.  Many of the open problems presented in the first edition have now been fully or partially solved; we discuss the progress on those problems, and we also pose some new ones.
}

\chapter*{Acknowledgments}
\markboth{\sffamily\slshape Acknowledgments}{\sffamily\slshape Acknowledgments}
\addcontentsline{toc}{chapter}{Acknowledgments}

\subsubsection*{From the first edition}

We are very grateful to Sidharth Jaggi, who made significant contributions to this project in its early stages and has provided helpful advice and feedback on its later drafts. 

We also thank our co-authors on various group testing papers whose results are surveyed in this monograph: Leonardo Baldassini, Volkan Cevher, Karen Gunderson, Thomas Kealy, and Dino Sejdinovic. 

We have discussed this monograph and received helpful feedback on it from many people, including George Atia, Mahdi Cheraghchi, Teddy Furon, Venkatesh Saligrama, and Ashley Montanaro. We received many helpful comments from two anonymous reviewers. Of course, any remaining errors and omissions are our responsibility.

M.~Aldridge was supported in part by the Heilbronn Institute for Mathematical Research.  J.~Scarlett was supported by an NUS Early Career Research Award. O.~Johnson would like to thank Maria, Emily, and Becca for their support.

Finally, we would like to thank the editorial board and staff of {\em Foundations and Trends in Communications and Information Theory} for their support with this monograph.

\subsubsection*{Added in the second edition}

\noindent \added{We are grateful to Daniel McMorrow for reading the new additions to the second edition and providing detailed feedback, and also to Sidharth Jaggi, Lena Krieg, Pavlos Nikolopoulos, Olga Scheftelowitsch, Mahdi Soleymani, Nelvin Tan, I-Hsiang Wang, and Ilias Zadik for their feedback on various specific parts.  

In addition, it is a pleasure to thank more colleagues who have worked with us on group testing problems: Wei Heng Bay, Steffen Bondorf, Thach Bui, Binbin Chen, Junren Chen, Lorenzo Ciampiconi, Fraser Daly, David Ellis, Oliver Gebhard, Bishwamittra Ghosh, Max Hahn-Klimroth, Ivan Lau, Philipp Loick, Daniel McMorrow, Kuldeep Meel, Vivekanand Paligadu, Olaf Parczyk, Pablo Pascual Cobo, Manuel Penschuck, Eric Price, Maurice Rolvien, Yang Sun, Nelvin Tan, Way Tan, Bernard Teo, Lan Truong, Ramji Venkataramanan, Haifeng Yu, Letian Yu, and Yuda Zhao.  }







\chapter*{Notation}
\markboth{\slshape NOTATION}{\slshape NOTATION}
\addcontentsline{toc}{chapter}{Notation}
\newcommand{\see}{} 
\begin{center}
\begin{tabular}{cp{9.6cm}}
$n$ & number of items (\see Definition \ref{def:nkbasic}) \\
$k$ & number of defective items (\see Definition \ref{def:nkbasic}) \\
$\K$ & defective set (\see Definition \ref{def:nkbasic}) \\
$\vc{u} = (u_i)$ & defectivity vector: $u_i = \II( i \in \K)$, shows if item $i$ is defective (\see Definition \ref{def:udef}) \\
$\alpha$ & sparsity parameter in the sparse regime $k = \Theta (n^\alpha)$ (\see Remark \ref{rem:sparse}) \\
$\beta$ & sparsity parameter in the linear regime $k  = \beta n$ (\see Remark \ref{rem:sparse}) \\
$T$ & number of tests (\see Definition \ref{def:testpool}) \\
$\mat X = (x_{ti})$ & test design matrix: $x_{ti} = \one$ if item $i$ is in test $t$; $x_{ti} = \zero$ otherwise (\see Definition \ref{def:testpool}) \\
$\vec y = (y_t)$ & test outcomes (\see Definition \ref{def:outcome}) \\
$\vee$ & Boolean inclusive \texttt{OR} (\see Remark \ref{rem:vee}) \\
$\widehat{\K}$ & estimate of the defective set (\see Definition \ref{def:decoding}) \\
$\Perr$ & average error probability (\see Definition \ref{def:error}) \\
$\Psuc $ & success probability $= 1 - \Perr$ (\see Definition \ref{def:error}) \\
$\text{rate}$ & $\log_2 \binom nk/T$ (\see Definition \ref{def:rate}) \\
$O$, $o$, $\Theta$ & asymptotic `Big O' notation \\
$R$ & an achievable rate (\see Definition \ref{def:achievable}) \\
$\olR$ & maximum achievable rate (\see Definition \ref{def:achievable}) \\
$S(i)$ & the support of column $i$ (\see Definition \ref{def:support}) \\
$S(\L)$ & the union of supports $ \bigcup_{i \in \L} S(i)$ (\see Definition \ref{def:support}) \\
$p$ & parameter for Bernoulli designs: each item is in each test independently with probability $p$ (\see Definition \ref{def:berndesign}) \\
$L$ & parameter for near-constant tests-per-item designs: each item is in $L$ tests sampled randomly with replacement (\see Definition \ref{def:const_col}) \\
$\nu$ &  test design parameter: for Bernoulli designs, $p = \nu/k$
(\see Definition \ref{def:berndesign}); for near-constant tests-per-item designs, $L = \nu T/k$ (\see Definition \ref{def:const_col}) 
\end{tabular}
\end{center}

\newpage 
\begin{center}
\begin{tabular}{cp{8.1cm}}
$h(x)$ & binary entropy function: $\displaystyle h(x) = - x \log_2 x - (1-x) \log_2 (1-x)$ (\see Theorem \ref{SSSub}) \\
$\qq$ & proportion of defectives  
(\see Appendix to Chapter \ref{ch:introduction}) \\
$\overline{k}$ & average number of defectives 
(\see Appendix to Chapter \ref{ch:introduction}) \\
$p(y \mid m, \ell)$  & probability of observing outcome $y$ from a test containing $\ell$ defective items and $m$ items in total (\see Definition \ref{def:noisyprob}) \\
$\rho$, $\varphi$, $\vartheta$, $\xi$   & noise parameters in binary symmetric (Example \ref{ex:bsc}), addition (Example \ref{ex:addition}), dilution/Z channel (Example \ref{ex:dilution}, \ref{ex:Z}), and erasure (Example \ref{ex:erasure}) models  \\
$\overline{\theta}$, $\underline{\theta}$   & threshold parameters in threshold group testing model (Example \ref{ex:threshold})  \\
$\Delta$ & decoding parameter for NCOMP (Section \ref{sec:NCOMP}) \\
$\gamma$ & decoding parameter for separate decoding of items (Section \ref{sec:separate}) and information-theoretic decoder (Section \ref{sec:pf_ach}) \\
$C_{\rm{chan}}$ & Shannon capacity of communication channel (\see Theorem \ref{thm:capcap}) \\
$\bpnohat{m}{r}{i}(u_i) $, $  \bphat{m}{r}{t}(u_i) $ & 
item-to-test and test-to-item messages (\see Section \ref{sec:belprop}) \\ 
$\mathcal{N}(i)$, $\mathcal{N}(t)$ &  neighbours of an item node and test node (\see Section \ref{sec:belprop}) \\
${\mat X}_{\K}$ & submatrix of columns of ${\mat X}$ indexed by $\K$ (\see Section \ref{eq:it_prelim}) \\
${\vec X}_{\K}$ & a single row of ${\mat X}_{\K}$ (\see Section \ref{eq:it_prelim}) \\
$V = V({\vec X}_\K)$ & random number of defective items in the test indicated by ${\vec X}$ (\see Section \ref{eq:it_prelim})\\
$P_{Y|V}$ & observation distribution depending on the test design only through $V$ (Equation \eqref{eq:Y_noisy}) \\
$S_0$, $S_1$ & partition of the defective set (Equation \eqref{eq:S0S1}) \\
$\imath$ & information density (\see Equation \eqref{eq:info_dens}) \\
${\mat X}_{0,\tau}$, ${\mat X}_{1,\tau}$ & submatrices of $\mat X$ corresponding to $(S_0,S_1)$ with $|S_0| = \tau$ (Equation \eqref{eq:simplify_ach}) \\
${\vec X}_{0,\tau}$, ${\vec X}_{1,\tau}$ & subvectors of ${\vec X}_{\K}$ corresponding to $(S_0,S_1)$ with $|S_0| = \tau$ (Equation \eqref{eq:I_tau}) \\
$I_\tau$ & conditional mutual information $I({\vec X}_{0,\tau}; Y | {\vec X}_{1,\tau})$ (Equation \eqref{eq:I_tau}) \\
$D_i$ & event that item $i$ is masked/disguised (Section \ref{sec:gen_conv}) \\
\end{tabular}

\end{center}

\chapter{Introduction to Group Testing} \label{ch:introduction}

\section{What is group testing?} \label{sec:whatis}

The `group testing' problem arose in the United States in the 1940s, when large numbers of men were being conscripted into the army and needed to be screened for syphilis. Since an accurate blood test (the Wassermann test) exists for syphilis, one can take a sample of blood from each soldier, and test it. However, since it is a rare disease, the vast majority of such tests will come back negative. From an information-theoretic point of view, this testing procedure seems inefficient, because each test is not particularly informative.

Robert Dorfman, in his seminal paper of 1943 \cite{dorfman}, founded the subject of group testing by noting that, for syphilis testing, the total number of tests needed could be dramatically reduced by pooling samples. That is, one can take blood samples from a `pool' (or `group') of many soldiers, mix the samples, and perform the syphilis test on the pooled sample. If the test is sufficiently precise, it should report whether or not any syphilis antigens are present in the combined sample. If the test comes back negative, one learns that all the soldiers in the pool are free of syphilis, whereas if the test comes back positive, one learns that \emph{at least one} of the soldiers in the pool must have syphilis. One can use several such tests to discover which soldiers have syphilis, using fewer tests than the number of soldiers.
(The origins of the problem, including the contributions of David Rosenblatt, are described in detail in \cite[Section 1.1]{du-hwang}.)

Of course, this idealized testing model is a mathematical convenience; in practice, a more realistic model could account for sources of error -- for example, that a large number of samples of negative blood could dilute syphilis antigens below a detectable level. However, the idealization results in a useful and interesting problem, which we will refer to as \defn{standard noiseless group testing}. 

Generally, we consider $n$ \defn{items} (in the above example, soldiers) of which $k$ are \defn{defective} (have syphilis). A test on a subset of items returns positive if at least one of the items is defective, and returns negative is all of the items are nondefective. The central problem of group testing is then the following: Given the number of items $n$ and the number of defectives $k$, how many such tests $T$ are required to accurately discover the defective items, and how can this be achieved?

As we shall see, the number of tests required depends on various assumptions on the mathematical model used.  An important distinction is the following:
\begin{description}
\item[Adaptive vs. nonadaptive] Under adaptive testing, the test pools are designed sequentially, and each one can depend on the previous test outcomes. Under nonadaptive testing, all the test pools are designed in advance, making them  amenable to being implemented in parallel. Most of the focus of this survey is on nonadaptive testing, though we present adaptive results in Section \ref{sec:adaptive} for comparison purposes, and consider the intermediate case of algorithms with a fixed number of stages in Section \ref{sec:two-stage}.  \added{In Section \ref{sec:linear} we consider adaptive testing in regimes with a relatively larger number of defective items.}
\end{description}

For nonadaptive testing, it is often useful to separate the \defn{design} and \defn{decoding} parts of the group testing problem. The design problem concerns the question of how to choose the testing strategy -- that is, which items should be placed in which tests. The decoding (or {\em detection}) problem consists of determining which items are defective given the test designs and their outcomes, ideally in a computationally efficient manner.

The number of tests required to achieve `success' depends on our criteria for declaring success:
\begin{description}
\item[Zero error probability vs.\ small error probability] Under a zero-error probability criterion, require being certain we will recover the defective set. Under a small error probability criterion, it suffices to recover the defective set with high probability. In particular, the set of defective items and/or the recovery algorithm may be random, and we may allow 
a tolerance $\epsilon > 0$ on the error probability with respect to this randomness. 
(In other sparse recovery problems, a similar distinction is sometimes made using the terminology `for-each setting' and `for-all setting' \cite{gilbert2007one}.) We will mostly focus on the small error probability criterion, but we give zero-error results in Section \ref{sec:zero-error} for comparison purposes.

\item[Exact recovery vs.\ approximate recovery] With an exact recovery criterion, we require that every defective item is correctly classified as defective, and every nondefective item is correctly classified as nondefective. With approximate recovery, we may tolerate having some small number of incorrectly classified items -- perhaps with different demands for false positives (nondefective items incorrectly classified as defective) and false negatives (defective items incorrectly classified as nondefective). For the most part, this survey focuses on the exact recovery criterion; some variants of approximate recovery are discussed in Section \ref{sec:partial}.
\end{description}

When considering more realistic settings, it is important to consider group testing models that do not fit into the standard noiseless group testing idealization we began by discussing. Important considerations include:
\begin{description}
\item[Noiseless vs.\ noisy testing] Under noiseless testing, we are guaranteed that the test procedure works perfectly: We get a negative test outcome if all items in the testing pool are nondefective, and a positive outcome if at least one item in the pool is defective. Under noisy testing, errors can occur, either according to some specified random model or in an adversarial manner. The first two chapters of this survey describe results in the noiseless case for simplicity, before the noisy case is introduced and discussed from Chapter \ref{ch:algorithms_noisy} onwards.
\item[Binary vs.\ nonbinary outcomes] Our description of standard group testing involves tests with binary outcomes -- that is, positive or negative results. In practice, we may find it useful to consider tests with a wider range of outcomes, perhaps corresponding to some idea of weak and strong positivity, according to the numbers of defective and nondefective items in the test, or even the `strength of defectivity' of individual items.  \rev{In such settings, the test matrix may even be nonbinary to indicate the `amount' of each item included in each test.} We discuss these matters further in Sections \ref{sec:noisy_models} and \ref{sec:noisy}. 
\end{description}


Further to the above distinctions, group testing results can also depend on the assumed distribution of the defective items among all items, and the decoder's knowledge (or lack of knowledge) of this distribution. In general, the true defective set could have an arbitrary prior distribution over all subsets of items. However, the following are important distinctions:
\begin{description}
\item[Combinatorial vs.\ i.i.d.~prior] For mathematical convenience, we will usually consider the scenario where there is a fixed number of defectives, and the defective set is uniformly random among all sets of this size. We refer to this as the \defn{combinatorial} prior (the terminology \defn{hypergeometric group testing} is also used). 
Alternatively (and perhaps more realistically), one might imagine that each item is defective independently with the same fixed probability $\qq$, which we call the \defn{i.i.d.~prior}.  \rev{In the appendix to this chapter, we discuss how results can be transferred from one prior to the other under suitable assumptions.}  Furthermore, in Section \ref{sec:prior}, we discuss a variant of the i.i.d.~prior in which each item has a different prior probability of being defective, \added{as well as more sophisticated priors in which the defectives are known to exhibit certain correlations or clustering structure.}
\item[Known vs.\ unknown number of defectives]  We may wish to distinguish between algorithms that require knowledge of the true number of defectives, and those that do not. An intermediate class of algorithms may be given bounds or approximations to the true number of defectives (see Remark \ref{rem:misspecify}). In Section \ref{sec:universal}, we discuss procedures that use pooled tests to estimate the number of defective items.
\end{description}
In this survey, we primarily consider the combinatorial prior. Furthermore, for the purpose of proving mathematical results, we will sometimes make the convenient assumption that $k$ is known. However, in our consideration of practical decoding methods in Chapters \ref{ch:algorithms} and \ref{ch:algorithms_noisy}, we focus on algorithms that do not require knowledge of $k$.

\section{About this survey}

The existing literature contains several excellent surveys of various aspects of group testing. The paper by Wolf \cite{wolf} gives an overview of the early history, following Dorfman's original work \cite{dorfman}, with a particular focus on adaptive algorithms. The textbooks of Du and Hwang \cite{du-hwang, du-hwang2} give extensive background on group testing, especially on adaptive testing, zero-error nonadaptive testing, and applications.  \rev{The lecture notes of D'yachkov \cite{dyachkov_lectures} focus primarily on the zero-error nonadaptive setting, considering both fundamental limits and explicit constructions.}  For the small-error setting, significant progress was made in  the Russian literature in the 1970s and 80s, particularly for nonadaptive testing in the very sparse regime where $k$ is constant as $n$ tends to infinity -- the review paper of Malyutov \cite{malyutov} is a useful guide to this work (see also \cite[Ch.~6]{dyachkov_lectures}). \added{A more recent survey by Aldridge and Ellis \cite{aldridge2022pooled} covers the use of group testing in the Covid-19 pandemic.}

The focus of this survey is distinct from these previous works.  In contrast with the adaptive and zero-error settings covered in \cite{du-hwang, dyachkov_lectures, wolf}, here we concentrate on the fundamentally distinct nonadaptive setting with a small (nonzero) error probability.  While this setting was also the focus of \cite{malyutov}, we survey a wide variety of recent algorithmic and information-theoretic developments not covered there, as well as considering a much wider range of sparsity regimes (that is, scaling behaviour of $k$ as $n \to \infty$).  We focus in particular on the `sparse regime' where $k = \Theta(n^{\alpha})$ for some $\alpha \in (0,1)$, which comes with a variety challenges compared to the `very sparse regime' in which $k$ is constant.
Another key feature of our survey is that we consider not only order-optimality results, but also quantify the performance of algorithms in terms of the precise constant factors, as captured by the {\em rate} of group testing. (While much of \cite{du-hwang2} focuses on the zero-error setting, a variety of probabilistic constructions are discussed in its fifth chapter, although the concept of a rate is not explicitly considered.)

Much of the work that we survey was inspired by the re-emergence of group testing in the information theory community following the paper of Atia and Saligrama \cite{atia-saligrama}. \nocite{atia-corr} However, to the best of our knowledge, the connection between group testing and information theory was first formally made by Sobel and Groll \cite[Appendix A]{sobel}, and was used frequently in the works surveyed in \cite{malyutov}. 

An outline of the rest of the survey is as follows. In the remainder of this chapter, we give basic definitions and fix notation (Section \ref{sec:basicdefs}), introduce the information-theoretic terminology of rate and capacity that we will use throughout the monograph (Section \ref{sec:counting_bound}), and briefly review results for adaptive (Section \ref{sec:adaptive}) and zero-error (Section \ref{sec:zero-error}) group testing algorithms, to provide a benchmark for other subsequent results. In Section \ref{sec:applications}, we discuss some applications of group testing in biology, communications, information technology, and data science. In a technical appendix to this chapter, we discuss the relationship between two common models for the defective set.

In Chapter \ref{ch:algorithms}, we introduce  a variety of nonadaptive algorithms for noiseless group testing, and discuss their performance.  Chapter \ref{ch:algorithms_noisy} shows how these ideas can be extended to various noisy group testing models.

Chapter \ref{ch:achievability} reviews the fundamental information-theoretic limits of group testing. This material is mostly independent of Chapters \ref{ch:algorithms} and \ref{ch:algorithms_noisy} and could be read before them, although readers may find that the more concrete algorithmic approaches of the earlier chapters provide helpful intuition.

In Chapter \ref{ch:other-topics}, we discuss a range of variations and extensions of the standard group testing problem.  The topics considered are approximate recovery of the defective set (Section \ref{sec:partial}), adaptive testing with limited stages (Section \ref{sec:two-stage}), counting the number of defective items (Section \ref{sec:universal}), decoding algorithms that run in sublinear time (Section \ref{sec:sublinear}), the linear sparsity regime $k = \Theta(n)$ (Section \ref{sec:linear}), group testing with more general prior distributions on the defective set (Section \ref{sec:prior}), explicit constructions of test designs (Section \ref{sec:explicit}), group testing under constraints on the design (Section \ref{sec:constrained}), computational considerations (Section \ref{sec:computation}), and more general group testing models (Section \ref{sec:other_models}). Each of these sections gives a brief outline of the topic with references to more detailed work, and can mostly be read independently of one another.  Finally, we conclude in Chapter \ref{ch:conclusion} with a partial list of interesting open problems, \added{as well as a summary of previously open problems from the first edition that have since been fully or partially solved.}

\rev{For key results in the survey, we include either full proofs or proof sketches (for brevity).  For other results that are not our main focus, we may omit proofs and instead provide pointers to the relevant references.}

\section{Basic definitions and notation} \label{sec:basicdefs}

We now describe the group testing problem in more formal mathematical language.

\begin{definition} \label{def:nkbasic}
We write $n$ for the number of items, which we label as $\{1,2,\dots, n\}$. We write $\K \subset \{1,2,\dots,n\}$ for the set of defective items (the \emph{defective set}), and write $k = |\K|$ for the number of defectives.
\end{definition}

\begin{definition} \label{def:udef}
We write $u_i = \one$ to denote that item $i \in \K$ is defective, and $u_i = \zero$ to denote that $i \not\in\K$ is nondefective. In other words, we define $u_i$ as an indicator random variable $u_i = \II \{ i \in \K \}$. We then write $\vc{u} = (u_i) \in \zo^n$ for the \emph{defectivity vector}. (In some contexts, an uppercase $\vc{U}$ will denote a \emph{random} defectivity vector.)
\end{definition}

\begin{remark} \label{rem:sparse}
We are interested in the case that the number of items $n$ is large, and accordingly consider asymptotic results as $n \to \infty$.  \rev{We use the standard `Big O' notation $O(\cdot)$, $o(\cdot)$, and $\Theta(\cdot)$ to denote asymptotic behaviour in this limit, and we write $f_n \sim g_n$ to mean that $\lim_{n\to\infty} \frac{f_n}{g_n} = 1$.}  We consider three scaling regimes for the number of defectives $k$:
\begin{description}
  \item[The very sparse regime:]  $k$ is constant (or bounded as $k = O(1)$) as $n \to \infty$;
  \item[The sparse regime:] $k$ scales sublinearly as $k = \Theta(n^\alpha)$ for some \emph{sparsity parameter} $\alpha \in [0,1)$  as $n \to \infty$. 
  \item[The linear regime:] $k$ scales linearly as $k \sim \beta n$ for $\beta \in (0,1)$ as $n \to \infty$.
\end{description}
\end{remark}

We are primarily interested in the case that defectivity is rare, where group testing has the greatest gains, though the linear regime is still of significant interest and will be considered in Section \ref{sec:linear}. Much of the early work on group testing considered only the very sparse regime, which is now quite well understood -- see, for example, \cite{malyutov} and the references therein -- so we will concentrate primarily on the wider sparse regime.

The case $\alpha = 0$, which covers the very sparse regime, usually behaves the same as small $\alpha$ but sometimes requires slightly different analysis to allow for the fact that $k$ may not tend to $\infty$. Hence, for brevity, we typically only explicitly deal with the cases $\alpha \in (0,1)$.

We assume for the most part that the true defective set $\K$ is uniformly random from the $\binom nk$ sets of items of size $k$ (the `combinatorial prior'). The assumption that $k$ is known exactly is often mathematically convenient, but unrealistic in most applications. For this reason, in Chapters \ref{ch:algorithms} and \ref{ch:algorithms_noisy} we focus primarily on decoding algorithms that do not require knowledge of $k$. However, there exist nonadaptive algorithms that can accurately estimate the number of defectives using $O( \log n)$ tests (see Section \ref{sec:universal} below), which could form the first part of a two-stage algorithm, if permitted.

\begin{definition} \label{def:testpool}
We write $T = T(n)$ for the \defn{number of tests} performed, and label the tests $\{1,2,\dots,T\}$.  To keep track of the design of the test pools, we write $x_{ti} = \one$ to denote that item $i \in \{1,2,\dots,n\}$ is in the pool for test $t \in \{1,2,\dots,T\}$, and $x_{ti} = \zero$ to denote that item $i$ is not in the pool for test $t$. We gather these into a matrix $\mat X \in \zo^{T \times n}$, which we shall refer to as the \defn{test matrix} or \defn{test design}.
\end{definition}

\begin{figure}[t] 
\begin{center}
\includegraphics[width=0.9\textwidth]{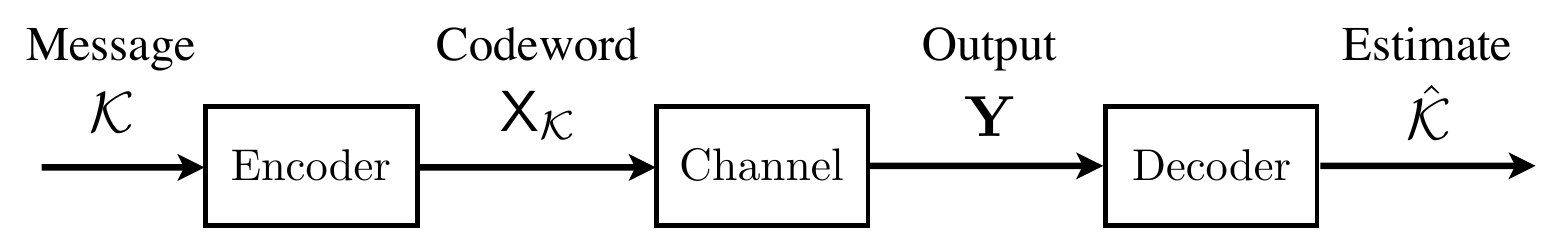}
\end{center}
\caption{Group testing interpreted as a channel coding problem.  The notation ${\mat X}_{\K}$ denotes the $T \times k$ submatrix of $\mat X$ obtained by keeping only the $k$ columns indexed by $\K$, and the output $\vec Y$ is the `OR' of these $k$ columns.} \label{fig:channel_diagram} 
\end{figure}

To our knowledge, this matrix representation was  introduced by Katona  \cite{katona1973combinatorial}. It can be helpful to think of group testing in a channel coding framework, where the particular defective set $\K$ acts like the source message, finding the defective set can be thought of as  decoding, and the matrix $\mat X$ acts like the codebook.   See Figure \ref{fig:channel_diagram} for an illustration.  See also \cite{malyutov} for a related interpretation of group testing as a type of {\em multiple-access channel}.

Similarly to the channel coding problem, explicit deterministic matrix designs often fail to achieve even order-optimal performance (see Sections \ref{sec:zero-error} and \ref{sec:explicit} for discussion). Following the development and successes of randomized channel codes (discussed further below), it is therefore a natural development to consider randomized test designs.  We will use a capital $X_{ti}$ to denote the entries of a random testing matrix. Some designs of particular interest include the following:
\begin{description}
\item[Bernoulli design] In the Bernoulli design, each item is included in each test independently at random with some fixed probability $p$. That is, we have $\PP(X_{ti} = \one) = p$ and $\PP(X_{ti} = \zero) = 1-p$, i.i.d.~over $i \in \{1,2, \dots, n\}$ and $t \in \{1,2,\dots, T\}$. Typically the parameter $p$ is chosen to scale as $p = \Theta(1/k)$. See Definition \ref{def:berndesign} for more details.
\item[Constant tests-per-item] In the constant tests-per-item design, each item is included in some fixed number $L$ of tests, with the $L$ tests for each item chosen uniformly at random, independent from the choices for all other items. In terms of the testing matrix $\mat{X}$, we have independent columns of constant weight. Typically the parameter $L$ is chosen to scale as $L = \Theta(T/k)$.
In fact, it is often more mathematically convenient to analyse the similar {\em near-constant tests-per-item} (near-constant column weight) design, where the $L$ tests for each item are chosen uniformly at random \emph{with replacement} -- see Definition \ref{def:const_col} for more details.
\item[Doubly regular design] One can also consider designs with both a constant number $L$ of tests-per-item (column weight) and also a constant number $m$ of items-per-test (row weight), with $nL = mT$.  For example, the matrix may be picked uniformly at random according to these constraints, \added{or alternatively, $L$ random matrices with column weight 1 and row weight $m$ may be stacked vertically}. Certain `near-constant' variants may also be considered. 
Again, $L = \Theta(T/k)$ (or equivalently, $m = \Theta(n/k)$) is a useful scaling. \added{We will only briefly touch on these designs in this monograph, in parts of Sections \ref{sec:linear} and \ref{sec:constrained}}, but mention that they were studied in the papers \cite{mezard,wadayama17,aldridge2020conservative,Tan2022}, \nocite{wadayama2} among others.
\end{description}

As hinted above, these random constructions can be viewed as analogous to random coding in channel coding, which is ubiquitous for proving achievability bounds.  However, while standard random coding designs in channel coding are impractical due to the exponential storage and computation required, the above designs can still be practical, since the random matrix only contains $T \times n$ entries.  In this sense, the constructions are in fact more akin to random linear codes, random LDPC codes, and so on.

We write $y_t \in \zo$ for the \defn{outcome} of test $t \in \{1,2,\dots, T\}$, where $y_t = \one$ denotes a positive outcome and $y_t = \zero$ a negative outcome. Recall that in the standard noiseless model, we have $y_t = \zero$ if all items in the test are nondefective, and $y_t = \one$ if at least one item in the test is defective. Formally, we have the following.

\begin{definition} \label{def:outcome}
Fix $n$ and $T$. Given a defective set $\K \subset \{1,2,\dots,n\}$ and a test design $\mat{X} \in \zo^{T \times n}$, the \defn{standard noiseless group testing model} is defined by the outcomes
  \begin{equation} \label{eq:outcome}
    y_t = \begin{cases} \one & \text{if there exists $i \in \K$ with $x_{ti} = \one$,} \\
             \zero & \text{if for all $i \in \K$ we have $x_{ti} = \zero$.} \end{cases}
  \end{equation}
We write $\vec y = (y_t) \in \zo^T$ for the vector of test outcomes.
\end{definition}

\begin{remark} \label{rem:vee}
A concise way to write \eqref{eq:outcome} is using the Boolean inclusive \texttt{OR} (or disjunction) operator $\vee$, where $\zero\vee\zero = \zero$ and $\zero \vee \one = \one \vee\zero = \one\vee\one = \one$. Then,
\begin{equation} y_t = \bigvee_{i \in \K} x_{ti} ; \label{eq:vee}
\end{equation}
or, with the understanding that \texttt{OR} is taken component-wise, 
\begin{equation} \vec y = \bigvee_{i \in \K} \vec x_i . \label{eq:veevec}
\end{equation}
Using the defectivity vector notation of Definition \ref{def:udef}, we can rewrite \eqref{eq:vee} in
analogy with matrix multiplication as
\begin{equation} y_t = \bigvee_{i} x_{ti} u_i. \label{eq:vee2}
\end{equation}
Note that the nonlinearity of the $\vee$ operation is what gives group testing its specific character.  
Indeed, we can consider \eqref{eq:vee2} to be a nonlinear `Boolean counterpart' to the well-known compressed sensing problem in which linear models are widely adopted \cite{gilbert}.
\end{remark} 
 
We illustrate a simple group testing procedure in Figure \ref{fig:construction}, where the defective items are represented by filled icons, and so it is clear that the positive tests are those containing at least one defective item. 

\begin{figure}
\begin{center}
\begin{tabular}{cccccccc|c}
\nondefman & \nondefman & \defman & \nondefman & \defman & \nondefman & \nondefman & \nondefman &  Outcome  \\
\hline
 $\one$ & $\one$ & \bfone & $\one$ & $\zero$ & $\zero$ & $\zero$ & $\zero$ & \mbox{ Positive} \\
$\zero$ & $\zero$ & $\zero$ & $\zero$ & \bfone & $\one$ & $\one$ & $\one$ & \mbox{ Positive} \\
 $\one$ & $\one$ & $\zero$ & $\zero$ & $\zero$ & $\zero$ & $\zero$ & $\zero$ & \mbox{ Negative} \\
 $\zero$ & $\zero$ & \bfone & $\zero$ & $\zero$ & $\zero$ & $\zero$ & $\zero$ & \mbox{ Positive} \\
$\zero$ & $\zero$ & \bfone & $\zero$ & \bfone & $\one$ & $\zero$ & $\zero$ & \mbox{ Positive} \\
$\zero$ & $\zero$ & $\zero$ & $\zero$ & \bfone & $\zero$ & $\zero$ & $\zero$ & \mbox{ Positive} \\
\end{tabular}
\end{center}
\caption{Example of a group testing procedure and its outcomes. Icons for defective individuals (items 3 and 5) are filled, and icons for nondefective individuals are unfilled. The testing matrix $\mat{X}$ is shown beneath the individuals, where elements $x_{ti}$ are circled for emphasis if $x_{ti} = \one$ and individual $i$ is defective. Hence, a test is positive if and only if it contains at least one circled $\one$. \label{fig:construction}}
\end{figure}

\begin{figure}
\begin{center}
\begin{tabular}{cccccccc|c}
\quest & \quest & \quest & \quest & \quest  & \quest & \quest & \quest &   $\vc{y}$ \\
\hline
 $\one$ & $\one$ & $\one$ & $\one$ & $\zero$ & $\zero$ & $\zero$ & $\zero$ & $\one$ \\
$\zero$ & $\zero$ & $\zero$ & $\zero$ & $\one$ & $\one$ & $\one$ & $\one$ & $\one$ \\
 $\one$ & $\one$ & $\zero$ & $\zero$ & $\zero$ & $\zero$ & $\zero$ & $\zero$ & $\zero$ \\
 $\zero$ & $\zero$ & $\one$ & $\zero$ & $\zero$ & $\zero$ & $\zero$ & $\zero$ & $\one$ \\
$\zero$ & $\zero$ & $\one$ & $\zero$ & $\one$ & $\one$ & $\zero$ & $\one$ & $\one$ \\
$\zero$ & $\zero$ & $\zero$ & $\zero$ & $\one$ & $\zero$ & $\zero$ & $\zero$ & $\one$ \\
\end{tabular}
\end{center}
\caption{Group testing inference problem. We write $\one$ for a positive test and $\zero$ for a negative test, but otherwise the matrix $\mat{X}$ is exactly as in Figure \ref{fig:construction} above. The defectivity status of the individuals is now unknown, and we hope to infer it from the outcomes $\vc{y}$ and matrix $\mat{X}$. \label{fig:inference}}
\end{figure}

Given the test design $\mat X$ and the outcomes $\vec y$, we wish to find the defective set.  Figure \ref{fig:inference} represents the inference problem we are required to solve -- the defectivity status of particular individuals is hidden, and we are required to infer it from the matrix $\mat{X}$ and the vector of outcomes $\vc{y}$. In Figure \ref{fig:inference}, we write $\one$ for a positive test and $\zero$ for a negative test. In general we write  $\hat\K = \hat\K(\mat X, \vec y)$ for our estimate of the defective set.

\begin{definition} \label{def:decoding}
A \defn{decoding} (or detection) \defn{algorithm} is a (possibly randomized) function $\hat\K \, \colon \zo^{T\times n} \times \zo^T \to \mathcal P \left( \{1,2,\dots,n\} \right)$, where the power-set $\mathcal P \left( \{1,2,\dots,n\} \right)$ is the collection of subsets of items.
\end{definition}

Under the exact recovery criterion, we succeed when $\hat\K = \K$, while under approximate recovery, we succeed if $\hat \K$ is close to $\K$ in some predefined sense (see Section \ref{sec:partial}).  Since we focus our attention on the former, we provide its formal definition as follows.

\begin{definition} \label{def:error}
Under the exact recovery criterion, the (\defn{average}) \defn{error probability} for noiseless group testing with a combinatorial prior is
\begin{equation} \label{eq:errordef} \Perr := \frac{1}{\binom nk} \sum_{\K \,:\, |\K| = k} \PP \big( \hat\K(\mat X, \vec y) \neq \K \mid \K \big), \end{equation}
where $\vec y$ is related to $\mat X$ and $\K$ via the group testing model and the probability $\PP$ is over the randomness in the test design $\mat X$ (if randomized), the group testing model (if random noise is present), and the decoding algorithm $\hat\K$ (if randomized). We call $\Psuc := 1 - \Perr$ the \defn{success probability}.
\end{definition}

We note that this average error probability refers to an average over a uniformly distributed choice of defective set $\K$, where we can think of this randomness as being introduced by nature. Even in a setting where the true defective set $\K$ is actually deterministic, this can be a useful way to think of randomness in the model. Since the outcomes of the tests only depend on the columns of the test matrix $\mat{X}$ corresponding to $\K$, the same average error probability is achieved even for a fixed $\K$ by any exchangeable matrix design (that is, one where the distribution of $\mat{X}$ is invariant under uniformly-chosen column permutations). This includes Bernoulli, near-constant tests-per-item, and doubly regular designs, as well as any deterministic matrix construction acted on by uniformly random column permutations.

\section{Counting bound and rate} \label{sec:counting_bound}

Recall that the goal is, given $n$ and $k$, to choose $\mat X$ and $\hat\K$ such that $T$ is as small as possible, while keeping the error probability $\Perr$ small.

Supposing momentarily that we were to require  an error probability of {\em exactly} zero, a simple counting argument based on the pigeonhole principle reveals that we require $T \geq \log_2 \binom nk$:  There are only $2^T$ combinations of test results, but there are $\binom nk$ possible defective sets that each must give a different set of results.   This argument is valid regardless of whether the test design is adaptive or nonadaptive.


The preceding argument extends without too much difficulty to the nonzero error probability case. For example, Chan \etal\ \cite{chan-etal-1} used an argument based on Fano's inequality to prove that
\begin{equation} \label{eq:psucweak} \Psuc \leq \frac{T}{\log_2 \binom nk},\end{equation}
which they refer to as `folklore', while Baldassini, Johnson, and Aldridge gave the following tighter bound on the success probability \cite[Theorem 3.1]{baldassini-johnson-aldridge} (see also \cite{johnsonconverse})

\begin{theorem}[Counting bound] \label{thm:bjaconverse}
Any algorithm (adaptive or nonadaptive) for recovering the defective set with $T$ tests has success probability satisfying
\begin{equation} \label{eq:psucstrong} \Psuc \leq \frac{2^T}{\binom nk}.\end{equation}
In particular, $\Psuc \to 0$ as $n \to \infty$ whenever $T \le (1-\eta)\log_2{\binom nk}$ for arbitrarily small $\eta > 0$.
\end{theorem}

From an information-theoretic viewpoint, this result essentially states that since the prior uncertainty is $\log_2 \binom nk$ for a uniformly random defective set, and each test is a yes/no answer revealing at most $1$ bit of information, we require at least $\log_2 \binom nk$ tests.  Because the result is based on counting the number of defective sets, we refer to it as the {\em counting bound}, often using this terminology for both the asymptotic and non-asymptotic versions when the distinction is clear from the context.

With this mind, it will be useful to think about how many bits of information we learn (on average) per test. Using an analogy with channel coding, we shall call this the \emph{rate} of group testing. In general, if the defective set $\K$ is chosen from some underlying random process with entropy $H$, then for a group testing strategy with $T$ tests, we define the rate to be $H/T$.  \added{(See Section \ref{sec:prior} for further discussion of this general version.)} In particular, under a combinatorial prior, where the defective set is chosen uniformly from the $\binom{n}{k}$ possible sets, the entropy is $H = \log_2 \binom{n}{k}$, leading to the following definition.

\begin{definition} \label{def:rate}
Given a group testing strategy under a combinatorial prior with $n$ items, $k$ defective items, and $T$ tests, we define the \defn{rate} to be
  \begin{equation} \text{rate} := \frac{\log_2 \binom nk}T . \end{equation}
\end{definition}

This definition was proposed for the combinatorial prior by Baldassini, Aldridge and Johnson \cite{baldassini-johnson-aldridge}, and extended to the general case (see Definition \ref{def:rate2}) in \cite{kealy-johnson-piechocki}. 
This definition generalizes a similar earlier definition of rate by Malyutov \cite{malyutov-1,malyutov}, which applied only in the very sparse (constant $k$) regime.

We note the following well-known bounds on the binomial coefficient
(see for example \cite[p.~1186]{cormen}):
\begin{equation} \bigg(\frac{n}{k}\bigg)^k \leq \binom nk \leq \bigg(\frac{\ee n}{k}\bigg)^k . \label{eq:binomcoeffbounds} \end{equation}
Thus, we have the asymptotic expression
\begin{equation} \log_2 \binom nk = k \log_2 \frac nk + O(k), \end{equation}
and in the sparse regime $k = \Theta(n^\alpha)$ for $\alpha \in [0,1)$, we have the asymptotic equivalence
\begin{equation} \label{eq:bincoeffequiv}
\log_2 \binom nk \asym k \log_2 \frac nk \asym (1 - \alpha) k \log_2 n
= \frac{(1-\alpha)}{\ln 2} k \ln n. \end{equation}
Thus, to achieve a positive rate in this regime, we seek group testing strategies with $T = O(k \log n)$ tests. In contrast, in Section \ref{sec:linear}, we will observe contrasting behaviour of the binomial coefficient in the linear regime $k \sim \beta n$, expressed in \eqref{eq:bincoeffequiv2}.

\begin{definition} \label{def:achievable}
Consider a group testing problem, possibly with some aspects fixed (for example, the random test design or the decoding algorithm), in a setting where the number of defectives scales as $k = k(n)$ according to some function (e.g., $k(n) = \Theta(n^{\alpha})$ with $\alpha \in (0,1)$).
\begin{enumerate}
\item We say a rate $R$ is \emph{achievable} if, for any $\delta, \epsilon > 0$, 
for $n$ sufficiently large there exists a group testing strategy with a number of tests $T = T(n)$ such that the rate satisfies
\begin{equation} \text{rate} = \frac{ \log_2 \binom nk}{T} > R - \delta, \label{eq:rate} \end{equation}
and the error probability $\Perr$ is at most $\epsilon$. 
\item  We say a rate $R$ is \emph{zero-error achievable} if, for any $\delta > 0$, for $n$ sufficiently large, there exists a group testing strategy with a number of tests $T = T(n)$ such that the rate exceeds $R - \delta$, and $\Perr = 0$. 
\item Given a random or deterministic test matrix construction (design), we define the \defn{maximum achievable rate} to be the supremum of all achievable rates that can be achieved by any decoding algorithm.  We sometimes also use this terminology when the decoding algorithm is fixed.  For example, we write $\olR_{\Bern}$ for the maximum rate achieved by Bernoulli designs and any decoding algorithm, and $\olR_{\Bern}^{\COMP}$ for the maximum rate achieved by Bernoulli rates using the COMP algorithm (to be described in Section \ref{sec:COMP}).
\item Similarly, the \defn{maximum zero-error achievable rate} is the supremum of all zero-error achievable rates for a particular design.
\item We define the \defn{capacity} $C$ to be the supremum of all achievable rates, and the \defn{zero-error capacity} $C_0$ to be the supremum of all zero-error achievable rates.  Whereas the notion of {\em maximum achievable rate} allows test design and/or decoding algorithm to be fixed, the definition of capacity optimizes over both.
\end{enumerate}
\end{definition}

Note that these notions of rate and capacity may depend on the scaling of $k(n)$. In our achievability and converse bounds for the sparse regime $k = \Theta(n^{\alpha})$, the maximum rate will typically vary with $\alpha$, but will not depend on the implied constant in the $\Theta(\cdot)$ notation.

\begin{remark} \label{rem:universal}
Note that the counting bound (Theorem \ref{thm:bjaconverse}) gives us a universal upper bound $C \leq 1$ on the capacity. In fact, it also implies the so-called strong converse:  The error probability $\Perr$ tends to $1$ when $T \le (1-\eta)\log_2{\binom nk}$ for arbitrarily small $\eta > 0$, which corresponds to a rate $R \geq 1/(1-\eta) > 1$.
\end{remark}

We are interested in determining when the upper bound $C=1$ can or cannot be achieved, as well as determining how close practical algorithms can come to achieving it. (We discuss what we mean by `practical' in this context in Section \ref{sec:algsummary}.)

\begin{figure}[t] 
\begin{center}
\includegraphics[width=0.95\textwidth]{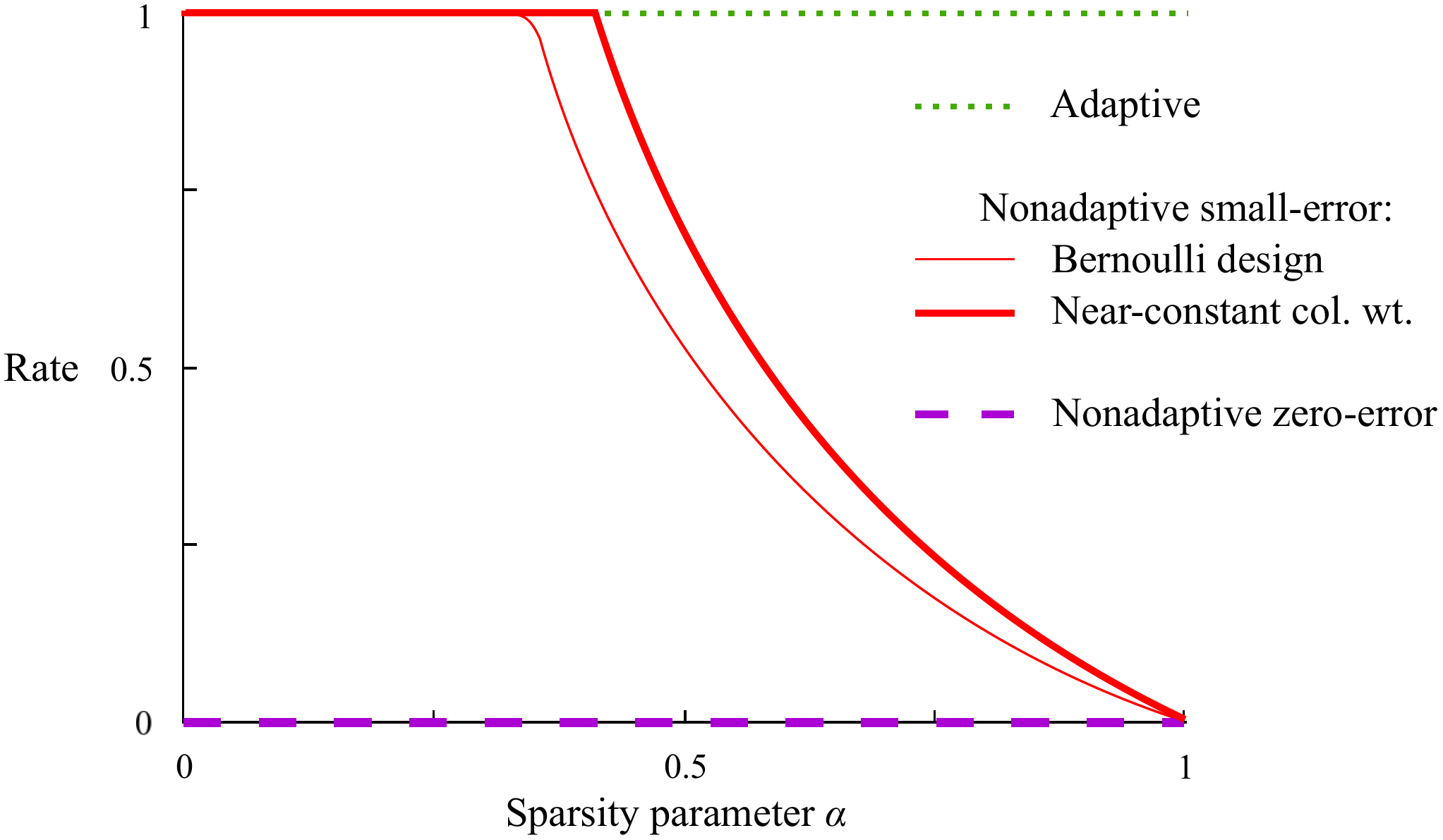}
\end{center}
\caption{\rev{Achievable rates for noiseless group testing with $k = \Theta(n^\alpha)$ for a sparsity parameter $\alpha \in (0,1)$: the adaptive capacity is $C =1$; the nonadaptive zero-error capacity is $C_0 = 0$; and the achievable rates for nonadaptive small-error group testing are given in Theorem \ref{thm:ch2main} for Bernoulli designs and Theorem \ref{thm:ch2nearconst} for near-constant column weight designs.  \added{(The rate of the near-constant column weight design is in fact the best possible.)}
}} \label{ch1-rates} 
\end{figure}

We will observe the following results for noiseless group testing in the sparse regime $k = \Theta(n^\alpha)$, which are illustrated in Figure \ref{ch1-rates}:
\begin{description}
\item[Adaptive testing] is very powerful, in that both the zero-error and small-error capacity equal $C_0 = C = 1$ for all $\alpha \in [0,1)$ (see Section \ref{sec:adaptive}).
\item[Zero-error nonadaptive testing] is much weaker, in the sense that the zero-error capacity is $C_0 = 0$ for all $\alpha \in (0,1)$ (see Section \ref{sec:zero-error}).
\item[Small-error nonadaptive testing] is more complicated. \rev{The capacity is $C = 1$ for $\alpha \in [0,0.409]$; this is achievable with a Bernoulli design for $\alpha < 1/3$ (Theorem \ref{thm:ch2main}), and with a (near-)constant tests-per-item design for the full interval (Theorem \ref{thm:ch2nearconst}).
\added{The capacity for $\alpha \in (0.409,1)$ turns out to also be attained by the latter design, and steadily decreases from 1 to 0 as $\alpha$ increases within this range.} 
We discuss these results further in Chapter \ref{ch:achievability}, and discuss rates for practical algorithms in Chapter \ref{ch:algorithms}.}
\end{description}

This survey is mostly concerned with nonadaptive group testing with small error probability, starting with the noiseless setting (Chapter \ref{ch:algorithms}).  Later in the monograph, we will expand our attention to the noisy nonadaptive setting (Chapter \ref{ch:algorithms_noisy}), approximate recovery criteria (Section \ref{sec:partial}), adaptive testing with limited stages (Section \ref{sec:two-stage}), and the linear regime $k = \Theta(n)$ (Section \ref{sec:linear}), among others.

It will be useful to compare the results to come with various well-established results for adaptive testing and for zero-error nonadaptive testing (in the noiseless setting). The next two sections provide a brief review of these two models.

\section[A brief review of noiseless adaptive group testing]{A brief review of noiseless adaptive group testing\sectionmark{A review of noiseless adaptive group testing}} \label{sec:adaptive}
\sectionmark{A review of noiseless adaptive group testing}

Much of the early group testing literature focused on adaptive procedures. Dorfman's original paper \cite{dorfman} proposed a simple procedure where items were partitioned into pools that undergo primary testing: A negative test indicates that all the items in that set are definitely nondefective, whereas for within the positive tests, all items are subsequently tested individually. It is easily checked (see, for example, \cite{li1962}, \cite[Ex. 26, Section IX.9]{feller}) that the optimal partition (assuming that $k$ is known) comprises $\sqrt{nk}$ pools, each of size $\sqrt{n/k}$. Dorfman's strategy therefore requires requires at most
  \begin{equation} \label{eq:dorfmanegg}
  T = \sqrt{nk} + k \sqrt{\frac nk} = 2\sqrt{nk} \end{equation}
tests, since the worst case is that all $k$ items are in different pools.

Sterrett \cite{sterrett} showed that improvements can be made by testing items in a positive pool individually until a defective item is found, and then re-testing all remaining items from the pool together. Li \cite{li1962} and Finucan \cite{finucan} provided variants of Dorfman's scheme based on multi-stage adaptive designs.  \added{We will revisit Dorfman's strategy in Section \ref{sec:linear}, where we will consider the linear regime $k \sim \beta n$ and a broader class of strategies called `conservative two-stage group testing' \cite{aldridge2020conservative}.}

The work of Sobel and Groll \cite{sobel} introduced the crucial idea of recursively splitting the set, with their later paper \cite{sobel2} showing that such a procedure performs well even if the number of defectives is unknown. We will describe the procedure of \defn{binary splitting}, which lies at the heart of many adaptive algorithms. Suppose we have a set $A$ of items. We can test whether $A$ contains any defectives, and, if it does, discover a defective item through binary splitting as follows.

\begin{algorithm}[Binary splitting] \label{alg:bin} Given a set $A$:
\begin{enumerate}
\item Initialize the algorithm with set $A$. Perform a single test containing every item in $A$.
\item If the preceding test is negative, $A$ contains no defective items, and we halt. If the test is positive, continue.
\item If $A$ consists of a single item, then that item is defective, and we halt. Otherwise, pick half of the items in $A$, and call this set $B$. Perform a single test of the pool $B$.
\item If the test is positive, set $A := B$. If the test is negative, set $A := A \setminus B$. Return to Step 3.
\end{enumerate}
\end{algorithm}

The key idea is to observe that even if the test in Step 3 is negative, we still gain information from it; since $A$ contained at least one defective (as confirmed by Steps 1 and 2), and $B$ contained no defective, we can be certain that $A \setminus B$ contains at least one defective.

In Step 3, when picking the set $B$ to be half the size of $A$, we can round $|A|/2$ in either direction. Since the size of the set $A$ essentially halves on each loop through the algorithm, we see that binary splitting finds a defective item in at most $\lceil \log_2 |A| \rceil$ adaptive tests, or confirms there are no defective items in a single test. We conclude the following.

\begin{theorem} \label{lem:rounds}
We can find all $k$ defectives in a set of $n$ items by repeated rounds of Algorithm \ref{alg:bin}, using a total of $k \log_2 n + O(k)$ adaptive tests, even when $k$ is unknown.  In the sparse regime $k = \Theta(n^\alpha)$ with $\alpha \in [0,1)$, this gives an achievable rate of $1-\alpha$.
\end{theorem}

\begin{proof}
In the first round, we initialize the binary splitting algorithm using $A = \{ 1, 2, \ldots, n \}$, and find the first defective (denoted by $d_1$) using at most $\lceil \log_2 n \rceil$ tests.

In subsequent rounds, if we have found defectives $\{ d_1, \ldots, d_r \}$ in the first $r$ rounds, then the $(r+1)$-th round of Algorithm \ref{alg:bin} is initialized with $A = \{ 1, 2, \ldots, n\} \setminus \{ d_1, d_2, \ldots, d_r \}$. We perform one further test to determine whether $\{ 1, 2, \ldots, n\} \setminus \{ d_1, d_2, \ldots, d_r \}$ contains at least one defective. If not, we are done. If it does, we find the next defective item using at most $\lceil \log_2 (n-r) \rceil \le \lceil \log_2 n \rceil$ tests. We repeat the procedure until no defective items remain, and the result follows.
\end{proof}

Note that for $\alpha > 0$, this rate $1 - \alpha$ fails to match the counting bound of $1$. However, we can reduce the number of tests required to $k \log_2 (n/k) + O(k)$, thus raising the rate to $1$ for all $\alpha \in [0,1)$, by using a variant of Hwang's generalized binary splitting algorithm \cite{hwang}. The key idea is to notice that, unless there are very few defectives remaining, the first tests in each round of the repeated binary splitting algorithm are overwhelmingly likely to be positive, and are therefore very uninformative. A refined procedure is as follows:

\begin{algorithm} \label{alg:hwang}
Divide the $n$ items into $k$ subsets of size $n/k$ (rounding if necessary), and apply Algorithm \ref{alg:bin} to each subset in turn. 
\end{algorithm}
Note that each of these subsets contains an average of one defective. Using the procedure above, if the $i$-th subset contains $k_i$ defectives,  taking $k = k_i$ and $n = n/k$ in Theorem \ref{lem:rounds}, we can find them all using $k_i \log_2(n/k) + O(k_i)$ tests, or confirm the absence of any defectives with one test if $k_i = 0$. Adding together the number of tests over each subset, we deduce the result.

Combining this analysis with the upper bound $C \leq 1$ (Remark \ref{rem:universal}), we deduce the following.

\begin{theorem} \label{thm:adaptcap}
Using Algorithm \ref{alg:hwang}, we can find the defective set with certainty using $k \log_2 (n/k) + O(k)$ adaptive tests. Thus, the capacity of adaptive group testing in the sparse regime $k = \Theta(n^\alpha)$ is $C_0 = C = 1$ for all $\alpha \in [0,1)$.
\end{theorem}

This theorem follows directly from the work of Hwang \cite{hwang}, and it was explicitly noted that such an algorithm attains the capacity of adaptive group testing by Baldassini, Johnson, and Aldridge \cite{baldassini-johnson-aldridge}.

The precise form of Hwang's generalized binary splitting algorithm \cite{hwang} used a variant of this method, with various tweaks to reduce the $O(k)$ term. For example, the set sizes are chosen to be powers of $2$ at each stage, so the splitting step in Algorithm \ref{alg:bin} is always exact. Further, items appearing at any stage in a negative test are removed completely, and the values $n$ and $k$ of remaining items are updated as the algorithm progresses. Some subsequent work further reduced the implied constant in the $O(k)$ term in the expression $k \log_2(n/k) + O(k)$ above; for example, Allemann \cite{allemann} reduced it to $0.255k$ plus lower order terms. 

We see that algorithms based on binary splitting  are very effective when the problem is sparse, with $k$ much smaller than $n$. For denser problems, the advantage may be diminished; for instance, when $k$ is a large enough fraction of $n$, it turns out that adaptive group testing offers no performance advantage over the simple strategy of individually testing every item once. For example, for adaptive zero-error combinatorial testing, Riccio and Colbourn \cite{riccio} proved that no algorithm can outperform individual testing if $k \geq 0.369n$, while the Hu--Hwang--Wang conjecture \cite{hu} suggests that such a result remains true for $k \geq n/3$. We further discuss adaptive (and nonadaptive) group testing in the linear regime $k = \Theta(n)$ in Section \ref{sec:linear}.
We re-iterate that the focus of this survey is on the sparse regime, $k = \Theta (n^\alpha)$ with $\alpha \in [0,1)$, where group testing techniques have the greatest gains over individual testing.

\section[A brief review of zero-error nonadaptive group testing]{A brief review of zero-error nonadaptive \\group testing\sectionmark{A brief review of zero-error group testing}} \label{sec:zero-error}
\sectionmark{A brief review of zero-error group testing}

In this section, we discuss nonadaptive group testing with under the zero error criterion -- that is, we must be certain of recovering the defective set (of a given size), without any error probability.  In particular, we examine the important concepts of separable and disjunct matrices. The literature in this area is deep and wide-ranging, and we shall barely scratch the surface here. The papers of Kautz and Singleton \cite{kautz} and D'yachkov and Rykov \cite{dyachkov-rykov} are classic early works in this area, while the textbook of Du and Hwang \cite{du-hwang} provides a comprehensive survey. 

We first introduce some simple notation regarding the support sets of binary vectors.

\begin{definition} \label{def:support}
Given a test matrix $\mat X = (x_{ti}) \in \zo^{T\times n}$, we write $S(i) := \{t : x_{ti} = \one \}$ for the support of column $i$. Further, for any subset $\L \subseteq \{1,2,\dots,n\}$ of columns, we write $S(\L) = \bigcup_{i \in \L} S(i)$ for the union of their supports. (By convention, $S(\emptyset) = \emptyset$.)
\end{definition}

Observe that $S(i)$ is the set of tests containing item $i$, while $S(\K)$ is the set of positive tests when the defective set is $\K$. 

We now proceed to introduce two definitions associated with the test matrices, which are widely studied -- see for example \cite[Chapter 7]{du-hwang} -- and are fundamental to zero-error nonadaptive group testing. 

\begin{definition} \label{def:separable}
A matrix $\mat X$ is called \emph{$k$-separable} if the support unions $S(\L)$ are distinct over all subsets $\L \subseteq \{1,2,\dots,n\}$ of size $|\L| = k$.

A matrix $\mat X$ is called \emph{$\bar k$-separable} if the support unions $S(\L)$ are distinct over all subsets $\L \subseteq \{1,2,\dots,n\}$ of size $|\L| \leq k$.
\end{definition}

Clearly, using a $k$-separable matrix as a test design ensures that group testing will provide different outcomes for each possible defective set of size $k$; thus, provided that there are exactly $k$ defectives, it is certain that the true defective set can be found (at least given enough computation -- we discuss what it means for an algorithm to be `practical' in Section \ref{sec:algsummary}). In fact, 
it is clear that $k$-separability of the test matrix is also a necessary condition for zero-error group testing to be possible: If the matrix is not separable, then there must be two sets $\L_1$ and $\L_2$ with $S(\L_1) = S(\L_2)$ which cannot be distinguished from the test outcomes. Similarly, a $\bar k$-separable test design ensures finding the defective set provided that there are \emph{at most} $k$ defectives.

Thus, given $n$ and $k$, we want to know how large $T$ must be for a $k$-separable $(T \times n)$-matrix to exist.

An important related definition is that of a disjunct matrix.

\begin{definition} \label{def:disjunct}
A matrix $\mat X$ is called \emph{$k$-disjunct} if for any subset $\L \subseteq \{1,2,\dots,n\}$ of size $|\L| = k$ and any $i \not\in \L$, the set $S(i) \setminus S(\L)$ is non-empty (i.e., $S(i)$ is not a subset of $S(\L)$).
\end{definition}

In group testing language, this ensures that no nondefective item appears only in positive tests. This not only guarantees that the defective set can be found, but also reveals how to do so easily: Any item that appears in a negative test is nondefective, while an item that appears solely in positive tests is defective. (We will study this simple algorithm under the name COMP in Chapter \ref{ch:algorithms}.)

We briefly mention that the notions of $k$-separability, $\bar k$-separability, and $k$-disjunctness often appear in the literature under different names.  In particular, the columns of a $k$-disjunct matrix are often said to form a {\em $k$-cover free family}, and the terminology {\em superimposed code} is often used to refer to the columns of either a $\bar k$-separable matrix or a $k$-disjunct matrix (see, for example, \cite{kautz,du-hwang,dyachkov1983survey}). 

It is clear that the implications
\begin{equation} \label{imps}
  k\text{-disjunct} \ \Rightarrow \ \bar k\text{-separable} \ \Rightarrow \ k\text{-separable}
\end{equation}
hold.
Furthermore, Chen and Hwang \cite{chen-hwang} showed that the number of tests $T$ required for separability and disjunctness in fact have the same order-wise scaling, proving the following.

\begin{theorem} \label{sep-dis}
Let $\mat X$ be $2k$-separable. Then there exists a $k$-disjunct matrix formed by adding at most one row to $\mat X$.
\end{theorem}

Because of this, attention is often focused on bounds for disjunct matrices, since \rev{such bounds are typically easier to derive}, and these results can be easily converted to statements on separable matrices using \eqref{imps} and Theorem \ref{sep-dis}.

\added{We now proceed to consider impossibility (converse) results, showing that attaining the $k$-disjunct property is impossible when the number of tests $T$ is too small.} 
The following result, which D'yachkov and Rykov \cite{dyachkov-rykov} attribute to Bassalygo, 
was an important early lower bound on the size of disjunct matrices.

\begin{theorem} \label{BDR}
Suppose there exists a $k$-disjunct $(T \times n)$-matrix. Then
  \begin{equation} \label{BDReq}
    T \geq \min \left\{ \frac{1}{2}(k+1)(k+2), n \right\} . 
  \end{equation}
\end{theorem}

There have been many improvements to this lower bound on $T$, of which we mention a few examples. Shangguan and Ge \cite{shangguan} improve the constant $1/2$ in front of the $k^2$ term of Theorem \ref{BDR} with the bound
  \begin{equation} \label{ge}
    T \geq \min \left\{ \frac{15+\sqrt{33}}{24} (k+1)^2, n \right\} \approx \min \left\{ 0.864 (k+1)^2, n \right\} .
  \end{equation}
Ruszink\'o\ \cite{ruszinko} proves the bound
  \begin{equation} \label{rusz}
    T \geq \frac18 \,k^2 \,\frac{\log n}{\log k} 
  \end{equation}
for $n$ sufficiently large, provided that $k$ grows slower than $\sqrt{n}$,%
\footnote{The first line of the proof in \cite{ruszinko} 
assumes $k^2$ divides $n$; this is not true when $k^2 > n$, but can be accommodated with a negligible increase in $n$ if $k$ grows slower than $\sqrt n$.}
 while F\"uredi \cite{furedi} proves a similar bound with $1/8$ improved to $1/4$.
In the sparse regime $k = \Theta(n^\alpha)$ with $\alpha \in (0,1)$, we have $\log n/\log k \to 1/\alpha$, which means that \eqref{rusz} and F\"uredi's improvement give improved constants compared to Theorem \ref{BDR} and \eqref{ge} for sufficiently small $\alpha$.
In the very sparse regime $k = O(1)$, \eqref{rusz} gives roughly a $\log n$ factor improvement, which D'yachkov and Rykov \cite{dyachkov-rykov} improve further, replacing $1/8$ by a complicated expression that is approximately $1/2$ for large (but constant) values of $k$.
  
In the case that $k = \Theta(n^\alpha)$ with $\alpha > 1/2$, the bound $T \geq n$ of Theorem \ref{BDR} can be achieved by the identity matrix (that is, testing each item individually), and the resulting number of tests $T = n$ is optimal. 

For $\alpha < 1/2$, the $T \geq \Omega(k^2)$  lower bound of Theorem \ref{BDR} (and the related results) are complemented by achievability results of the form $T \leq O(k^2 \log n)$, which is only a logarithmic factor larger. For example, considering a Bernoulli random design with $p = 1/(k+1)$, one can prove the existence of a $k$-disjunct $(T \times n)$-matrix with
  \[ T \leq (1+\delta)\ee(k+1) \ln \left( (k+1) \binom{n}{k+1} \right) \sim (1+\delta)\ee(k+1)^2 \ln n  \]
for any $\delta > 0$ \cite[Theorem 8.1.3]{du-hwang}. (Du and Hwang \cite[Section 8.1]{du-hwang} attribute this result to unpublished work by Busschbach \cite{busschbach}.) Kautz and Singleton \cite{kautz} give a number of constructions of separable and disjunct matrices, notably including  a construction based on Reed--Solomon codes that we discuss further in Section \ref{sec:explicit}.  Porat and Rothschild \cite{porat} give a construction with $T = O(k^2 \log n)$ using linear codes.

Note that in the sparse regime, the lower bound from Theorem \ref{BDR} is on the order of $\min\{\Omega(k^2),n\}$  which is much larger than the order $k \log n$ of the counting bound. Thus, nonadaptive zero-error group testing has rate $0$ according to Definition \ref{def:rate}.

\begin{theorem} \label{thm:nonadaptcap}
The capacity of nonadaptive group testing with the zero-error criterion is $C_0 = 0$ in the case that $k = \Theta(n^\alpha)$ with $\alpha \in (0,1)$.  \added{In other words, the number of tests must have a growth rate strictly higher than $k \log\frac{n}{k}$.}
\end{theorem}

\begin{remark}
    \rev{In the context of zero-error communication \cite{shannon-zero}, a memoryless channel having a zero-error capacity of zero is a very negative result, as it implies that not even two distinct codewords can be distinguished with zero error probability.  We emphasize that when it comes to group testing, the picture is very different: A result stating that $C_0 = 0$ by no means implies that attaining zero error probability is a hopeless task; rather, it simply indicates that it is insufficient to take $O\big(k \log \frac{n}{k}\big)$ tests.  As discussed above, there is an extensive amount of literature establishing highly valuable results in which the number of tests is $O(k^2 \log n)$ or similar.}
\end{remark}

In contrast with Theorem \ref{thm:nonadaptcap}, in Chapters \ref{ch:algorithms} and \ref{ch:achievability} of this survey, we will see that under the small-error criterion (i.e., asymptotically vanishing but nonzero error probability), we can achieve nonzero rates for all $\alpha \in [0,1)$, and can even attain a rate of $1$ for \rev{$\alpha \in [0,0.409]$}. This demonstrates the significant savings in the number of tests permitted by allowing a small nonzero error probability. 

An interesting point of view is provided by Gilbert \etal\ \cite{gilbert2012}, who argue that zero-error group testing can be viewed as corresponding to an adversarial model on the defective set; specifically, the adversary selects $\K$ as a function of $\mat X$ in order to make the decoder fail.  Building on this viewpoint, \cite{gilbert2012} gives a range of models where the adversary's choice is limited by computation or other factors, effectively interpolating between the zero-error and small-error models.

\section{Applications of group testing} \label{sec:applications}

Although group testing was first formulated in terms of testing for syphilis \cite{dorfman}, it has been abstracted into a combinatorial and algorithmic problem, and subsequently been applied in many contexts. The early paper of Sobel and Groll \cite{sobel} lists some basic applications to unit testing in industrial processes, such as the detection of faulty containers, capacitors, or Christmas tree lights. Indeed, solutions based on group testing  have been proposed more recently for quality control in other manufacturing contexts, such as integrated circuits \cite{kahng2006new} and molecular electronics \cite{stan} (though the latter paper studies a scenario closer to the linear model discussed in Section \ref{sec:other_models}).

We review some additional applications here;  this list is certainly not exhaustive, and is only intended to give a flavour of the wide range of contexts in which group testing has been applied. Many of these applications motivate our focus on nonadaptive algorithms. This is because in many settings, adaptive algorithms are impractical, and it is preferable to fix the test design in advance -- for example, to allow a large number of tests to be run in parallel.  

\added{
\subsection*{Medical testing}

As discussed at the start of this chapter, medical testing was the very first proposed application of group testing, where expensive tests could be reduced when testing for a rare disease -- in that case, syphilis.

This idea re-emerged more recently in the Covid-19 pandemic, when many researchers and public figures proposed the use of pooled testing. This was particularly relevant at the beginning of the pandemic, when PCR tests were scarce and expensive, so there would be a large benefit to being able to make use of the test savings inherent in the group testing paradigm. Since group testing is most beneficial when infection is rare, it was most suited to the screening of asymptomatic people -- if an individual is displaying symptoms, they may be sufficiently likely to be infected that group testing would not be beneficial.

Aldridge and Ellis \cite{aldridge2022pooled} wrote a survey on group testing and the pandemic, covering several relevant mathematical models and giving an account of how and where group testing was actually used. We only give a brief overview here, and direct readers to \cite{aldridge2022pooled} for more details.

The most widely used method in practice was Dorfman's algorithm (see Sections \ref{sec:adaptive} and \ref{sec:two-stage}). This was preferred for several reasons including the simplicity of the method, the requirement of only two stages of testing (which is important since PCR testing is quite slow), and the fact that this approach is quite effective in the `uncommon but not super-rare' region where the prevalence is roughly between 2\% and 25\% (see Section \ref{sec:conservative} and \cite{aldridge2022pooled} for details).  One prominent and well documented use of Dorfman's algorithm was testing students at the University of Cambridge, UK, where pools were `social bubbles', typically of size around 8 -- see \cite[Section 6.2]{aldridge2022pooled} for a more detailed description of this and other well publicized uses of group testing during the pandemic.

Ultimately, group testing was not used particularly extensively for Covid-19, despite being championed by prominent figures in the media (see, for example, \cite{nyt}). As time went on, rapid at-home lateral flow tests became widely and cheaply available. Although these are less effective than PCR tests in detecting the presence of disease, the benefits of having fast and easy-to-obtain results ultimately proved to be more important.  Naturally, taking a quick test at home is not conducive to the sort of organised process needed to carry out group testing.

Nevertheless, there was an understandable explosion in interest in the topic of group testing. Much of this seemingly amounted to the rediscovery of known ideas.  On the other hand, there were also plenty of innovations in the mathematics of group testing too.  Several of the additions to the second edition of this survey are thanks to this upsurge in interest in the topic that occurred at that time.  In particular, some of the new research directions that arose are specifically motivated by applications in testing for pandemic diseases, such as community-aware group testing (see Section \ref{sec:community}) and certain non-binary testing models (see Section \ref{sec:other_models}). 
}

\subsection*{Biology}

With medicine and biology being closely related, it is no  surprise that group testing has found many  uses in biology in addition to medical testing, as summarized, for example, in  \cite{balding,chen10, du-hwang2}. We list some examples as follows:

\paragraph{DNA testing}

As described in \cite[Chapter 9]{du-hwang}, \cite{sham} and \cite{shental}, modern sequencing methods search for particular subsequences of a genome in relatively short fragments of DNA. Samples from individuals can easily be mixed, and accordingly, group testing can lead to significant reductions in the number of tests required to isolate individuals with rare genetic conditions -- see, for example, \cite{balding, curnow, gille}. In this context, it is typical to use nonadaptive methods (as in \cite{du-hwang2, erlich, erlich2, macula, shental}), since it is preferable not to stop the testing machines in order to rearrange the sequencing strategy. Furthermore, the physical design of modern DNA testing plates means that it can often be desirable to use exactly $T=96$ tests (see \cite{erlich2}). Macula \cite{macula} describes combinatorial constructions that are robust to errors in testing.

\paragraph{Counting defective items}

Often we do not need to estimate the defective set itself, but rather wish to efficiently estimate the proportion of defective items. This may be because we have no need to distinguish individuals (for example, when dealing with insects \cite{thompson,walter}), or wish to preserve confidentiality of individuals (for example, monitoring prevalence of diseases).    References \cite{chen9, sobel3, swallow} were early works showing that group testing offers an efficient way to estimate the proportion of defectives, particularly when defectivity is rare. This testing paradigm continues to be used in recent medical research, where pooling can provide significant reductions in the cost of DNA testing  -- see for example \cite{khan}, \cite{tilghman}. 
  
Specific applications are found in  works such as \cite{katholi,swallow,thompson,walter}, in which the proportion of insects carrying a disease is estimated; and in \cite{gastwirth,tu}, in which the proportion of the population with HIV/AIDS is estimated while preserving individual privacy. Many of these protocols require nonadaptive testing, since tests may be time-consuming -- for example, one may need to place a group of possibly infected insects with a plant, and wait to see if the plant becomes infected. 
We review the question of counting defectives using group testing in more detail in Section \ref{sec:universal}.

\paragraph{Other biological applications}

We briefly remark that group testing has also been used in many other biological contexts -- see \cite[Section 1.3]{du-hwang2} for a review. For example, this includes the design of protein--protein interaction experiments \cite{mourad}, high-throughput drug screening \cite{kainkaryam}, \added{catalyst discovery \cite{Sak2025}}, and efficient learning of Immune--Defective graphs in drug design \cite{ganesan}.

\subsection*{Commmunication and signal processing}

Group testing has been applied in a number of communications scenarios, including the following:

\paragraph{Multiple access channels}

We refer to a channel where several users can communicate with a single receiver as a \defn{multiple access channel}. Wolf \cite{wolf} describes how this can be formulated in terms of group testing: At any one time, a small subset of users (active users) will have messages to transmit, and correspond to defective items in this context. Hayes \cite{hayes} introduced adaptive protocols based on group testing to schedule transmissions, which were further developed by many authors (see \cite{wolf} for a review). In fact, Berger \etal~\cite{berger2} argue for the consideration of a ternary group testing problem with outcomes `idle', `success', and `collision', corresponding to no user, one user, or multiple users broadcasting simultaneously, and develop an adaptive transmission protocol.

These adaptive group testing protocols for multiple access channels are complemented by corresponding nonadaptive protocols developed in works such as \cite{komlos} (using random designs) and \cite{debonis} (using designs based on superimposed code constructions). Variants of these schemes were further developed in works such as \cite{debonis2}, \cite{wu5} and \cite{wu4}. The paper \cite{varanasi} uses a similar argument for the related problem of Code-Division Multiple Access (CDMA), where decoding can be performed for a group of users simultaneously transmitting from constellations of possible points.  \added{See also \cite{inan2018energy,inan2019group} for further group testing approaches in short-packet or energy-limited random access settings.  

Along similar lines as some of the preceding works, in \cite{cohen2020efficient} a group testing based method is proposed for data collection over multiple-access wireless sensor networks.  See also \cite{cohen2021serial} for analogous methods in the context of sampling sparse time series signals.}

\paragraph{Cognitive radios}

A related communication scenario is that of cognitive radio networks, where `secondary users' can opportunistically transmit on frequency bands which are unoccupied by primary users. We can scan combinations of several bands at the same time and detect if any signal is being transmitted across any of them, and use procedures based on group testing to determine which bands are unoccupied -- see for example \cite{atia2, sharma2}.

\paragraph{Network tomography and anomaly discovery}

Group testing has been used to perform network tomography; that is, to detect faults in a computer network only using certain end-to-end measurements.  In this scenario, users send a packet from one machine to another, and check whether it successfully arrives. For example, we can view the edges of the network as corresponding to items, with items in a test corresponding to the collection of edges along which the packet travelled.  If (and only if) a packet arrives safely, we know that no edge on that route is faulty (no item is defective), which precisely corresponds to the OR operation of the standard noiseless group testing model.

As described in several works including \cite{ cheraghchi-etal, harvey2007, ma, xu}, and discussed briefly in Section \ref{sec:constrained}, this leads to a scenario where arbitrary choices of tests cannot be taken, since each test must correspond to a connected path in the graph topology. This motivates the study of graph-constrained group testing, which is a problem of interest in its own right.

Goodrich and Hirschberg \cite{goodrich} describes how an adaptive algorithm for ternary group testing can be used to find faulty sensors in networks, and a nonadaptive algorithm (combining group testing with Kalman filters) is described in \cite{lo}.

\subsection*{Information technology}

The discrete nature of the group testing problem makes it particularly useful for various problems in computing, such as the following:

\paragraph{Data storage and compression}

Kautz and Singleton \cite{kautz} describe early applications of superimposed coding strategies to efficiently searching punch cards and properties of core memories. Hong and Ladner \cite{hong} describe an adaptive data compression algorithm for images, based on the wavelet coefficients. In particular, they show that the standard Golomb algorithm for data compression is equivalent to Hwang's group testing algorithm \cite{hwang} (see Section \ref{sec:adaptive}). These ideas have been extended, for example in \cite{hong2} in the context of compressing correlated data from sensor networks, using ideas related to the multiple access channel described above.  \added{See also \cite{agarwal2020group} for a group testing problem with runlength constraints based on applications in topological DNA-based data storage.}

\paragraph{Cybersecurity}

An important cybersecurity problem is to efficiently determine which computer files have changed, based on a collection of hashes of various combinations of files (this is sometimes referred to as the `file comparison problem'). Here the modified files correspond to defective items, with the combined hash acting as a testing pool.  References \cite{goodrich2} and \cite{madej}    demonstrate methods to solve this problem using nonadaptive procedures based on group testing. 

Khattab \etal\ \cite{khattab} and Xuan \etal\ \cite{xuan} describe how group testing can be used to detect denial-of-service attacks, by dividing the server into a number of virtual servers (each corresponding to a test), observing which ones receive large amounts of traffic (interpreted as a positive test) and hence deducing which users are providing the greatest amount of traffic.

\paragraph{Database systems}

In order to manage databases efficiently, it can be useful to classify items as `hot' (in high demand), corresponding to defectivity in group testing language. Cormode and Muthukrishnan \cite{cormode} show that this can be achieved using both adaptive and nonadaptive group testing, even in the presence of noise. A related application is given in \cite{wang2018}, which considers the problem of identifying `heavy hitters' (high-traffic flows) in Internet traffic, and provides a solution using a linear version of group testing, in which each test gives the number of defective items in the testing pool (see Section \ref{sec:other_models}).

\paragraph{Bloom filters}

A \defn{Bloom filter} \cite{bloom} is a data structure that allows one to test if a given item is in a special set of distinguished items extremely quickly, with no possibility of false negatives and very rare false positives.

The Bloom filter uses $L$ hash functions, each of which maps items to $\{1, 2, ..., T\}$. For each of the items in the distinguished set, one sets up the Bloom filter by hashing the item using each of the $L$ hash functions, and setting the corresponding bits in a $T$-bit array to $\one$. (If the bit is already set to $\one$, it is left as $\one$.)
To test if another item is in the distinguished set, one hashes the new item with each of the $L$ hash functions and looks up the corresponding bits in the array. If any of the bits are set to $\zero$, the item is not in the distinguished set; while if the bits are all set $\one$, one assumes the item is in the set, although there is some chance of a false positive.

The problem of deciding how many hash functions $L$ to use, and how large the size of the array $T$ is, essentially amounts to a group testing problem. For instance, for suitably-chosen $L$, the analysis is almost identical to that of the COMP algorithm with a near-constant tests-per-item design (see Section \ref{sec:near_constant}).  We also mention that \cite{zhanghuang} makes a connection between Bloom filters and coding over an \texttt{OR} multiple-access channel, which is also closely related to group testing.

\added{
\subsection*{Machine learning}

Group testing has played a role in several aspects of machine learning, including the following.

\paragraph{Classification and clustering} 
Two of the most fundamental problems in machine learning are classification (a supervised problem) and clustering (an unsupervised problem).   
Ubaru and Mazumdar \cite{ubaru2017multilabel} use group testing to solve classification problems with many labels using combinations of binary problems, each corresponding to the {\tt OR} of relatively fewer labels.  This was further refined by Ubaru \etal~\cite{ubaru2020multilabel} using methods based on hierarchical partitioning and data-dependent grouping.  
Turning to clustering problems, a `near neighbour' search technique was proposed by Engels, Coleman, and Shrivastava~\cite{engels2021practical} based on treating near neighbours as positives (defectives), non-neighbours as negatives (nondefectives), and using hashing techniques to test groups of points and return, within some accuracy, whether any point in that group is a near neighbour of a particular query point.  A different clustering problem was studied by Black \etal~\cite{black2024clustering} in which queries can be performed on subsets, with the result returning how many clusters intersect the subset.  This was tackled using group testing techniques and shown to require significantly fewer queries compared to previously-studied `same-cluster queries'.  

\paragraph{Neural networks} In recent years, the field of machine learning has drastically evolved following the rise of deep neural networks, and group testing has continued to play a role in certain aspects. 
With the goal of detecting certain objects in images, Liang and Zou \cite{liang2021neural}, 
`pool' multiple images (by averaging pixels or aggregating features) and then pass them through a neural network, with group testing methods then used for detection.  Pooling sizes up to at least 16 are shown to be feasible experimentally.  A counterpart based on non-binary test results was proposed by Ghosh, Saxena, and Rajwade~\cite{ghosh2023efficient}, who further used binary group testing for outlier detection based on testing pools of data points.

\paragraph{Learning interpretable rules} \rev{In the work of Malioutov \etal~\cite{Malioutov2017}, group testing is used to learn classification rules that are interpretable by practitioners. For example, in medicine we may wish to develop a rule based on training data that can diagnose a condition or identify high-risk groups from a number of pieces of measured medical data (features). However, standard machine learning approaches such as support vector machines or neural networks can lead to classification rules that are complex, opaque, and hard to interpret for a clinician. For reasons of simplicity, it can be preferable to use suboptimal classification rules based on a small collection of {\tt AND} clauses or a small collection of {\tt OR} clauses. In  \cite{Malioutov2017},  the authors show how such rules can be obtained using a relaxed noisy linear programming formulation of group testing (to be described in Section \ref{sec:noisy_LP}). 
The work of Emad, Varshney, and Malioutov \cite{emad2015semiquantitative} further studied such problems under a semi-quantitative group testing model, in which the outcomes are non-binary.}

\paragraph{Distributed learning}  Jain, Cardone, and Mohajer~\cite{jain2023probabilistic} considered distributed computation of matrix--vector products, and proposed group testing methods to identify which machines are `attacked' (that is, these machines output adversarially chosen unreliable answers).  A similar kind of problem of identifying malicious clients using group testing was studied by Xhemrishi \etal~\cite{xhemrishi2023fedgt}, in the context of federated learning.

}
 
\subsection*{Other theoretical domains}

Finally, group testing has been applied to a number of theoretical problems in statistics, data science, and theoretical computer science.

\paragraph{Search problems}

Du and Hwang \cite[Part IV]{du-hwang} give an extensive review of applications of group testing to a variety of search problems, including the famous problem of finding a counterfeit coin and membership problems.  This can be seen as a generalization of group testing; a significant early contribution to establish order-optimal performance was made by Erd\H{o}s and R\'{e}nyi \cite{erdos-renyi}.  \added{Further search problems related to group testing include locating targets via region queries \cite{Kaspi2015}, `20 questions' problems \cite[Section 6.2]{zhou2025twenty}, and finding similar items in high-dimensional spaces \cite{shi}.}

\paragraph{Sparse inference and learning}

Gilbert, Iwen and Strauss \cite{gilbert} discuss the relationship between group testing and compressed sensing, and show that group testing can be used in a variety of sparse inference problems, including streaming algorithms and learning sparse linear functions. Malioutov and Varshney \cite{malioutov} build on this idea by showing how group testing can be used to perform binary classification of objects, and Emad and Milenkovic \cite{emad2014poisson} develop a framework for testing arrivals with decreasing defectivity probability.

\paragraph{Theoretical computer science}

Group testing has been applied to  classical problems in theoretical computer science, including pattern matching \cite{clifford,indyk,macula2004} and the estimation of high degree vertices in hidden bipartite graphs \cite{wang11}.

\rev{
In addition, generalizations of the group testing problem are studied in this community in their own right, including the `$k$-junta problem' (see for example \cite{blais,bshouty2,mossel}). A binary function $f$ is referred to as a $k$-junta if it depends on at most $k$ of its inputs, and we wish to investigate this property using a limited number of input--output pairs $(x, f(x))$. 

It is worth noting that {\em testing} $k$-juntas only requires determining whether a given $f$ has this property or is far from having this property \cite{blais}, which is distinct from {\em learning} $k$-juntas, i.e., either determining the $k$ inputs that $f$ depends on or estimating $f$ itself.  Further studies of the $k$-junta problem vary according to whether the inputs $x$ are chosen by the tester (`membership queries') \cite{bshouty}, uniformly at random by nature \cite{mossel}, or according to some quantum state \cite{ambainis,atici}.
In view of the above description, group testing with a combinatorial prior is closely related to the $k$-junta learning problem, but with the additional knowledge that the function is an {\tt OR} of the $k$ inputs. 
}

\section*{Appendix: Comparison of combinatorial and\\ i.i.d.~priors} \label{sec:comb_iid}
\markright{\slshape APPENDIX: COMPARISON OF PRIORS}
\addcontentsline{toc}{section}{Appendix: Comparison of combinatorial and i.i.d.~priors}

In this technical appendix, we discuss the relationship between combinatorial and i.i.d.~priors for the defective set. We tend to use the combinatorial prior throughout this survey, so new readers can safely skip this appendix on first reading.

Recall the two related prior distributions on the defective set $\K$: 
\begin{itemize}
    \item Under the {\em combinatorial prior}, there are exactly $k$ defective items, and the defective set $\K$ is uniformly random over the $\binom nk$ possible subsets of that size.
    \item Under the {\em i.i.d.~prior}, each item is defective independently with a given probability $\qq \in (0,1)$, and hence the number of defectives $k = |\K|$ is distributed as $k \sim {\rm Binomial}(n,\qq)$, with $\EE( k ) = n\qq$.  For brevity, we adopt the notation $\overline{k} = n\qq$.
\end{itemize}

Intuitively, when the (average) number of defectives is large, one should expect the combinatorial prior with parameter $k$ to behave similarly to the i.i.d.~prior with a matching choice of $\overline{k}$, since in the latter case we have $k = \overline{k} (1+o(1))$ with high probability due to standard binomial concentration bounds. 

To formalize this intuition, first consider the definition $\text{rate} := \frac{1}{T} \log_2 \binom nk$ for the combinatorial prior (see Definition \ref{def:rate}), along with the following analogous definition for the i.i.d.~prior:
\begin{equation}
     \text{rate} := \frac{nh(\qq)}{T},
\end{equation}
where $h(\qq) = -\qq \log_2 {\qq} - (1-\qq)\log_2{(1-\qq)}$ is the binary entropy function.  Using standard estimates of the binomial coefficient \cite[Sec.~4.7]{ash}, the former is asymptotically equivalent to $\frac{1}{T} n h(k/n)$, which matches $\frac{1}{T} nh(\qq) = \frac{1}{T} nh(\overline{k} / n)$ with $k$ in place  of $\overline{k}$.  Consistent with our focus in this monograph, in this section we focus on scaling laws of the form $k \to \infty$ and $k = o(n)$ (or similarly with $\overline{k}$ in place of $k$), in which case the preceding rates are asymptotically equivalent to $\frac{1}{T} k\log_2\frac{n}{k}$ and $\frac{1}{T} \overline{k}\log_2\frac{n}{\overline{k}}$.  With some minor modifications, the arguments that we present below also permit extensions to the linear regime $k = \Theta(n)$ (discussed in Section \ref{sec:other_models}).

Having established that the rate expressions asymptotically coincide, further arguments are needed to transfer achievability and converse results from one prior to the other.  In the following, we present two results for this purpose.  In both results, any statement on the existence or non-existence of a decoding rule may refer to decoders that have perfect knowledge of the number of defectives $k$, or only partial knowledge (such as high-probability bounds), or no knowledge at all -- but the assumed decoder knowledge must remain consistent throughout the entire theorem.  Note that having exact knowledge of $k$ in the i.i.d.~setting is a particularly unrealistic assumption, since in that setting it is a random quantity.  We further discuss the issue of known vs.~unknown $k$ at the end of the section.

\subsubsection{From combinatorial to i.i.d.}

The following theorem describes how to transfer achievability bounds from the combinatorial prior to the i.i.d.~prior.

\begin{theorem}  \label{thm:relation_ach}
    Consider a sequence of (possibly randomized or adaptive) test designs $\mat{X}$ (indexed by $n$) attaining $\Perr \to 0$ under the combinatorial prior whenever $k = k_0 (1+o(1))$ for some nominal number of defectives $k_0$, with $k_0 \to \infty$ and $k_0 = o(n)$ as $n \to \infty$.  Then the same $\mat{X}$ and decoding rule also attains $\Perr \to 0$ under the i.i.d.~prior with $\qq = k_0 / n$ (i.e., $\overline{k} = k_0$).  In particular, if a given rate $R_0$ is achievable under the combinatorial prior whenever $k = k_0 (1+o(1))$, then it is also achievable under the i.i.d.~prior with $\overline{k} = k_0$.
\end{theorem}
\begin{proof}
    Since $\overline{k} = k_0$ grows unbounded as $n \to \infty$ by assumption, we have by standard binomial concentration that $k = \overline{k}(1+o(1)) = k_0(1+o(1))$ with probability approaching one under the i.i.d.~prior.  Letting $\mathcal{I}$ denote the corresponding set of `typical' $k$ values, we deduce that
    \begin{align}
        \PP({\rm err}) 
            &= \sum_{k} \PP(k) \PP({\rm err} \,|\, k) \\
            &\le \sum_{k \in \mathcal{I}} \PP(k) \PP({\rm err} \,|\, k) + \PP( k \notin \mathcal{I} ) \\
            &\to 0,
    \end{align}
    since $\PP({\rm err} \,|\, k) \to 0$ for all $k \in \mathcal{I}$ by assumption, and we established that $\PP( k \notin \mathcal{I} ) \to 0$.  This establishes the first claim.  
    The additional claim on the rate follows since, as discussed above (and using $k = k_0 (1+o(1))$), the achievable rates for both priors are asymptotically equivalent to $\frac{1}{T} k_0\log_2\frac{n}{k_0}$.
\end{proof}

Analogously, the following theorem describes how to transfer converse bounds from the combinatorial prior to the i.i.d.~prior.

\begin{theorem} \label{thm:relation_conv}
    Fix $R_0 > 0$, and suppose that under the combinatorial prior with some sequence of defective set sizes $k = k_0$ (indexed by $n$) satisfying $k_0 \to \infty$ and $k_0 = o(n)$, there does not exist any algorithm achieving rate $R_0$.  Then for any arbitrarily small constant $\epsilon > 0$, under the the i.i.d.~prior with $\qq = k_0 (1+\epsilon) / n$, there does not exist any algorithm achieving rate $R_0 (1+\epsilon)$.
\end{theorem}
\begin{proof}
    Since $k_0 \to \infty$ and the average number of defectives under the i.i.d.~prior is $\overline{k} = k_0(1+\epsilon)$, we deduce via binomial concentration that $k \in [k_0,k_0(1+2\epsilon)]$ with probability approaching one.  For $k$ outside this range, any contribution to the overall error probability is asymptotically negligible.
    
    On the other hand, when $k$ does fall in this range, we can consider a genie argument in which a uniformly random subset of $k_1 = k - k_0$ defectives is revealed to the decoder.  The decoder is then left to identify $k_0$ defectives out of $n_1 = n-k_1$ items.  Hence, the problem is reduced to the combinatorial prior with slightly fewer items.

    The condition $k \le k_0(1+2\epsilon)$ implies that $k_1 \le 2\epsilon k_0$ and hence $n_1 \ge n - 2\epsilon k_0$, which behaves as $n(1-o(1))$ since $k_0 = o(n)$.  As discussed at the start of this section, in the sparse regime, the asymptotic rate is equal to the asymptotic value of $\frac{1}{T} k\log_2\frac{n}{k}$ (combinatorial prior) or $\frac{1}{T}\overline{k}\log_2\frac{n}{\overline{k}}$ (i.i.d.~prior), and we observe that (i) replacing $n$ by $n(1-o(1))$ does not impact the asymptotic rate; and (ii) replacing $\overline{k} = k_0(1+\epsilon)$ by $k_0$ reduces the rate by a factor of $1/(1+\epsilon)$ asymptotically.

    We conclude the proof via a contradiction argument: If $R_0(1+\epsilon)$ were achievable with $\qq = k_0(1+\epsilon) / n$ under the i.i.d.~prior, the preceding reduction would imply that $R_0$ is achievable with $k = k_0$ under the combinatorial prior, which was assumed to be impossible.
\end{proof}

Unlike Theorem \ref{thm:relation_ach}, this result involves scaling the (average) number of defectives and the rate by $1+\epsilon$.  However, since $\epsilon$ is arbitrarily small, one can think of this scaling as being negligible.  In fact, for all achievability and converse results that we are aware of, in the case that $k = \Theta( n^{\alpha} )$ for some $\alpha \in (0,1)$, the asymptotic rate depends only on $\alpha$ and not on the implied constant in the $\Theta(\cdot)$ notation.

\subsubsection{From i.i.d.~to combinatorial}

It is also of interest to transfer results in the opposite direction, i.e., to infer achievability or converse bounds for the combinatorial prior based on those for the i.i.d.~prior.  For this purpose, the contrapositive statements of Theorems \ref{thm:relation_ach} and \ref{thm:relation_conv} read as follows:
\begin{itemize}
    \item (Theorem \ref{thm:relation_ach}) If there does not exist any test design and decoder achieving $\Perr \to 0$ under the i.i.d.~prior when $\qq = k_0 / n$ with $k_0 \to \infty$ and $k_0 = o(n)$, then there also does not exist any test design and decoder that simultaneously achieves $\Perr \to 0$ under the combinatorial prior for all $k$ such that $k = k_0(1+o(1))$.
    \item (Theorem \ref{thm:relation_conv}) Again assuming $k_0 \to \infty$ and $k_0 = o(n)$, if the rate $R_0$ is achievable with $\qq = k_0/n$ under the i.i.d.~prior, then for arbitrarily small $\epsilon > 0$ the rate $R_0/(1+\epsilon)$ is achievable with $k = k_0/(1+\epsilon)$ under the combinatorial~prior.
\end{itemize}
It is worth noting that the former of these statements does not directly provide a converse result for any particular value of $k$, but rather, only does so for the case that several $k$ must be handled simultaneously.

\subsubsection{Knowledge of the number of defective items} 

A somewhat more elusive challenge is to transfer results from the case that the decoder knows the number of defective items $k$ to the case that it does not (or the case that it only knows bounds on $k$), and vice versa.  In particular, it is sometimes convenient to prove achievability results for the case that $k$ is known exactly (see for example Chapter \ref{ch:achievability}), and to prove converse results for the case that $k$ is unknown (see for example Section \ref{sec:sss}).  This potentially poses a `gap' in the achievability and converse bounds even when the associated rates coincide.

\added{
While we are not aware of any completely general results allowing one to close such a gap, we highlight a number of techniques that have addressed this issue and provided evidence that the best possible rates tend to be the same regardless of whether $k$ is known exactly or to within a multiplicative $1+o(1)$ factor (as is the case under the i.i.d.~prior).
\begin{itemize}
    \item In Section \ref{sec:algcon}, we will describe an approach that combines the failure of two algorithms called SSS (Section \ref{sec:sss}) and COMP (Section \ref{sec:COMP}) to deduce the failure of an arbitrary algorithm.  Neither of these algorithms has knowledge of $k$.  It is argued that if both of them fail, then there must exist some $\K' \ne \K$ with $|\K'| = |\K| = k$ such that $\K'$ is also consistent with the outcomes.  Since the decoder cannot do better than randomly guess between the two (or more) consistent size-$k$ sets even when $k$ is known, we deduce that $\Perr$ cannot tend to zero.
    \item In Section \ref{sec:gen_conv}, we will outline a converse technique based on \emph{masking events}, in which many items appear only in tests containing other items that are defective, leading to decoding errors.  Such techniques are more naturally suited to the i.i.d.~prior, but a simple trick can be used under the combinatorial prior.  Letting $k$ be the number of defectives in the latter, one considers the i.i.d.~prior with $\bar k$ slightly below $k$, so that the random number of defectives is below $k$ with high probability.  Assuming this high-probability event holds, further items can be marked as defective to produce exactly $k$ in total, effectively recovering the combinatorial prior.  The idea is then that if masking occurred before marking the extra items as defective, it certainly still occurs after doing so.
    \item Certain achievability results that assume exact knowledge of $k$ have been adapted to only require approximate knowledge: for example, \cite[Appendix B]{Sca18}, which applies to both noiseless and noisy settings.
\end{itemize}
}

\chapter{Algorithms for Noiseless Group Testing} \label{ch:algorithms}

\section{Summary of algorithms} \label{sec:algsummary}

In this chapter, we discuss decoding algorithms for noiseless nonadaptive group testing. That is, we are interested in methods for forming the estimate $\hat\K$ of the defective set given the test matrix $\mat X$ and the outcomes $\vec y$.  In addition, we present performance bounds for these algorithms under random test designs.

We are particularly interested in algorithms that are practical in the following two senses:
\begin{enumerate}
 \item The algorithm does not require knowledge of the number of defectives $k$, other than general imprecise knowledge such as `defectivity is rare compared to nondefectivity', which we implicitly assume throughout.
 \item The algorithm is computationally feasible. Specifically, we seek algorithms that run in time and space polynomial in $n$, and preferably no worse than the $O(nT)$ time and space needed to read and store an arbitrary test design $\mat X$.  Certain algorithms exist having even faster `sublinear' decoding times, but their discussion is deferred to Section \ref{sec:sublinear}.
\end{enumerate}

All algorithms in this chapter are practical in the first sense \added{and, with one exception, in the second sense.  The exception is an algorithm called SSS, which has (worst-case) runtime higher than polynomial and can typically only be run on small-to-medium size problems}, but we include it here as a benchmark. Since several algorithms are based on approximating the SSS algorithm in faster time, the analysis of SSS is useful for understanding these other more efficient algorithms.

An important concept that we will use is that of a satisfying set.

\begin{definition} \label{def:sat}
Consider the noiseless group testing problem with $n$ items, using a test design $\mat X$ and producing outcomes $\vec y$. A set $\L \subset \{1,2,\dots,n\}$ is called a \defn{satisfying set} if
\begin{itemize}
  \item every positive test contains at least one item from $\L$;
  \item no negative test contains any item from $\L$.
\end{itemize}
\end{definition}

Thus, when performing group testing with test design $\mat X$, had the true defective set been $\L$, we would have observed the outcome $\vec y$.  Clearly the true defective set $\K$ is itself a satisfying set.

While we postpone complete definitions of algorithms until the relevant sections, it is worth providing a quick outline here:
\begin{description}
\item[SSS (smallest satisfying set)] chooses the smallest satisfying set.  This is based on the idea that the defective set $\K$ is a satisfying set, but since defectivity is rare, $\K$ is likely to be small in size.  \added{This algorithm faces computational challenges when $n$ is large, since it is equivalent to solving an integer program.}  See Sections \ref{sec:sss} and \ref{sec:lp} for details.
\item[COMP (combinatorial orthogonal matching pursuit)] assumes \\
that each item is defective unless there is a simple proof that it is nondefective, namely, that the item is in at least one negative test. See Section \ref{sec:COMP} for details.
\item[DD (definite defectives)] assumes that each item is nondefective unless there is a certain simple proof that it is defective. See Section \ref{sec:DD} for details.
\item[SCOMP (sequential COMP)] attempts to find the smallest satisfying set by beginning with the DD set of definite defectives, and sequentially adding items to the estimated defective set until a satisfying set is obtained. See Section \ref{sec:SCOMP} for details.
\item[Linear programming relaxations] solve a relaxed form of the smallest satisfying set problem, and use this to attempt to find the smallest satisfying set. See Section \ref{sec:lp} for details.
\end{description}
In Chapter \ref{ch:algorithms_noisy}, we will present additional algorithms that can be specialized to the noiseless case, particularly \textit{belief propagation} (see Section \ref{sec:belprop}) and \textit{separate decoding of items} (Section \ref{sec:separate}).  Since these algorithms are primarily of interest for noisy settings, we omit them from this chapter to avoid repetition.  \added{We also briefly mention an additional technique called \emph{spatial coupling}; in Chapter \ref{ch:achievability}, we will discuss that this technique has polynomial runtime and provably achieves optimal rates.   The reason for deferring this approach to Chapter \ref{ch:achievability} is that its development is strongly aligned with the theoretical developments of that chapter, and less so with the algorithmic developments of the present chapter.}

We will shortly see numerical evidence that SSS performs the best when the problem is sufficiently small that it can be implemented, while the SCOMP and linear programming approaches perform best among the more practical algorithms.  As another means to compare these algorithms, in the upcoming sections, we will mathematically analyse the rates achieved by these algorithms in the case that the matrix $\mat X$ has a Bernoulli design, formally stated as follows.

\begin{definition} \label{def:berndesign}
In a Bernoulli design, each item is included in each test independently at random with some fixed probability $p = \nu/k$. In other words, independently over $i \in \{1,2, \dots, n\}$ and $t \in \{1,2,\dots, T\}$, we have $\PP(X_{ti} = \one) = p = \nu/k$ and $\PP(X_{ti} = \zero) = 1-p = 1 - \nu/k$.
\end{definition}

\rev{The parametrization $p = \nu / k$ is chosen because such scaling with constant $\nu$ will be seen to be optimal as $k$ grows large.  Intuitively, since the average number of defectives in each test is $\nu$, one should avoid the cases $\nu \to 0$ or $\nu \to \infty$ because they lead to uninformative tests (that is, tests returning a given outcome with high probability).}
 In Section \ref{sec:near_constant}, we turn to the near-constant tests-per-item design, formally introduced in Definition \ref{def:const_col} below. We will see that the rates of all of the above algorithms are slightly improved when used with this alternative design.
 
Later, in Section \ref{sec:algcon}, we will see how the analysis of the SSS and COMP algorithms allows us to establish algorithm-independent lower bounds on the number of tests required under Bernoulli and near-constant test-per-item designs.

The result of this chapter are further complemented by those of Chapter \ref{ch:achievability}, which looks further into information-theoretic achievability and converse bounds.  A collective highlight of these two chapters is the following result on the information-theoretic optimality of some practical algorithms when certain random test designs are adopted.

\begin{theorem} \label{thm:highlight}
Consider group testing with $n$ items and $k = \Theta(n^\alpha)$ defectives, for $\alpha > 1/2$. Suppose that the test design is Bernoulli (with the parameter $p$ chosen optimally) or constant tests-per-item (with the parameter $L$ chosen optimally), and suppose that the decoding algorithm is SSS, DD, SCOMP, or linear programming. Then we can achieve the rate
  \[ R^* = \begin{cases} \displaystyle \frac{1}{\ee \ln 2} \, \frac{1-\alpha}{\alpha} \approx 0.531 \frac{1-\alpha}{\alpha}  & \text{for Bernoulli design,} \\[0.4cm]
        \displaystyle \ln 2 \, \frac{1-\alpha}{\alpha} \approx 0.693 \frac{1-\alpha}{\alpha}  & \text{for near-constant tests-per-item,} \end{cases} \]
which is the optimal rate for $\alpha > \frac{1}{2}$ under these random test designs regardless of the decoding algorithm used.  \added{This rate corresponds to a number of tests satisfying the following for arbitrarily small $\eta > 0$: $T \ge (1+\eta) \ee k \ln k$ (Bernoulli design), or $T \ge (1+\eta) \frac{1}{\ln 2} k \log_2 k$ (near-constant tests-per-item).\footnote{\added{When stating a rate, we will often also state the corresponding condition to $T$ if it is sufficiently simple/compact.  If the condition on $T$ is more complicated, then we will tend to only state the rate $R$, with the understanding that this corresponds to $T \sim \frac{1}{R} k \log_2\frac{n}{k}$.}}}

\added{Moreover, for the near-constant tests-per-item design, this rate is optimal for $\alpha > \frac{1}{2}$ among all nonadaptive designs.}
\end{theorem}

\begin{figure}[!t] 
\begin{center}
\includegraphics[width=\textwidth]{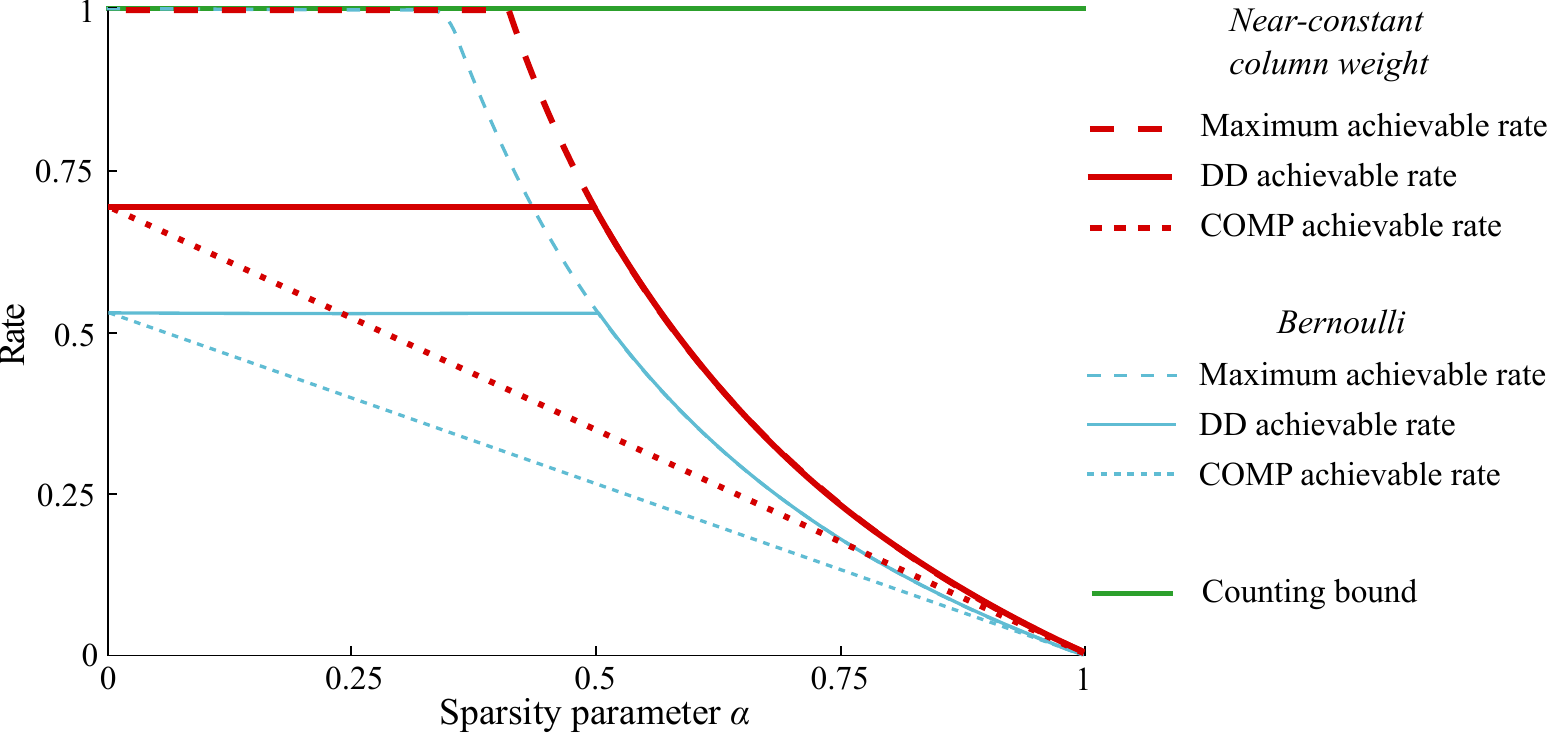}
\end{center}
\caption{Rates of various algorithms for nonadaptive group testing in the sparse regime with a Bernoulli design and with a near-constant column weight design.} \label{fig:algrates} 
\end{figure}

This and the other main results from this chapter are illustrated in Figure \ref{fig:algrates}.  
%
%
%
Table \ref{fig:summary} summarizes the algorithms. Whether or not the linear programming approach is guaranteed to give a satisfying set depends on the rule used to convert the LP solution to an estimate of the defective set -- see Section \ref{sec:lp} for further discussion.

\begin{table}[!t]
\begin{center}
\begin{tabular}{cccccc}
\hline
 \mbox{}  & Optimal rate & Fast & SS & No false $+$ & No false $-$ \\
\hline
\textbf{SSS}  & $\alpha\in[0,1)$ & no & yes & no & no \\
\textbf{COMP} & no & yes & yes & no & yes \\
\textbf{DD}   & $\alpha\in[1/2,1)$ & yes & no & yes & no \\
\textbf{SCOMP}& $\alpha\in[1/2,1)$ & yes & yes & no & no \\
\textbf{LP}   & $\alpha\in[1/2,1)$ & yes & depends & no & no \\
\hline
\end{tabular}
\end{center}
\caption{Summary of features of algorithms: (i) range of $\alpha$ (if any) for which the optimal rate is attained under randomized testing; (ii) whether an efficient algorithm is known; (iii) whether the output is guaranteed to be a satisfying set; (iv)-(v) guarantees on the false positives and false negatives in the reconstruction. \rev{The algorithms labelled `fast' can be implemented in time $O(nT)$, except possibly LP, whose complexity depends on the solver used (see Section \ref{sec:lp} for discussion).}  \label{fig:summary}}
\end{table}

\begin{figure}[!t] 
\begin{center}
\includegraphics[width=0.86\textwidth]{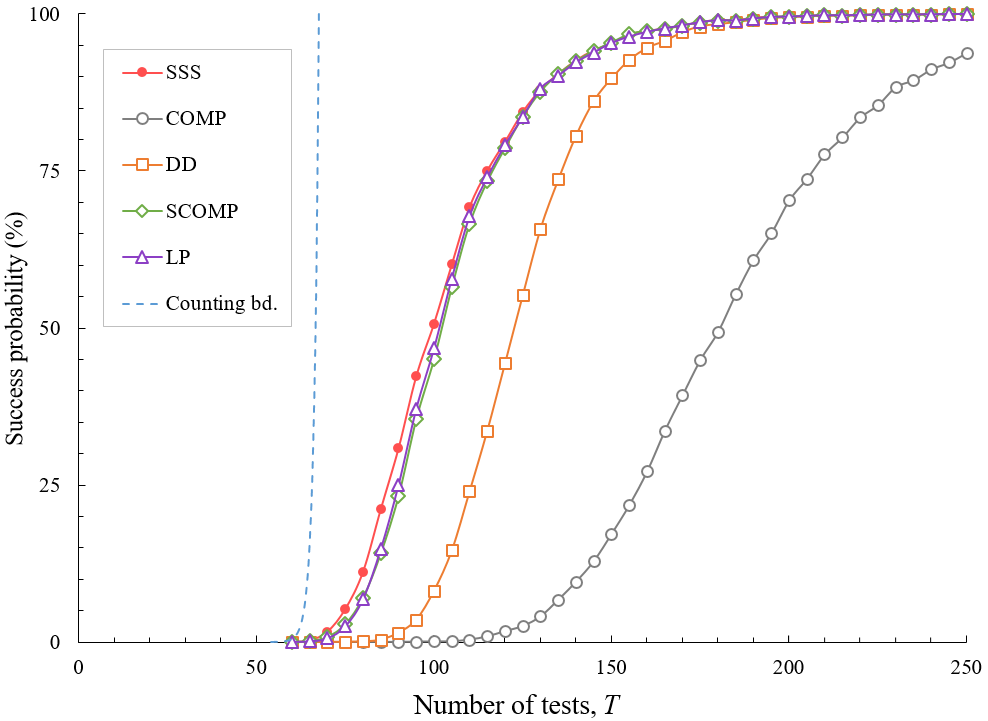}
\end{center}
\caption{Empirical performance through simulation of five algorithms with a Bernoulli design. The parameters are $n = 500$, $k=10$, and $p = 1/(k+1) = 0.0909$. For comparison purposes, we plot the theoretical upper bound on $\Psuc$ from Theorem \ref{thm:bjaconverse} as `counting bd'.} \label{fig:allfive} 
\end{figure}

While this chapter focuses on mathematical results, we can also examine the behaviour of the algorithms presented here in simulations. Figure \ref{fig:allfive} compares the five algorithms with a Bernoulli design. We see that at least in this example, SCOMP and linear programming are almost as good as SSS (which would be infeasible for larger problems), while DD is not much worse, and COMP lags behind more significantly.

\begin{figure}[!t] 
\begin{center}
\includegraphics[width=\textwidth]{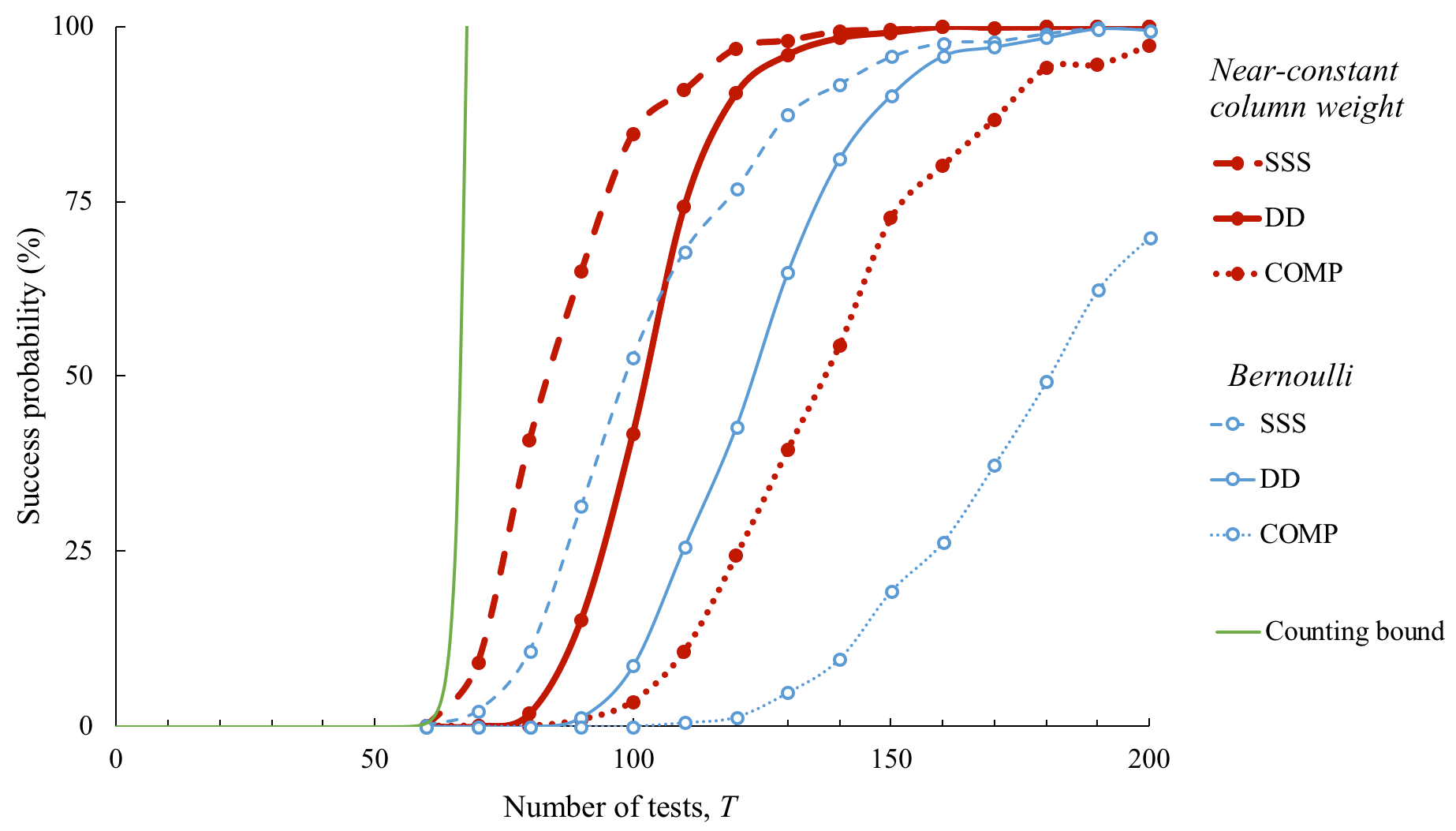}
\end{center}
\caption{Simulation of the COMP, DD and SSS algorithms with Bernoulli and near-constant column weight designs. The problem parameters are $n = 500$, and $k=10$. The Bernoulli parameter is $p = 1/k = 0.1$; the near-constant column weight parameter $L$ is the nearest integer to $(\ln 2)T/k \simeq 0.0693T$.} \label{fig:500-10} 
\end{figure}

Figure \ref{fig:500-10} compares the performance of the  COMP, DD and SSS algorithms under Bernoulli and near-constant column weight matrix designs. We see that the near-constant column weight design provides a noticeable improvement for all three algorithms.

\section{SSS: Smallest satisfying set} \label{sec:sss}

We first discuss the smallest satisfying set (SSS) algorithm. While SSS does not require knowledge of the number of defectives, it \added{faces computational challenges when $n$ is large.}  This is due to the highly combinatorial nature of the algorithm, which amounts to solving an integer program (see Section \ref{sec:lp}). 

Recall from Definition \ref{def:sat} that a set $\L$ is satisfying if every positive test contains at least one item from $\L$, and no negative test contains any item from $\L$.  Moreover, since the true defective set $\K$ is satisfying, a smallest satisfying set definitely exists -- though it may not be unique.

The SSS algorithm is based on the idea that since $\K$ is satisfying, and since defectivity is rare, it seems plausible that $\K$ should be the smallest satisfying set.

\begin{algorithm}
The \defn{smallest satisfying set} (\defn{SSS}) algorithm is defined by setting $\hat\K_\SSS$ to be the smallest satisfying set, with ties broken arbitrarily if such a set is not unique.
\end{algorithm}

\begin{figure}
\begin{center}
\begin{tabular}{ccccccc|c}
\quest & \quest & \quest & \quest & \quest & \quest & \quest & \yspace \\
\hline
 $\one$ & $\zero$ & $\one$ & $\zero$  & $\zero$ & $\one$ & $\zero$ & $\zero$ \\
  $\one$ & $\one$ & $\zero$ & $\one$  & $\zero$ & $\zero$ & $\one$  & $\one$ \\
 $\one$ & $\zero$ & $\zero$ & $\zero$  & $\one$ & $\zero$ &  $\zero$ & $\zero$ \\
 $\zero$ & $\one$ & $\one$ & $\zero$  & $\one$ & $\one$ & $\zero$  & $\one$ \\
$\one$ & $\zero$ & $\one$ & $\one$  & $\zero$ & $\one$ &  $\zero$ & $\one$ \\
\end{tabular}
\caption{Group testing example. \label{fig:sssexample} We describe the outcome of the SSS algorithm in Example \ref{ex:sss}, and that of the COMP algorithm in Example \ref{ex:comp}.}
\end{center}
\end{figure}

\begin{example}\label{ex:sss}
In Figure \ref{fig:sssexample}, we give an example of a group testing matrix and its outcome. (This example will reappear throughout this chapter.) Since we have only $n = 7$ items, it is, in this small case, practical to check all $2^7 = 128$ subsets. It is not difficult to check that the sets
$\{2,4\}$ and $\{2,4,7\}$ are the only satisfying sets. Of these, $\{2,4\}$ is the smallest satisfying set. Thus, we have $\hat\K_\SSS = \{2, 4\}$ as the output of the SSS algorithm.
\end{example}

\begin{remark} \label{rem:setcover}
    A naive implementation of SSS requires an exhaustive search over $\sum_{k' \le k}\binom{n}{k'}$ putative defective sets, and in general, we do not regard SSS as being practical in the sense described at the beginning of this chapter.  To make this intuition more precise, we describe a connection to the \emph{set cover} problem (see \cite{vazirani}). Given a universe $U$ and a family $\mathcal S$ of subsets of $U$, the set cover problem is to find the smallest family of subsets in $\mathcal S$ such that its union is the whole of $U$. Suppose that we let $U$ be the set of positive tests, and the subsets in $\mathcal S$ list the tests in which each possibly defective item is included. (An item is `possibly defective' if it appears in no negative tests; this definition will play a key role in the COMP and DD algorithms later.) Then the minimal set cover is exactly the smallest satisfying set.    The set cover problem is known to be NP-hard to solve \cite{Karp1972}, or even to verify that a given putative solution is optimal.

    \added{While the above-mentioned search time  $\sum_{k' \le k}\binom{n}{k'}$ may seem overly pessimistic, even a more careful analysis suggests high worst-case decoding time in general.  In particular, a significant reduction is attained by replacing $n$ by the number $n'$ of `possibly defective' items as mentioned above.  However, for rates exceeding a certain threshold in $\big[\frac{1}{2},\ln 2\big]$ depending on the test design, it can be shown we still have $n' = \omega(k)$ (e.g., see \cite{coja2020optimal,coja2022statistical,lovig2024mcmc}), meaning that even if $k$ were known, the number of possibilities $\binom{n'}{k}$ would still scale as $\exp(\omega(k))$. }
\end{remark}

Since SSS is in some sense the `best possible' (albeit possibly impractical) algorithm, we are interested in upper bounds on its rate.  In Section \ref{sec:algcon}, we make the term `best possible' more precise, and show that in fact these upper bounds are achievable in the information-theoretic sense. \rev{In fact, if we switch from the combinatorial prior to the i.i.d.~prior on the defective set (see the Appendix to Chapter \ref{ch:introduction}), then it can be shown that SSS is equivalent to {\em maximum a posteriori} (MAP) estimation of $\K$, and in this case its optimality is immediate.}

\begin{theorem} \label{SSSub}
Consider noiseless nonadaptive group testing with exact recovery and the small error criterion, with $k = \Theta(n^{\alpha})$ for some $\alpha \in (0,1)$, and using the SSS algorithm for decoding. The maximum rate achievable by SSS with a Bernoulli design is bounded above by
\begin{equation} \label{SSSBern}
  \olR^{\SSS}_{\Bern} \leq \max_{\nu > 0} \min \left\{ h( \ee^{-\nu}),\, \frac{\nu}{\ee^\nu \ln 2} \frac{1-\alpha}{\alpha} \right\} .
\end{equation}
Here and throughout, we write $h(x) = - x \log_2 x - (1-x) \log_2 (1-x)$ for the binary entropy function, measured in bits. 
\end{theorem}

\begin{proof}
The first term follows using an argument from \cite{aldridge}, and can be seen as a strengthening of the Fano-based counting bound \eqref{eq:psucweak} provided by Chan \etal\ \cite{chan-etal-1}. The idea is to observe that \cite{chan-etal-1} considers tests for which the probability of being negative could take any value, so the entropy is upper bounded by the maximum possible value, $H(Y_t) \leq h(1/2) = 1$. However, under Bernoulli testing with a general value of $p= \nu/k$, the probability that a test is negative is $(1- \nu/k)^k \simeq \ee^{-\nu}$, meaning that $H(Y_t) \simeq h(\ee^{-\nu})$. The full argument, which is justified by a typical set argument, is given in \cite[Lemma 1]{aldridge}.

We explain how the second term follows using a simplified version of the argument from \cite{aldridge-baldassini-johnson}. 
We say that a defective item is \emph{masked} if every time it is tested, it appears in a test with at least one other defective item.
We observe that if some $i \in \K$ is masked then SSS fails, since the set $\K \setminus \{ i \}$ forms a smaller satisfying set.
In other words, we know that
\begin{equation}
 \Perr \geq \PP \left( \bigcup_{i \in \K} \{ \mbox{$i$ masked} \} \right).\end{equation}
Hence, using the Bonferroni inequalities (see for example \cite[Chapter IV, eq.~(5.6)]{feller}), we can bound
\begin{align}
\Psuc & \leq 1- \PP \left( \bigcup_{i \in \K} \{ \mbox{$i$ masked} \} \right)  \notag \\
& \leq 1- \sum_{i \in \K} \PP \left(  \{ \mbox{$i$ masked} \} \right) + \sum_{i < j \in \K} \PP \left( \{ \mbox{$i$ and $j$ masked} \} \right) . \label{eq:genform}
\end{align}
Now, any particular defective $i$ is masked if all the tests it is included in also contain one or more other defective items.  Using a Bernoulli test design with item probability $p= \nu/k$, the probability that item $i$ appears in a particular test with no other defectives is $p (1-p)^{k-1}$, so the probability it is masked is $(1- p (1-p)^{k-1})^T$.  Similarly, the probability that  items $i$ and $j$ are both masked is $(1- 2 p(1-p)^{k-1})^T$, since we need to avoid the two  events `item $i$ and no other defective tested' and `item $j$ and no other defective tested', which are disjoint. 

Overall, then, in \eqref{eq:genform} we can deduce that 
\begin{align}
\Psuc & \leq 1 - k (1- p (1-p)^{k-1})^T + \frac{k^2}{2} (1- 2 p (1-p)^{k-1})^T  . \label{eq:SSSPsuc}
\end{align}
Since $p = \nu/k$, we can write $r = p(1-p)^{k-1} \sim \nu/(k \ee^{\nu})$. Taking
\begin{equation}
    T = \left\lceil \frac{(1-r) \ln k}{r} \right\rceil \label{eq:T_choice_SSS}
\end{equation}
and using the fact that 
\begin{align}
k (1- r)^T - \frac{k^2}{2} (1-2r)^T 
& \geq \left( k \ee^{-r T/(1-r)} \right) \left( 1 - \frac{k}{2} \ee^{-r T/(1-r)} \right)
\end{align}
(see \cite[eq. (41)]{aldridge-baldassini-johnson}), we can deduce that \eqref{eq:SSSPsuc} is bounded away from 1.  \rev{Note that proving this converse for the choice \eqref{eq:T_choice_SSS} also proves it for all larger values of $T$, since additional tests can only lead to fewer masked items.}

In Definition \ref{def:achievable}, using the binomial coefficient approximation \eqref{eq:bincoeffequiv}, this means that if the rate exceeds
\begin{equation} \frac{\log_2 \binom nk}{T} \asym \frac{(1-\alpha)}{\ln 2} \frac{k \ln n}{T} =
\frac{(1-\alpha) k \ln n}{\ln 2} \frac{r}{(1-r) \ln k} \sim \frac{\nu}{e^{\nu} \ln 2} \frac{1-\alpha}{ \alpha}, \end{equation}
then the success probability is bounded away from 1.
\end{proof}

\rev{In Chapter \ref{ch:achievability}, we will survey a result of Scarlett and Cevher \cite{scarlett-cevher-1} showing that the rate \eqref{SSSBern} is achievable for all $\alpha \in (0,1)$, and deduce that this is the maximum achievable rate for the Bernoulli design.}

\section{COMP: Combinatorial orthogonal matching pursuit} \label{sec:COMP}

The COMP algorithm was the first practical group testing algorithm shown to provably achieve a nonzero rate for all $\alpha < 1$.  The proof was given by Chan \etal~\cite{chan-etal-1,chan-etal-3}. 

COMP is based on the simple observation that any item in a negative test is definitely nondefective. COMP makes the assumption that the other items are defective.

\begin{algorithm} \label{def:comp}
The \defn{COMP} algorithm is defined as follows. We call any item in a negative test \defn{definitely nondefective} (\defn{DND}), and call the remaining items \defn{possibly defective} (\defn{PD}). Then the COMP algorithm outputs $\hat\K_\COMP$ equalling the set of possible defectives.
\end{algorithm}

The basic idea behind the COMP algorithm has appeared many times under many names -- the papers \cite{kautz,malyutov,chan-etal-1,chan-etal-3,chen10, luo} are just a few examples. The first appearance of the COMP idea that we are aware of is by Kautz and Singleton \cite{kautz}. 
We use the name `COMP' (for Combinatorial Orthogonal Matching Pursuit) following Chan \etal\ \cite{chan-etal-1}, who named it based on connections to a related compressed sensing algorithm.

\begin{example} \label{ex:comp}
Recall our worked example from Figure \ref{fig:sssexample}, previously discussed in Example \ref{ex:sss}.  We consider each the negative tests. Test 1 is negative, so items 1, 3, and 6 are definitely nondefective (DND). Test 3 is negative, so items 1 and 5 are definitely nondefective. Putting this together, we deduce that the remaining items 2, 4, 7 are the possible defective (PD) items, so we choose to mark them as defective. In other words, $\hat\K_\COMP = \{2, 4, 7 \}$.
\end{example}

The following lemma shows that COMP can be interpreted as finding the largest satisfying set, in stark contrast with SSS.  Here and subsequently, we write $\K^\comp = \{1,\dotsc,n\} \setminus \K$ for the items not in $\K$.

\begin{lemma}  \label{lem:COMPsat}
The estimate $\hat\K_\COMP$ generated by the COMP algorithm is a satisfying set (in the sense of Definition \ref{def:sat}) and contains no false negatives. Every satisfying set  is a subset of $\hat\K_\COMP$, so $\hat\K_\COMP$ is the unique largest satisfying set. \end{lemma}
\begin{proof} Observe that since every DND item appears in a negative test, and so must indeed be nondefective, the COMP algorithm outputs no false defectives, and the true defective set satisfies $\K \subseteq \hat\K_\COMP$. Furthermore, since every positive test contains an element of $\K$, and hence of $\hat\K_\COMP$, we deduce that $\hat\K_\COMP$ is a satisfying set.

Fix a satisfying set $\L$ and consider item $i \not\in \hat\K_\COMP$. By construction $i$ must be a DND, meaning that it appears in a negative test and therefore (see the second part of Definition \ref{def:sat}) cannot be in $\L$. In other words $\hat\K_\COMP^\comp \subseteq \L^\comp$, or reversing the inclusion, $\L \subseteq \hat\K_\COMP$. 
\end{proof}


In the remainder of this section, we will establish the rate achievable by the COMP algorithm with a Bernoulli design. These results are due to \cite{chan-etal-1} and \cite{aldridge}. We should expect that COMP is suboptimal, since it does not make use of the positive tests. 
\begin{theorem} \label{COMPub1}
Consider noiseless nonadaptive group testing with exact recovery and the small error criterion, with $k = \Theta(n^{\alpha})$ for some $\alpha \in (0,1)$, and using the COMP algorithm for decoding. With a Bernoulli design and an optimized parameter $p = 1/k$, the following rate is achievable:
\begin{equation} \label{COMPBern}
 R^{\COMP}_{\Bern} = \frac{1}{\ee \ln 2} (1-\alpha) \approx 0.531 (1-\alpha).
\end{equation}
\added{This rate corresponds to a number of tests satisfying the following for arbitrarily small $\eta > 0$: $T \ge (1+\eta) \ee k \ln n$.}
\end{theorem}
\begin{proof}
For any given nondefective item,  the probability that it appears in a particular test, and that such a test is negative, is $p(1-p)^k$. This follows from the independence assumption in the Bernoulli design; the test is negative with probability $(1-p)^k$, and the given item appears with probability $p$. Hence, the probability that this given nondefective appears in no negative tests is $(1- p(1-p)^k)^T$.

The COMP algorithm succeeds precisely when every nondefective item appears in a negative test, so the union bound gives
\begin{align}
\Perr & = \PP \left( \bigcup_{i \in \K^\comp} \{\mbox{item $i$ does not appear in a negative test} \} \right) 
\nonumber \\
& \leq | \K^\comp | \left( 1- p(1-p)^k \right)^T \nonumber \\
& \leq n \exp( - T p (1-p)^k). \label{eq:COMPprob}
\end{align}
The expression $p(1-p)^k$ is maximized at $p = 1/(k+1) \asym 1/k$, so we take $p = 1/k$ (or equivalently $\nu =1$), meaning that $(1-1/k)^k \to \ee^{-1}$. Hence, taking $T = (1+ \delta) \ee k \ln n$ means that $T p (1-p)^k \sim (1+ \delta) \ln n$.

Using the binomial coefficient approximation \eqref{eq:bincoeffequiv}, we have the asymptotic expression
\[  \frac{\log_2 \binom nk}{T} \asym \frac{(1-\alpha)}{\ln 2} \frac{k \ln n}{T} . \]
Then following Definition \ref{def:achievable}, taking $(1+ \delta) \ee k \ln n$  gives a rate arbitrarily close to  $ (1-\alpha)/(\ee \ln 2)$,
as desired.
\end{proof}

\begin{remark} \label{rmk:tightness}
    Using a similar but slightly more involved argument, one can show that the expression $R^{\COMP}_{\Bern}$ from \eqref{COMPBern} gives the maximum achievable rate when COMP is used in conjunction with Bernoulli testing.  The argument is outlined as follows for a general parameter $p = \nu/k$.
    
    First, we use binomial concentration to show that the number of negative tests $T_0$ is tightly concentrated around its mean, yielding $T_0 \simeq T \ee^{-\nu}$.  Conditioned on $T_0$, the probability of a given nondefective failing to be in any negative test is
    \[ q := (1-p)^{T_0} \simeq \exp( -p T_0 ) \simeq \exp\left( -\frac{\nu}{k}  T e^{-\nu}  \right).  \]
    The error events are (conditionally) independent for different nondefective items, so the total error probability is $1 - (1-q)^{n-k}$.  Substituting $q \simeq \exp( -T\nu e^{-\nu} /k )$, applying some simple manipulations, and noticing that $\nu \ee^{-\nu}$ is maximized at $\nu = 1$, we find that
    \begin{gather*}  
        T > (1 + \eta) \ee k \ln n  \quad \implies \quad \Perr \to 0 \\
        T < (1 - \eta) \ee k \ln n  \quad \implies \quad \Perr \to 1,
    \end{gather*}
    for arbitrarily small $\eta > 0$.  This matches the choice of $T$ in the proof of Theorem \ref{COMPub1} above.  In fact, this argument not only shows that \eqref{COMPBern} the highest achievable rate, but also that the error probability tends to one when this rate is exceeded.
    
    The preceding argument essentially views COMP as a coupon-collecting algorithm, gradually building up a list of nondefectives using negative tests.  We say that a nondefective item is `collected' if it appears in a negative test.  A result dating back to Laplace (see also \cite{erdos-renyi2}) states that in order to collect all coupons in a set of size $m$, it suffices to have $m \ln m$ trials. Here, we need to collect $m = n-k \asym n$ coupons, and each of the $\ee^{-1}T$ negative tests (assuming $\nu = 1$) contains on average $pn = n/k$ such coupons. Thus, we require $\frac{n}{k} \ee^{-1}T \asym n \ln n$, which rearranges to $T \asym \ee k \ln n$.  
\end{remark}


\begin{remark} \label{rem:misspecify}
The analysis above also allows us to consider the extent to which COMP and other algorithms require us to know the exact number of defectives $k$. While, for a given test matrix, the decisions taken by COMP do not require this knowledge, clearly to form a Bernoulli test matrix with $p = 1/k$ requires the value of $k$ itself.

However, COMP is reasonably robust to misspecification of $k$, in the following sense. Suppose that for a fixed $c$ we base our matrix on the erroneous assumption that there are $\widehat{k} = k/c$ defectives, and hence use $p= 1/\widehat{k} = c/k$ instead. Repeating the analysis above, we can see that \eqref{eq:COMPprob} behaves like $n \exp( - T c \ee^{-c}/k)$, so we should use $T = (1+\delta) k \ln n/(c \ee^{-c})$ tests, corresponding to a rate of $(1-\alpha) c \ee^{-c}/\ln 2$.

In other words, even with a multiplicative error in our estimate of $k$, COMP will achieve a nonzero rate, and if the multiplicative factor $c$ is close to 1, COMP will achieve a rate close to that given in Theorem \ref{COMPub1} above. Although the analysis would be more involved, we anticipate that similar results should hold for other algorithms. We further discuss the question of uncertainty in the number of defectives, including how group testing can provide estimates of $k$ which are accurate up to a multiplicative factor, in Section \ref{sec:universal}.
\end{remark}

\section{DD: Definite defectives} \label{sec:DD}

The DD (Definite Defectives) algorithm, due to Aldridge, Baldassini and Johnson \cite{aldridge-baldassini-johnson}, was the first practical group testing algorithm to provably achieve the optimal rate for Bernoulli designs for a range of values of $\alpha$ (namely, $\alpha \ge \frac{1}{2}$).

Recall the definitions of `definite nondefective' and `possible defective' from Definition \ref{def:comp}. DD is based on the observation that if a (necessarily positive) test contains exactly one possible defective, then that item is in fact definitely defective.

\begin{algorithm} \label{def:dd}
The \defn{definite defectives} (\defn{DD}) algorithm is defined as follows. 
\begin{enumerate}
\item We say that any item in a negative test is \defn{definitely nondefective} (\defn{DND}), and that any remaining item is a \defn{possible defective} (\defn{PD}). 
\item If any PD item is the only PD item in a positive test, we call that item \defn{definitely defective} (DD). 
\item The DD algorithm outputs $\hat\K_\DD$, the set of definitely defective items.
\end{enumerate}
\end{algorithm}
One justification for DD is the observation that removing nondefective items from a test does not affect the outcome of the test, so the problem is the same as if we use the submatrix with columns in PD.  In addition, the principle `assume nondefective unless proved otherwise' (used by DD) should be preferable to the rule `assume defective unless proved otherwise' (used by COMP) under the assumption that defectivity is rare.
We illustrate this by continuing Example \ref{ex:comp}.
\begin{example} \label{ex:dd}
We present a worked example of DD. From Example \ref{ex:comp}, we know that items 2, 4, and 7 are the possible defectives (PD). Now consider the submatrix with the corresponding columns, illustrated in Figure \ref{fig:ddexample}. Notice that tests 4 and 5 are positive, and only contain one (possible defective) item, so we can deduce that items
$2$ and $4$ must be defective. The defectivity status of item 7 is still unclear, but the DD algorithm marks it as nondefective.
Thus, we have $\hat\K_\DD = \{2, 4 \}$.
\end{example}

\begin{figure}
\begin{center}
\begin{tabular}{ccccccc|c}
\nondefman & \quest & \nondefman & \quest & \nondefman & \nondefman & \quest & \yspace \\
\hline
  & $\zero$ &  & $\zero$  &  &  & $\zero$ & $\zero$ \\
   & $\one$ &  & $\one$  &  &  & $\one$  & $\one$ \\
  & $\zero$ &  & $\zero$  &  &  &  $\zero$ & $\zero$ \\
  & $\one$ &  & $\zero$  &  & & $\zero$  & $\one$ \\
 & $\zero$ &  & $\one$  &  &  &  $\zero$ & $\one$ \\
\end{tabular}
\caption{Example of the DD algorithm.  We describe the inferences that we make in Example \ref{ex:dd}, but give the submatrix of PD columns only, having marked the DND items discovered in Example \ref{ex:comp} by COMP as nondefective (replacing question marks by outlined figures to represent our knowledge). \label{fig:ddexample}}
\end{center}
\end{figure}
\begin{lemma} The estimate $\hat\K_\DD$ generated by the DD algorithm has no false positives.
\end{lemma}
\begin{proof} Since (as in COMP), all DND items are indeed definitely nondefective, the first stage of DD makes no mistakes. Furthermore, we know that each DD item is indeed defective, so the second stage of inference in DD is also certainly correct.  Hence, DD can only makes an error in the final step, by marking a defective item as nondefective. In other words, DD has no false positives, and $\hat\K_\DD \subseteq \K$.  \end{proof}

\begin{remark} \label{rem:ddnotss}
Unlike COMP, the DD algorithm does not necessarily produce an estimate which is a satisfying set. Figure \ref{fig:ddexample2} contains a simple example that illustrates this; since no test is negative, all items are marked as PDs, but no test contains a single item, so $\hat\K_{\DD} = \emptyset$. 

\begin{figure}[t]
\begin{center}
\begin{tabular}{ccc|c}
\quest & \quest & \quest & \yspace \\
\hline
$\one$ & $\zero$ & $\one$ & $\one$ \\
$\zero$ & $\one$ & $\one$ & $\one$ \\
$\one$ & $\one$ & $\zero$ & $\one$ \\
\end{tabular}
\caption{Example of the DD algorithm finding a set which is not satisfying, to illustrate Remark \ref{rem:ddnotss}. In this case
$\hat\K_{\DD} = \varnothing$. \label{fig:ddexample2}}
\end{center}
\end{figure}

However, when DD does produce a satisfying set, it must necessarily be the smallest satisfying set -- that is, $\hat\K_{\DD} = \hat\K_{\SSS}$. We prove this  by contradiction as follows:  Assume that $\hat\K_{\DD}$ is a satisfying set, but not the smallest one.  Then $\hat\K_{\DD}$ must have more elements than $\hat\K_{\SSS}$, meaning there exists some item $i \in \hat\K_{\DD} \cap (\hat\K_{\SSS})^\comp$.  By the definition of the DD algorithm, this item $i$ appears in some non-empty collection of positive tests, say indexed by $\{ t_1, \ldots, t_m \} \subseteq \{1,\dotsc,T\}$, with no other element of $\hat\K_{\DD}$. However, each such test $t_j$ must also contain some item $i_j \in \hat\K_{\SSS}$, because $\hat\K_{\SSS}$ is satisfying.  On the other hand, by definition, no element in $\hat\K_{\DD} \cup \hat\K_{\SSS}$ appears in any negative tests, so $\{ i, i_1, \ldots i_m \}$ would all be counted as PD by Stage 1 of the DD algorithm. As a result, item $i$ would never appear as a lone PD item, and so would never be marked as DD, giving a contradiction.
\end{remark}

We now discuss how to calculate the rate of DD under a Bernoulli test design.  It turns out that DD outperforms the COMP rate given in Theorem \ref{COMPub1} above, as is immediately deduced from the following result due to \cite{aldridge-baldassini-johnson}.

\begin{theorem} \label{DDub}
Consider noiseless nonadaptive group testing with exact recovery and the small error criterion, with $k = \Theta(n^{\alpha})$ for some $\alpha \in (0,1)$, and using the DD algorithm for decoding.  Under a Bernoulli design and an optimized parameter $p = \frac{1}{k}$, the following rate is achievable:
\begin{equation} \label{DDBern}
  R^{\DD}_{\Bern} = \frac{1}{\ee \ln 2} \min \left\{ 1, \frac{1-\alpha}{\alpha} \right\} \approx 0.531 \min \left\{ 1, \frac{1-\alpha}{\alpha} \right\} .
\end{equation}
\added{This rate corresponds to a number of tests satisfying the following for arbitrarily small $\eta > 0$: $T \ge (1+\eta) \max\big\{\ee k \ln \frac{n}{k}, \ee k \ln k\big\}$.}

Moreover, for $\alpha \geq 1/2$, this matches the maximum achievable rate for SSS given in Theorem \ref{SSSub}.
\end{theorem}

Note that in \eqref{DDBern}, the first minimand dominates for $\alpha \leq 1/2$ and the second minimand dominates for $\alpha \geq 1/2$.

\added{The original proof in \cite{aldridge-baldassini-johnson} is rather technical, so we instead outline a proof that can be inferred from the more recent work of Scarlett and Johnson \cite{scarlett-johnson} (who focused on noisy settings), several ingredients of which came from \cite{aldridge-baldassini-johnson}.  The intuition behind the two terms in \eqref{DDBern} is fairly straightforward:
\begin{enumerate}
    \item Call a nondefective item \emph{intruding} if it appears only in positive tests, so cannot be ruled out as definitely nondefective. We first want the number of `intruding' nondefective items to be small; specifically, to scale like $o(k)$. This ensures that almost all of the PD items, $1 - o(1)$ of them, really are defective. This can be achieved for rates below $\frac{1}{\ee \ln 2}$, which is the first term in \eqref{DDBern}. (In comparison, COMP requires the number of intruding nondefectives to not merely be small, but to be $0$; this requires an even lower rate.)
    \item Second, we want every defective item to be the only PD in at least one test, which will allow it to be confirmed as definitely defective. When the $o(k)$ scaling from the first point holds and the rate is below $\frac{1}{\ee \ln 2}\frac{1-\alpha}{\alpha}$, then this is achieved, with high probability. This is the second term in term in \eqref{DDBern}. (This threshold is exactly that appearing in the analysis of SSS in Section \ref{sec:sss}; this is because the masking event therein is very similar to the event `failure to be a unique PD in any test' when the number of PD items is $k+o(k)$.)
\end{enumerate}
We proceed to outline how this intuition can be made rigorous, though we still skip some details and refer the reader to \cite{aldridge-baldassini-johnson,scarlett-johnson} for the full details.

Studying the nondefective items is relatively straightforward, and is similar to the analysis of COMP.  A simple calculation shows that each test outcome satisfies $\PP(Y = 0) \sim \ee^{-\nu}$, so the number of negative tests $M_0$ satisfies $\EE M_0 \sim T \ee^{-\nu}$, and a standard binomial concentration bound shows that $M_0 \sim T \ee^{-\nu}$ with high probability. Conditioned on the number of negative tests $M_0 = m_0$ for a value $m_0 \sim T \ee^{-\nu}$ satisfying this high-probability behaviour, we have for any nondefective item $i \notin \K$ that
\begin{equation*}
    \PP(i \in {\rm PD} \mid M_0 = m_0) = \Big(1 - \frac{\nu}{k}\Big)^{m_0} \sim \ee^{-T \nu \ee^{-\nu}/k},
\end{equation*}
and hence, the number $G$ of intruding nondefective items satisfies
\begin{equation}
    \EE(G \mid M_0 = m_0) \sim (n-k) \,\ee^{-T \nu \ee^{-\nu}/k}. \label{eq:G_avg}
\end{equation}
A simple calculation shows that this scales as $o(k)$ for rates below $\frac{\nu \ee^{-\nu}}{\ln 2}$, the optimal value of which is $\frac{1}{\ee \ln 2}$ when $\nu = 1$.  Hence, by Markov's inequality, we have $G = o(k)$ with probability tending to one.

Studying the defective items is somewhat more complicated, partly because conditioning on $G$ is difficult given that $G$ is not independent of the placements of defective items.  Before attempting such conditioning, we consider a fixed defective item $i \in \K$ and an associated triplet $(M_0, M_i, M_{\rm other})$ corresponding to a useful partitioning of the $T$ tests: $M_0$ is the number of negative tests, $M_i$ is the number of tests containing item $i$ but no other defectives, and $M_{\rm other} = T - M_0 - M_i$ counts the remaining tests.  Due to the i.i.d.~nature of the Bernoulli design, this triplet has a multinomial distribution:
\[ (M_0, M_i, M_{\rm other}) \sim \Mn(T; p_0,p_i,p_{\rm other}), \]
where $p_0 \sim \ee^{-\nu}$, $p_i \sim \frac{\nu \ee^{-\nu}}{k}$, and $p_{\rm other} = 1-p_0-p_i$.  These parameters come from the above calculation for $M_0$ and a similar calculation for $M_i$

We now consider conditioning on a fixed pair $(G,M_0) = (g,m_0)$ satisfying $m_0 \sim T \ee^{-\nu}$ and $g = o(k)$; recall that we showed above that these properties hold with high probability.  Under such conditioning, we consider `splitting' the $M_i$ tests in the second category into a further pair $(L_i, M_i-L_i)$, where $L_i$ counts the tests in which $i$ is the only PD item, and $M_i-L_i$ counts the tests containing both $i$ and a nondefective item from PD.  Note that the DD algorithm succeeds if and only if $L_i > 0$ for all $i \in \K$.

While the effect of the above-mentioned conditioning on $(G,M_0)$ may appear to be complicated, a simple analysis of the conditional probability can be performed (see \cite[Lemma 2]{scarlett-johnson}, which is related to \cite[Lemma C.2]{aldridge-baldassini-johnson}) to deduce that 
\begin{multline*}
  (L_i, M_i - L_i, M_{\rm other}) \mid \{G = g, M_0 = m_0\} \\
  \sim \Mn\Big( T- m_0; \frac{p'_i}{1 - p_0}, \frac{p''_i}{1 - p_0}, \frac{p_{\rm other}}{1-p_0} \Big),
\end{multline*}
where $p'_i = p_i (1-\psi)$ and $p''_i = \psi$ with $\psi$ being the conditional probability that at least one nondefective item from PD is included in a given test.  A simple calculation shows that the condition $g = o(k)$ implies $\psi = o(1)$, which in turn implies $p'_i \sim p_i$.  Combining this with $p_0 \sim \ee^{-\nu}$, $p_i \sim \frac{\nu \ee^{-\nu}}{k}$, and $m_0 \sim T \ee^{-\nu}$, it follows that the conditional distribution of $L_i$ is roughly binomial with $T(1-\ee^{-\nu})$ trials and probability parameter $\frac{\nu \ee^{-\nu}/k}{1-\ee^{-\nu}}$.  This implies that the event $L_i = 0$ has probability roughly 
$$ \Big(1 - \frac{\nu \ee^{-\nu} / k}{1-\ee^{-\nu}}\Big)^{T(1-\ee^{-\nu})} \sim \exp\Big( - \frac{T}{k} \nu \ee^{-\nu} \Big).$$
Hence, by a union bound over the $k$ defectives, the conditional probability of any defective item satisfying $L_i = 0$ (thus causing DD to fail) is roughly
\[ \PP(\text{error} \mid G = g, M_0 = m_0) \sim k \exp\Big( - \frac{T}{k} \nu \ee^{-\nu} \Big). \]
A simple calculation shows that this approaches $0$ as $n \to \infty$ for rates below
\[ \frac{\nu \ee^{-\nu}}{\ln 2}\frac{1-\alpha}{\alpha} = \frac{1}{\ee \ln 2}\frac{1-\alpha}{\alpha} , \]
where the right-hand side has taken the optimal $\nu = 1$. This completes the argument.
}

\begin{remark} We conclude this section with the following  remarks.
\begin{enumerate}
    \item As shown in \cite{scarlett-johnson}, Theorem \ref{DDub} provides the best possible achievable rate for DD with Bernoulli testing, so any further improvements require changing either the detection algorithm or the test design, not just the analysis.  For $\alpha \ge 1/2$, such a claim is trivially deduced from an observation that no algorithm can do better under Bernoulli testing.  For $\alpha < 1/2$, taking $C = \max \{ \alpha, 1 - \alpha \} - \epsilon = 1-\alpha - \epsilon$ in \eqref{eq:G_avg} gives $\EE(G) \gg k$, which leads to a large number of true defectives getting `drowned out' by the nondefectives that are marked as possible defectives.
    \item \added{With the exception of SSS, the DD algorithm achieves the highest rate under the Bernoulli test design among the simple algorithms surveyed in this chapter.  This will also be the case under the improved near-constant tests-per-item design in Section \ref{sec:near_constant}.   For the Bernoulli design, we note that an algorithm called \emph{separate decoding of items} achieves slightly higher rates when $\theta$ is very small, in particular approaching $\ln 2 \simeq 0.693$ as $\theta \to 0$.  Since this algorithm is primarily targeted at noisy settings, we defer its description to Section \ref{sec:separate}.  }
\end{enumerate}
\end{remark}

\section{SCOMP: Sequential COMP} \label{sec:SCOMP}

SCOMP is an algorithm due to Aldridge, Baldassini, and Johnson \cite{aldridge-baldassini-johnson} that builds a satisfying set by starting from the set of definite defectives (DD) and sequentially adding new items until a satisfying set is reached. The name comes from `Sequential COMP', as it can be viewed as a sequential version of the COMP algorithm.

\begin{algorithm} \label{alg:scomp}
The \defn{SCOMP} algorithm is defined as follows.
\begin{enumerate}
\item Initialize $\hat{\K}$ as the estimate $\hat\K_{\DD}$ produced by the DD algorithm (Algorithm \ref{def:dd}), and declare any definitely nondefective items (items appearing in a negative test) to be nondefective. The other possible defectives are not yet declared either way.
  \item Any positive test is called \emph{unexplained} if it does not contain any items from $\hat\K$. Add to $\hat\K$ the possible defective not in $\hat\K$ that appears in the most unexplained tests, and mark the corresponding tests as no longer unexplained. (Ties may be broken arbitrarily.)
  \item Repeat step 2 until no tests remain unexplained. The estimate of the defective set is $\hat\K$.
\end{enumerate}
\end{algorithm}

Note that a satisfying set leaves no unexplained tests, and any set containing no definite nondefectives and leaving no unexplained tests is satisfying. Note also that the set of all possible defectives is satisfying, so the SCOMP algorithm does indeed terminate.

The following result \cite{aldridge-2} is relatively straightforward to prove.   

\begin{theorem} \label{SCOMPub}
For any given test design, any rate achievable by DD is also achievable by SCOMP. In particular, 
with a Bernoulli design and optimal choice of the parameter $p$, SCOMP can achieve the rate given in \eqref{DDub} above.
Moreover, for $\alpha \geq 1/2$, this matches the best achievable
rate obtained using SSS (Theorem \ref{SSSub}).
\end{theorem}

\begin{proof}
The simple idea is that for each particular test design $\mat{X}$ and defective set $\K$, whenever DD succeeds, SCOMP also succeeds. More specifically, if $\hat\K_{\DD}  = \K$, then the initial choice $\hat\K = \K$ in step 1 of SCOMP (Algorithm \ref{alg:scomp}) is already a satisfying set, so there are no unexplained tests to consider in step 2, and the algorithm terminates. \end{proof}

\added{It is natural to ask} whether SCOMP has a larger achievable rate than DD for Bernoulli testing.  We note from simulations such as Figure \ref{fig:allfive} that SCOMP appears to perform strictly better than DD experimentally.  \added{However, small experimental improvements at fixed problem sizes may not necessarily translate to strictly improved achievable rates as $n \to \infty$.}  In fact, as described in Section \ref{sec:near_constant} below, Coja-Oghlan \etal\, \cite{coja} have proved that SCOMP provides no such improvement in the rate for the near-constant column weight design.

There is an analogy between SCOMP and Chvatal's approximation algorithm for the set cover problem (see Remark \ref{rem:setcover}). At each stage, Chvatal's algorithm \cite{chvatal} greedily chooses a subset that covers the largest number of currently uncovered elements in the universe of elements. 
Similarly, SCOMP makes a greedy choice of possibly defective items that explain as many currently unexplained positive tests as possible.
For a universe of $|U|=m$ items, Chvatal's algorithm produces a solution that is at most $H_m$ times larger than the optimal set cover, where $H_m \asym \ln m$ is the $m$-th harmonic number. This can be shown to be the best possible approximation factor for a polynomial-time algorithm for set cover (in the worst case) \cite[Theorem 29.31]{vazirani}.  This means that for certain test matrices, we can view SCOMP as outputting the `tightest possible polynomial-time approximation to the smallest satisfying set'. {\bf }  However, this does not preclude the possibility of improved efficient approximations under other well-chosen test matrices.

\section{Linear programming relaxations} \label{sec:lp}

\added{We saw in Section \ref{sec:sss} that SSS can be cast as an integer program (IP).  In discrete optimization, a prominent approach to solving an IP is to solve its {\em linear programming (LP) relaxation}, in which the integer variables are replaced by real-valued random variables to produce an LP.  The LP can then be solved in polynomial time, and its solution can be rounded to be integer-valued; one hopes that this will solve (exactly or approximately) the original IP.}

\added{An LP relaxation of SSS was proposed by Malioutov and Malyutov \cite{malioutov-malyutov}.}
Recalling that $x_{ti}$ indicates whether item $i$ is in test $t$, and $y_t$ is the outcome of test $t$, a smallest satisfying set corresponds to an optimal solution to the integer program 
  \begin{align*} \text{minimize}_{\mathbf{z}} \ &\sum_{i=1}^n z_i \\
    \qquad \text{subject to } \  &\sum_{i=1}^n x_{ti} z_i \geq 1 \quad \text{when } y_t = \one, \\
                                   &\sum_{i=1}^n x_{ti} z_i = 0 \quad \text{when } y_t = \zero, \\
                                   &z_i \in \{0,1\} . \end{align*}
We hope that the optimal $\vec{z}$ will be close to the true defectivity vector $\vc{u}$ introduced in Definition \ref{def:udef}, since taking $\vc{z} = \vc{u}$ will satisfy the constraints.
In general, we think of each $0$--$1$ vector $\vec{z}$ as the indicator function of some putative defective set $\L$, with $\L = \L(\vec{z}) := \{ i: z_i = 1 \}$. The first two constraints on $\vec{z}$ ensure that $\L(\vec{z})$ is satisfying in the sense of Definition \ref{def:sat}, by considering the positive and negative tests respectively. Hence, each $\vec{z}$ that achieves the minimal value of the linear program is the indicator function of a satisfying set of minimal size, i.e., $\hat\K =  \{i : z_i = 1 \}$ is a smallest satisfying set.    

The LP approach attempts to estimate the defective set via a relaxed version of the $0$--$1$ problem, where each $z_i$ can be any nonnegative real number.  That is, the optimization formulation is exactly as above, but with each constraint $z_i \in \{0,1\}$ replaced by $z_i \geq 0$. \added{(It is unnecessary to add the extra constraints $z_i \leq 1$, as it is straightforward to check that an optimal solution will never contain any entries exceeding 1.)}

Linear programs of this form can be solved efficiently; for example, the ellipsoid algorithm is guaranteed to find a solution in polynomial time, though it is typically outperformed in practice by the simplex algorithm. (See, for example, \cite[p.~897]{cormen} for a discussion of the running times of linear programming algorithms.) 

There are various heuristics for how to turn an optimal solution $\vec z = (z_i)$ to the relaxed program into an estimate of the defective set. 
For example, one could consider the following crude method: If there is any $i$ with $z_i \notin \{0, 1\}$, declare a global error; otherwise, estimate $\hat{\mathcal K} = \{i : z_i = 1 \}$ to be the defective set.  Malioutov and Malyutov \cite{malioutov-malyutov} suggest an estimate $\hat{\mathcal K} = \{i : z_i > 0 \}$, and show strong performance on simulated problems. Note that this rule will always provide a satisfying set, since each positive test will have some possible defective $i$ with $z_i$ that is declared defective.
Alternatively, the estimate $\hat{\mathcal K} = \{i : z_i \geq 1/2 \}$ appears to be (very) slightly better in simulations, but does not guarantee a satisfying set.

For the purposes of the following theorem, it suffices that in the event that all $z_i$ are $0$ or $1$, the heuristic chooses $\hat{\mathcal K} = \{i : z_i = 1 \}$, as any sensible heuristic surely must.
This theorem is due to \cite{aldridge-2}, and shows that the above LP approach, like SCOMP, is at least as good as DD.

\begin{theorem} \label{LPub}
For any given test design, any rate achievable by DD is also achievable by LP. In particular, 
with a Bernoulli design and optimal choice of the parameter $p$, LP can achieve the rate given in \eqref{DDub} above.
Moreover, for $\alpha \geq 1/2$, this matches the best achievable
rate obtained using SSS (Theorem \ref{SSSub}).
\end{theorem}

\begin{proof} We argue that for any given test design $\mat{X}$ and defective set $\K$, whenever DD succeeds LP also succeeds. To be precise, any item $i$ that appears in some negative test $t$  must have $z_i$ = 0 in order  to satisfy the second constraint of the linear program, $\sum_{i=1}^n x_{ti} z_i = 0$. Furthermore, if a positive test $t$ contains only one possible  defective $i$, the LP solution must have $z_i \geq 1$ to ensure the first constraint $\sum_{i=1}^n x_{ti} z_i \geq 1$ holds, and it will choose $z_i$ = 1 to minimize $\sum_i z_i$. Finally, if DD succeeds then these definite defectives form a satisfying set, so all constraints are satisfied, and the algorithm will set all other $z_i = 0$, to minimize $\sum_i z_i$.  
\end{proof}

As with SCOMP, while simulation evidence suggests that LP often outperforms DD (e.g., see Figure \ref{fig:allfive}), it is unclear whether one can formally establish that LP attains a strictly larger rate than that of DD for $\alpha < 1/2$. 

We briefly mention an earlier result establishing that a (different) linear program attains a nonzero rate for all $\alpha$, albeit a much lower rate than that of Theorem \ref{LPub}.  Specifically, the \defn{LiPo} algorithm of Chan \etal\ \cite{chan-etal-3} is based on relaxing a similar integer program to the one we surveyed, and further assumes the decoder knows $k$ exactly, so the linear program can be phrased as a feasibility problem. They showed that LiPo achieves the rate 
\begin{equation} R^{\mathrm{LiPo}}_{\mathrm{Bern}} = \frac{1}{\frac83 \, \text{e}^2 \ln 2} \, \frac{1-\alpha}{1+\alpha} \approx 0.073 \, \frac{1-\alpha}{1+\alpha} . \end{equation}

\section{Improved rates with near-constant tests-per-item} \label{sec:near_constant}

Throughout this chapter, we have focused on Bernoulli testing designs, where each item is independently placed in each test with a given probability, and hence $\mat X$ contains i.i.d.~Bernoulli entries.  Such a design is conceptually simple, is typically the easiest to analyse mathematically, and is known to be information-theoretically optimal for $k = O(n^{1/3})$ (see Chapter \ref{ch:achievability}).

However, it turns out that the Bernoulli design is typically not the best choice.  Below, we will see that an alternative random design based on \emph{near-constant tests-per-item} can improve the COMP and DD rates by a factor of $\mathrm{e}(\ln 2)^2 \simeq 1.306$, leading to two key implications. First, for $k = \Theta(n^{\alpha})$ with $\alpha$ sufficiently close to one, this combination outperforms Bernoulli testing used in conjunction with with {\em any} decoder. Second, for small $\alpha$, this combination improves on the best known rate for Bernoulli testing under any {\em practical} decoder.  \added{In Chapter \ref{ch:achievability}, we will see even stronger results establishing that the near-constant tests-per-item design, paired with an optimal decoder, attains the optimal rate among all nonadaptive designs.} The results of this section are due to Johnson, Aldridge, and Scarlett \cite{johnson-aldridge-scarlett} and Coja-Oghlan \etal \cite{coja}. 

The following definition formally introduces the random design that provides the above-mentioned improved rates.

\begin{definition} \label{def:const_col}
	The {\em near-constant column weight} (or {\em near-constant tests-per-item}) design with parameter $\nu > 0$ forms a group testing matrix $\mat X$ in which $L = \nu T/k$ entries of each column are selected uniformly at random with replacement and set to one, with independence between columns.  The remaining entries of $\mat{X}$ are set to zero.
  \end{definition}

(When we write $L = \nu T/k$, we ignore rounding issues, and noting that the results are unaffected by whether we set $L = \lfloor \nu T/k \rfloor$ or $L = \lceil \nu T/k \rceil$)

Since we sample with replacement, some items may be in fewer than $L$ tests, but typically only slightly fewer, hence the terminology `near-constant'. This is a mathematical convenience that turns out to simplify the analysis.  \rev{The parametrization $L = \nu T/k$ is chosen because such scaling with $\nu = \Theta(1)$ turns out to be optimal, analogously to the scaling $p = \nu/k$ in Definition \ref{def:berndesign}. (In Section \ref{sec:constrained}, we will briefly survey a setting in which the number of tests per item is constrained to be much smaller than $O(T/k)$.) An intuitive reason as to why the above design may be preferable to the Bernoulli design is that it prevents any item from being included in too few tests.}

While Definition \ref{def:const_col} suffices for our purposes, it is worth mentioning that it is one of a variety of related randomized designs that have appeared in the literature.  Indeed, a variety of works have considered {\em (exactly) constant tests-per-item} (see for example \cite{macula,mezard}).  \added{We refer the reader to \cite{johnson-aldridge-scarlett} for a more detailed overview of the history, dating back at least as far as the early work of Kautz and Singleton \cite{kautz}.
Other works have also considered constant items-per-test (row weight) designs (for example, \cite{chan-etal-1}) and doubly-regular designs with both constant row weight and constant column weight (for example, \cite{Tan2022}).  Such designs are not needed in this chapter, but they can provide additional benefits in other contexts; see Sections \ref{sec:linear} and \ref{sec:sparse} for examples.
}

We first present the rate achieved by  the simple COMP algorithm (see Section \ref{sec:COMP}).

\begin{theorem} \label{COMPub2}
Consider noiseless nonadaptive group testing with exact recovery and the small error criterion, with $k = \Theta(n^{\alpha})$ for some $\alpha \in (0,1)$, and using the COMP algorithm for decoding. With a near-constant column weight design and an optimized parameter $\nu = \ln 2$, the maximum achievable rate using COMP is
\begin{equation} \label{COMPNCC}
  \olR^{\COMP}_{\mathrm{NCC}} = \ln 2 \cdot (1-\alpha) \approx 0.693 (1-\alpha) .
\end{equation}
\added{This rate corresponds to a number of tests satisfying the following for arbitrarily small $\eta > 0$: $T \ge (1+\eta) \frac{1}{(\ln 2)^2} k \ln n = (1+\eta) \frac{1}{\ln 2} k \log_2 n$.}
\end{theorem}
\begin{proof}[Proof sketch] 
We omit the details of the proof for brevity, and refer the reader to \cite{johnson-aldridge-scarlett}.  The key idea is again to formulate a coupon-collector problem (see Remark \ref{rmk:tightness}).

First, we consider the total number of positive tests $T_1$. A given test $t$ is negative if, for each of the $k$ defective items, none of the $L = \nu T/k$ choices of column entries equal $t$. Since these choices take place independently with replacement, this is the same as choosing $kL = \nu T$ entries in total, all independently with replacement. Hence, the probability that test $t$ is negative is $(1- 1/T)^{\nu T} \sim \ee^{-\nu}$, so the expected number of positive tests $\EE T_1 \sim T (1-\ee^{-\nu})$.

Now, since changing one choice of column entry changes the number of positive tests by at most 1, the random variable $T_1$ satisfies the bounded difference property in the sense of McDiarmid \cite{mcdiarmid}, which allows us to prove a standard concentration bound. Specifically we can deduce by McDiarmid's inequality that $T_1$ is close to its expected value, $T_1 \approx T (1- \ee^{-\nu})$, with high probability.

Conditioned on $T_1$, each nondefective item appears in some negative test with probability $1- (T_1/T)^L$, independently of one another. Hence, assuming the concentration result holds (replacing $T_1$ by its mean), we find that
$$ \Psuc \sim \big(1 - (T_1/T)^L\big)^{n-k} \sim \big(1 - (1- \ee^{-\nu})^L\big)^n = \big(1 - (1- \ee^{-\nu})^{\nu T/k}\big)^n
.$$
It is easy to check that $(1-\ee^{-\nu})^\nu$ is maximized at $\nu = \ln 2$, where it takes the value $\ee^{-(\ln 2)^2}$.
Thus, choosing $T = (1+\delta) (k \ln n) /(\ln 2)^2$ gives us that 
$(1- \ee^{-\nu})^{\nu T/k} = \ee^{-(1+\delta) \ln n}$. This allows us to deduce that
$\Psuc \sim (1- n^{-1+\delta})^n \sim \exp(-n^{-\delta})$, which tends to 1.

In terms of rates, again using Definition \ref{def:rate} and \eqref{eq:bincoeffequiv}, we can deduce that this equates to a rate of $(1-\alpha) (k \ln n)/(T \ln 2)$ which approaches $\ln 2 (1-\alpha) $ as required.
\end{proof}

Comparing with Theorem \ref{COMPub1}, we see that for the COMP algorithm, the near-constant column weight design provides an improvement of roughly $30.6\%$ over Bernoulli testing.  In addition, the rate of Theorem \ref{COMPub2} improves even over that of the DD algorithm with Bernoulli testing, both for sufficiently small $\alpha$ and sufficiently high $\alpha$.  See  Figure \ref{fig:algrates} for an illustration.

We now turn to the DD algorithm (see Section \ref{sec:DD}), which  strictly improves on Theorem \ref{COMPub2} for all $\alpha \in (0,1)$.  

\begin{theorem} \label{DDub2}
Consider noiseless nonadaptive group testing with exact recovery and the small error criterion, with $k = \Theta(n^{\alpha})$ for some $\alpha \in (0,1)$, and using the DD algorithm for decoding.  Under a near-constant column weight design with an optimized parameter $\nu = \ln 2$, the following rate is achievable:
\begin{equation} \label{DDCCW}
  R^{\mathrm{DD}}_{\mathrm{NCC}} = (\ln 2) \,\min \left\{ 1, \frac{1-\alpha}{\alpha} \right\} \approx 0.693 \min \left\{ 1, \frac{1-\alpha}{\alpha} \right\}.
\end{equation} 
\added{This rate corresponds to a number of tests satisfying the following for arbitrarily small $\eta > 0$: $T \ge (1+\eta) \max\big\{ \frac{1}{(\ln 2)^2} k \ln \frac{n}{k}, \frac{1}{(\ln 2)^2} k \ln k \big\}$.}

Moreover, for $\alpha \geq 1/2$, this achieves the maximum achievable
rate for SSS using this design, $\olR^{\SSS}_{\NCC}$ (see Theorem \ref{SSSub2} below).
\end{theorem}

The proof of this result bears some similarity to that of Bernoulli testing, but is more technically challenging.  The interested reader is referred to \cite{johnson-aldridge-scarlett} (see also \cite{gebhard2021improved}).

Comparing Theorem \ref{DDub2} to Theorem \ref{DDub}, we see that the achievable rate with the near-constant column weight design is roughly 30.6\% higher than the Bernoulli design -- the same gain as that observed for COMP (see Figure \ref{fig:algrates}).  

As in the case of Bernoulli testing, one immediately deduces (by Theorem \ref{SCOMPub} and Theorem \ref{LPub} respectively) that the SCOMP and LP algorithms (see Sections \ref{sec:SCOMP} and \ref{sec:lp}) also achieve the rate \eqref{DDCCW}.  \rev{ Moreover, in \cite{coja}, it was shown that \eqref{DDCCW} is the {\em maximum} achievable rate for the SCOMP (and DD) algorithm under the near-constant column weight design, meaning that asymptotically SCOMP does not outperform DD.  The fact that the rate cannot exceed $(\ln 2) \frac{1-\alpha}{\alpha}$ comes from Theorem \ref{SSSub2} below.  As for the $\ln 2$ term, the idea is to show that for rates above $\ln 2$ there exist many nondefectives that explain the maximum possible number $L$ of tests, and hence even the first iteration of SCOMP fails.  It remains an open problem as to whether a similar phenomenon holds for other test designs, such as Bernoulli testing.  }

Next, we present a converse bound \cite[Theorem 4]{johnson-aldridge-scarlett} for SSS, which notably matches the rate of the DD algorithm when $\alpha$ is above a threshold of roughly $0.409$.

\begin{theorem} \label{SSSub2}
  Consider noiseless nonadaptive group testing with exact recovery and the small error criterion, with $k = \Theta(n^{\alpha})$ for $\alpha \in (0,1)$, and using the SSS algorithm for decoding. With a near-constant column weight design, the maximum achievable rate is bounded above by
  \begin{equation} \label{SSSCCW}
    \olR^{\SSS}_{\NCC} \leq  \min \left\{1,\, \ln 2 \,\frac{1-\alpha}{\alpha} \right\}.
  \end{equation}
  \added{This rate upper bound corresponds to the condition that any algorithm must have a number of tests satisfying the following for arbitrarily small $\eta > 0$: $T \ge (1-\eta) \max\big\{ k \log_2 \frac{n}{k}, \frac{1}{(\ln 2)^2} k \ln k \big\}$.}
\end{theorem}

This result is analogous to Theorem \ref{SSSub}.  \rev{In Chapter \ref{ch:achievability}, we will survey a recent result of Coja-Oghlan \etal~\cite{coja} showing that the rate \eqref{SSSCCW} is achievable for all $\alpha \in (0,1)$, and deduce that \eqref{SSSCCW} gives the maximum achievable rate for the near-constant column weight design.}

\chapter{Algorithms for Noisy Group Testing} \label{ch:algorithms_noisy}

\section{Noisy channel models} \label{sec:noisy_models}

In Chapter \ref{ch:algorithms}, we focused on noiseless group testing algorithms and their theoretical guarantees.  From both a theoretical and practical perspective, these algorithms (as presented) rely strongly on the assumption that there is no noise.  In this chapter, we give an overview of some algorithms that are designed to handle noisy scenarios, most of which build on the ideas from the noiseless setting.  We initially present heuristic approaches, and then move on to techniques with theoretical guarantees.

For many of the applications described in Section \ref{sec:applications}, it is clearly an unrealistic modelling assumption that the tests would be able to perfectly identify whether any defective item is present in the pool. There are a variety of ways of modelling the noise, which affect the algorithms and their performance in different ways. We proceed by giving several illustrative examples.

Recall that standard noiseless group testing can be formulated component-wise using the Boolean \texttt{OR} operation as $y_t = \bigvee_{i \in \K} x_{ti}$ (see \eqref{eq:vee}). One of the simplest noise models simply considers the scenario where these values $\bigvee_{i \in \K} x_{ti}$ are flipped independently at random with a given probability.

\begin{example}[Binary symmetric noise] \label{ex:bsc}
In the binary symmetric noise model, the $t$-th test outcome is given by
\begin{equation} \label{eq:bsc}
Y_t = \begin{cases}
    \bigvee_{i \in \K} X_{ti} & \text{with probability } 1-\rho \\
    1 - \bigvee_{i \in \K} X_{ti} & \text{with probability } \rho.
\end{cases}
\end{equation}
This is, each test is flipped independently with probability $\rho$.
\end{example}

While the binary  noise model is an interesting one, many applications in Section \ref{sec:applications} suggest that false positive tests and false negative tests may occur with different probabilities.  We proceed by presenting some examples, maintaining the standard assumption that distinct test outcomes are conditionally independent given $\mat X$. Furthermore, we assume that each test has the same probability distribution specifying its outcome, and that this distribution depends on the test design $\mat{X}$ only through the number of defective items in the test and the total number of items in the test.  
For reasons of generality, we no longer insist that the test outcomes $y_t$ can only take values in $\zo$, but rather consider the case of $y_t \in \YY$ for some finite alphabet $\YY$. We follow in part the notation of \cite[Section 6.3]{aldridge-thesis}.

\begin{definition} \label{def:noisyprob}
\rev{
We define the probability transition function $p( \cdot \;|\; m, \ell)$ such that for a test  containing $m$ items, $\ell$ of which are defective, for each outcome $y \in \YY$ we have
\begin{equation}
\PP \left( Y_t = y \;\left|\; \sum_{i=1}^n X_{ti} = m, \sum_{i \in \K} X_{ti} = \ell \right. \right)  =  p( y \mid m, \ell),
\end{equation}
independently of all other tests.}
\end{definition}

In other words, $p(y \mid m, \ell)$ is the probability of observing outcome $y$ from a test containing $\ell$ defective items and $m$ items in total. Note that $\sum_{y \in \YY} p(y \mid m, \ell) = 1$ for all $m$ and $\ell$.

For example, the standard noiseless group testing model has probability
transition function
\begin{equation} \label{eq:ptfnoiseless}
\begin{array}{lllll}
p( \one \mid m, \ell) = 1 & \mbox{ if $\ell \geq 1$,} & \hspace*{0.5cm} &
p( \zero \mid m, \ell) = 0 & \mbox{ if $\ell \geq 1$,} \\
p( \one \mid m, \ell) = 0 & \mbox{ if $\ell = 0$,} & \hspace*{0.5cm} &
p( \zero \mid m, \ell) = 1 & \mbox{ if $\ell = 0$,} \\
\end{array}
\end{equation}
independent of $m$.
Similarly, the binary symmetric noise model of Example \ref{ex:bsc} has probability transition function
\begin{equation} \label{eq:ptfbsc}
\begin{array}{lllll}
\!\!\! p( \one \mid m, \ell) = 1-\rho & \mbox{if $\ell \geq 1$,} & \hspace*{0.0cm} & p( \zero \mid m, \ell) = \rho & \mbox{if $\ell \geq 1$,} \\
\!\!\! p( \one \mid m, \ell) = \rho & \mbox{if $\ell = 0$,} & \hspace*{0.0cm} & p( \zero \mid m, \ell) = 1 - \rho & \mbox{if $\ell = 0$.} \\
\end{array}
\end{equation}

Definition \ref{def:noisyprob} captures a variety of other noise models, one of which is the addition noise model of \cite{atia-saligrama}. Here false negative tests never occur, but false positive tests occur independently with a given probability $\varphi$. 

\begin{example}[Addition noise] \label{ex:addition}
In the \emph{addition noise} model, the probability transition function is given by
\begin{equation} \label{eq:ptfaddition}
\begin{array}{lllll}
p( \one \;|\; m, \ell) = 1 & \mbox{ if $\ell \geq 1$,} & \hspace*{0.3cm} & p( \zero \;|\; m, \ell) = 0 & \mbox{ if $\ell \geq 1$,} \\
p( \one \;|\; m, \ell) = \varphi & \mbox{ if $\ell = 0$,} & \hspace*{0.3cm} & p( \zero \;|\; m, \ell) = 1 - \varphi & \mbox{ if $\ell = 0$,} \\
\end{array}
\end{equation}
where $\varphi \in (0,1)$ is a noise parameter.
\end{example}

We note that the noise processes described in Examples \ref{ex:bsc} and \ref{ex:addition} can both be thought of as  sending the outcome of standard noiseless group testing through a noisy communication channel (see Definition \ref{def:ndc} below). Another interesting model, which cannot be represented in this way, is the dilution model of \cite{atia-saligrama}. This captures the idea that in some scenarios (such as DNA testing), the more defectives are present, the more likely we are to observe a positive test. 

In this model, the outcome of a test containing $\ell \geq 1$ defectives will be positive if and only if a Binomial$(\ell,1-\vartheta)$ random variable is at least one. Equivalently, this can be thought of as a scenario where every defective item included in the test only `behaves as a defective' with probability $1-\vartheta$, whereas with probability $\vartheta$ it is `diluted'.

\begin{example}[Dilution noise] \label{ex:dilution}
In the \emph{dilution noise} model, the probability transition function is given by
\begin{equation} \label{eq:ptfdilution}
\begin{array}{llll}
p( \one \;|\; m, \ell) = 1 - \vartheta^\ell, & \hspace*{0.3cm} & p( \zero \;|\; m, \ell) = \vartheta^\ell, & \mbox{ for all $\ell \geq 0$,} 
\end{array}
\end{equation}
where $\vartheta \in (0,1)$ is a noise parameter.
\end{example}

An analogous model to the dilution noise model is the Z channel noise model, in which tests containing defective items are erroneously negative with some fixed probability.

\begin{example}[Z channel noise] \label{ex:Z}
    In the \emph{Z channel} noise model, the probability transition function is given by
    \begin{equation} \label{eq:ptfZ}
    \begin{array}{lllll}
    p( \one \;|\; m, \ell) = 1 - \vartheta & \mbox{ if $\ell \geq 1$,} & \hspace*{0.3cm} & p( \zero \;|\; m, \ell) = \vartheta & \mbox{ if $\ell \geq 1$,} \\
    p( \one \;|\; m, \ell) = 0 & \mbox{ if $\ell = 0$,} & \hspace*{0.3cm} & p( \zero \;|\; m, \ell) = 1 & \mbox{ if $\ell = 0$,} \\
    \end{array}
    \end{equation}
    where $\vartheta \in (0,1)$ is a noise parameter.
\end{example}

By analogy, the addition noise channel (Example \ref{ex:addition}) can also be viewed as `reverse Z channel' noise.  \added{Examples \ref{ex:bsc}, \ref{ex:addition}, and \ref{ex:Z} can also all be viewed as special cases of the following general noise model.

\begin{example}[Binary channel noise] \label{ex:binary_channel}
    In the \emph{general binary channel noise} model, the probability transition function is given by
    \begin{equation} \label{eq:ptfZ_binary}
    \begin{array}{lllll}
    p( \one \;|\; m, \ell) = 1 - \rho_{10} & \mbox{ if $\ell \geq 1$,} & \hspace*{0.3cm} & p( \zero \;|\; m, \ell) = \rho_{10} & \mbox{ if $\ell \geq 1$,} \\
    p( \one \;|\; m, \ell) = \rho_{01} & \mbox{ if $\ell = 0$,} & \hspace*{0.3cm} & p( \zero \;|\; m, \ell) = 1-\rho_{01} & \mbox{ if $\ell = 0$,} \\
    \end{array}
    \end{equation}
    where $\rho_{01},\rho_{10} \in [0,1]$ are noise parameters.
\end{example}}

An example to illustrate the fact that the alphabet $\YY$ need not be $\zo$ is the erasure noise model, where each test may fail to give a conclusive result. We represent such an outcome by a question mark $\question$. In this case, $\YY = \zoq$, and the noise model is defined as follows.

\begin{example}[Erasure noise] \label{ex:erasure}
In the \emph{erasure noise} model, the probability transition function is given by
\begin{equation} \label{eq:ptferasure}
\begin{array}{lllll}
p( \one \;|\; m, \ell) = 1 - \xi & \mbox{ if $\ell \geq 1$,} & \hspace*{0.3cm} & p( \question \;|\; m, \ell) = \xi & \mbox{ if $\ell \geq 1$,} \\
p( \question \;|\; m, \ell) = \xi & \mbox{ if $\ell = 0$,} & \hspace*{0.3cm} & p( \zero \;|\; m, \ell) = 1 - \xi & \mbox{ if $\ell = 0$,} \\
\end{array}
\end{equation} 
where $\xi \in (0,1)$ is a noise parameter, and all other values of $p( \,\cdot\, \mid m, \ell)$ are zero.
\end{example}

Next, we provide another example of interest from \cite{laarhoven-1}, falling under the broad category of {\em threshold group testing} (see, for example, \cite{cheraghchi2013improved,damaschke2006threshold}).  In this example, a positive result is attained when the proportion of items in the test exceeds some threshold $\overline{\theta}$, a negative result is obtained when the proportion is below another threshold $\underline{\theta}$ (with $\underline{\theta} \le \overline{\theta}$), and positive and negative outcomes are equally likely when the proportion is in between these thresholds.

\begin{example}[Threshold group testing] \label{ex:threshold}
In the \emph{probabilistic threshold group testing noise} model, the probability transition function is given by
\begin{equation} \label{eq:ptfthresh}
\begin{array}{lllll}
p( \one \;|\; m, \ell) = 1  & \mbox{ if $\frac{\ell}{m} \ge \overline{\theta}$,} & \hspace*{0.1cm} & p( \zero \;|\; m, \ell) = 0 & \mbox{ if $\frac{\ell}{m} \ge \overline{\theta}$,} \\
p( \one \;|\; m, \ell) = 0 & \mbox{ if $\frac{\ell}{m} \le \underline{\theta}$,} & \hspace*{0.1cm} & p( \zero \;|\; m, \ell) = 1  & \mbox{ if $\frac{\ell}{m} \le \underline{\theta}$,} \\
p( \one \;|\; m, \ell) = \frac{1}{2} & \mbox{ if $\underline{\theta} < \frac{\ell}{m} < \overline{\theta}$,} & \hspace*{0.1cm} & p( \zero \;|\; m, \ell) = \frac{1}{2}  & \mbox{ if $\underline{\theta} < \frac{\ell}{m} < \overline{\theta}$ },
\end{array}
\end{equation}
where $\underline{\theta} \le \overline{\theta}$ are thresholds.
\end{example}

Another variation in \cite{laarhoven-1} instead assumes that the probability of a positive test increases from $0$ to $1$ in a linear fashion in between the two thresholds, rather than always equalling $\frac{1}{2}$.  It is worth noting that, while our focus is on random noise models, most works on threshold group testing have focused on {\em adversarial} noise \cite{cheraghchi2013improved,damaschke2006threshold}.

The noise models described in \eqref{eq:ptfnoiseless}, \eqref{eq:ptfbsc}, \eqref{eq:ptfaddition}, \eqref{eq:ptfdilution}, \eqref{eq:ptfZ}, and \eqref{eq:ptferasure} above share the property that $p( \cdot \;|\; m, \ell)$ does not  depend on $m$. Of course, this need not be the case in general, as Example \ref{ex:threshold}  shows. However, this property is sufficiently useful that we follow \cite[Definition 6.11]{aldridge-thesis} in explicitly naming it.

\begin{definition}
\label{def:odm}
We say that a noise model satisfies the \emph{only defects matter} property if the probability transition function is of the form
\begin{equation} \label{eq:odm} p( y \;|\; m, \ell) = p( y \;|\; \ell).
\end{equation}
\end{definition}

Properties of this type have been exploited in general sparse estimation problems beyond group testing (see for example  \cite{aksoylar,malyutov,scarlett-cevher-1}). In these cases, this property means that only the columns of a measurement matrix that correspond to the nonzero entries of a sparse vector impact the samples, and that the corresponding output distribution is permutation-invariant with respect to these columns.

While the only defects matter property, Definition \ref{def:odm}, does not hold in general, it plays a significant role in many proofs of noisy group testing results. For example, this assumption will be used throughout Chapter \ref{ch:achievability} to provide information-theoretic achievability and converse results. Some further evidence for the value of Definition \ref{def:odm} is that Furon \cite{furon} gives examples where `only defects matter' does not hold and a nonzero rate cannot be achieved.

A further interesting special case of Definition \ref{def:odm} is when the noisy group testing process can be thought of as sending the outcome of standard noiseless group testing through a noisy `communication' channel. 
\begin{definition}[Noisy defective channel] \label{def:ndc}
If we can express 
\begin{equation} p( y \mid m, \ell) = p(y \mid \II\{ \ell \geq 1 \} ), \label{eq:noisychan}
\end{equation}
where $p( y \mid \II\{ \ell \geq 1 \} )$ is the transition probability function of a noisy binary-input communication channel, then we say that the \emph{noisy defective channel property} holds.
\end{definition} 

\added{In the typical scenario where $y \in \{\zero, \one\}$ is also binary, this definition coincides with the binary channel noise of Example \ref{ex:binary_channel}. }
In the case that this property holds, the following result is stated in \cite{baldassini-johnson-aldridge}.

\begin{theorem} \label{thm:capcap}
If the noisy defective channel property (Definition \ref{def:ndc}) holds
then the group testing capacity $C$ (in the sense of Definition
\ref{def:achievable}) satisfies the following, regardless of whether the test design is adaptive or nonadaptive:
\begin{equation} \label{eq:noisychanbds}
C \leq C_{\rm{chan}},
\end{equation}
where $C_{\rm{chan}}$ is the Shannon capacity of the corresponding noisy communication channel $p( y \mid \II\{ \ell \geq 1 \} )$.
\end{theorem}

In fact, a similar result holds more generally even when the channel $p(y|\ell)$ has a nonbinary input indicating the number of defectives in the test; however, it is primarily the form stated in Theorem \ref{thm:capcap} that has been useful when comparing to achievability results.

One may be tempted to conjecture that for $k=o(n)$, equality holds in \eqref{eq:noisychanbds} for {\em adaptive} group testing.  This conjecture was shown to be true in \cite{Sca18} for the Z channel noise model (Example \ref{ex:Z}), but false when $k = \Theta(n^{\alpha})$ for $\alpha$ sufficiently close to one under the binary symmetric noise and addition noise models (Examples \ref{ex:bsc} and \ref{ex:addition}).

The argument given in \cite{baldassini-johnson-aldridge} to prove Theorem \ref{thm:capcap} uses the fact that the test outcome vector ${\vec y} = (y_1,\dotsc,y_T)$ acts like the output of the channel whose input codeword is indexed by the defective set.  Since the transmission of information is impossible at rates above capacity, it certainly remains impossible in the presence of the extra constraints imposed by the group testing problem.

The noisy defective channel property of Definition \ref{def:ndc} is satisfied by the models described in \eqref{eq:ptfnoiseless}, \eqref{eq:ptfbsc}, \eqref{eq:ptfaddition}, \eqref{eq:ptfZ}, and \eqref{eq:ptferasure} (though not the dilution model \eqref{eq:ptfdilution} or threshold model \eqref{eq:ptfthresh}). For example, we can deduce that the binary symmetric model Definition \ref{eq:ptfbsc} has group testing capacity $C \leq 1 - h(\rho)$, where $h(\cdot)$ is the binary entropy function.  It remains an open problem to determine under what conditions this bound is sharp; in Section \ref{sec:noisy}, we will see that it is sharp in the sparse regime $k = O(n^{\alpha})$ when $\alpha$ is sufficiently small. 

A simple case in which we can determine the adaptive group testing capacity is the erasure model; the following result is from \cite[Theorem 1.3.1]{baldassini-johnson-aldridge}.
\begin{theorem} \label{thm:erasure}
The capacity of adaptive group testing is $C = 1 - \xi$ for the erasure model of Example \ref{ex:erasure} when $k = o(n)$.
\end{theorem}
\begin{proof} This is achieved by simply using a noiseless adaptive group testing scheme (see Section \ref{sec:adaptive}), and repeating tests for which $y_t = \question$. Standard concentration bounds reveal that, for any $\epsilon > 0$, no more than $T(\xi + \epsilon)$ tests will need repeating (with probability approaching one as $n \to \infty$), and the result follows from Theorem \ref{thm:adaptcap}. \end{proof}

A similar argument can be used to determine bounds on the rates of nonadaptive algorithms under Bernoulli designs for the erasure noise model of Example \ref{ex:erasure}. Again, with high probability, given $T$ tests, we know there should be at least $T(1-\xi-\epsilon)$ tests that are not erased. Hence, simply ignoring the tests which return a $\question$, it is as if we have been given a Bernoulli design matrix with at least $T(1-\xi-\epsilon)$ rows. 

Hence, for example, if there are $k = \Theta(n^\alpha)$ defectives, then building on Theorem \ref{DDub}, we can achieve a rate of 
\begin{equation} \label{eq:erasurerate}
\frac{1 - \xi}{\ee \ln 2} \min \left\{ 1, \frac{1-\alpha}{\alpha} \right\}
\end{equation}
using the DD algorithm and a Bernoulli test design. Similarly, building on Theorem \ref{SSSub}, we know that even the SSS algorithm cannot achieve a rate greater than
\begin{equation} \label{eq:sssrate}
 (1-\xi) \max_{\nu > 0} \min \left\{ h(\mathrm \ee^{-\nu}),\, \frac{\nu}{\ee^\nu \ln 2} \frac{1-\alpha}{\alpha} \right\} .
\end{equation}
for Bernoulli designs.

We also briefly mention that since the addition noise channel (Example \ref{ex:addition}) satisfies the property that a negative outcome is definitive proof of no defectives being present, we can easily extend the analysis of the COMP algorithm to deduce a counterpart to Theorem \ref{COMPub1} (see \cite[Lemma 1]{scarlett-johnson} for details). Specifically, since a proportion $\varphi$ of the negative tests are flipped at random, we can achieve a rate of $\frac{1-\varphi}{\ee \ln 2} (1-\alpha)$ using the COMP algorithm (see \cite[eq. (149)]{scarlett-johnson}).

In the remainder of the chapter, we describe a variety of algorithms that can be used to solve noisy group testing problems in the presence of both false positive tests and false negative tests (e.g., for the binary symmetric noise model).  Most of these are extensions of the noiseless algorithms presented in Chapter \ref{ch:algorithms}, and like that chapter, we initially focus our attention on nonadaptive Bernoulli test designs, the small-error recovery criterion, and the scaling $k = \Theta(n^{\alpha})$ with $\alpha \in (0,1)$.  \added{See also Section \ref{sec:near_const_noisy} for improved rates via near-constant tests-per-item.}

\section{Noisy linear programming relaxations} \label{sec:noisy_LP}

Recall the linear programming relaxation for the noiseless setting in Section \ref{sec:lp}.  A similar idea can be used in the noisy setting by introducing {\em slack variables}, which leads to a formulation allowing `flipped' test outcomes but paying a penalty in the objective function for doing so.  Using this idea, the following formulation was proposed in \cite{malioutov-malyutov}: 
\begin{align*}
  \mathrm{minimize}_{\vc{z},\boldsymbol{\xi}} ~~~& \sum_{i=1}^n z_i + \lambda \sum_{t=1}^T \xi_j  \\
  \text{subject to } ~~~& z_i \ge 0 \\
  & \xi_t \ge 0  \\
  & \xi_t \le 1 \hspace*{13.7ex} \text{when }y_t = 1 \\
  & \sum_{i=1}^n x_{ti}z_i = \xi_t \hspace*{6ex}\, \text{when }y_t = 0 \\
  &  \sum_{i=1}^n x_{ti}z_i + \xi_t \ge 1 \quad \text{when }y_t = 1.
\end{align*}
As in the noiseless setting, $\vc{z}$ represents an estimate of the defectivity indicator vector $\vc{u}$ (see Definition \ref{def:udef}), \rev{whereas here we also have a vector of $T$ slack variables $\boldsymbol{\xi} = (\xi_1,\dotsc,\xi_T)$}. The parameter $\lambda$ controls the trade-off between declaring a small number of items to be defective (sparsity) and the degree to which the test outcomes are in agreement with the decoded $z_i$ (most slack variables being zero).  \rev{Observe that if we were to further constrain each $z_i$ and $\xi_t$ to be binary-valued ($0$ or $1$), then the above formulation would be minimizing a weighted combination of the number of (estimated) defectives and the number of `flipped' tests.  Such a binary-valued minimization problem, with a suitable choice of $\lambda$, can also be shown to be equivalent to {\em maximum a posteriori} (MAP) decoding under an i.i.d.~defectivity model (see the Appendix to Chapter \ref{ch:introduction}) and symmetric noise (see Example \ref{ex:bsc}); \added{the details can be found in \cite[Section~5.3]{ciampiconi2020maxsat}}.}

The above formulation treats false positive tests and false negative tests equally. 
However, it can also be modified to weigh the two differently; in the extreme case, if it is known that a test with no defectives {\em definitely} results in a negative outcome (e.g., dilution noise of Example \ref{ex:dilution}, or Z channel noise of Example \ref{ex:Z}), then we could replace all of the slack variables corresponding to negative tests by zero.  An analogous statement holds true when a test with at least one defective {\em definitely} results in a positive outcome (e.g., addition noise of Example \ref{ex:addition}).

To the best of our knowledge, no theoretical results are known for the above noisy linear program relaxation.  However, this method has been seen to provide excellent performance in numerical experiments \cite{malioutov-malyutov}; see Section \ref{sec:noisy_rates} for an illustration. 

A related noisy linear program relaxation using {\em negative tests only} was proved to achieve positive rates in \cite{chan-etal-3}.  However, there are two notable limitations.  First, from a theoretical view, the constants were not optimized in the proofs, and so the rates are far from optimal.  Second, from a practical view, ignoring the tests with positive outcomes can significantly worsen the performance.  

\section{Belief propagation} \label{sec:belprop}

A decoding algorithm based on belief propagation was described in \cite[Section III]{sejdinovic-johnson}. Although there was no attempt to calculate performance bounds or rates, there was some numerical evidence presented to show that this approach can work very well (see also Section \ref{sec:noisy_rates}, as well as \cite{coja2022efficient} for an extensive empirical study).  The success of belief propagation experimentally is perhaps unsurprising, since it has enjoyed considerable success for the decoding of LDPC codes over noisy channels, a problem that shares characteristics with group testing. 

Recall from Definition \ref{def:udef} that we write $u_i = \II \{i \in \K\}$ to indicate whether or not item $i$ is defective. The idea is to estimate the defective set by working with the marginals of the posterior distribution, and for each $i$, seek to estimate $u_i$ as 
\begin{equation} \label{eq:map}
\widehat{u}_i := \argmax_{u_i \in \zo} \PP(u_i \mid \vc{y}),
\end{equation}
where $\vc{y}$ is the vector of test outcomes. Clearly, we would prefer to optimize this posterior probability as a function  of all the $(u_i)_{i \in \{ 1, \ldots, n \}}$, but this would be computationally infeasible due to the size of the search space.

\begin{figure}[t] 
\begin{center}
\includegraphics[width=0.7\textwidth]{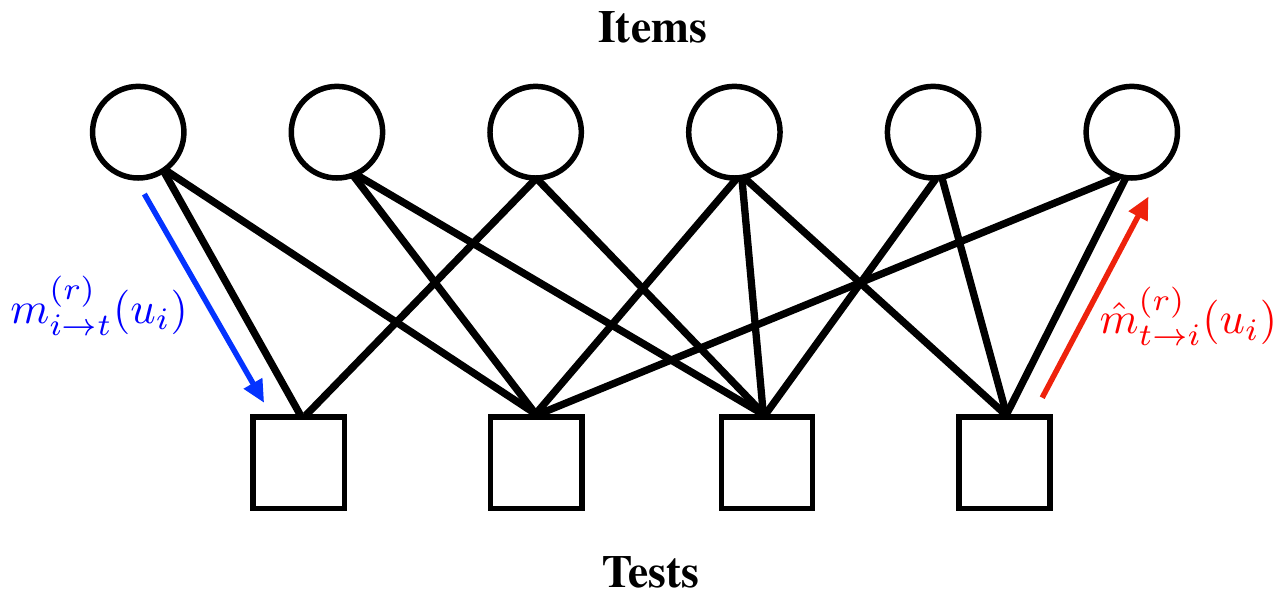}
\end{center}
\caption{Example bipartite graph used in belief propagation decoding.  Edges represent the inclusion of items in tests, and messages are passed in both directions.} \label{fig:bp_graph} 
\end{figure}

While exactly computing the probability $\PP(u_i \mid \vc{y})$ appearing in \eqref{eq:map} is also difficult, we can approximately compute it using loopy belief propagation.  To understand this, we set up a bipartite graph with $n$ nodes on one side corresponding to items, and $T$ nodes on the other side corresponding to tests.  Each test node is connected to all of the nodes corresponding to items included in the test.  See Figure \ref{fig:bp_graph} for a simple example.

Assuming that $k$ out of $n$ items are defective, a natural prior is given by
\begin{equation} \label{eq:lambdadef}
\PP(U_i = \one) = \frac kn =: \qq,
\end{equation}
and for analytical tractability, an independent prior $\PP(\vc{u}) = \prod_{i=1}^n \PP(u_i)$ is adopted.  Even for a combinatorial prior where $\K$ is uniform over $\binom{n}{k}$ possible defective sets, \eqref{eq:lambdadef} yields a good approximation for large $k$ due to concentration of measure.  In either case, this method (as described here) requires at least approximate knowledge of $k$. 

In accordance with general-purpose techniques for loopy belief propagation (e.g., see \cite[Ch.~26]{mackay}), messages are iteratively passed from items to tests and tests to items.  Letting $\mathcal{N}(i)$ and $\mathcal{N}(t)$ denote the neighbours of an item node and test node respectively, the item-to-test and test-to-item message are given as follows \cite{sejdinovic-johnson}: 
\begin{gather}
    \bpnohat{m}{r}{i}(u_i) \propto \big( \qq \II\{ u_i = 1 \} + (1-\qq) \II\{ u_i = 0 \}\big) \prod_{t' \in \mathcal{N}(i) \setminus \{t\}} \bphat{m}{r}{t'}(u_i) \label{eq:bp1} \\
    \bphat{m}{r}{t}(u_i) \propto \sum_{\{u_{i'}\}_{i' \in \mathcal{N}(t) \setminus \{i\}}} \PP(y_t \mid u_{[t]} ) \prod_{i' \in \mathcal{N}(t) \setminus \{i\}} \bpnohat{m}{r}{i'}(u_i), \label{eq:bp2}
\end{gather}
where $r$ indexes the round of message passing, $\propto$ denotes equality up to a normalizing constant, and $u_{[t]}$ denotes the subvector of $\vc{u}$ corresponding to the items in test $t$, which are the only ones that impact $y_t$.  These messages amount to updating beliefs of the test outcomes $y_t$ in terms of $u_i$, and beliefs of $u_i$ in terms of the test outcomes $y_t$.  By iterating these steps, we hope to converge to a sufficiently good approximation of the posterior.  

The sum over $\{u_{i'}\}_{i' \in N(t) \setminus \{i\}}$ in \eqref{eq:bp2} grows exponentially in the number of items in the test, so these messages are still expensive to compute if the computation is done naively.  Fortunately, at least for certain noise models, it is possible to rewrite the messages in a form that permits efficient computation.  In \cite{sejdinovic-johnson}, this was shown for the following model that combines the addition and dilution models of Examples \ref{ex:addition} and \ref{ex:dilution}:
\begin{equation} \label{eq:ptfadditiondilution}
p( \one \;|\; m, \ell) = 1 - (1-\varphi)\vartheta^\ell, \quad p( \zero \;|\; m, \ell) = (1-\varphi)\vartheta^\ell, \quad \mbox{for all $\ell \geq 0$.}
\end{equation}
Observe that setting $\varphi= 0$ recovers the dilution model, whereas setting $\vartheta = 0$ recovers the addition model.

For this model, it is convenient to work with log-ratios of the messages, defined as
\begin{equation}
\bpnohat{L}{r}{i} = \ln \frac{\bpnohat{m}{r}{i}(\one)}{\bpnohat{m}{r}{i}(\zero)} \mbox{ \;\;\; and \;\;\;}
\bphat{L}{r}{t} = \ln \frac{\bphat{m}{r}{t}(\one)}{\bphat{m}{r}{t}(\zero)}.
\end{equation}
The natural prior $\PP(u_i = \one) = \qq$ mentioned above means that $\bpnohat{L}{r}{i}$ should be initialized as $\bpnohat{L}{0}{i} = \ln \frac{\qq}{1-\qq}$. Then the item-to-test updates in subsequent rounds easily follow from \eqref{eq:bp1}:
\begin{equation}
\bpnohat{L}{r+1}{i} =  \ln \left( \frac{\qq}{1-\qq} \right)
+ \sum_{t' \in {\mathcal{N}}(i) \setminus{\{t\}}} \bphat{L}{r}{t'}.
\end{equation}

The test-to-item messages require a bit more effort to derive, but the analysis is entirely elementary.  If the test $t$ is positive ($y_t = 1$), we obtain \cite{sejdinovic-johnson}
\begin{equation*}
\bphat{L}{r}{t} = \ln \left(  \vartheta +  
\frac{1-\vartheta}{1 - (1-\varphi) \prod\limits_{{\scriptscriptstyle j \in {\mathcal{N}}(t) \setminus \{i \} }} \Big( \vartheta + \frac{1-\vartheta}{1 + \exp( \bpnohat{L}{r}{j} )} \Big)} \right),
\end{equation*}
and if the test $t$ is negative ($y_t = 0$), we simply have $\bphat{L}{r}{i} = \ln \vartheta$ \cite{sejdinovic-johnson}.

We are not aware of any works simplifying the messages (or their log-ratios) for general noise models.  Since the binary symmetric noise model of Example \ref{ex:bsc} (with parameter $\rho$) is particularly widely-adopted, we also state such a simplification here without proof. If $y_t = 0$, then
 \begin{equation*}
\hat{m}^{(r+1)}_{t \to i}(u_i) \propto
\begin{cases}
\; \rho\displaystyle\prod_{i'\in N(t) \setminus \{i\}}\big(m^{(r)}_{i'\to t}(0)+m^{(r)}_{i'\to t}(1)\big)  & \quad u_i = 1 \vspace*{2ex}\\
\begin{aligned}
&\rho\displaystyle\prod_{i'\in N(t) \setminus \{i\}}\big(m^{(r)}_{i'\to t}(0)+m^{(r)}_{i'\to t}(1)\big) \\
&\qquad {}+(1-2\rho)\displaystyle\prod_{i'\in N(t) \setminus \{i\}}m^{(r)}_{i'\to t}(0)
\end{aligned} 
&  \quad u_i = 0,
\end{cases} 
\end{equation*}
while if $y_t = 1$, then
\begin{equation*}
\hat{m}^{(r+1)}_{t \to i}(u_i) \propto
\begin{cases}
\; (1-\rho)\displaystyle\prod_{i'\in N(t) \setminus \{i\}}\big(m^{(r)}_{i'\to t}(0)+m^{(r)}_{i'\to t}(1)\big)  & \quad u_i = 1 \vspace*{2ex} \\
\begin{aligned}
&(1-\rho)\displaystyle\prod_{i'\in N(t) \setminus \{i\}}\big(m^{(r)}_{i'\to t}(0)+m^{(r)}_{i'\to t}(1)\big) \\
&\qquad {}- (1-2\rho)\displaystyle\prod_{i'\in N(t) \setminus \{i\}}m^{(r)}_{i'\to t}(0) \end{aligned}
&  \quad u_i = 0 .
\end{cases} 
\end{equation*}
Here we found it more convenient to work directly with the messages rather than their log-ratios; the two are equivalent in the sense that either can be computed from the other.

In the case that $k$ is known exactly, instead of declaring $\hat{u}_i$ to be zero or one according to \eqref{eq:map}, one can sort the estimates of $\PP(u_i = 1 \mid \vc{y})$ in decreasing order and declare the resulting top $k$ items to be the defective set.  Moreover, while \eqref{eq:map} amounts to declaring an item defective if the estimate of $\PP(u_i = 1 \mid \vc{y})$ exceeds $\frac{1}{2}$, one could threshold at values other than $\frac{1}{2}$.  This would be of interest, for example, in scenarios where false positives and false negatives in the reconstruction are not considered equally bad.

\subsection{Other algorithms based on Monte Carlo and message passing} \label{sec:monte_carlo}

\rev{
A distinct but related approach to belief propagation is based on generating samples from $\PP(\K \mid \vc{y})$ via Markov chain Monte Carlo (MCMC).  To our knowledge, the MCMC approach to group testing was initiated by Knill, Schliep, and Torney \cite{knill1996mcmc}; see also \cite{schliep2003dna} and \cite{furon2012mcmc} for related follow-up works.  These papers use the notion of Gibbs sampling: A randomly-initialized set $\K_0 \subseteq \{1,\dotsc,n\}$ is sequentially updated by choosing an item in $\{1,\dotsc,n\}$ (e.g., uniformly at random) and deciding whether it should be added or removed (or unchanged) from the set.  Specifically, this decision is made based on a posterior calculation using Bayes rule, analogously to the belief propagation updates.

The Gibbs sampling procedure is designed to produce a Markov chain with stationary distribution $\PP(\K \mid \vc{y})$, so that after sufficiently many iterations, the set being maintained is also approximately distributed according to $\PP(\K \mid \vc{y})$.  After taking numerous samples of sets from this distribution, the most commonly-occurring items are taken to be the final estimate $\Khat$.} \added{Similarly to belief propagation, most studies on MCMC have lacked theory and been based on empirical evaluation.  However, a notable exception is the recent work of Lovig and Zadik \cite{lovig2024mcmc} studying the theory of MCMC in the noiseless setting; the details are deferred to Section \ref{sec:computation}.

Yet another related decoding strategy is \emph{generalized approximate message passing} (GAMP), which has had considerable success in related areas such as compressed sensing.  Its use in group testing was explored in \cite{cao2023gamp,goenka2021contact,karimi2022noisy}, all of which allowed the incorporation of additional prior information (see Section \ref{sec:prior} for further discussion).  These works again focused on empirical evaluation rather than theoretical guarantees.
}

\section{Noisy COMP} \label{sec:NCOMP}

In Section \ref{sec:COMP}, we discussed the analysis given by \cite{chan-etal-1,chan-etal-3} of the simple COMP algorithm in the noiseless case.  In the same works, the authors also introduced a noisy version of COMP, which we refer to as NCOMP.  The authors focused on the binary symmetric noise model (Example \ref{ex:bsc}) with parameter $\rho \in \big(0,\frac{1}{2}\big)$, and we do the same in this section, but it is also possible to handle other noise models (e.g., see \cite{gebhard2021improved}).

The idea of NCOMP is that for any item $i\in\{1,\dotsc,n\}$, if the item is defective, then among the tests where $i$ is included, we should expect roughly a fraction $1-\rho$ of the outcomes to be positive.  In contrast, if the item is nondefective, we should expect a smaller fraction of the outcomes to be positive.  Thus, the algorithm declares item $i$ to be defective or nondefective according to the following rule:
\begin{equation}
	\text{Declare $i$ defective} \iff \frac{ \sum_{t=1}^T \boldsymbol{1}\{ X_{ti} = 1 \cap y_t = 1 \} }{ \sum_{t=1}^T \boldsymbol{1}\{ X_{ti} = 1\}  } \ge 1 - \rho(1 + \Delta) \label{eq:ncomp_rule}
\end{equation}
for some parameter $\Delta > 0$.  Note that this rule requires knowledge of the noise level $\rho$.

It was shown in \cite{chan-etal-1,chan-etal-3} that with a suitable choice of $\Delta$, NCOMP achieves a positive rate for all $\alpha \in (0,1)$, albeit generally far from the information-theoretic limits of Section \ref{sec:noisy}.  \rev{The rates presented in  \cite{chan-etal-1,chan-etal-3} differ according to the choice of $\nu > 0$, but as discussed in \cite[Footnote 3]{Sca17b}, the best rate that can be ascertained {\em directly} from these works is
\begin{equation}
    \tilde{R}_{\rm Bern}^{\rm NCOMP} = \frac{(1-2\rho)^2 (1-\alpha)}{4.36(1+\sqrt{\alpha})^2}, \label{eq:ncomp_rate}
\end{equation}
and amounts to choosing $\nu = 1$.  

We provide only a high-level outline of the proof of \eqref{eq:ncomp_rate}, and refer the reader to \cite{chan-etal-1,chan-etal-3} for the details.  The analysis separately characterizes the probability of a given defective wrongly being declared as nondefective (by failing the threshold test in \eqref{eq:ncomp_rule}) and a given nondefective wrongly being declared defective.  The i.i.d.~nature of the test matrix and noise permits a concentration of measure argument, from which it can be shown that both of these error events decay exponentially in $T$ as long as $\Delta$ is not too high.  Applying the union bound leads to a multiplication of the preceding probabilities by $k$ and $n-k$ respectively, and the analysis is completed by choosing $T$ large enough to make the resulting bound decay to zero, as well as optimizing $\Delta$ and $\nu$.}

\added{
An improved rate for a slight variation of NCOMP can be inferred from the subsequent work of Scarlett and Johnson \cite{scarlett-johnson}, and is formally stated in a related paper by Gebhard \etal~\cite[Proposition E.1]{gebhard2021improved}.\footnote{Their expressions are written in terms of binary relative entropies; below we simplify these by replacing $k \,d\big(\frac{a}{k} \mmid \frac{b}{k}\big)$ by $a\log\frac{a}{b} + b - a$; these are equivalent up to lower-order terms.}  Specifically, instead of thresholding the ratio as stated in \eqref{eq:ncomp_rule}, this variant of NCOMP simply thresholds the number of negative tests containing the item:
\begin{equation}
	\text{Declare $i$ defective} \iff \sum_{t=1}^T \boldsymbol{1}\{ X_{ti} = 1 \cap y_t = 0 \} < \gamma 
 \cdot \frac{T\nu}{k}, \label{eq:ncomp_rule2}
\end{equation}
for some parameter $\gamma > 0$.  Recalling that we consider Bernoulli testing with $p = \frac{\nu}{k}$, we see that $\frac{T\nu}{k}$ is simply the \emph{average} number of tests each item is placed in, and using this average slightly simplifies the analysis compared to using the actual (random) value.  The resulting rate is stated in the following theorem, in which the noise model may be symmetric or asymmetric.

\begin{theorem} \label{thm:NCOMP_bern_improved}
    Consider the NCOMP decoding rule \eqref{eq:ncomp_rule2} under the Bernoulli test design with parameter $p = \frac{\nu}{k}$, and suppose that $k = \Theta(n^{\alpha})$ for some $\alpha \in (0,1)$.  Moreover, consider the noise model from Example \ref{ex:binary_channel} in which tests with no defectives (respectively, with defectives) are flipped to 1 (respectively, 0) with probability $\rho_{01}$ (respectively, $\rho_{10}$), where the noise parameters $(\rho_{01},\rho_{10})$ satisfy $\rho_{01} + \rho_{10} < 1$.  Then, the following rate is achieved:
    \begin{equation}
        R_{\rm Bern}^{\mathrm{NCOMP}} = \frac{(1-\alpha)\nu}{\ln 2} \max_{\gamma \in (\rho_{10},w)} \min\bigg\{ \frac{1}{\alpha} D_{\rho_{10}}(\gamma), D_{w}(\gamma) \bigg\}, \label{eq:NCOMP_improved}
    \end{equation}
    where we define $w = e^{-\nu}(1-\rho_{01}) + (1-e^{-\nu})\rho_{10}$ and $D_{a}(t) = t \log\frac{t}{a} + a - t$.
\end{theorem}

While this rate expression is slightly complicated, the idea of the analysis is broadly similar to before: Bound the probability that a given item leads to an incorrect decision in \eqref{eq:ncomp_rule2}, and then apply a union bound over all items.  For the defectives, a union bound over $k$ items yields the first term in the minimum in \eqref{eq:NCOMP_improved}, and for the nondefectives, a union bound over $n-k$ items yields the second term.  Importantly, the analysis of each item uses tight Chernoff bounds, as opposed to looser concentration inequalities that were used to derive rates such as \eqref{eq:ncomp_rate}.  The maximization over $\gamma$ in \eqref{eq:NCOMP_improved} arises from optimizing the choice of threshold.

The rate in Theorem \ref{thm:NCOMP_bern_improved}, along with those of other algorithms and test designs, will be compared visually in Section \ref{sec:noisy_rates}.
}

\section{Separate decoding of items} \label{sec:separate}

The NCOMP algorithm described above decodes each item individually: The decision on whether or not item $i$ is defective is based only on the $i$-th column of $\mat{X}$, along with $\vec{y}$.  This general principle of decoding items separately was in fact introduced in an early work of Malyutov and Mateev \cite{Mal80}, and shown to come with strong theoretical guarantees in the case that $k = O(1)$.  It was originally referred to as {\em separate testing of inputs}, but we adopt the terminology {\em separate decoding of items} to avoid possible confusion \added{with individual testing (choosing $\mat{X}$ to be the identity matrix)}.

Again, recall that $u_i = \II \{i \in \K\}$ indicates whether or not item $i$ is defective.  The decoding rule for item $i$ proposed in \cite{Mal80} is as follows: 
\begin{equation}
	\text{Declare $i$ defective} \iff \sum_{t=1}^T \log_2\frac{P_{Y|X_i,U_i}(y_t|x_{ti},1)}{P_{Y}(y_t)} \ge \gamma \label{eq:separate_dec}
\end{equation}
where $\gamma > 0$ is a threshold, and $P_{Y|X_i,U_i}$ is the conditional distribution of a test outcome given a single item's inclusion indicator and defectivity indicator.  This can be interpreted as the Neyman-Pearson test for binary hypothesis testing with hypotheses $H_0 \colon u_i = 0$ and $H_1 \colon u_i = 1$; note that $P_{Y|X_i,U_i}(y_t|x_{ti},0)$ is the same as $P_{Y}(y_t)$ regardless of the value of $x_{ti}$, since nondefective items do not impact the test outcome.

\rev{We briefly mention that the decoder \eqref{eq:separate_dec}, along with its analysis (outlined below), can be viewed as a simplified and computationally efficient counterpart to an intractable {\em joint} decoding rule based on thresholding.  The latter is surveyed in Chapter \ref{ch:achievability} as a means to deriving information-theoretic achievability bounds. See also \cite{laarhoven-1,huleihel} for further works comparing separate and joint decoding.}

The results of \cite{Mal80} indicate the following: When $k = O(1)$ and $n \to \infty$, the rate achieved by separate decoding of items for the noiseless model or binary symmetric noise model is {\em within an $\ln 2$ factor of the optimal (joint) decoder}.  For instance, in the noiseless setting, a rate of $\ln 2 \simeq 0.693$ bits/test is attained, thus being reasonably close to the optimal rate of one.

For more general noise models, under Bernoulli testing, a sufficient condition on the number of tests for vanishing error probability is \cite{Mal80}
\begin{equation*} 
    T \ge \frac{\log_2 p}{I_1} (1+o(1)), \label{eq:sep_dec_T}
\end{equation*}
where the {\em single-item mutual information} $I_1$ is defined as follows, with implicit conditioning on item 1 being defective, and $X_1$ denoting whether it was included in a given test that produced the outcome $Y$:
\begin{equation}
    I_1 = I(X_1;Y). \label{eq:I1_intro}
\end{equation}
In a follow-up work \cite{Mal98}, similar results were shown when the rule \eqref{eq:separate_dec} is replaced by a {\em universal} rule (one that does not depend on the noise distribution) based on the empirical mutual information.

In this monograph, we are primarily interested in the sparse regime $k = \Theta(n^{\alpha})$, as opposed to the very sparse regime $k = O(1)$.  Separate decoding of items was studied under the former setting by Scarlett and Cevher \cite{Sca17b}, with the main results for specific models including the following.

\begin{theorem} \label{thm:separate}
    Consider the separate decoding of items technique under i.i.d. Bernoulli testing with parameter $p = \frac{\ln 2}{k}$ (so $\nu = \ln 2$), with $k = \Theta(n^{\alpha})$ for some $\alpha \in (0,1)$. Then we have the following:
    \begin{itemize}
        \item Under the noiseless model, there exists a constant $c(\delta') > 0$ such that the following rate is achieved:
        \begin{equation}
        R_{\rm Bern}^{\mathrm{SD}} = \max_{\delta' > 0} \min\bigg\{ (\ln 2) (1-\alpha)(1-\delta'), \\ c(\delta') \frac{1-\alpha}{\alpha} \bigg\}.
        \end{equation}
        In particular, as $\alpha \to 0$, the rate approaches $\ln 2$ bits/test.
        \item Under the binary symmetric noise model \eqref{eq:bsc} with parameter $\rho \in \big(0,\frac{1}{2}\big)$, there exists a constant $c_{\rho}(\delta') > 0$ such that the following rate is achieved:
        \begin{align}
        \hspace*{-3ex}R_{\rm Bern}^{\mathrm{SD}}(\rho) = \max_{\delta' > 0} \min\bigg\{ (\ln 2)(1 - h(\rho)) (1-\alpha)(1-\delta'), c_{\rho}(\delta')\frac{1-\alpha}{\alpha} \bigg\}.
        \end{align}
        Hence, as $\alpha \to 0$, the rate approaches $(\ln 2)(1 - h(\rho))$. 
    \end{itemize}
\end{theorem}

The quantities  $c(\delta')$ and $c_{\rho}(\delta')$ are related to concentration bounds arising in the analysis, as we discuss in the proof outline below.   Explicit expressions for these quantities can be found in \cite{Sca17b}, but they are omitted here since they are somewhat complicated.  For both the noiseless and symmetric noise models, in the limit as $\alpha \to 0$, the rate comes within a $\ln 2$ factor of the channel capacity, which cannot be exceeded by any group testing algorithm (see Theorem \ref{thm:capcap}).  In \cite{laarhoven-1}, characterizations of the mutual information $I_1$ in \eqref{eq:I1_intro} were also given for a variety of other noisy group testing models.



\paragraph{Overview of proof of Theorem \ref{thm:separate}} As stated following \eqref{eq:separate_dec}, the decoder for a given item performs a binary hypothesis test to determine whether the item is defective.  As a result, analysing the error probability amounts to characterizing the probabilities of false positives and false negatives in the recovery.

We first consider false positives.  Letting $i$ represent a nondefective item, and letting $\vec{X}_{i} = [X_{1i},\dotsc,X_{Ti}]^T$ be the corresponding column of $\mat{X}$, the probability of being declared defective is
\begin{align}
	P_{\mathrm{fp}} &= \sum_{\vec{x}_{i},\vec{y}} \PP(\vec{x}_{i}) \PP(\vec{y}) \mathbf{1}\bigg\{ \sum_{t=1}^T \log_2\frac{P_{Y|X_i,U_i}(y_i\mid x_{ti},1)}{P_{Y}(y_t)} \ge \gamma \bigg\} \label{eq:sd_line1} \\
    	&\le \sum_{\vec{x}_{i},\vec{y}} \PP(\vec{x}_{i}) \bigg( \prod_{t=1}^T P_{Y|X_i,U_i}(y_i\mid x_{ti},1) \bigg) 2^{-\gamma} \label{eq:sd_line2} \\
        &= 2^{-\gamma}, \label{eq:sd_line3}
\end{align}
where \eqref{eq:sd_line1} substitutes the decoding rule \eqref{eq:separate_dec} and uses the fact that the column $\vec{X}_{i}$ and test outcomes $\vec{Y}$ are independent when $i$ is nondefective; \eqref{eq:sd_line2} follows by writing the sum of logarithms as the logarithm of a product and noting that $\PP(\vec{y}) = \prod_{t=1}^T P_{Y}(y_t)$, which means that the event in the indicator funtion can be re-arranged to $\PP(\vec{y}) \le \big( \prod_{t=1}^T P_{Y|X_i,U_i}(y_i\mid x_{ti},1) \big)2^{-\gamma}$; and \eqref{eq:sd_line3} follows since we are summing a joint probability distribution over all of its values.  Since there are $n-k$ nondefectives, we can use the union bound to conclude that for any $\delta > 0$, the choice
	$$ \gamma = \log_2\frac{n-k}{\delta} $$
suffices to ensure that the probability of {\em any} false positives is at most $\delta$.

With this choice of $\gamma$, the probability of any given defective item $i$ being declared as nondefective is given by
\begin{equation}
	P_{\mathrm{fn}} = \PP\bigg( \sum_{t=1}^T \log_2\frac{P_{Y|X_i,U_i}(Y_t|X_{ti},1)}{P_{Y}(Y_t)} \le \log_2\frac{n-k}{\delta} \bigg).
\end{equation}
Observe that the mean of the left-hand side inside the probability is exactly $T I_1$, with $I_1$ being the mutual information term defined in \eqref{eq:I1_intro}.  Moreover, the probability itself is simply the lower tail probability of an i.i.d.~sum, and hence, we should expect some degree of concentration around the mean.  To see this more concretely, we note that as long as
\begin{equation}	
	T \ge \frac{\log_2\frac{n-k}{\delta}}{I_1 (1-\delta')}
\end{equation}
for some $\delta' \in (0,1)$, we have
\begin{equation}
	P_{\mathrm{fn}} \le \PP\bigg( \sum_{t=1}^T \log_2\frac{P_{Y|X_i,U_i}(Y_t \mid X_{ti},1)}{P_{Y}(Y_t)} \le T I_1 (1-\delta') \bigg),
\end{equation}
which is the probability of an i.i.d.~sum being a factor $1-\delta'$ below its mean.

In the very sparse regime $k = O(1)$, establishing the required concentration is straightforward -- it suffices to apply Chebyshev's inequality to conclude that $P_{\mathrm{fn}} \to 0$ for arbitrarily small $\delta'$.  We can then apply a union bound over the $k$ defective items to deduce that the probability of {\em any} false negatives vanishes, and we readily deduce \eqref{eq:sep_dec_T}.

The sparse regime $k = \Theta(n^{\alpha})$ is more challenging, and the choice of concentration inequality can differ depending on the specific noise model. We omit the details, which are given in \cite{Sca17b}, and merely state that in Theorem \ref{thm:separate}, the second result makes use of a general bound based on Bernstein's inequality, whereas the first result uses a sharper bound specifically tailored to the noiseless model.  

\section{Noisy (near-)definite defectives} \label{sec:ndd}

We saw in Chapter \ref{ch:algorithms} that the Definite Defectives (DD) algorithm (Algorithm \ref{def:dd}) achieves the best known rates of any practical algorithm in the noiseless setting.  As a result, there is substantial motivation for developing analogous algorithms in noisy settings.  Here we present such an algorithm, developed by Scarlett and Johnson \cite{scarlett-johnson}, which is suitable for noise models satisfying the noisy defective channel property (Definition \ref{def:ndc}), and is again practical in the sense of having $O(nT)$ decoding time.

Under Bernoulli testing with parameter $\nu > 0$, the algorithm accepts two parameters $(\gamma_1,\gamma_2)$ and proceeds as follows:
    \begin{enumerate}
    \item For each $i \in \{1,\dotsc,n\}$, let $\Nneg(i)$ be the number of negative tests in which item $i$ is included.  In the first step, we construct the following set of items that are believed to be nondefective:
    \begin{equation}
    \NDhat = \bigg\{ i  \,:\, \Nneg(i) > \frac{\gamma_1 T \nu}{k} \bigg\} \label{eq:NDhat}
    \end{equation}
    for some threshold $\gamma_1$.
    The remaining items, $\PDhat = \{1,\dotsc,n\} \setminus \NDhat$, are believed to be `possible defective' items.
    \item For each $j \in \PDhat$, let $\Ntilpos(j)$ be the number of positive tests that include item $j$ and no other item from $\PDhat$.  In the second step, we estimate the defective set as follows:
    \begin{equation}
    \Khat = \bigg\{ i \in \PDhat \,:\, \Ntilpos(i) > \frac{\gamma_2 T \nu \ee^{-\nu}}{k} \bigg\} \label{eq:Shat}
    \end{equation}
    for some threshold $\gamma_2$.
\end{enumerate}
In the noiseless case, setting $\gamma_1 = \gamma_2 = 0$ recovers the standard DD algorithm, Algorithm \ref{def:dd}.  For the addition noise model (Example \ref{ex:addition}), since negative test outcomes are perfectly reliable, one can set $\gamma_1 = 0$.  Similarly, for the Z channel noise model Example \ref{ex:Z}, since positive test outcomes are perfectly reliable, one can set $\gamma_2 = 0$.  In fact, one of the main goals of \cite{scarlett-johnson} was to show that these two noise models can behave quite differently in group testing despite corresponding to channels with the same Shannon capacity.

Using concentration of measure results, it is possible to give exponential tail bounds for error events corresponding to particular values of $\gamma_1$ and $\gamma_2$.  \added{This gives rise to a rate resembling the NCOMP one in \eqref{eq:NCOMP_improved}, but with some important modifications:
\begin{itemize}
    \item The maximum is over both $\gamma_1$ and $\gamma_2$, as well as an extra parameter $\xi$ discussed below.
    \item There are four terms in the minimization instead of two, with the added terms corresponding to the additional thresholding done in the second step.  Accordingly, the first two terms depend on $\gamma_1$ and the second two terms depend on $\gamma_2$.
    \item An extra parameter $\xi \in (0,\alpha)$ is introduced to represent the existence of at most $n^{\xi}$ `false positives' in the first step, which is now permissible as long as the second step corrects them.
    \end{itemize}
By balancing these four terms, it is shown in \cite{scarlett-johnson} that at least in certain cases, one can find expressions for the optimal values of these parameters, with the help of the  Lambert W function.  The interested reader is referred to \cite{scarlett-johnson} for the rate expressions both before and after optimizing parameters.  We will provide a rate plot for the symmetric noise model (Example \ref{ex:bsc}) in Section \ref{sec:noisy_rates}, along with some numerical experiments.}

The strongest results among those in \cite{scarlett-johnson} are for the addition noise model (Example \ref{ex:Z}), in which the achievability curve matches an algorithm-independent converse for Bernoulli testing for a wide range of $\alpha \in (0,1)$.  In contrast, wider gaps are observed for the Z channel and symmetric noise models.  
For each of these models, the rate converges to the noiseless DD rate (Section \ref{sec:DD}) in the low noise limit, though the convergence can be rather slow, with visible gaps remaining even for low noise levels such as $0.001$.

\section{Improved rates with constant tests-per-item} \label{sec:near_const_noisy}

\added{Given the significant benefits of the near-constant tests-per-item design in the noiseless setting, it is natural to ask whether this design also outperforms the Bernoulli design in noisy settings.  This question was addressed by Gebhard \etal~\cite{gebhard2021improved}, though they chose to adopt the closely-related \emph{constant tests-per-item design}, defined as follows.

\begin{definition} \label{def:const_col_2}
    The {\em constant column weight} (or {\em constant tests-per-item}) design with parameter $\nu > 0$ forms a group testing matrix $\mat X$ in which $L = \nu T/k$ entries of each column are selected uniformly at random \emph{without} replacement and set to one, with independence between columns.  The remaining entries of $\mat{X}$ are set to zero.
  \end{definition}

The only difference between this and Definition \ref{def:const_col} is that the sampling here is done without replacement, so all columns have weight exactly $L$.  

We first state the improved NCOMP bound from \cite{gebhard2021improved}, which serves as a counterpart to Theorem \ref{thm:NCOMP_bern_improved}.

\begin{theorem} \label{thm:NCOMP_improved}
    Consider the NCOMP decoding rule \eqref{eq:ncomp_rule} under the constant column weight design with parameter $L = \nu T/k$, and suppose that $k = \Theta(n^{\alpha})$ for some $\alpha \in (0,1)$.  Moreover, consider the noise model from Example \ref{ex:binary_channel} in which
    tests with no defectives are flipped to $\one$ with probability $\rho_{01}$ and
    tests with defectives are flipped to $\zero$ with probability $\rho_{10}$,
    where the noise parameters $(\rho_{01},\rho_{10})$ satisfy $\rho_{01} + \rho_{10} < 1$.  Then, the following rate is achieved: 
    \begin{equation}
        R_{\rm CC}^{\mathrm{NCOMP}} = \frac{(1-\alpha)\nu}{\ln 2} \max_{\gamma \in (\rho_{10},w)} \min\bigg\{ \frac{1}{\alpha} \, d(\gamma \mmid \rho_{10}), \, d(\gamma \mmid w) \bigg\}, \label{eq:NCOMP_improved_NCC}
    \end{equation}
    where $w = e^{-\nu}(1-\rho_{01}) + (1-e^{-\nu})\rho_{10}$, and 
    \[ d(p \mmid q) = p\log\frac{p}{q} + (1-p)\log\frac{1-p}{1-q} \]
    is the relative entropy between a ${\rm Bernoulli}(p)$ and a ${\rm Bernoulli}(q)$ random variable.
\end{theorem}

The terms in \eqref{eq:NCOMP_improved_NCC} share exactly the same interpretation as those in \eqref{eq:NCOMP_improved}, but turn out to be different due to the different test design.  It is shown in \cite{gebhard2021improved} that 
\begin{equation}
    R_{\rm CC}^{\mathrm{NCOMP}} \ge R_{\rm Bern}^{\mathrm{NCOMP}},    
\end{equation}
meaning that this achievable rate is always at least as high as its counterpart for the Bernoulli design.

The main result of \cite{gebhard2021improved} is an achievable rate for the noisy DD algorithm, which is omitted here but again shares a similar form and interpretation to its Bernoulli counterpart discussed in Section \ref{sec:ndd}.  It is shown that the bound is at least as strong as its Bernoulli counterpart under the Z channel noise (Example \ref{ex:Z}).  It is not yet known whether such a result also holds for more general noise models (such as symmetric or addition noise), but numerical evaluations suggest that this may be the case.
}


\section{Rate comparisons and numerical simulations} \label{sec:noisy_rates}

In this section, we compare the achievable rates of the algorithms considered throughout this chapter, as well as comparing the algorithms numerically.
We focus here on the symmetric noise model (Definition \ref{ex:bsc}) with parameter $\rho \in \big(0,\frac12\big)$, since it has received the most attention in the context of proving achievable rates for noisy group testing algorithms. 

\begin{figure}[t] 
\begin{center}
\includegraphics[width=0.95\textwidth]{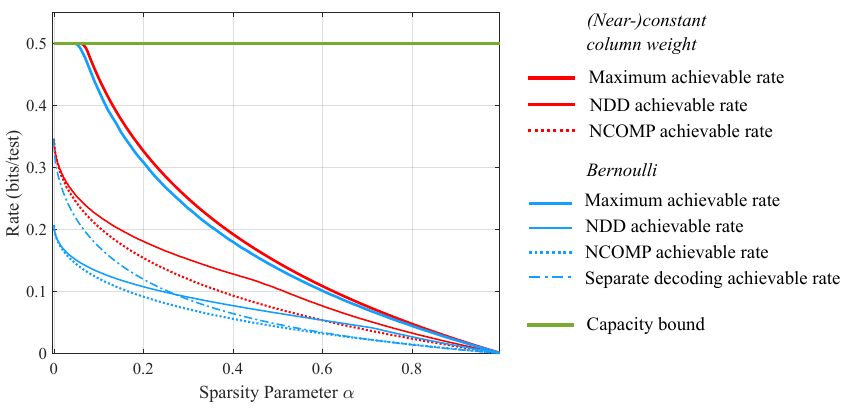}
\end{center}
\caption{Achievable rates for the symmetric noise model with noise level $\rho = 0.11$.  The `maximum achievable rate' curves are discussed in Section \ref{sec:high_k_ach}.} \label{fig:noisy_rates} 
\end{figure}

\paragraph{Rate comparisons} \added{In Figure \ref{fig:noisy_rates}, we plot the achievable rates of NCOMP (the improved rates from Theorems \ref{thm:NCOMP_bern_improved} and \ref{thm:NCOMP_improved}), separate decoding of items, and noisy DD with noise level $\rho = 0.11$, under both the Bernoulli and near-constant column weight designs (except separate decoding whose existing analysis is specific to the Bernoulli design).  We optimize the test design parameter $\nu > 0$ separately for each curve, except separate decoding since its rate was proved specifically for $\nu = (\ln 2)(1+o(1))$.  We also plot the best possible rates for these designs \cite{chen2024exact,coja2024noisy}, which we will cover in Section \ref{sec:high_k_ach}. 

We observe that at least in the setting shown, the rates for separate decoding of items and noisy DD are uniformly stronger than the rate proved for NCOMP.  For the Bernoulli design, separate decoding attains a higher rate for small $\alpha$, whereas noisy DD attains a higher rate for large $\alpha$.  While we only showed a single noise level here, we note that the benefits of noisy DD compared to other algorithms have been seen to increase at lower noise levels \cite{scarlett-johnson,gebhard2021improved}.}




 \begin{figure}[t] 
     \begin{center}
        \includegraphics[width=0.6\textwidth]{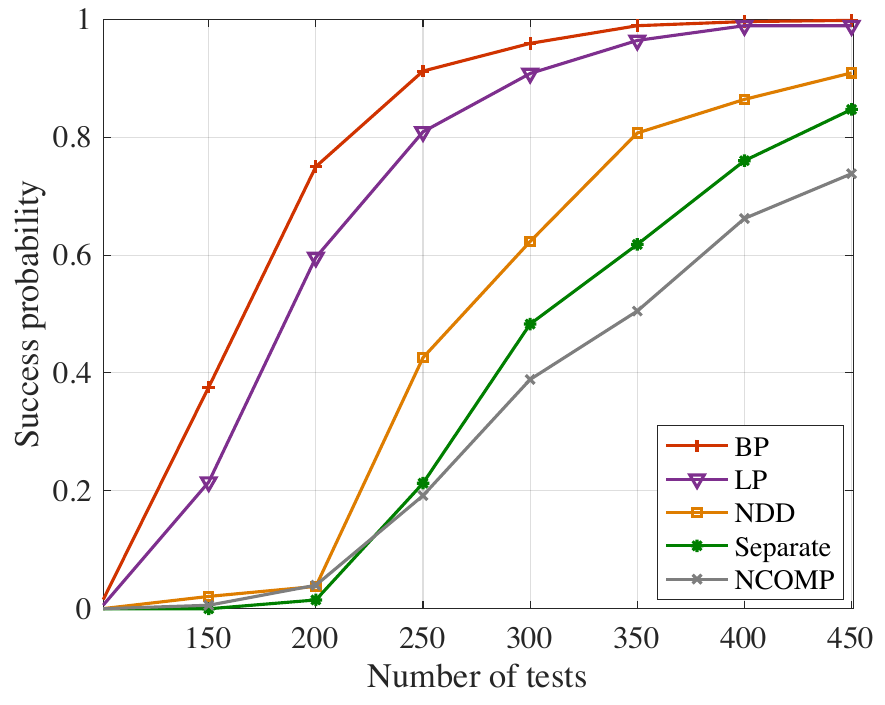}
    \end{center}
    \caption{Experimental simulations for the symmetric noise model under Bernoulli testing with parameter $\nu = \ln 2$, with $n=500$ items, $k=10$ defectives, and noise parameter $\rho = 0.05$.} \label{fig:noisy_sim1} 
\end{figure}

\begin{figure}[t] 
     \begin{center}
        \includegraphics[width=0.6\textwidth]{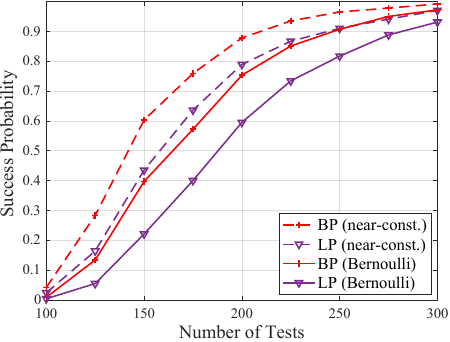}
    \end{center}
    \caption{Comparison of the Bernoulli and near-constant column weight designs under the same setup as that of Figure \ref{fig:noisy_sim1}.} \label{fig:ExperimentsNCC} 
\end{figure}

\begin{figure}[t] 
     \begin{center}
        \includegraphics[width=0.6\textwidth]{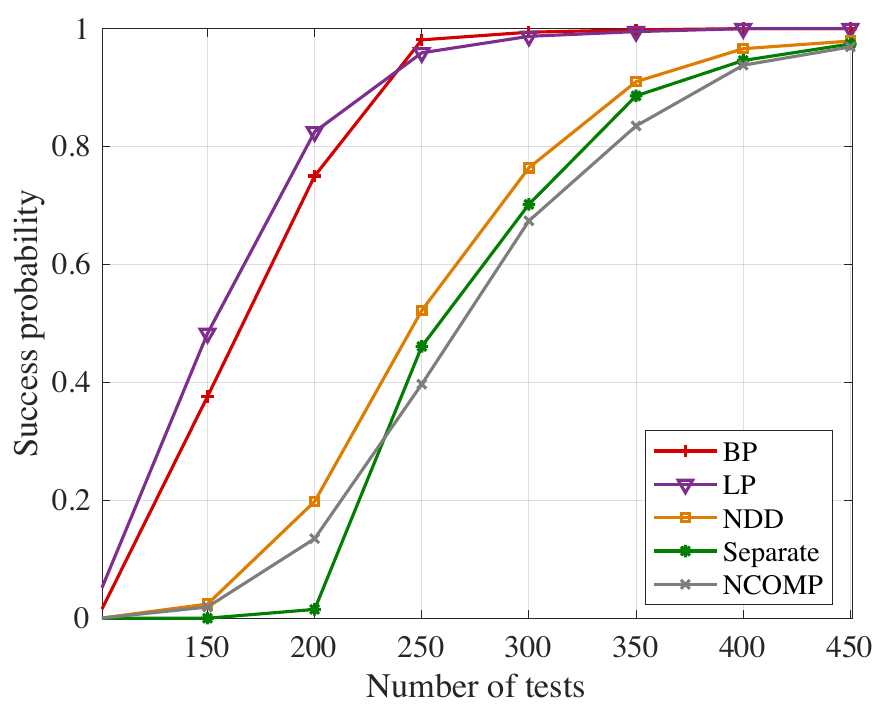}
    \end{center}
    \caption{Performance of oracle versions of the respective algorithms under the same setup as that of Figure \ref{fig:noisy_sim1}.} \label{fig:noisy_sim2} 
\end{figure}

\paragraph{Numerical simulations} 
In Figure \ref{fig:noisy_sim1}, we plot experimental simulation results under the symmetric noise model (Example \ref{ex:bsc}) with parameter $\rho = 0.05$, and with $n=500$ items and $k=10$ defectives.
We first consider i.i.d.~Bernoulli testing with parameter $\nu = \ln 2$, along with the following decoding rules:
\begin{itemize}
    \item Noisy linear programming (LP) as described in Section \ref{sec:noisy_LP}, with parameter $\lambda = 0.5$ and each $u_i$ rounded to the nearest integer in $\{0,1\}$;
    \item Belief propagation (BP) as described in Section \ref{sec:belprop}, with 10 message passing iterations;
    \item NCOMP as described in Section \ref{sec:NCOMP} (specifically, see \eqref{eq:ncomp_rule}), with $\Delta = \frac{0.1(1-2\rho)}{\rho}$ based on the theoretical choice in \cite{chan-etal-1} along with some manual tuning of the constant factor;
    \item Separate decoding of items as described in Section \ref{sec:separate}, with $\gamma = (1-\delta) I(X_1;Y)$ in accordance with the theoretical analysis, and $\delta$ chosen based on manual tuning to be $\frac{1}{3}$;
    \item Noisy DD as described in Section \ref{sec:ndd}, with parameters $\gamma_1 = \gamma_2 = 0.175$ based on manual tuning.
\end{itemize}
\added{(In Section \ref{sec:high_k_ach} we will discuss a spatially coupled test design and polynomial-time decoder that achieves the optimal rate \cite{coja2024noisy}; it is omitted here since, to our knowledge, it is currently a construction that exists for theoretical purposes but has not been implemented experimentally.)}

We observe that BP performs best in this setup, followed closely by LP.  There is then a larger gap to noisy DD and separate decoding, and finally NCOMP requires the most tests.  While our experiments are far from being an exhaustive treatment, these results indicate somewhat of a gap between the current theory and practice, with the best-performing methods (BP and LP) also being the least well-understood from a theoretical point of view.  Closing this gap remains an interesting direction for further research.

\added{Next, we compare the Bernoulli design to the near-constant column weight design. \added{(We found the near-constant column weight and exactly-constant column weight designs to have nearly identical performance, so we only show the former.)}
To avoid clutter, we focus on the two best-performing decoders, namely, LP and BP.  The results are shown in Figure \ref{fig:ExperimentsNCC}, in which we use the same parameters as those above.  Similarly to the noiseless setting, we see that the near-constant column weight design is able to provide clear improvements over the Bernoulli design.}

\paragraph{Oracle versions} Finally, to better understand the impact of knowledge of $k$ in noisy group testing, in Figure \ref{fig:noisy_sim2}, we repeat the experiment of Figure \ref{fig:noisy_sim1} with `oracle' versions of the algorithms for the case that the number of defectives $k$ is known:
\begin{itemize}
    \item Noisy LP includes the additional constraint that the estimated defectivity indicator variables $z_i$ sum to $k$.
    \item Instead of thresholding, BP chooses the $k$ items with the highest estimated probabilities of being defective.
    \item NCOMP takes the $k$ items for which the proportions of positive tests (relative to those the item is included in) are highest.
    \item Separate decoding of items chooses the $k$ items with the highest sum of log-probability ratios (see \eqref{eq:separate_dec}).
    \item Noisy DD estimates the `possible defectives' to be the set of $(1+\Delta)k$ items in the lowest number of negative tests, where we set $\Delta = \frac{1}{2}$ based on manual tuning.  The algorithm then estimates the defective set to be the set of $k$ items with the highest number of positive tests in which it is the unique possible defective.
\end{itemize} 
We observe that knowledge of $k$ brings the performance of NCOMP, separate decoding, and noisy DD closer together, but generally maintains their relative order.  On the other hand, the performance of LP improves more than that of BP, making it become the best performing algorithm for most values of $T$. 
    

\chapter{Information-Theoretic Limits} \label{ch:achievability}

In this chapter, we present information-theoretic achievability and converse bounds characterizing the fundamental limits of group testing regardless of the computational complexity.  We have already seen a few converse results in the previous chapters, including the counting bound (Theorem \ref{thm:bjaconverse}) in the noiseless setting, and a capacity-based bound for noisy settings (Theorem \ref{thm:capcap}).

The main results presented in this chapter are as follows:
\begin{itemize}
    \item an achievable rate for the noiseless setting under Bernoulli testing, which matches or improves on all the algorithms considered in Chapter \ref{ch:algorithms} (see the discussion in Section \ref{sec:achmain} and the details in Section \ref{sec:pf_ach});
    \item a matching converse bound for the noiseless setting establishing the exact maximum achievable rate of nonadaptive testing with a Bernoulli design (Section \ref{sec:algcon});
    \item \rev{an improved achievable rate for the noiseless setting under a near-constant column weight design (see the discussion in Section \ref{sec:achmain} and the details in \ref{sec:ach_near_const});}
    \item analogous achievability and converse bounds for noisy settings under the Bernoulli design, and applications to specific models (Section \ref{sec:noisy}).
\end{itemize}

\section{Overview of the standard noiseless model}\label{sec:achmain}

\rev{The first two major results in this chapter give achievable rates for noiseless nonadaptive group testing with two different designs.
Theorem \ref{thm:ch2main}, due to Scarlett and Cevher \cite{scarlett-cevher-1,scarlett-cevher-2}, concerns the Bernoulli design (see Definition \ref{def:berndesign}), and Theorem \ref{thm:ch2nearconst}, due to Coja-Oghlan \etal\ \cite{coja}, concerns the near-constant column weight design (see Definition \ref{def:const_col}).}

\begin{theorem} \label{thm:ch2main}
Consider noiseless nonadaptive group testing, under the exact recovery criterion in the small error setting, and $k = \Theta(n^\alpha)$ defectives with $\alpha \in (0,1)$. Then the rate 
  \begin{equation} \label{bern-ach}
    \olR_{\rm Bern} = \max_{\nu > 0} \min \left\{ h(\mathrm \ee^{-\nu}),\, \frac{\nu\ee^{-\nu}}{ \ln 2} \frac{1-\alpha}{\alpha} \right\} 
  \end{equation}
is achievable, and can be achieved by a Bernoulli test design.
\end{theorem}

\rev{
\begin{theorem} \label{thm:ch2nearconst}
Consider noiseless nonadaptive group testing, under the exact recovery criterion in the small error setting, and $k = \Theta(n^\alpha)$ defectives with $\alpha \in (0,1)$. Then the rate 
  \begin{equation} \label{ncc-ach}
    \olR_{\rm NCC} =  \min\bigg\{1,(\ln 2)\frac{1-\alpha}{\alpha}\bigg\}
  \end{equation}
is achievable, and can be achieved by a near-constant column weight design with $\nu = \ln 2$.  

\added{This rate corresponds to a number of tests satisfying the following for arbitrarily small $\eta > 0$: $T \ge (1+\eta) \max\big\{ k \log_2 \frac{n}{k}, \frac{1}{(\ln 2)^2} k \ln k \big\} .$}
\end{theorem}}

\begin{figure}[t] 
\begin{center}
\includegraphics[width=\textwidth]{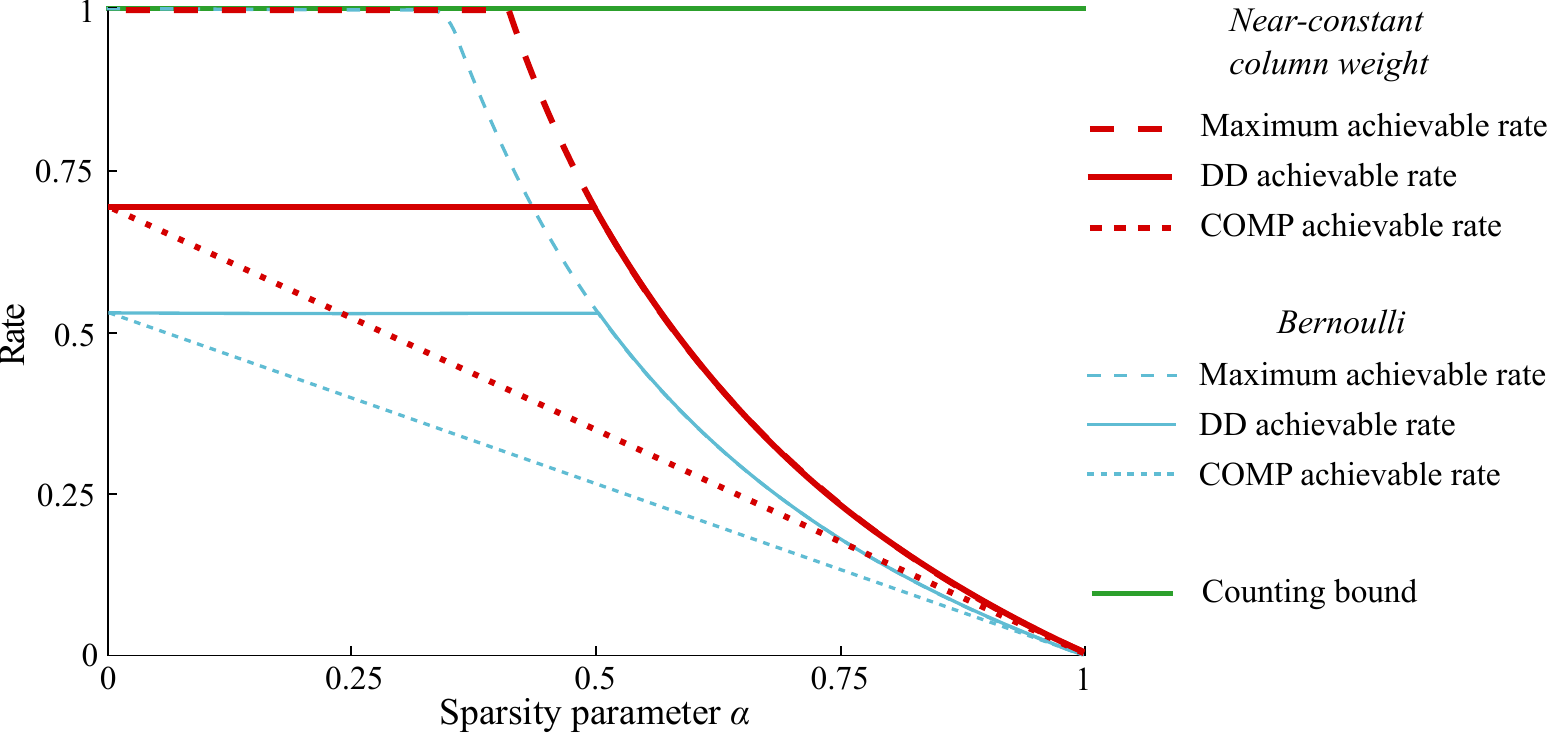}
\end{center}
\caption{Rate for nonadaptive group testing in the sparse regime with a Bernoulli design and with a near-constant column weight designs.} \label{fig:algrates2} 
\end{figure}

\added{
We will shortly see (in Figure \ref{fig:algrates2}) that $\olR_{\rm NCC} \ge \olR_{\rm Bern}$, with strict inequality when $\alpha > \frac{1}{3}$.  
Moreover, in Theorems \ref{thm:bern_conv} and \ref{thm:ncc_conv} later in the chapter, we will further see that these rates are the best possible for their respective designs, by establishing \emph{design-specific converse bounds}.  Thus, \eqref{bern-ach} and \eqref{ncc-ach} serve as exact thresholds giving the best possible rates for these designs.  

However, these results collectively still fall short of answering two important questions: (i) Are there other nonadaptive designs that can achieve even better rates? (ii) For $\alpha < \frac{1}{2}$, can the stronger rate $\olR_{\rm NCC}$ be achieved using a computationally efficient decoder?  (For $\alpha \ge \frac{1}{2}$ it is achieved by the DD algorithm; see Theorem \ref{DDub2}.) 
These questions were completely answered by Coja-Oghlan \etal~\cite{coja2020optimal}, with Theorem \ref{thm:gen_conv} below answering question (i) in the negative by showing that $\olR_{\rm NCC}$ is optimal, and Theorem \ref{thm:spatial} below answering question (ii) in the affirmative as long as a modified test design is allowed.  Section \ref{sec:computation} outlines the challenges of attaining optimal rates for $\alpha < \frac{1}{2}$ with efficient decoding using the Bernoulli and near-constant column weight designs.}

\added{
\begin{theorem} \label{thm:gen_conv}
    Consider noiseless nonadaptive group testing, under the exact recovery criterion in the small error setting, and $k = \Theta(n^\alpha)$ defectives with $\alpha \in (0,1)$.  Then regardless of the test design and decoding strategy, it is impossible to achieve any rate higher than $\olR_{\rm NCC}$ defined in \eqref{ncc-ach}. 
\end{theorem}

\begin{theorem} \label{thm:spatial}
    Consider noiseless nonadaptive group testing, under the exact recovery criterion in the small error setting, and $k = \Theta(n^\alpha)$ defectives with $\alpha \in (0,1)$.  Then there exists a test design and decoder that attains the rate $\olR_{\rm NCC}$ defined in \eqref{ncc-ach}, and whose decoding time is polynomial with respect to $n$.
\end{theorem}

}

\rev{Theorem \ref{thm:ch2main} is proved in Section \ref{sec:pf_ach} using information-theoretic methods akin to those used in studies of channel coding. We dedicate a large section of this chapter to the study of this proof, as the information theory approach is a powerful and flexible method that can be applied to other sparse inference problems (see \cite{scarlett-cevher-1}), and in particular to noisy group testing models (see Section \ref{sec:noisy}). The proof of Theorem \ref{thm:ch2nearconst} uses a more direct probabilistic method to show that there exists only one satisfying set (see Definition \ref{def:sat}) with high probability -- arguably a simpler strategy, but one that may be harder to generalize to other models. We discuss this proof in Section \ref{sec:ach_near_const}.} \added{The proof of Theorem \ref{thm:gen_conv} is outlined in Section \ref{sec:gen_conv}, and is based on the idea of identifying many items that are \emph{masked} in the sense that they only appear in tests that contain other defectives.  The proof of Theorem \ref{thm:spatial} is relatively complicated, so in Section \ref{sec:spatial} we primarily focus on describing the test design and decoding rule, only giving hints on the proof techniques.  The test design is based on the idea of \emph{spatial coupling}; while this is a well-known technique in error-correcting coding, its application to group testing is far from trivial.  }  

\paragraph{Discussion of the rates} \rev{The rates are shown in Figure \ref{fig:algrates2}, which is a repeat of Figure \ref{fig:algrates} included here for convenience. We see that for $\alpha \leq 1/3$, both theorems give an equal rate of $1$, while for $\alpha > 1/3$, the rate \eqref{ncc-ach} for near-constant column weight designs is slightly higher than the rate \eqref{bern-ach} for Bernoulli designs, in particular equalling $1$ for $\alpha \le 0.409$.}

The rate expression in \eqref{bern-ach} is somewhat complicated. It will become apparent in the forthcoming proof of Theorem \ref{thm:ch2main} that the parameter $\nu$ enters through the choice of the Bernoulli parameter as $p = \nu/k$. It is easy to see that the first minimand of \eqref{bern-ach} is maximized at $\nu = \ln 2$, and is the value of $p$ that corresponds to (asymptotically) half of the tests being positive. By differentiation, we see that the second minimand of \eqref{bern-ach} is maximized at $\nu = 1$, which corresponds to $p = 1/k$, and is the value of $p$ that corresponds to an average of one defective per test. Using these findings, we can check that the following simplification of \eqref{bern-ach} holds:
  \[ \olR_{\rm Bern} = \begin{cases} 1 & \text{for }\alpha \leq 1/3 \\
       \text{as in \eqref{bern-ach}} & \text{for }1/3 < \alpha < 0.359 \\
       \displaystyle 0.531\, \frac{1 - \alpha}{\alpha} & \text{for }\alpha \ge 0.359,\end{cases} \]
where it should be understood that the decimal values are non-exact (rounded to three decimal places).
       
\rev{The near-constant column weight design with $L = \nu T/k$ tests per item is always optimized with $\nu = \ln 2$, which corresponds to (asymptotically) half of the tests being positive. This makes the expression \eqref{ncc-ach} simpler, and we have
\[  \olR_{\rm NCC}  = \begin{cases} 1 & \text{for $\alpha \leq 0.409$,} \\ 0.693 \,\displaystyle\frac{1-\alpha}{\alpha} & \text{for $\alpha > 0.409$.} \end{cases} \]
       
Thus, we see that nonadaptive group testing achieves the rate $1$ of the counting bound (see Section \ref{sec:counting_bound}) and has the same rate as adaptive testing for $\alpha \leq 1/3$ with a Bernoulli design and for $\alpha \leq 0.409$ with a near-constant column weight design.
On the other hand, for $\alpha$ above these thresholds, the rates are strictly below the counting bound.}
\added{Since the counting bound can be achieved for all $\alpha \in (0,1)$ using adaptive designs (Section \ref{sec:adaptive}) and the near-constant column weight design attains the optimal rate among nonadaptive designs (Theorem \ref{thm:gen_conv}), we conclude that there is a strict gap between the optimal adaptive and nonadaptive rates for $\alpha > 0.409$.}

\paragraph{Other related work} Before continuing, we briefly review work on achievable rates for noiseless nonadaptive group testing that preceded \rev{Theorems \ref{thm:ch2main} and \ref{thm:ch2nearconst}} (although these papers did not necessarily phrase their results this way). We begin with results using Bernoulli test designs.

Freidlina \cite{friedlina} and Malyutov \cite{malyutov-1} showed that a rate of $1$ is achievable in the very sparse regime where $k$ is constant as $n \to \infty$. Malyutov used an information-theoretic approach based on a multiple access channel model with one input for each defective item.  Seb\H{o} \cite{sebo} also attained a rate of $1$ for constant $k$ using a more direct probabilistic method. 

Atia and Saligrama \cite{atia-saligrama} reignited interest in the use of information-theoretic methods for studying group testing. They used a model of channel coding with correlated codewords, where each potential defective set is a message (recall the channel coding interpretation of group testing shown in Figure \ref{fig:channel_diagram}). Atia and Saligrama showed that, in the limiting regime where $k \to \infty$ after $n \to \infty$, one can succeed with $T = O(k \log n)$ tests, although they did not specify the implicit constant. Effectively, in our notation, this shows a nonzero rate for $\alpha = 0$, but does not prove the  Freidlina--Malyutov--Seb\H o\ rate of $1$. They also showed that  $T = O(k \log n \, \log^2 k)$ suffices for any $k = o(n)$, though this falls short of proving a nonzero rate for $\alpha \in (0,1)$. They also gave order-wise results for some noisy models. A similar approach to Atia and Saligrama was taken by Scarlett and Cevher \cite{scarlett-cevher-1} (outlined below), but with tighter analysis and careful calculation of constants giving the better rates of Theorem \ref{thm:ch2main}.

Aldridge, Baldassini, and Gunderson \cite{aldridge-baldassini-gunderson} generalized Seb\H o's approach to all $\alpha \in [0,1)$, showing a nonzero rate for all $\alpha$ that achieves the rate of $1$ at $\alpha = 0$, but that is suboptimal compared to \eqref{bern-ach} for $\alpha \in (0,1)$.

In Chapter \ref{ch:algorithms} of this monograph, we saw some rates that can be achieved with practical algorithms. Chan \etal\ \cite{chan-etal-1, chan-etal-3} were the first to show a nonzero rate for all $\alpha \in (0,1)$, albeit one that is suboptimal compared to \eqref{bern-ach}, by analysing the COMP algorithm (Theorem \ref{COMPub1}). They also showed nonzero rates for some non-Bernoulli designs. The DD algorithm of Aldridge, Baldassini, and Johnson \cite{aldridge-baldassini-johnson} also achieves nonzero rates for all $\alpha \in (0,1)$ with Bernoulli testing, in particular matching \eqref{bern-ach} for $\alpha > 1/2$ (Theorem \ref{DDub}).  

\rev{We also saw in Section \ref{sec:near_constant} that the performance of these algorithms is improved when used with the near-constant column weight design. In particular, the DD algorithm achieves the same rate as \eqref{ncc-ach} for $\alpha \ge 1/2$, as shown by Johnson, Aldridge and Scarlett \cite{johnson-aldridge-scarlett}. We direct the reader back to Chapter \ref{ch:algorithms} for detailed discussions of these results and other algorithms.

M\'ezard, Tarzia and Toninelli \cite{mezard} had suggested that Theorem \ref{thm:ch2nearconst} should be true by appealing to heuristics from statistical physics -- the innovation of Coja-Oghlan \etal\ \cite{coja} was to prove this rigorously.}
  

\section{Proof of achievable rate for Bernoulli testing} \label{sec:pf_ach}

\subsection{Discussion of proof techniques} \label{sec:disc_info_dens}

Our proof follows Scarlett and Cevher \cite{scarlett-cevher-1}, who proved Theorem \ref{thm:ch2main} as a special case of a more general framework for noiseless and noisy group testing.  The analysis is based on {\em thresholding techniques} that are rooted in  early information-theoretic works, such as \cite{Fei54,Sha57}, as well as recent developments in information-spectrum methods \cite{Han03}.  In fact, we also saw a simpler version of this approach when studying separate decoding of items in Section \ref{sec:separate}.

To describe these methods in more detail, we momentarily depart from the group testing problem and consider a simple channel coding scenario where $M$ codewords are drawn from some distribution $P_{\vec{X}}$, and one of them is transmitted over a channel $P_{\vec{Y}|\vec{X}}$ to produce an output sequence $\vec{y}$.  The optimal (yet generally computationally intractable) decoding rule chooses the codeword $\vec{x}$ maximizing the likelihood $P_{\vec{Y}|\vec{X}}(\vec{y}|\vec{x})$, and the resulting error probability is upper bounded by the probability that the true codeword is the only one such that $\log_2\frac{ P_{\vec{Y}|\vec{X}}(\vec{y}|\vec{x}) }{ P_{\vec{Y}}(\vec{y}) }$ exceeds a suitably-chosen threshold.  Intuitively, we should expect $P_{\vec{Y}|\vec{X}}(\vec{y}|\vec{x})$ to be considerably larger than $P_{\vec{Y}}(\vec{y})$ when $\vec{x}$ is the true transmitted codeword, whereas if $\vec{x}$ is an incorrect codeword then this is unlikely to be the case.

More precisely, by a simple change of measure technique, the probability of $\log_2\frac{ P_{\vec{Y}|\vec{X}}(\vec{y}|\vec{x}) }{ P_{\vec{Y}}(\vec{y}) }$ exceeding any threshold $\gamma$ for a single non-transmitted codeword is at most $2^{-\gamma}$, and hence the union of these events across all $M-1$ non-transmitted codewords has probability at most $(M-1)2^{-\gamma}$.  Choosing $\gamma$ slightly larger than $\log_2 M$ ensures that this probability is small, and hence the error probability is roughly the probability that the true codeword $\vec{x}$ fails the threshold test, i.e., $\log_2\frac{ P_{\vec{Y}|\vec{X}}(\vec{y}|\vec{x}) }{ P_{\vec{Y}}(\vec{y}) } < \gamma \approx \log_2 M$.

Finally, for a memoryless channel taking the form $P_{\vec{Y}|\vec{X}}(\vec{y}|\vec{x}) = \prod_{i=1}^n P_{Y|X}(y_i|x_i)$, and an i.i.d.~codeword distribution of the form $P_{\vec{X}}(\vec{x}) = \prod_{i=1}^n P_{X}(x_i)$, the quantity $\log_2\frac{ P_{\vec{Y}|\vec{X}}(\vec{y}|\vec{x}) }{ P_{\vec{Y}}(\vec{y}) }$ concentrates about its mean $n I(X;Y)$.  As a result, we get vanishing error probability when the number of codewords satisfies $M \lesssim 2^{n I(X;Y)}$, and we can achieve any coding rate up to the mutual information $I(X;Y)$.

For group testing, we follow the same general idea, but with a notable change: the `codewords' (that is, the $T \times k$ submatrices $\mat{X}_{\K}$ of the test matrix for $\K$ of cardinality $k)$ are not independent. For example, the codewords corresponding to $\K_1 = \{1,2,3\}$ and $\{\K_2\} = \{1,4,7\}$ have a common first column.  To handle this issue, we treat different incorrect codewords separately depending on their amount of overlap with the true codeword: If there is no overlap then the analysis is similar to that of channel coding above, while if there is overlap then we consider probabilities of the form $\log_2\frac{P_1(\vec{y}|\cdot)}{P_2(\vec{y}|\cdot)}$, where $P_1$ conditions on the true codeword, and $P_2$ only conditions on the overlapping part.

We now proceed with the proof of Theorem \ref{thm:ch2main}.  We first introduce some notation that will allow the initial steps to be re-used for the noisy setting, then formally specify the decoder used, provide the non-asymptotic bound that forms the starting point of the analysis, and finally, outline the subsequent asymptotic analysis that leads to the final result.

\subsection{Information-theoretic notation} \label{eq:it_prelim}

While our focus is primarily on the noiseless setting, the initial steps of the analysis are just as easily done simultaneously for general noise models.  Specifically, we consider an arbitrary model studying the `only defects matter' property, given in Definition \ref{def:odm}.  Due to this property and the symmetry in the random construction of ${\mat X}$, the analysis will not be impacted by the realization of $\K$, and we will therefore set $\K = \{1,\dotsc,k\}$ without loss of generality.

We now introduce some notation.  We again consider the Bernoulli design (Definition \ref{def:berndesign}), in which each item is included in each test independently with probability $p$.  For convenience, we write $p = \nu/k$ for some $\nu > 0$, and as usual the $T \times n$ test matrix is denoted by $\mat X$.

The submatrix ${\mat X}_{\K}$ denotes only the columns of the matrix ${\mat X}$ indexed by $\K$, and ${\vec X}_{\K}$ denotes a single row of ${\mat X}_{\K}$.
We write $V = V({\vec X}_\K)$ for the random number of defective items in the test indicated by ${\vec X}$.

The observation $Y \in \{0, 1\}$ is generated according to some general distribution $P_{Y | {\vec X}_{\K}}$ depending on ${\vec X}_{\K}$ only through $V$:
    \begin{equation}
        (Y \mid \vec{X}, \K) \sim P_{Y|{\vec X}_{\K}} = P_{Y|V} . \label{eq:Y_noisy}
    \end{equation}
    This is precisely the only defects matter property of Definition \ref{def:odm}.
     The $T$-fold product of $P_{Y|{\vec X}_{\K}}$ gives the distribution of the overall test vector ${\vec Y} = (Y_1,\dotsc,Y_T)$ given ${\mat X}_{\K}$, and is denoted by $P_{{\vec Y} | {\mat X}_{\K}}$.

As discussed above, we consider separate error events according to how much an incorrect defective set $\K'$ overlaps with $\K$.  To facilitate this, for a given partition $(S_0,S_1)$ of $\K$, we write
\begin{equation}
    P_{Y|{\vec X}_{S_0}, {\vec X}_{S_1}}(y \mid {\vec x}_{S_0}, {\vec x}_{S_1}) = P_{Y|\vec{X}_{\K}}(y \mid {\vec x}_{\K}), \label{eq:S0S1}
\end{equation}
and use this to define the marginal distribution
\begin{equation}
    P_{Y \mid {\vec X}_{S_1}}(y | {\vec x}_{S_1}) = \sum_{ {\vec x}_{S_0} } P_{{\vec X}_{S_0}}( {\vec x}_{S_0} ) P_{Y|{\vec X}_{S_0}, {\vec X}_{S_1}}(y \mid {\vec x}_{S_0}, {\vec x}_{S_1}),
\end{equation}
where $({\vec x}_{S_0}, {\vec x}_{S_1}, y)$ is a specific realization of $({\vec X}_{S_0}, {\vec X}_{S_1}, Y)$.
In the analysis, $S_1$ will represent the intersection $\K \cap \K'$ between the defective set $\K$ and some incorrect set $\K'$, whereas $S_0$ will represent the set difference $\K \setminus \K'$.

Finally, in accordance with the techniques outlined in Section \ref{sec:disc_info_dens}, we define the {\em information density}
    \begin{equation}
        \imath( {\vec X}_{S_0}; Y \mid {\vec X}_{S_1} ) = \log_2 \frac{ P_{Y|{\vec X}_{S_0}, {\vec X}_{S_1}}(y \mid {\vec X}_{S_0}, {\vec X}_{S_1}) }{ P_{Y|{\vec X}_{S_1}}(Y \mid {\vec X}_{S_1}) }, \label{eq:info_dens}
    \end{equation}
    and let $\imath^T( {\mat X}_{S_0}; {\vec Y} | {\mat X}_{S_1} )$ be the $T$-letter extension obtained by summing \eqref{eq:info_dens} over the $T$ tests.  Since the tests are independent, writing the sum of logarithms as the logarithm of a product yields
    \begin{equation}
        \imath^T( {\mat X}_{S_0}; {\vec Y} \mid {\mat X}_{S_1} ) = \log_2 \frac{ P_{{\vec Y} | {\mat X}_{S_0}, {\mat X}_{S_1}}({\vec Y} \mid {\mat X}_{S_0}, {\mat X}_{S_1}) }{ P_{{\vec Y}|{\mat X}_{S_1}}({\vec Y} \mid {\mat X}_{S_1}) }. \label{eq:info_dens_T}
    \end{equation}
    We also note that the expectation of \eqref{eq:info_dens} is equal to the conditional mutual information $I({\vec X}_{S_0}; Y \mid {\vec X}_{S_1})$, and the expectation of \eqref{eq:info_dens_T} is equal to $T \cdot I({\vec X}_{S_0}; Y \mid {\vec X}_{S_1})$.

\subsection{Choice of decoder}

Inspired by classical information-theoretic works such as \cite{Fei54} (again see Section \ref{sec:disc_info_dens}), we consider a decoder that searches for a defective set $\K \subseteq \{1,\dotsc,n\}$ of cardinality $k$ such that
\begin{equation}
    \imath^T( {\mat X}_{S_0}; {\vec Y} \mid \mat X_{S_1} ) > \gamma_{|S_0|} \quad \text{for all $(S_0,S_1)$ partitioning $\K$ with $S_0 \ne \emptyset$} \label{eq:decoder}
\end{equation}
for suitable constants $\gamma_1,\dotsc,\gamma_K$ to be chosen later.   If no such set exists, or if multiple sets exist, then an error is declared.

The rule \eqref{eq:decoder} can be viewed as a weakened version of the maximum-likelihood (ML) rule -- that is, the decoder that chooses the set $\K$ maximizing $P_{{\vec Y} | {\mat X}_{\K}}$.  Specifically, if a unique set satisfies \eqref{eq:decoder}, it must be the ML choice, whereas sometimes the ML decoder might succeed where the above decoder fails -- for example, in cases where no $\K$ passes all $2^k - 1$ of its threshold tests.
    
The above decoder is unlikely to be computationally feasible in practice even for moderate problem sizes.  The focus in this section is on information-theoretic achievability regardless of such considerations.  Moreover, while this rule requires knowledge of $k$, we argue in Section \ref{sec:algcon} that at least in the noiseless setting, the resulting rate can be achieved even without this knowledge.

\subsection{Non-asymptotic bound}

Observe that in order for an error to occur, it must be the case that either the true defective set $\K = \{1,\dotsc,k\}$ fails one of the threshold tests in  \eqref{eq:decoder}, or some incorrect set $\K'$ passes all of the threshold tests.  As a result, the union bound gives
\begin{equation}
\begin{split}
  \Perr \le \PP\bigg( \bigcup_{(S_0, S_1)} &\big\{ \imath^T( {\mat X}_{S_0}; {\vec Y} \mid {\mat X}_{S_1} ) \le \gamma_{|S_0|} \big\} \bigg) \\ 
  &+ \sum_{\K' \ne \K} \PP\Big( \imath^T( {\mat X}_{\K' \setminus \K}; {\vec Y} \mid {\mat X}_{\K \cap \K'} ) > \gamma_{|\K' \setminus \K|} \Big),
\end{split}
\label{eq:nonasymp_init}
\end{equation}
where in the first term the union is implicitly subject to the conditions in \eqref{eq:decoder}, and in the second term, we upper bound the probability of passing all threshold tests by the probability of passing a single one (namely, the one with $S_1 = \K \cap \K'$).

Using the form of $\imath^T$ in \eqref{eq:info_dens_T}, we can upper bound any given summand of \eqref{eq:nonasymp_init} as follows with $S_0 = \K' \setminus \K$, $S_1 = \K \cap \K'$, and $\tau = |\K' \setminus \K|$:
\begin{align}
	\PP \Big( \imath^T ( &{\mat X}_{\K' \setminus \K}; {\vec Y} \mid {\mat X}_{\K \cap \K'} ) > \gamma_\tau \Big) \\ 
    \begin{split} & = \sum_{ {\mat X}_{S_0}, {\mat X}_{S_1}, {\vec y} } \PP( {\mat X}_{S_0}, {\mat X}_{S_1} ) P_{{\vec Y}|{\mat X}_{S_1}}({\vec y} \mid {\mat X}_{S_1}) \\ &\qquad\quad {}\times \boldsymbol{1}\bigg\{ \log_2 \frac{ P_{{\vec Y} | {\mat X}_{S_0}, {\mat X}_{S_1}}({\vec y} \mid {\mat X}_{S_0}, {\mat X}_{S_1}) }{ P_{{\vec Y}|{\mat X}_{S_1}}({\vec y} \mid {\mat X}_{S_1}) } > \gamma_{\tau} \bigg\}    \end{split} \label{eq:nonasymp_3_1} \\
    & \le \sum_{ {\mat X}_{S_0}, {\mat X}_{S_1}, {\vec y} } \PP( {\mat X}_{S_0}, {\mat X}_{S_1} ) P_{{\vec Y} | {\mat X}_{S_0}, {\mat X}_{S_1}}({\vec y} \mid {\mat X}_{S_0}, {\mat X}_{S_1}) 2^{-\gamma_{\tau}}  \label{eq:nonasymp_3_2}\\ 
    & = 2^{-\gamma_{\tau}}. \label{eq:nonasymp_3_3}
\end{align}
Here, \eqref{eq:nonasymp_3_1} follows since the observations depend on $\mat{X}_{\K'}$ only through the columns $S_1 = \K \cap \K'$ overlapping with $\K$, \eqref{eq:nonasymp_3_2} follows by upper bounding $P_{{\vec Y}|{\mat X}_{S_1}}({\vec y} \mid {\mat X}_{S_1})$ according to the event in the indicator function and then upper bounding the indicator function by one, and  \eqref{eq:nonasymp_3_3} follows from the fact that we are summing a joint distribution over all of its values.  

Combining \eqref{eq:nonasymp_init} and \eqref{eq:nonasymp_3_3}, and also applying the union bound in the first term of the former, we obtain
\begin{align*}
	\Perr \le \sum_{ (S_0,S_1) } \PP\Big( \imath^T( {\mat X}_{S_0}; {\vec Y} \mid {\mat X}_{S_1} ) \le \gamma_{|S_0|} \Big) + \sum_{\K' \ne \K} 2^{-\gamma_{|\K' \setminus \K|}}.
\end{align*}
where we have applied the definition $\tau = |\K' \setminus \K|$.   By counting the number of $S_0 \subset \K$ of cardinality $\tau \in \{1,\dotsc,k\}$, as well as the number of $\K' \ne \K$ such that $|\K' \setminus \K| = \tau \in \{1,\dotsc,k\}$, we can simplify the above bound to
\begin{align}
\Perr \le \sum_{\tau=1}^k {\binom{k}{\tau}}\PP\Big( \imath^T( {\mat X}_{0,\tau}; {\vec Y} \mid {\mat X}_{1,\tau} ) \le \gamma_{\tau} \Big) + \sum_{\tau=1}^k {\binom{k}{\tau}}{\binom{n-k}{\tau}} 2^{-\gamma_{\tau}}, \label{eq:simplify_ach}
\end{align}
where ${\mat X}_{0,\tau} = {\mat X}_{S_0}$ and  ${\mat X}_{1,\tau} = {\mat X}_{S_1}$ for an arbitrary partition $(S_0, S_1)$ of $\{1,\dotsc,k\}$ with $|S_0| = \tau$; by the i.i.d.~test design and model assumption \eqref{eq:Y_noisy}, the probability in \eqref{eq:simplify_ach} is the same for any such partition.

Finally, choosing 
  \[ \gamma_{\tau} = \log_2\frac{k\binom{k}{\tau}\binom{n-k}{\tau}}{\delta} \]
for some $\delta > 0$, we obtain the non-asymptotic bound
\begin{equation}
	\Perr \le \sum_{\tau=1}^k \binom{k}{\tau} \PP\bigg( \imath^T( {\mat X}_{0,\tau}; {\vec Y} \mid {\mat X}_{1,\tau} ) \le \log_2\frac{k\binom{k}{\tau}\binom{n-k}{\tau}}{\delta} \bigg) + \delta. \label{eq:non_asymp}
\end{equation}

\subsection{Characterizing the tail probabilities}

The next step is to characterize the probability appearing on the right-hand side of \eqref{eq:non_asymp}.  The idea is to note that this is the tail probability of an i.i.d.~sum, and hence we should expect some concentration around the mean.  Recall from Section \ref{eq:it_prelim} that the mean of the information density is the conditional mutual information:
\begin{equation}
    \EE\big[ \imath^T( {\mat X}_{0,\tau}; {\vec Y} \mid {\mat X}_{1,\tau} ) \big] = T \cdot I({\vec X}_{0,\tau}; Y \mid {\vec X}_{1,\tau}) =: T \cdot I_\tau, \label{eq:I_tau}
\end{equation}
where $({\vec X}_{0,\tau},{\vec X}_{1,\tau})$ correspond to single rows in $({\mat X}_{0,\tau},{\mat X}_{1,\tau})$, and $Y$ is the corresponding entry of $\vec Y$.
The following lemma characterizes $I_\tau$ for the noiseless model; we return to the noisy setting in Section \ref{sec:noisy}.

\begin{lemma} \label{lem:mi_asymp}
	Under the noiseless group testing model using Bernoulli testing with probability $p=\nu/k$ for some  fixed $\nu > 0$, the conditional mutual information $I_{\tau}$ behaves as follows as $k \to \infty$:
\begin{enumerate}
  \item If $\tau/k \to 0$, then
    $$ I_{\tau} \asym \ee^{-\nu} \nu \frac{\tau}{k}\log_2\frac{k}{\tau}  . $$
  \item If $\tau/k \to \psi \in (0,1]$, then     $$ I_{\tau} \asym \ee^{-(1-\psi)\nu} h(\ee^{-\psi\nu}) , $$
    where $h(\psi)$ 
    is the binary entropy function.
\end{enumerate}
\end{lemma}
\begin{proof}
    In the noiseless setting, we have $I({\vec X}_{0,\tau};Y \mid {\vec X}_{1,\tau}) = H(Y \mid {\vec X}_{1,\tau})$.  If ${\vec X}_{1,\tau}$ contains any ones, then the conditional entropy of $Y$ is zero, and otherwise, the conditional entropy is the binary entropy function evaluated at the conditional probability of $Y=1$.  Evaluating these probabilities explicitly, we obtain
    \[ I_{\tau} = (1-p)^{k-\tau} h\big((1-p)^\tau\big) = \Big(1 - \frac{\nu}{k}\Big)^{k - \tau} h\bigg( \Big( 1 - \frac{\nu}{k} \Big)^{\tau} \bigg). \]
    
	In the case that $\tau/k \to 0$, the lemma now follows from the asymptotic expressions $(1-\nu/k)^{k-\tau} \to \ee^{-\nu}$ and $(1-\nu/k)^{\tau} \asym 1 - \nu\tau/k$, as well as $h(1-a) \asym -a\log_2 a$ as $a \to 0$.
	
	In the case that $\tau/k \to \psi \in (0,1]$, the lemma follows from the limits $(1-\nu/k)^{k-\tau} \to \ee^{-(1-\psi)\nu}$ and  $(1-\nu/k)^{\tau} \to \ee^{-\psi\nu}$, as well as the continuity of entropy.
\end{proof}

We now fix a set of constants $\delta'_{\tau}$ for $\tau=1,\dots,k$, and observe that as long as
\begin{equation}
	T \ge \frac{ \log_2{\binom{n-k}{\tau}} + \log_2\big(\frac{k}{\delta}\binom{k}{\tau}\big) }{ (1-\delta'_{\tau}) I_{\tau}  }, \label{eq:n_cond_Il}
\end{equation}
we can upper bound the probability in \eqref{eq:non_asymp} by
\begin{equation}
	\PP\big( \imath^T( {\mat X}_{0,\tau}; {\vec Y} \mid {\mat X}_{1,\tau} ) < (1-\delta'_{\tau}) T I_{\tau}   \big). \label{eq:pe_simplified}
\end{equation}
As mentioned above, $\imath^T$ is a sum of $T$ i.i.d.~random variables having mean $I_{\tau}$, and as a result, we can bound \eqref{eq:pe_simplified} using concentration inequalities.  

In fact, in the case that $k$ is constant (that is, not growing with $n$), it suffices to use Chebyshev's inequality to show that each term of the form \eqref{eq:pe_simplified} vanishes for arbitrarily small $\delta'_{\tau}$ \cite{scarlett-cevher-2}.  Since each such term vanishes, then so does the weighted sum of all such terms in \eqref{eq:non_asymp} (using the fact that $k$ is constant), and we are left only with the sufficient condition in \eqref{eq:n_cond_Il} for $\PP(\mathrm{err}) \le \delta + o(1)$.  By taking $\delta \to 0$ sufficiently slowly, we are left with the condition
  \[ T \ge \max_{\tau = 1,\dotsc,k} \frac{ \log_2{\binom{n-k}{\tau}}}{ I_{\tau} } (1+o(1)) \]
for $\PP(\mathrm{err}) \to 0$.
 
However, our main interest is not in the fixed-$k$ regime, but in the regime $k = \Theta(n^{\alpha})$ for $\alpha \in (0,1)$.  In this case, more sophisticated concentration bounds are needed, and these turn out to introduce extra requirements on $T$ beyond \eqref{eq:n_cond_Il} alone.

\begin{lemma} \label{lem:ach_concentration}
	Set $\tau^* = k/{\log_2 k}$. Under Bernoulli group testing with probability $p=\nu/k$, the quantities $\imath_{\tau,T} := \imath^T( {\mat X}_{0,\tau}; {\vec Y} \mid {\mat X}_{1,\tau} )$ satisfy the following concentration bounds provided that the quantities $\delta'_{\tau}$ are uniformly bounded away from zero and one:
\begin{enumerate}
\item For $\tau \le \tau^*$, we have
    \begin{multline*} \PP\big( \imath_{\tau,T} < T I_{\tau} (1-\delta'_{\tau}) \big) \\
    \le  \exp\bigg(-T \frac{\tau}{k} \ee^{-\nu}\nu \big( (1-\delta'_{\tau})\log_2(1-\delta'_{\tau}) + \delta'_{\tau}\big) (1+o(1)) \bigg). \end{multline*}
\item  For $\tau > \tau^*$, we have
    $$ \PP\big( \imath_{\tau,T} < T I_{\tau} (1-\delta'_{\tau}) \big) \le 2\exp\bigg(- \frac{(\delta'_{\tau} I_{\tau})^2 T}{4(8+\delta'_{\tau} I_{\tau})} \bigg).$$ 
\end{enumerate}
\end{lemma}
\begin{proof}
	The first bound is proved by lower bounding $i_{\tau,T}$ by a scaled binomial random variable and applying a well-known concentration bound specific to the binomial distribution. The second bound is proved using Bernstein's inequality.  The details can be found in \cite{scarlett-cevher-1}.
\end{proof}

The remainder of the proof amounts to rather tedious yet elementary algebraic manipulations, and we therefore provide only an outline.
We start by choosing choose $\delta'_{\tau} = 1-\epsilon$ for $\tau \le \tau^*$, and $\delta'_{\tau} = \epsilon$ for $\tau > \tau^*$, where $\epsilon > 0$ is arbitrarily small.

The first requirement on $T$ is that it satisfies \eqref{eq:n_cond_Il} for all $\tau = 1,\dotsc,k$.  Using Lemma \ref{lem:mi_asymp} and the preceding choices of $\delta'_{\tau}$, one can show that the value of $\tau$ that gives the most stringent requirement on $T$ is $\tau = k$, at least in the asymptotic limit.  As a result, we get the condition
    \begin{equation}
    T \ge \Big( k\log_2\frac{n}{k}\Big) \big(1 + O(\epsilon) + o(1) \big). \label{eq:ach_final_1}
    \end{equation}
This arises from the fact that the numerator in \eqref{eq:n_cond_Il} with $\tau = k$ is dominated by $\log_2{\binom{n-k}{k}}$, which behaves as $\big(k \log_2 \frac{n}{k}\big)(1+o(1))$ whenever $k = o(n)$.

The second requirement on $T$ is that, upon substituting the bounds of Lemma \ref{lem:ach_concentration} into \eqref{eq:non_asymp} and taking $\delta \to 0$, the resulting summation on the right-hand side vanishes.  For this to be true, it suffices that both the summations over $\tau\in\{1,\dotsc,\tau^*\}$ and $\tau\in\{\tau^* + 1,\dotsc,k\}$ vanish.  The second of these (the `small overlap' case) turns out to already vanish under the condition in \eqref{eq:ach_final_1}.  On the other hand, after some rearranging and asymptotic simplifications, we find that the first of these summations (the `large overlap' case) vanishes provided that
    \begin{equation}
    T \ge \Bigg( \frac{\frac{\alpha}{1-\alpha} k\log_2\frac{n}{k} }{\nu \ee^{-\nu}} \Bigg) \big(1 + O(\epsilon) + o(1) \big). \label{eq:ach_final_2}
    \end{equation}
Theorem \ref{thm:ch2main} follows by combining these bounds and taking $\epsilon \to 0$.

\section{Converse bound for Bernoulli testing} \label{sec:algcon}

We have seen that nonadaptive Bernoulli matrix designs achieve a rate of $1$ bit per test whenever $\alpha \le \frac{1}{3}$, thus matching the counting bound and proving their asymptotic optimality.  On the other hand, for $\alpha \in \big(\frac{1}{3},1\big)$, there remains a gap between the two, with the gap growing larger as $\alpha$ approaches one.  

A priori, there are several possible reasons for the remaining gaps: the analysis in Section \ref{sec:pf_ach} could be loose, the use of Bernoulli tests could be suboptimal, or the counting bound itself could be loose.  The following result, due to Aldridge \cite{aldridge}, rules out the first of these, showing that Theorem \ref{thm:ch2main} provides the best rate that one could hope for given that Bernoulli designs are used.

\begin{theorem} \label{thm:bern_conv}
	Consider noiseless group testing in the sparse regime $k=\Theta(n^{\alpha})$, with the exact recovery criterion, and Bernoulli testing.  If the rate exceeds $\olR_{\rm Bern}$ defined in \eqref{bern-ach}, then the error probability averaged over the testing matrix is bounded away from zero, regardless of the decoding algorithm.
\end{theorem}
\begin{proof}
	The idea of the proof is as follows. 
	First, we argue that if both the COMP and SSS (see Chapter \ref{ch:algorithms}) algorithms fail, then any algorithm fails with a certain probability. Second, we argue that for rates above $\olR_{\rm Bern}$, both COMP and SSS fail.
    
    Let $\hat\K_\COMP$ and $\hat\K_\SSS$ be the sets returned by COMP and SSS, respectively.  Recall that these are respectively the largest (see Lemma \ref{lem:COMPsat}) and smallest satisfying sets, where a satisfying set is any putative set of defective items that could have produced the observed output (see Definition \ref{def:sat}).  The key observation is that if $|\hat\K_\COMP| > k$ and $|\hat\K_\SSS| < k$, then there exist at least two satisfying sets $\L$ with $|\L| = k$; any such set can be found by adding elements of $\hat\K_\COMP$ to $\hat\K_\SSS$ until reaching size $k$.  Then, even if $k$ is known, the best one can do given multiple such sets is to choose one arbitrarily, yielding an error probability of at least $1/2$.  Thus,
    \begin{align}
    	\Perr &\ge \frac{1}{2} \,\PP\big( |\hat\K_\COMP| > k \,\cap\, |\hat\K_\SSS| < k \big) \notag \\
        	&\ge \frac{1}{2} \,\big( 1 - \PP(|\hat\K_\COMP| = k) - \PP(|\hat\K_\SSS| = k) \big) \label{twoterms}
    \end{align}
    by the union bound.
    
    We handle the two above terms separately. First, we observe that $\PP(|\hat\K_\COMP| = k)$ is precisely the probability of COMP succeeding, since the largest satisfying set is necessarily unique (see Lemma \ref{lem:COMPsat}).  In Remark \ref{rmk:tightness}, it was shown that the success probability of COMP tends to zero for rates above the value $R_\mathrm{Bern}^\COMP$ defined in \eqref{COMPBern}, which is strictly less than the rate $\olR_{\rm Bern}$ that we consider here. 
    Second, $\{|\hat\K_\SSS| = k\}$ cannot occur when a defective item is masked, and such a masking event was shown to occur with probability bounded away from zero in the proof of Theorem \ref{SSSub}.  Combining these two results, we find that \eqref{twoterms} is bounded away from $0$, which completes the proof.
\end{proof} 


\section{Improved rates with near-constant tests-per-item} \label{sec:ach_near_const}

\rev{
In Section \ref{sec:near_constant}, we saw that the near-constant column weight (or near-constant tests-per-item) design introduced in Definition \ref{def:const_col} gives improved rates for the COMP and DD algorithms, and that the SSS algorithm cannot attain a higher rate than  $\min\big\{1,\ln 2 \, \frac{1-\alpha}{\alpha}\big\}$.

Similarly to the Bernoulli design, the converse of $\min\big\{1,\ln 2 \, \frac{1-\alpha}{\alpha}\big\}$ bits per test for SSS under the near-constant column weight design matches the achievability result for the DD algorithm stated in Theorem \ref{DDub2} when $\alpha \ge 1/2$.  In this section, we describe a recent development that extends the achievability of the preceding rate to {\em all} $\alpha \in (0,1)$, albeit at the expense of (potentially considerably) increased computation compared to DD.

Formally, the main result of Coja-Oghlan \etal\ \cite{coja} proves that the near-constant column weight design has a maximum achievable rate of
  \[   
    \olR_{\rm NCC} =  \min\bigg\{1,(\ln 2)\frac{1-\alpha}{\alpha}\bigg\},
  \]
as we stated in Theorem \ref{thm:ch2nearconst} above.
We refrain from presenting the full details of the somewhat lengthy proof of Theorem \ref{thm:ch2nearconst}, but we sketch the main steps. 

On the whole, the analysis is less based on tools from information theory, and more based on direct probabilistic arguments. (Coja-Oghlan \etal\ note that similar arguments have been successful in the theory of constraint satisfaction problems.)  Nevertheless, some similarities do exist between this approach and the information-theoretic analysis of Section \ref{sec:pf_ach}. Notably, the error events associated with incorrect defective sets are handled separately according to the amount of difference with the true defective set. The error events corresponding to a small overlap (that is, an incorrect set having relatively few items in common with $\K$) have low probability for rates up to $1$, and the error events corresponding to a large overlap (a large number of items in common) have low probability for rate up to $(\ln 2)\frac{1-\alpha}{\alpha}$ after the application of a tight concentration bound.

\begin{proof}[Proof sketch of Theorem \ref{thm:ch2nearconst}]
Consider the number of satisfying sets $\Khat$ of the correct size $|\Khat| = k$ that have a set difference with the true defective set of size $|\K \setminus \Khat| = |\Khat \setminus \K| = \tau$.  Clearly there is one such set with $\tau = 0$, namely, the true defective set.  If it can be shown that with high probability there are no others, then the true defective set is the only satisfying set, and can -- at least given enough computation time -- be found reliably.  Different bounds are used depending on whether $\tau \geq \tau^*$ or $\tau < \tau^*$, where we choose $\tau^* = k/\log_2 n$.

Similarly to the Bernoulli design, we may assume that $\K$ is fixed, say $\K = \{1,\dotsc,k\}$, without loss of generality.  We first consider the `small overlap' (or `large difference') case, where $\tau \geq \tau^*$. Let $\mathcal S$ be the event that there exists a satisfying set corresponding to such a $\tau$. Using the union bound, we have
  \begin{equation} \label{eg:curlyS}
      \mathbb P (\mathcal S ) \leq \sum_{\tau = \tau^*}^{k} \binom{n - k}{\tau} \binom{k}{\tau} \mathbb P\big(\Khat_\tau \text{ is satisfying}\big)
  \end{equation}  
where $\Khat_\tau$ is any set of size $k$ containing $\tau$ nondefectives and $k-\tau$ defectives. In \cite{coja}, a concentration result and a coupling argument is used to show that the bound \eqref{eg:curlyS} for the near-constant column weight design is very close to the analogous bound for the Bernoulli design.  This means that we can treat this case as though we were using the Bernoulli$(p)$ design, where $p = 1 - \ee^{-\nu/k} \sim \nu/k$.

Pick $\nu = \ln 2$, so $1 - p = 2^{-1/k}$. (Here we follow an argument from \cite{aldridge-baldassini-gunderson}.) A test has a different result under $\Khat_\tau$ compared to $\K$ if no item in $\K$ is tested but an item in $\Khat_\tau \setminus \K$ is tested, or vice versa. This has probability
\[ 2(1 - p)^k \big(1 - (1 - p)^{\tau}\big) = 2(1 - p)^k - 2(1 - p)^{k+\tau} . \]
Hence,
\[ \mathbb P\big(\Khat_\tau \text{ is satisfying}\big) = \big( 1 - 2(1 - p)^k + 2(1 - p)^{k+\tau}\big)^T , \]
and, using $(1-p)^k = \frac12$, we have
\begin{align*} \mathbb P (\mathcal S ) &\leq \sum_{\tau = \tau^*}^{k} \binom{n - k}{\tau} \binom{k}{\tau} \big( 1 - 2(1 - p)^k + 2(1 - p)^{k+\tau}\big)^T \\
&= \sum_{\tau = \tau^*}^{k} \binom{n - k}{\tau} \binom{k}{\tau} \left( 1 - 2\cdot\tfrac12 + 2\cdot\tfrac12\big(2^{-1/k}\big)^\tau\right)^T \\
&= \sum_{\tau = \tau^*}^{k} \binom{n - k}{\tau} \binom{k}{\tau} 2^{-\tau T/k} . \end{align*}
One can check that the summands here are decreasing, so the largest term is that for $\tau = \tau^*$, and for any $\delta > 0$ we have $\tau^* < \delta k$ for $n$ sufficiently large. Hence, we have 
\begin{align*}
\mathbb P (\mathcal S ) &\leq k \binom{n - k}{\delta k} \binom{k}{\delta k} 2^{-\delta k T/k} \\
&\leq k \left(\frac{\ee n}{\delta k}\right)^{\delta k} \left(\frac{\ee }{\delta }\right)^{\delta k} 2^{-\delta  T} \\
&= k 2^{-\delta(T - k \log_2(n/k) -2 (\log_2 \ee - \log_2 \delta) k)}.
\end{align*}
We see for $T > (1+\eta) k \log_2 (n/k)$, which corresponds to any rate up to $1$, that $\mathbb P(\mathcal S)$ can be made arbitrarily small.

Next, we consider the `large overlap' (or `small difference') case, where $\tau < \tau^*$. The union bound argument above is too weak here, as `rare solution-rich instances drive up the expected number of solutions' \cite{coja}. Instead, we first show that a certain property $\mathcal R$ holds with high probability, and then use the expansion properties of the near-constant column weight design to show that, with high probability, no large-overlap solutions exist when $\mathcal R$ holds. 

Property $\mathcal R$ is the event that every defective item $i \in \K$ is the unique defective item in at least $\delta L$ tests, for some $\delta > 0$. We need to show that $\mathcal R$ holds with high probability. To simplify this proof sketch, we present the analysis as though each item $i$ were included in exactly $L = \nu T/k$ tests chosen {\em without} replacement (rather than with replacement). Each of these $L$ tests contains no other defective items with probability
\[ \left(1 - \frac 1T \right)^{L(k-1)} = \left(1 - \frac 1T \right)^{\nu T (1 - 1/k)} \to \ee^{-\nu} . \]
For further simplification here (with the full details given in \cite{coja}), we make another non-rigorous approximation and suppose that each such test contains no other defectives with probability exactly $\ee^{-\nu}$, and that this event is independent across tests. Write $M_i \stackrel{\rm d}{\approx} \text{Bin}(L, \ee^{-\nu})$ for the number of tests in which $i$ is the unique defective. Then the probability this is fewer than $\delta L$ tests is (approximately)
\[ \mathbb P( M_i < \delta L) \approx \mathbb P\big( \text{Bin}(L, \ee^{-\nu}) < \delta L\big) \leq 2^{-L d(\delta \,\|\, \ee^{-\nu})} , \]
where $d(p \,\|\, q)$ is the relative entropy between a Bernoulli$(p)$ and a Bernoulli$(q)$ random variable, and we have used the standard Chernoff bound for the binomial distribution. It is clearly advantageous to take $\delta$ as small as possible, and doing so yields
\begin{align*} d(\delta \,\|\, \ee^{-\nu}) &= h(\delta) - \big(\delta \log_2 \ee^{-\nu}  + (1-\delta) \log_2(1 - \ee^{-\nu})\big) \\  &\to - \log_2 (1 - \ee^{-\nu}) . \end{align*}
We can then use a union bound to write
\begin{align*}
\mathbb P(\mathcal R) &= \mathbb P \left( \bigcap_{i \in \mathcal K} M_i \geq \delta L \right) \\
&\leq  1 - k\mathbb P( M_i < \delta L ) \\
&\lesssim 1 - k 2^{L\log_2(1 - \ee^{-\nu})} \\
&= 1 - 2^{-(-\nu\log_2(1 - \ee^{-\nu})T/k - \log_2 k)},
\end{align*}
where we substituted $L = \nu T / k$.
The preceding bound can be made to approach $1$ provided that
\[ T > (1 + \eta) \frac{1}{-\nu\log_2(1 - \ee^{-\nu})} k \log_2 k \]
for some small $\eta > 0$.  The term $-\nu\log_2(1 - \ee^{-\nu})$ is maximized at $\nu = \ln 2$, where it takes the value $\ln 2$.  Hence, the preceding condition reduces to
\[ T > (1 + \eta) \frac{1}{\ln 2} k \log_2 k ,\]
which corresponds to rates no larger than $(\ln 2) \frac{1-\alpha}{\alpha}$.

It remains to argue that, conditioned on the event $\mathcal R$, we have no small-difference satisfying sets with high probability. The key observation of \cite{coja} is that switching even a single item from defective to nondefective would change the result of a large number (at least $\delta L$) of formerly positive tests. But turning these tests back positive requires switching many items from nondefective to defective, because the expansion properties of the design imply that it is unlikely that any item will be able to cover many of these tests. These switches in turn change the result of many more tests, requiring more switches, and so on. Hence, to get another satisfying set, one must switch the status of many items, and no `large overlap' set can exist.  While this is only an intuitive argument, it is formalized in \cite{coja}.

Together, we have establishing vanishing probability for the existence of satisfying sets with either a small overlap or a large overlap (with `small' and `large' collectively covering all cases), and we are done.
\end{proof}

The use of the near-constant column weight design was crucial in checking that property $\mathcal R$ holds with high probability. Suppose that we instead use a Bernoulli$(\nu/k)$ design; then, the probability that a defective item is the unique defective in a given tests is
\[ p(1-p)^{k-1} = \frac{\nu}{k} \left(1 - \frac{\nu}{k}\right)^{k-1} \sim \frac{\nu \ee^{-\nu}}{k} . \]
Hence, the probability of being the unique defective item in
fewer than $\delta L = \delta\nu T/k$ tests is
\[
\mathbb P\big( \text{Bin}(T, \nu\ee^{-\nu}/k) < \delta \nu T/k) \leq 2^{-T d(  \delta \nu /k \,\|\, \nu\ee^{-\nu}/k )} .
\]
Again, taking $\delta$ as small as possible, we have
\begin{align*} d( \delta \nu /k \,\|\,  \nu\ee^{-\nu}/k ) \sim \log_2(1 - \nu\ee^{-\nu}/k) \sim \frac{1}{\ln 2} \frac{\nu \ee^{-\nu}}{k} . \end{align*}
Following the same argument leads to the conclusion that we avoid large-overlap errors with rates up to $\frac{\nu \ee^{-\nu}}{\ln 2}\frac{1-\alpha}{\alpha}$, as in \eqref{bern-ach}. Thus, we see that the achievable rate for Bernoulli designs (Theorem \ref{thm:ch2main}) can also be proved using the approach of \cite{coja}.

We can match the achievability result of Theorem \ref{thm:ch2nearconst} with a converse for near-constant column weight designs. In particular, the proof of the corresponding result for Bernoulli designs (Theorem \ref{thm:bern_conv}) extends easily to the near-constant column weight design once Theorem \ref{SSSub2} on the SSS algorithm is in place. This extension is stated formally as follows for completeness.

\begin{theorem} \label{thm:ncc_conv}
	Consider noiseless group testing in the sparse regime $k=\Theta(n^{\alpha})$, with the exact recovery criterion, and the near-constant column weight design (Definition \ref{def:const_col}).  If the rate exceeds $\olR_{\rm NCC}$ defined in \eqref{bern-ach}, then the error probability averaged over the testing matrix is bounded away from zero, regardless of the decoding algorithm.
\end{theorem}

This result readily establishes that the achievable rate for the DD algorithm, stated in Theorem \ref{DDub2}, is optimal (with respect to the random test design) when $\alpha \ge 1/2$.  Recall that these rates are plotted in Figure \ref{fig:algrates2}.}

\added{
\section{General converse and polynomial-time strategies} \label{sec:optimal_gt}

Having established the exact rates attained by the Bernoulli and near-constant column weight design, we now turn to the question of whether the latter design's rate can be improved further, and whether the optimal rate can be attained in a computationally efficient manner (i.e., polynomial time) for all $\alpha \in (0,1)$ (recall that the DD algorithm suffices for $\alpha \ge \frac{1}{2}$).  As stated in Theorem \ref{thm:gen_conv} and \ref{thm:spatial}, the near-constant column weight design's rate $\olR_{\rm NCC} = \min\big\{1,(\ln 2)\frac{1-\alpha}{\alpha}\big\}$ is in fact optimal among \emph{all designs}, and it can be attained with a computationally efficient method when a modified test design based on spatial coupling is used.  These results are due to Coja-Oghlan \etal~\cite{coja2020optimal}, and we proceed to outline their techniques for the converse and achievability parts separately.


\subsection{General converse} \label{sec:gen_conv}

Here we outline the proof Theorem \ref{thm:gen_conv}, which states that no nonadaptive strategy can improve on the rate $\olR_{\rm NCC}$ stated in \eqref{ncc-ach}.  The converse of 1 is already known from the counting bound, so it remains to establish a converse of $(\ln 2)\frac{1-\alpha}{\alpha}$.

Our presentation combines aspects of Coja-Oghlan \etal~\cite{coja2020optimal} and Bay, Price, and Scarlett~\cite{bay2022optimal}; the latter built on the former but considered general sparsity regimes rather than only $k = \Theta(n^{\alpha})$.  Like with the analysis of SSS in Section \ref{sec:sss}, the idea is to identify \emph{masked items}; that is, items that only appear in tests with one or more other defectives.  The details here are quite different due to the need to consider arbitrary test designs.

We first highlight a number of useful reductions or simplifications that will make the analysis easier.
\begin{itemize}
    \item It turns out that once the result is established with a sparsity parameter $\alpha$ arbitrarily close to 1, it readily implies the same for all smaller values of $\alpha$.  The reasoning behind this is that the rate $(\ln 2)\frac{1-\alpha}{\alpha}$ corresponds to having $T = \frac{1 + o(1)}{\ln 2} k \log_2 k$ tests, which does not directly depend on $n$.  Then, supposing that this rate is achievable for some smaller $\alpha = \alpha_{\rm small}$, we can consider an algorithm that handles some larger $\alpha = \alpha_{\rm large}$ by first adding `dummy' nondefective items (specifically, enough to induce an overall sparsity level of $\alpha_{\rm small}$).  This means that achievability for smaller $\alpha$ implies achievability for larger $\alpha$; the contrapositive statement is then that a converse for higher $\alpha$ implies a converse for smaller $\alpha$.
    \item It is more convenient to establish the converse for the i.i.d.\ prior with $q = k/n$, rather than the combinatorial prior.  The former can then be translated to the latter by considering a slightly reduced value of $q$ and then adding some extra defectives to make the total exactly $k$. (If masking occurs before these additions, it still occurs after them.)
    \item It is useful to impose upper bounds on the number of tests-per-item and items-per-test.  We can assume there are at most $\frac{n}{k}\ln n$ items per test without loss of generality, since all tests with more items could be `guessed' as being positive without even performing them, and the guesses would all be correct with probability approaching one.  Once this items-per-test bound is imposed, a simple degree counting argument further shows that almost all of the items -- namely, $n(1-o(1))$ of them -- appear in at most $(\ln n)^3$ tests.  To simplify the exposition, we will assume that this holds for all $n$ items; the general case only requires minor modifications.
    \item We also assume that every test contains at least two items.  This is justified by the fact that tests containing no items are useless, while tests containing one item simply reveal that item's defectivity status.  Since we are in the regime $T = o(n)$, only an asymptotically negligible fraction of items' statuses can be learned in this way.
\end{itemize}
With these reductions in place, we proceed to study the i.i.d.~prior, $\alpha$ arbitrarily close to one, at most $\frac{n}{k}\ln n$ items per test, at most $\ln^3 n$ tests per item, and at least two items per test.

For $i=1,\dotsc,n$, let $D_i$ be the event that item $i$ is masked (also referred to as \emph{disguised}, hence the use of the symbol $D$).  Under the i.i.d.~prior, it is straightforward to show that the probability of $i$ being defective given $D_i$ is equal to the prior defectivity probability $q$.  A single item being masked does not immediately imply failure: we have $q = o(1)$, so guessing that item to be nondefective would be the correct decision with high probability.  However if we can establish that \emph{many} items are masked, then guessing they will \emph{all} be nondefective will be likely to fail.

The key result for identifying masked items is due to Aldridge \cite{aldridge-linear} (see also Section \ref{sec:lin_nonada}). 
Under the i.i.d.$(q)$ prior, if every test contains at least two items, then
\[ \PP(D_i) \ge \prod_{t:x_{ti}=1} \big(1-(1-q)^{w_t-1}\big) , \] where $w_t$ is the number of items in test $t$; this can be proved using the FKG inequality. 
Taking logarithms and rearranging, we get
\[ \frac{1}{n}\sum_{i=1}^n \ln \PP(D_i) \ge \frac{T}{n} \mathcal{L}(q),\] where \[ \mathcal{L}(q) = \min_{w=2,3,\dotsc,n} w\ln\big(1-(1-q)^{w-1}\big) . \]

The value of $w$ attaining the minimum in $\mathcal{L}(q)$ is roughly $(\ln 2)/q$, which gives 
\[ -\mathcal{L}(q) \sim \frac{(\ln 2)^2}{q} = (\ln 2)^2 \, \frac{n}{k} .\]
Thus, 
\begin{equation}
    \frac{1}{n}\sum_{i=1}^n \ln \PP(D_i) \gtrsim - \frac{T}{n} \, \frac{(\ln 2)^2}{ q }.
\end{equation}
Since the maximum is at least as high as the average, this shows that there exists a specific item $i_0$ such that
\begin{equation}
    \PP(D_{i_0}) \gtrsim \exp \left(-\frac{T}{n} \, \frac{(\ln 2)^2}{q} \right). \label{eq:i0_bound}
\end{equation}
One can then use similar ideas to establish the existence of \emph{many} items whose masking probability $\PP(D_i)$ satisfies a similar lower bound.  When doing so, it is convenient to choose the various $i$ values such that the $D_i$ events are independent of one another.  To do so, we note that if the placements of items into tests are represented by a bipartite graph (see, for example, Figure \ref{fig:bp_graph}), then a masking event for a given item $i$ only depends on the defectivity status of items at distance 2 in the graph; that is, items sharing a test with $i$.  Thus, items separated by distance exceeding 4 have independent masking events.  Note that items $i,i'$ having a distance exceeding 4 means that (i) $i$ and $i'$ do not share any test, and (ii) the set of items sharing a test with $i$ is disjoint from the set of items sharing a test with $i'$.

In view of this independence observation, we can perform an iterative procedure as follows:
\begin{itemize}
    \item Identify some $i_0$ whose masking probability is at least the bound from \eqref{eq:i0_bound}, as described above.
    \item Remove all items and tests within distance 4 of $i_0$ in the bipartite graph, and further remove all tests of degree at most 1 and their associated items.
    \item Proceed back to the first step on the reduced graph and continue this procedure until some stopping criterion is met.
\end{itemize}
Note that if an item is masked in the reduced graph, it is also masked in the original graph.  While modifying the graph can slightly change $n$ and $T$, it turns out that, thanks to the reduction bounding the number of tests-per-item and items-per-test, their changes can be made negligible while still extracting enough items; namely, the ratio $T/n$ can be kept within a $1+\eta$ factor of its original value for arbitrarily small $\eta > 0$.  This implies, via \eqref{eq:i0_bound}, that all of the extracted items satisfy
\[ \PP(D_i) \gtrsim \exp \left(-(1+\eta)\frac{T}{n} \,\frac{(\ln 2)^2}{q} \right) . \]

The number of extracted items comes out to be `nearly linear' -- say, at least $n^{1-\eta}$ -- and we know that the associated masking events are independent of one another.
Thus, the average number of masking events is roughly lower bounded by
\begin{equation}
    n^{1-\eta} \, \exp\bigg(-(1+\eta)\frac{T}{n} \,\frac{(\ln 2)^2}{q}\bigg).\label{eq:num_masked}    
\end{equation}
Due to the independence property, we can readily establish concentration around this average.  Moreover, if this number is significantly higher than $n/k$, then with high probability there are both defectives and nondefectives among these masked items, implying that reliable recovery is impossible.

Since we reduced to $\alpha$ arbitrarily close to 1, we may further assume that $n/k \le n^{\eta}$.  Putting this all together with \eqref{eq:num_masked}, substituting $q = k/n$, and solving for $T$, we find that failure indeed occurs when $T$ is slightly below 
\[ \frac{1}{(\ln 2)^2}\,k \ln k = \frac{1}{\ln 2}\,k \log_2 k , \]
which corresponds to a rate of $(\ln 2)\frac{1-\alpha}{\alpha}$ and establishes the desired converse.  See \cite{coja2020optimal} (or alternatively \cite{bay2022optimal}) for the full details.

\subsection{Achievability via spatial coupling} \label{sec:spatial}

Here we outline the test design and decoding algorithm used for proving Theorem \ref{thm:spatial}; that is, a computationally efficient strategy that achieves the rate $\olR_{\rm NCC}$ stated in \eqref{ncc-ach} for all $\alpha \in (0,1)$.

\paragraph{Spatially coupled test design}

The test design groups items and tests into equal-size compartments that can be decoded one at a time, with the information learned about each compartment being propagated to a small number of subsequent compartments.  The number of compartments is $\ell = \sqrt{\ln n}$, and we view them as being arranged in a circular structure so that the `subsequent compartment' after the $\ell$th one is the first one.  See Figure \ref{fig:spatial} for an illustration.

\begin{figure}[t]
\begin{center}
\includegraphics[width=\textwidth]{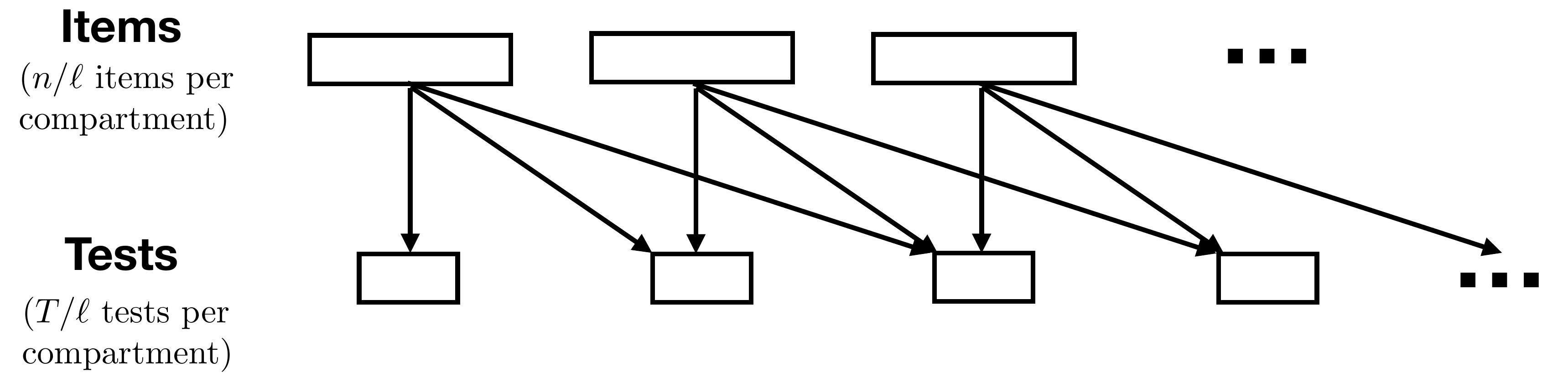}
\end{center}
\caption{\added{Illustration of the spatially coupled test design.  Each arrow corresponds to performing a certain number of random placements of items (from the item compartment) into tests (from the test compartment).  For the final item compartments (not shown), these arrows would wrap around to the first test compartments.}} \label{fig:spatial}
\end{figure}

Within the $r$th compartment of items, each item joins a random number of tests not only in the $r$th compartment of tests, but also in the subsequent $s-1$ compartments, with $s = \ln \ln n$.  In order to `kick start' the recovery procedure, the first $s$ compartments of items are given some additional tests, namely, sufficiently many for DD decoding to succeed on them.

\paragraph{Decoding strategy} The decoding algorithm, termed \emph{spatial inference vertex cover} (SPIV), operates in several phases:
\begin{itemize}
    \item In Phase 1, the first $s$ compartments of items are decoded using the DD algorithm, as hinted above.
    \item In Phase 2, the subsequent compartments of items are decoded iteratively using information from previous compartments.  Within each decoding subroutine, any item in a negative test is first marked as nondefective.  After doing so, the remaining items are categorized by looking for tests containing that item but not containing items previously declared as defective; see below for further details.
    \item In Phase 3, a `clean-up' procedure is performed to correct for a possible small number of mistakes from Phase 2; see below for further details.
\end{itemize}

We first provide the additional details on Phase 2.  Suppose that item $i$ lies in compartment $r+1$, meaning it is involved in tests in compartments $\{r+j\}_{j=1}^s$.  For each $j$, let $W_{i,j}$ be the number of positive tests in compartment $r+j$ that contain item $i$ but none of the items that were previously declared defective (in the compartments up to $r$).  Then, an natural strategy would be to decide defectivity by thresholding $\sum_{j=1}^s W_{i,j}$, which is indeed higher on average when $i$ is defective.  However, the degree of concentration around the average turns out to be insufficient for such an approach to be asymptotically optimal.  To get the required improvement, the idea is to consider a \emph{weighted combination} of the form $\sum_{j=1}^s w_j W_{i,j}$ for suitably-chosen $\{w_j\}_{j=1}^s$.  The idea is that the counts of $W_{i,j}$ are more informative for smaller values of $j$, since the earlier compartments are able to receive information from a larger number of `already decoded' compartments.  The precise choice of $\{w_j\}_{j=1}^s$ is omitted here, but it turns out to be sufficient for correctly classifying $n - o(k)$ out of the $n$ items.

The purpose of Phase 3 is to correct for the $o(k)$ mistakes.  To do this, for each item currently believed to be defective (except those in the first $s$ compartments), we count the number of positive tests in which it is the only item estimated as being defective.  It can be shown that if the item is truly defective, this number should be $\Theta(\log n)$, whereas for nondefectives it is much lower.  Accordingly, the item's defectivity status is updated by thresholding the count at $(\ln n)^{1/4}$.  Using expansion properties of the bipartite graph representing the test design, it is shown that the number of misclassifications reduces by a factor of at least 3 each time this step is performed, meaning it can be iterated $O(\log n)$ times to correct all mistakes.

We again refer the reader to \cite{coja2020optimal} for the full details of this outline.  To our knowledge, no attempt has been made to implement the test design and decoding strategy experimentally (or a modified version of it); accordingly, it may be of interest in future work to establish whether the theoretical asymptotic optimality guarantee can indeed translate to being beneficial at practical problem sizes. 
}

\section{Noisy settings} \label{sec:noisy}

We now turn our attention to noisy settings, considering general noise models of the form \eqref{eq:Y_noisy} -- that is, those that satisfy the only defects matter property of Definition \ref{def:odm}.  As we discussed previously, our initial achievability analysis leading to the non-asymptotic bound \eqref{eq:non_asymp} is valid for any such model, and hence, a reasonable approach is to follow the subsequent steps of the noiseless model.  The main difficulty in doing so is establishing suitable concentration inequalities analogous to Lemma \ref{lem:ach_concentration}. 

We proceed by presenting a general achievability result for the very sparse regime $k = O(1)$, where establishing the desired concentration is straightforward.  We also give a matching converse bound that remains valid for the sparse regime $k = \Theta(n^{\alpha})$.  Achievability bounds and refined converse bounds in the sparse regime are more difficult, and are postponed to Section \ref{sec:high_k_ach}.

\subsection{General noise models in the very sparse regime}

The following theorem provides a general characterization of the required number of tests in terms of suitable conditional mutual information quantities.  This result was given in the works of Malyutov \cite{malyutov-1} and Atia and Saligrama \cite{atia-saligrama}; \rev{see also \cite{dyachkov_lectures} for a survey paying finer attention to the error exponent (that is, the exponential rate of decay of the error probability) and considering universal decoding rules (where the noise distribution is not known). }

\begin{theorem} \label{thm:general_noise}
	Consider any noiseless group testing setup of the form \eqref{eq:Y_noisy}, with $\mathrm{Bernoulli}(p)$ testing and $k = \Theta(n^{\alpha})$ with $\alpha \in [0,1)$.  Then in order to achieve vanishing error probability as $n \to \infty$, it is necessary that
    \begin{equation}
    	T \ge \max_{\tau=1,\dotsc,k} \frac{\tau \log_2 \frac{n}{\tau} }{ I({\vec X}_{0,\tau} ; Y \mid {\vec X}_{1,\tau} ) } (1 - o(1)), \label{eq:general_noise}
    \end{equation}
    where the mutual information is with respect to the independent random vectors $({\vec X}_{0,\tau},{\vec X}_{1,\tau})$ of sizes $(\tau,k-\tau)$ containing independent $\mathrm{Bernoulli}(p)$ entries, along with the noise model $P_{Y|V}$ in \eqref{eq:Y_noisy}.  Moreover, in the case that $\alpha = 0$ ($k = O(1)$) a matching achievability bound holds, and the maximum achievable rate is given by
    \begin{align}
        \olR_{\mathrm{Bern}}^{\mathrm{noisy}} 
            &= \min_{\tau=1,\dotsc,k} \frac{1}{\tau} I({\vec X}_{0,\tau} ; Y \mid {\vec X}_{1,\tau} ) \label{eq:general_noise_ach} \\
            &= I({\vec X}_{0,k} ; Y). \label{eq:general_noise_ach2}
    \end{align}
\end{theorem}

Observe that the equality \eqref{eq:general_noise_ach2} states that the minimum in \eqref{eq:general_noise_ach} is achieved by $\tau = k$, and the capacity reduces to a single unconditional mutual information term.  Moreover, if the noisy defective channel property holds (Definition \ref{def:ndc}) and the Bernoulli testing parameter is optimized, this mutual information term reduces to the corresponding channel capacity -- for example, $I(U;Y) = 1 - h(\rho)$ for the symmetric noise model.

However, the capacity equalling \eqref{eq:general_noise_ach2} crucially relies on two assumptions: (i) the observation model \eqref{eq:Y_noisy} is symmetric, in the sense of depending only on the number of defectives in the test, and not the specific defectives included; and (ii) the number of defectives is bounded, i.e., $k = O(1)$.  Counterexamples to \eqref{eq:general_noise_ach2} in cases that the former condition fails can be found in  \cite{malyutov-1}.  As for the latter condition, we observe that the term $\log_2 \frac{n}{\tau}$ can range from $\log_2\frac{n}{k}$ to $\log_2 n$, and these two terms can have a non-negligible difference when $k$ scales with $n$.  For instance, one can infer from the analysis of \cite{scarlett-cevher-2} that the term corresponding to $\tau = 1$ can dominate in the regime $k = \Theta(n^{\alpha})$ when $\alpha \in (0,1)$ is sufficiently close to one.

The assumption $k = O(1)$ in the achievability part is highly restrictive; we discuss this point further in Section \ref{sec:high_k_ach}.  Another limitation of Theorem \ref{thm:general_noise} is that the converse part is specific to Bernoulli testing; however, analogous results for arbitrary nonadaptive test designs have also been established in \cite{malyutov,scarlett-cevher-3}, the details of which are omitted here.

\subsubsection{Discussion of achievability proof}

Proofs of the achievability part of Theorem \ref{thm:general_noise} can be found in \cite{Mal80,atia-saligrama,scarlett-cevher-2}; continuing the earlier analysis, we discuss the approach of \cite{scarlett-cevher-2}.

As mentioned above, the bound \eqref{eq:non_asymp} remains valid in the noisy setting, and the main step in the subsequent analysis is establishing the concentration of $\imath^T( {\mat X}_{0,\tau}; {\vec Y} | {\mat X}_{1,\tau} )$. In general, this is a challenging task, and may introduce extra conditions on $T$, as we saw in the proof of Theorem \ref{thm:ch2main}.  However, it turns out that when $k = O(1)$, the concentration bound given in the second part of Lemma \ref{lem:ach_concentration} (which extends immediately to general noise models \cite{scarlett-cevher-1}) is sufficient.  Indeed, assuming bounded $k$ greatly simplifies matters, since it means that the combinatorial term $\binom{k}{\tau}$ in \eqref{eq:non_asymp} is also bounded.

The equality \eqref{eq:general_noise_ach2} follows from elementary information-theoretic arguments, which we outline here.  Assuming without loss of generality that $\K = \{1,\dotsc,k\}$, writing the entries of ${\vec X}_{\K}$ as $(X_1,\dotsc,X_k)$ accordingly, and letting ${\vec X}_{j}^{j'}$ denote the collection $(X_{j},\dotsc,X_{j'})$ for indices $1 \le j \le j' \le k$, we have
\begin{align}
    \frac{1}{\tau} I({\vec X}_{0,\tau} ; Y \mid {\vec X}_{1,\tau} )
    &= \frac{1}{\tau} I({\vec X}_{k-\tau+1}^{k};Y\mid {\vec X}_{1}^{k-\tau}) \label{eq:ratio1} \\
    &= \frac{1}{\tau} \sum_{j=k-\tau+1}^k I( X_j; Y \mid {\vec X}_{1}^{j-1} ) \label{eq:ratio2} \\
    &= \frac{1}{\tau} \sum_{j=k-\tau+1}^k \big( H(X_j) - H(X_j\mid Y,{\vec X}_{1}^{j-1}) \big), \label{eq:ratio3}
\end{align}
where \eqref{eq:ratio1} follows from the definition of $({\vec X}_{0,\tau},{\vec X}_{1,\tau})$ and the symmetry of the noise model in \eqref{eq:Y_noisy}, \eqref{eq:ratio2} follows from the chain rule for mutual information, and \eqref{eq:ratio3} follows since $X_j$ is independent of ${\vec X}_{1}^{j-1}$.  We establish the desired claim by observing that \eqref{eq:ratio3} is decreasing in $\tau$: The term $H(X_j)$ is the same for all $j$, whereas the term $H(X_j|Y,{\vec X}_{1}^{j-1})$ is smaller for higher values of $j$ because conditioning reduces entropy.

\subsubsection{Discussion of converse proof} 

In light of the apparent connection between channel coding and group testing (see Figure \ref{fig:channel_diagram}), a natural starting point is to apply Fano's inequality, which states that in order to achieve an error probability of $\delta$, it is necessary that
\begin{equation}
	I(\K;\vec{Y} \mid \mat{X}) \ge \log_2{ \binom{n}{k}} (1-\delta) - 1. \label{eq:Fano_init}
\end{equation}
Note that $\vec{Y}$ depends on $\K$ only through $\mat{X}_{\K}$, which corresponds to $\mat{X}_{0,k}$ in the above notation.  We can therefore replace $I(\K;\vec{Y} \mid \mat{X})$ by $I(\mat{X}_{0,k};\vec{Y})$, which in turn equals $T I(\vec{X}_{0,k};Y)$ since the tests are independent. Substituting into \eqref{eq:Fano_init} and rearranging, we obtain the necessary condition
\begin{equation}
	T \ge \frac{ \log_2{\binom{n}{k}} }{ I(\vec{X}_{0,k};Y) } \bigg(1 - \delta - \frac{1}{ \log_2{\binom{n}{k}} }\bigg). \label{eq:Fano_2}
\end{equation}
This bound matches \eqref{eq:general_noise} whenever the maximum therein is achieved by $\tau = k$.  However, as discussed above, this is not always the case.

The key to overcoming this limitation is to use a `genie argument' \cite{atia-saligrama}, in which a subset of $\K$ is revealed to the decoder, and it only remains to estimate the non-revealed part.  This clearly only makes the recovery problem easier, so any converse for this genie-aided setting remains valid in the original setting.  Note that since $\mat{X}$ is generated in a symmetric i.i.d.~manner and the assumed model \eqref{eq:Y_noisy} is invariant to relabelling, it makes no difference precisely which indices are revealed; all that matters is the number revealed. (However, the revealed indices must not depend on $\mat X$ or $\vec y$.)  Letting $\tau$ denote the number of defectives left to estimate, the number revealed is equal to $k - \tau$.

In the genie-aided setting, the number of possible defective sets reduces from $\binom{n}{k}$ to $\binom{n - k + \tau}{\tau}$.  Moreover, the relevant mutual information in Fano's inequality is not $I(\K;\vec{Y}|\mat{X})$, but instead $I(\K_{0,\tau};\vec{Y}|\K_{1,\tau},\mat{X})$, where $\K_{0,\tau}$ (respectively, $\K_{1,\tau}$) denotes the non-revealed (respectively, revealed) defective item indices.  Upon upper bounding the mutual information via the data processing inequality, we obtain the following analogue of \eqref{eq:Fano_2}:
\begin{equation}
	T \ge \frac{ \log_2{ \binom{n-k+\tau}{\tau}} }{ I(\vec{X}_{0,\tau};Y|\vec{X}_{1,\tau}) } \bigg(1 - \delta - \frac{1}{ \log_2{\binom{n-k+\tau}{\tau}} }\bigg). \label{eq:Fano3}
\end{equation}
We then recover \eqref{eq:general_noise} by maximizing over $\tau=1,\dotsc,k$ and noting that
\begin{equation}
    \log_2 \binom{n-k+\tau}{\tau} = \Big(\tau\log_2\frac{n}{\tau}\Big)(1+o(1)). \label{eq:log2}
\end{equation}

We mention that an alternative approach was taken in \cite{scarlett-cevher-2}, bearing a stronger resemblance to the above achievability proof and again relying on change-of-measure techniques from the channel coding literature.  The proof of \cite{scarlett-cevher-2} has the advantage of recovering the so-called `strong converse' (see Remark \ref{rem:universal}), but it requires additional effort in ensuring that the suitable sums of information densities concentrate around the corresponding conditional mutual information.

\subsubsection{Examples: Symmetric, addition, and dilution noise}

We briefly discuss the application of Theorem \ref{thm:general_noise} to one symmetric noise model and two asymmetric noise models introduced above:
\begin{itemize}
    \item \added{Recall from Example \ref{ex:bsc} that in the binary symmetric noise model, each test outcome gets flipped with some probability $\rho$.  By bounding the mutual information in Theorem \ref{thm:general_noise}, we find that the optimal number of tests with $k=O(1)$ behaves as \[ \frac{1+o(1)}{1-h(\rho)} \,k \log_2 n , \] where $h(\rho)$ is the binary entropy function in bits (see, for example, \cite{Mal80,scarlett-cevher-2,atia-saligrama}).  This is natural, since $1-h(\rho)$ is the capacity of the binary symmetric channel.  We note that that $1-h(\rho) = \Theta( (1-2\rho)^2 )$ has a quadratic dependence in $\rho$ for $\rho \approx \frac{1}{2}$. }
	\item Recall from Example \ref{ex:addition} that the \emph{addition noise} model takes the form $Y_t = \big(\bigvee_{i \in \K} {\mat X}_{ti}\big) \vee Z_t$, where $Z_t \sim \mathrm{Bernoulli}(\varphi)$.  By bounding the mutual information in Theorem \ref{thm:general_noise}, it was shown in \cite{atia-saligrama} that the optimal number of tests with $k=O(1)$ behaves as $O\big( \frac{ k \log n }{ 1 - \varphi } \big)$.  Hence, we have a simple linear dependence on the addition noise parameter.
    \item Recall from Example \ref{ex:dilution} that the \emph{dilution noise} model takes the form $Y_t = \bigvee_{i \in \K} \big( {\mat X}_{ti} \wedge Z_{ti} \big)$, where $Z_{ti} \sim \mathrm{Bernoulli}(\vartheta)$.  By bounding the mutual information in Theorem \ref{thm:general_noise}, it was shown in \cite{atia-saligrama} that the optimal number of tests $k=O(1)$ behaves as $O\big( \frac{ k \log n }{ (1-\vartheta)^2 } \big)$.  Hence, the dependence in the denominator is quadratic.
\end{itemize}
Additional expressions for the relevant mutual information terms for various noise models, often including precise constant factors, can be found in \cite{laarhoven-1}.

\subsection{Binary channel noise models in general sparse regimes} \label{sec:high_k_ach}

\added{
In Section \ref{sec:pf_ach}, we outlined the techniques of Scarlett and Cevher \cite{scarlett-cevher-1} for establishing the achievability threshold for Bernoulli testing in the noiseless setting.  These techniques also naturally apply to noisy settings, and, accordingly, they were applied to the symmetric noise model in \cite{scarlett-cevher-1} to obtain an achievable rate that notably equals $1-h(\rho)$ (with $h(\rho)$ being the binary entropy function) when $\alpha$ is sufficiently small in the scaling $k = \Theta(n^{\alpha})$.  However, for larger $\alpha$ values, significant gaps still remained following their work, and the understanding of the noisy setting had generally lagged behind the noiseless setting.

In recent concurrent works by Chen and Scarlett \cite{chen2024exact} and Coja-Oghlan \etal~\cite{coja2024noisy}, these gaps have been substantially reduced.  The main results of these papers are highly technical even to state (let alone prove), so we only provide an outline conveying the main ideas and findings.

We first highlight the differences in the results in these two works:
\begin{itemize}
    \item Chen and Scarlett \cite{chen2024exact} focus on the binary symmetric noise model (see Example \ref{ex:bsc}), in which each test outcome is flipped with probability $\rho \in (0,\frac{1}{2})$.  
    Under this model, they attain exact information-theoretic thresholds for both the Bernoulli design and the near-constant column weight design, with the former design establishing that the achievable rates in \cite{scarlett-cevher-1} are suboptimal (except for small $\alpha$), and the latter design providing further improvements.
    \item Coja-Oghlan \etal~\cite{coja2024noisy} consider arbitrary binary channel noise models (see Example \ref{ex:binary_channel}) that need not be symmetric, and show that a design based on spatial coupling attains a threshold matching that of \cite{chen2024exact} when specialized to symmetric noise, while having polynomial-time decoding.  They also show that the same threshold serves as a converse for the near-constant column weight design.
\end{itemize}

To simplify the exposition, we will focus only on symmetric noise below.  Collectively, the results outlined above establish the exact rates achieved by the Bernoulli design and the near-constant column weight design, and show that the latter can be achieved with polynomial-time decoding using spatial coupling.  Thus, they bring the understanding to a level that nearly matches the noiseless setting, except that it remains unknown whether the near-constant column weight design is optimal among \emph{all} nonadaptive designs.

The following theorem reveals the general form of the thresholds attained in these works, as well as the weaker threshold in \cite{scarlett-cevher-1}.

\begin{theorem} \label{thm:symm_high_k}
    Consider the symmetric noise model with parameter $\rho$ in the sparse regime $k = \Theta(n^{\alpha})$, $\alpha \in (0,1)$. With the Bernoulli design or the near-constant column weight design with an optimized parameter $\nu$, there exists a decoder achieving the rate
    \begin{equation}
        R^{\mathrm{Bern}}_{\mathrm{sym}} = \min\big\{  1 - h(\rho), \,c(\rho,\alpha)  \big\}, \label{eq:symm_rate}
    \end{equation}
    where $h(\rho)$ is the binary entropy function in bits, and $c(\rho,\alpha)$ is a continuous function with $c(\rho, 0) > 1 - h(\rho)$.  In particular, for sufficiently small $\alpha$, the capacity is given by
    \begin{equation}
        C_{\mathrm{sym}} = 1 - h(\rho). \label{eq:symm_capacity}
    \end{equation}
\end{theorem}

While the result of \cite{scarlett-cevher-1} for Bernoulli testing was in this form, it was given with a highly suboptimal expression for $c(\rho,\alpha)$, whereas \cite{chen2024exact,coja2024noisy} established precise expressions for the respective designs.  These precise expressions for $c(\rho,\alpha)$ are complicated, so are omitted here, but the resulting rate plots for $\rho =0.01$ and $\rho = 0.11$ are shown in Figure \ref{fig:rates_noisy}.

\begin{figure}[t]
\begin{center}
\includegraphics[width=0.47\textwidth]{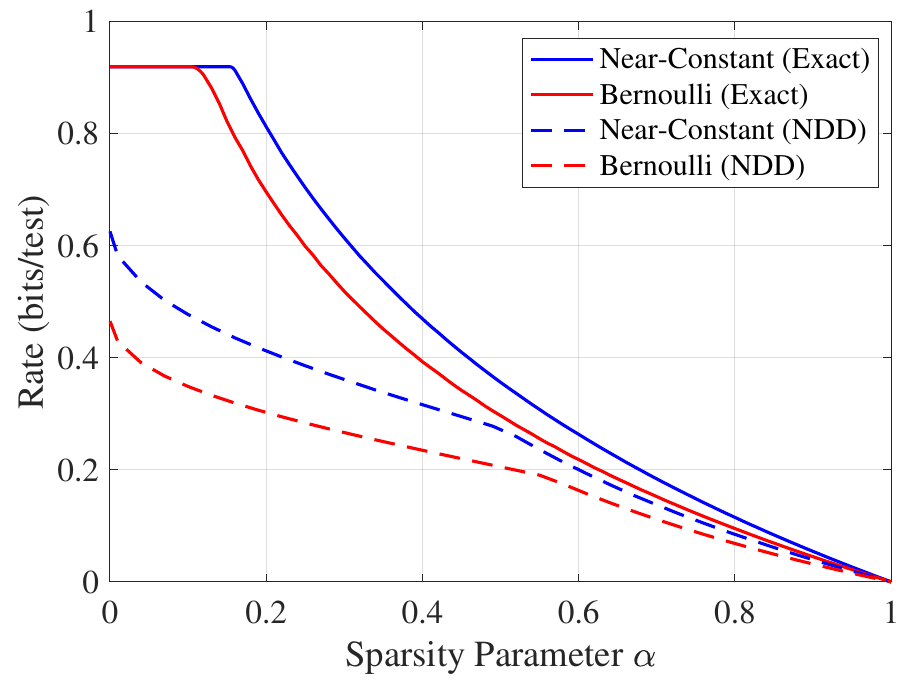} 
\includegraphics[width=0.455\textwidth]{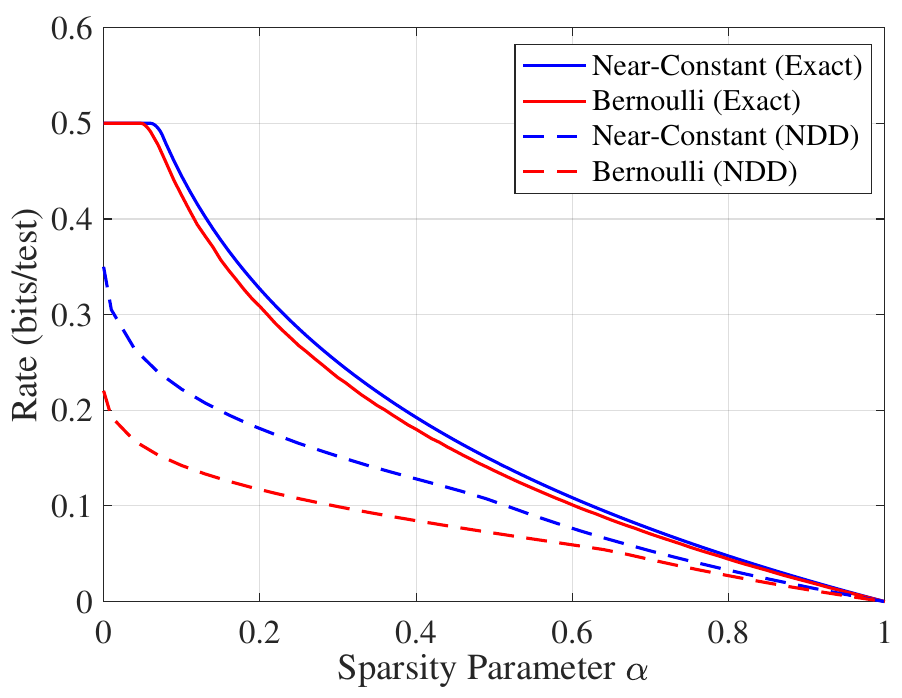}
\end{center}
\caption{Rates of noisy group test with $\rho = 0.01$ (left) and $\rho = 0.11$ (right).  The curves labelled NDD represent the achievable rates of the near-definite defectives algorithm (Section \ref{sec:ndd}).} \label{fig:rates_noisy}
\end{figure}

A consequence of these results is that existing approaches such as noisy DD (see Section \ref{sec:ndd}) are strictly suboptimal under symmetric noise.  This is in contrast with certain asymmetric models, where noisy DD can be optimal for $\alpha$ close to $1$.  Another curious finding in these works is that for the near-constant column weight design, the optimal test design parameter $\nu$ is not always $\ln 2$; rather, it is $\ln 2$ for small $\alpha$, then quickly jumps to a higher value, and then steadily decreases again.

We give a very brief outline of the proof techniques in \cite{chen2024exact,coja2024noisy} as follows, mainly focusing on the differences to the noiseless setting:
\begin{itemize}
    \item To establish converse results, the dominant error event is no longer \emph{complete masking} (see Sections \ref{sec:sss} and \ref{sec:gen_conv}), but instead the following: (i)~a defective $i \in \K$ has relatively few tests in which it is the only defective, and not too few of those tests are flipped by the noise; (ii) a nondefective $j \notin \K$ appears in sufficiently many tests without any defectives, and not too few of those tests are flipped by the noise.  Under suitably-defined events of this type, we can deduce that $\K \cup \{j\} \setminus \{i\}$ is favoured over $\K$ by the optimal maximum-likelihood decoder, thus leading to an error.
    \item In \cite{chen2024exact}, it is discussed that, perhaps surprisingly, extending the techniques of \cite{scarlett-cevher-1} leads to a suboptimal rate for Bernoulli testing.  The looseness stems from the set difference parameter $\tau$ (see \eqref{eq:general_noise}) sometimes being dominant at $\tau = 1$ rather than $\tau = k$, particularly as $\alpha$ increases. (Another source of weakness in \cite{scarlett-cevher-1} is using a suboptimal concentration bound for the information density when $\tau$ is small, but the limitation being discussed here applies even when an optimal concentration bound is used.)  
    To circumvent this weakness, a \emph{hybrid} decoder is proposed in \cite{chen2024exact} that reduces the analysis to information-theoretic thresholding (as in Section \ref{sec:pf_ach}) for incorrect sets `far' from the true one, but to maximum-likelihood decoding for those `close' to the true one.  The maximum-likelihood decoding analysis then follows similar ideas to the converse analysis.
    \item In \cite{chen2024exact} the spatial coupling approach is taken.  The general ideas are similar to those outlined in Section \ref{sec:spatial} but with non-minor technical differences, whose details we omit here.
\end{itemize}

We conclude this section by reiterating that neither \cite{chen2024exact} nor \cite{coja2024noisy} attempted to establish a matching converse bound for \emph{arbitrary} nonadaptive test designs, and this remains an important gap in this line of works.
}

\chapter{Other Topics in Group Testing} \label{ch:other-topics}

In this chapter, we explore extensions of group testing beyond the settings considered in the previous chapters, which were primarily focused on nonadaptive randomized designs, the exact recovery criterion, and sublinear scaling in the number of defectives. 
Many of the extensions considered below have natural analogues in classical information theory, 
and we attempt to draw such parallels when they arise naturally.


\section{Approximate recovery} \label{sec:partial}

In many group testing situations, one might be satisfied with an estimate of the defective set $\Khat$ being `sufficiently close' to the true defective set $\K$, without demanding the exact recovery criterion we have considered throughout this survey. That is, it might suffice for the number of \emph{false negative items} $|\K \setminus \Khat|$ and the number of \emph{false positive items} $|\Khat \setminus \K|$ to be small, 
without them necessarily being zero. For example, when screening for diseases, a small number of false positives might lead to slightly more medical attention for those who did not need it, a cost which might be small compared to performing many more pooled tests. 

We briefly mention that moving to approximate recovery is known to significantly help under the zero-error recovery criterion, allowing one to break the $\Omega(k^2)$ barrier discussed in Section \ref{sec:zero-error} and instead use only $O(k \log n)$ tests, even in the presence of adversarial test errors \cite{cheraghchi2009noise}.  In the following,  we focus only on the small-error criterion.

\subsection{Achievable rates}

Permitting a limited number of mistakes in the reconstruction is analogous to the notion of {\em rate--distortion theory} in information theory, where one only requires a source to be reconstructed approximately (see \cite[Chapter 10]{cover} for a review of this topic).  A natural performance criterion that treats both error types equally is
  \[ \Perrd = \PP\big( |\K \setminus \Khat| > \dmax \text{ or } |\Khat \setminus \K| > \dmax \big),  \]
which declares an error if either the number of false negatives or false positives exceeds a common threshold $d$.

The following result, due to Scarlett and Cevher \cite{scarlett-cevher-2}, characterizes how this relaxed criterion affects the required number of tests when this threshold is a constant fraction of the number of defectives.  

\begin{theorem} \label{thm:partial_noiseless}
	For nonadaptive group testing with $k = o(n)$ and $\dmax = \gamma k$ for some $\gamma \in (0,1)$, under the Bernoulli test design with parameter $p$, we have the following:
\begin{itemize}
  \item With $p = \frac{\ln 2}{k}$, there exists an algorithm such that $\Perrd \to 0$ provided that $T > (1+\eta)T^*$ for arbitrarily small $\eta > 0$, where
    $$ T^* = k \log_2 \frac{n}{k}.$$
  \item For any $p$ and any algorithm, in order to achieve $\Perrd \to 0$ it is necessary that $T > (1-\eta)T^*_\gamma$ for arbitrarily small $\eta > 0$, where
    \[ T^*_\gamma =  (1 - \gamma)k \log_2 \frac{n}{k} = (1-\gamma)T^*. \]
 \end{itemize}
\end{theorem}

The achievability part can be proved using the argument used to prove Theorem \ref{thm:ch2main} (see Section \ref{sec:pf_ach}).  The difference is that we only need to consider incorrect sets having small overlap with the true defective set, since those with large overlap still suffice for approximate recovery.  The converse part is also proved in a similar manner to the exact recovery result.

Theorem \ref{thm:partial_noiseless} implies both positive and negative results on the extent to which approximate recovery reduces the number of tests under the Bernoulli design compared with the bounds discussed in Chapter \ref{ch:achievability}.  Letting $\alpha$ be the exponent such that $k = \Theta(n^{\alpha})$ as usual, we observe the following:
\begin{itemize}
	\item For $\alpha \le 1/3$, the gain is very limited, amounting to at most a reduction by the multiplicative factor $1 - \gamma$, which vanishes as $\gamma \to 0$.  This is because nonadaptive Bernoulli testing achieves a rate of $1$ in this regime.
    \item For $\alpha > 1/3$, the gain is more significant -- the number of tests remains $\big(k \log_2 \frac{n}{k}\big)(1+o(1))$ under the approximate recovery criterion, whereas for exact recovery the rate tends to zero as $\alpha \to 1$.
\end{itemize}
Extensions of Theorem \ref{thm:partial_noiseless} are given in \cite{scarlett-cevher-4} to a {\em list decoding} setting, in which the decoder outputs a list of length $L \ge k$ and it is only required that the list contains at least $(1-\gamma)k$ defectives.  \rev{(The concept of list decoding for group testing also appeared much earlier under the zero-error criterion, e.g., see \cite{indyk2010}).}  If $L$ is much larger than $k$, this means that we are potentially allowing a large number of false positives.  However, a finding of \cite{scarlett-cevher-4} is that this relaxation only amounts to a replacement of $k\log_2\frac{n}{k}$ by $k\log_2\frac{n}{L}$ in the required number of tests (asymptotically), which is a rather minimal gain.  

\added{We additionally note some follow-up works giving further related results:
\begin{itemize}
    \item A counterpart to Theorem \ref{thm:partial_noiseless} for the near-constant tests-per-item design can be inferred from \cite{coja}, though they focused on exact recovery (as we surveyed in Section \ref{sec:ach_near_const}).
    \item The achievability part can also be proved for a spatially coupled test design that permits polynomial time decoding \cite[Sec.~5]{coja2020optimal} (as we surveyed for exact recovery in Section \ref{sec:spatial}).
    \item The necessary condition $T \ge (1-\gamma)k \log_2 \frac{n}{k}$ (and its proof) applies to general nonadaptive designs, and there is a simple nonadaptive design that gives a matching achievability result \cite{truong2020all}.
\end{itemize}
\added{To establish the last claim, the idea is to ignore a $\gamma - \eta'$ fraction of the items for some small $\eta' > 0$, i.e., do not test them, and simply guess that they are nondefective.  Among the remaining $(1-\gamma + \eta')n$ items, there are roughly $(1-\gamma + \eta')k$ defectives, and we test these items using a Bernoulli design in a similar manner to Theorem \ref{thm:partial_noiseless}, but with suitably modified parameters to ensure fewer than $\eta' k$ false positives and false negatives (rather than $\gamma k$).  Since the step of ignoring items incurs roughly $(\gamma - \eta')k$ false negatives, we can ensure at most $\gamma k$ false negatives overall (and at most $\eta' k < \gamma k$ false positives), as desired.  Taking $\eta' \to 0$ then gives the desired result with a $1-\gamma$ factor saving in the number of tests.}  The full details can be found in \cite{truong2020all}. }

It is also of interest to understand the achievable rates of the practical algorithms studied in Chapter \ref{ch:algorithms} and \ref{ch:algorithms_noisy} under the approximate recovery criterion.  In the noiseless case, it is in fact straightforward to extend the exact recovery analysis of COMP and DD; starting with the Bernoulli design, we note the following:
\begin{itemize}
    \item The COMP algorithm (Algorithm \ref{def:comp}) always has no false negatives, and the analysis of Section \ref{sec:COMP} shows that when $T \ge (\ee k \ln n)(1+\eta)$ so (i.e., the rate is below $(1-\alpha)/(\ee \ln 2)$), the average number of false positives tends to zero, and therefore the probability of having one or more false positives also tends to zero.  By a nearly identical analysis, one finds that when $T \ge \big(\ee k \ln \frac{n}{k} \big)(1+\eta)$, the average number of false positives behaves as $o(k)$, and therefore the probability of having more than $\gamma k$ false positives tends to zero for any fixed $\gamma \in (0,1)$, by Markov's inequality.
    \item The DD algorithm (Algorithm \ref{def:dd}) always has no false positives, and the analysis of Section \ref{sec:DD} shows that when $T \ge \big(\ee k \ln \frac{n}{k}\big) (1+\eta)$, any given defective item is the unique `possible defective' (PD) in some test, with probability approaching one.  For exact recovery, an additional condition $T \ge (\ee k \ln k)(1+\eta)$ arises from a union bound over the $k$ defective items.  In the case of approximate recovery, however, we can instead use the fact that the average number of defective items failing to be the unique PD in some test behaves as $o(k)$, and therefore, the probability of having more than $\gamma k$ false negatives tends to zero for any fixed $\gamma \in (0,1)$, by Markov's inequality.
\end{itemize}
Hence, using Definition \ref{def:rate} and \eqref{eq:bincoeffequiv}, a rate of $R = \frac{1}{\ee \ln 2} \approx 0.531$ is achieved by COMP with no false negatives, and by DD with no false positives.  For the near-constant column weight design one can follow similar ideas and attain analogous results with the rates improved to $\ln 2 \approx 0.693$.  \added{The details can be found in the recent work of McMorrow and Scarlett \cite{mcmorrow}, who further established the information-theoretically optimal rates under both one-sided approximate recovery criteria (false positives only and false negatives only).}

In fact, the rate $\ln 2$ for approximate recovery can also be obtained under the Bernoulli design.  Specifically, using similar arguments based on avoiding the union bound and instead applying Markov's inequality, it has been shown that separate decoding of items (see Section \ref{sec:separate}) achieves a rate of $\ln 2 \approx 0.693$ in the noiseless setting when both $\gamma k$ false positives and $\gamma k$ false negatives are allowed \cite{Sca17b}.  This rate is higher than that of COMP and DD for the Bernoulli design, but comes with the caveat of requiring both false positives and false negatives.  \rev{The results are summarized in Table \ref{tbl:partial}.}

\begin{table}
\begin{center}
\begin{tabular}{ccccc}
\hline
 \mbox{}  & Rate  & Rate & No false $+$ & No false $-$ \\
 \mbox{}  & (Bern.) & (near-const.) & \\
\hline
\textbf{Optimal} & $1$ & 1 & no & no \\
\textbf{COMP} & $\frac{1}{\ee \ln 2}$ & $\ln 2$ & no & yes \\
\textbf{DD}   & $\frac{1}{\ee \ln 2}$ & $\ln 2$ & yes & no \\
\textbf{Separate Dec.} & $\ln 2$ & -- & no & no \\
\hline
\end{tabular}
\end{center}
\caption{\rev{Summary of achievable rates (in bits/test) for approximate recovery under Bernoulli testing.  Each achievable rate holds for all $\alpha \in (0,1)$ and an arbitrarily small (but constant) fraction of mistakes in the reconstruction.  The final two columns indicate whether the algorithm is guaranteed to have no false positives/negatives. \label{tbl:partial}}}
\end{table}

    Analogous results have also been given in noisy settings.  For instance, under the symmetric noise model with parameter $\rho \in (0,1/2)$, the information-theoretic rate given in Theorem \ref{thm:partial_noiseless} naturally becomes $1 - h(\rho)$, and the rate based on separate decoding of items becomes $(1 - h(\rho)) \ln 2$.  The interested reader is referred to \cite{Sca17b,scarlett-cevher-2}.

\subsection{All-or-nothing thresholds} \label{sec:allnothing}

\added{As discussed following Theorem \ref{thm:partial_noiseless}, the $k \log_2 \frac{n}{k}$ threshold for Bernoulli design can be improved to $(1-\gamma)k \log_2 \frac{n}{k}$ when the approximate recovery parameter is $d = \gamma k$, by moving to a slightly modified design that ignores some items and declares them as nondefective.  This raises the question of whether modifying the design is actually necessary to break the $k \log_2 \frac{n}{k}$ barrier while still attaining the approximate recovery guarantee for some $\gamma \in (0,1)$.

A result of Niles-Weed and Zadik \cite{niles2023all}, which strengthened earlier results of Truong, Aldridge, and Scarlett~\cite{truong2020all}, shows that the threshold $T^* = k \log_2 \frac{n}{k}$ in fact serves as a very strong threshold for the Bernoulli design whenever $k = o(n)$:  If the number of tests is reduced even slightly (by a multiplicative factor $1-\eta$ for arbitrarily small $\eta > 0$), then not even a small fraction of the defectives will be recovered correctly.  Specifically, we cannot achieve approximate recovery with $d = \gamma k$ for \emph{any} fixed $\gamma \in (0,1)$; even a large choice such as $\gamma = 0.99$.

Combining this result with Theorem \ref{thm:partial_noiseless}, we see that under the Bernoulli design, `near-perfect' recovery is possible when $T > (1+\eta)T^*$, but `near-useless' recovery is unavoidable when $T < (1-\eta)T^*$.  This phenomenon is known as an \emph{all-or-nothing threshold}, and had previously been observed for other statistical estimation problems.  An analogous result was subsequently given for the constant column weight design by Coja-Oghlan \emph{et al.}~\cite{coja2022statistical}.  Collectively, these results establish the following.

\begin{theorem}
    Consider the noiseless nonadaptive group testing problem in the sparse regime $k = \Theta(n^{\alpha})$ with $\alpha \in (0,1)$, and let $\K$ be the true defective set of size $k$.  Then, under either the Bernoulli design or the constant column weight design with the parameter $\nu$ chosen such that half the tests are positive on average (up to suitable rounding in the latter design), we have the following:
    \begin{itemize}
        \item (`All') If the rate is below $1$, in that $T \ge (1+\eta)k\log_2\frac{n}{k}$ for some $\eta > 0$, then there exists a decoding strategy recovering an estimate $\Khat$ such that $|\K \setminus \Khat| = o(k)$ and $|\Khat \setminus \K| = o(k)$ with probability approaching one.
        \item (`Nothing') If the rate exceeds $1$, in that $T \le (1-\eta)k\log_2\frac{n}{k}$ for some $\eta > 0$, then for any decoding strategy producing an estimate $\Khat$ of size $k$, it holds with probability approaching one that $|\K \cap \Khat| = o(k)$.
    \end{itemize}
\end{theorem}

For the Bernoulli design with parameter $p = \nu/k$, to have half the tests be positive on average, one should choose $\nu$ to satisfy $\big(1 - \frac{\nu}{k}\big)^k = \frac{1}{2}$, which implies $\nu = \ln 2+o(1)$.  Similarly, one can show for the near-constant column weight design with parameter $L = \nu T/k$ that one should also set $\nu = \ln 2+o(1)$.  For the Bernoulli design, the more general case of $(1 - \frac{\nu}{k})^k$ equalling a constant in $(0,\frac{1}{2})$ was also handled in \cite{niles2023all}, though the case that this constant lies in $(\frac{1}{2},1)$ was skipped for technical reasons.
}




\section{Adaptive testing with limited stages} \label{sec:two-stage}

\subsection{Noiseless testing}

We saw in Section \ref{sec:adaptive} that adaptive testing permits the exact zero-error identification of $\K$ with an information-theoretically optimal rate $R= 1$.  This offers two key advantages over nonadaptive testing: replacing the small-error criterion by the zero-error criterion, and achieving $R= 1$, which is only possible for $k = O(n^{0.409})$ in the nonadaptive case.  On the other hand, adaptive testing schemes may come with considerable overhead compared to nonadaptive testing, since it is no longer possible to perform all of the tests in parallel.

An interesting variation that potentially attains the benefits of both worlds is {\em two-stage testing}, in which a very limited amount of adaptivity is allowed; namely, one can only perform two stages of testing, in which the tests in the second stage can depend on the outcomes in the first stage.  The binary splitting algorithm (Algorithm \ref{alg:bin}) described in Section \ref{sec:adaptive} does not fall into this category, and in fact uses $O(\log n)$ stages.

A variety of algorithms and results have been proposed for the two-stage setting \cite{berger2002,damaschke2012randomized,deBonis2005optimal,macula1998two,mezard2011two}.  Here we present a result of M{\'e}zard and Toninelli \cite{mezard2011two}, which improves on the earlier bounds of \cite{berger2002}.  Note that here the notion of `rate' is defined with respect to the {\em average} number of tests for a random defective set (i.e., using $\frac{\log_2 \binom{n}{k}}{ \EE T }$ in \eqref{eq:rate} instead of $\frac{\log_2 \binom{n}{k}}{ T }$); we refer to this as the {\em variable-$T$ setting}.  

\begin{theorem} \label{thm:two-stage}
    Consider the problem of two-stage group testing in the variable-$T$ setting with zero error.  
    When $k = \Theta(n^{\alpha})$ for some $\alpha \in (0,1)$, the following rate is achievable:
        $$ R_{2} = 
        \begin{cases}
            \dfrac{1}{\ee \ln 2} \approx 0.531 & \alpha \le \frac{1}{2} \\
            \ln 2 \approx 0.693 & \alpha > \frac{1}{2}.
        \end{cases} $$
\end{theorem}

\rev{This result was stated in \cite{mezard2011two} under the i.i.d.~prior defectivity model (see the Appendix to Chapter \ref{ch:introduction}), but the proof transfers easily to the combinatorial prior.  
A converse of $\ln 2$ for all $\alpha \in (0,1)$ is also given in \cite{mezard2011two} under the i.i.d.~prior (in particular matching the achievability part when $\alpha > \frac{1}{2}$).  
}


\added{The idea behind achieving the zero-error recovery criterion in Theorem \ref{thm:two-stage} is to individually test any item whose status is not determined by the first stage.  Once the structure of the second stage is fixed in this way,}
 the freedom in the test design is entirely in the first stage.  For $\alpha > 1/2$, this stage is based on the standard i.i.d.~Bernoulli testing procedure considered throughout this monograph, whereas for $\alpha \leq 1/2$, an alternative construction is used in which $\mat{X}$ has both constant row weight and constant column weight. 
We observe that two-stage testing with zero error probability requires {\em considerably} fewer tests compared to the nonadaptive case, in particular avoiding the $\Omega(k^2)$ barrier (see Section \ref{sec:zero-error}).  \added{The benefits of two-stage testing will further be explored for the linear regime $k \sim \beta n$ in Section \ref{sec:linear}, using similar ideas to those of Theorem \ref{thm:two-stage}.}


At this point, it is natural to question whether there exists a more general trade-off between the number of stages and the rate.  This question was addressed by Damaschke and Muhammad \cite{damaschke2012randomized} under both the small-error and zero-error recovery criteria.  Among other things, it was proved that even in the zero-error setting, four stages is enough to attain a rate of one.

\begin{theorem} \label{thm:four-stage}
	For any $k = o(n)$, the capacity of four-stage adaptive group testing in the variable-$T$ setting with zero error is $C_4 = 1$.
\end{theorem}

A high-level description of the four-stage algorithm establishing this result is as follows:
\begin{itemize}
    \item The $n$ items are split into $\frac{k}{\Delta}$ `cells' of size $\frac{\Delta n}{k}$, where $\Delta \to 0$ sufficiently slowly so that $\Delta \log\frac{n}{k} \to \infty$.
    \item In the first stage, each test (one per cell) consists of all the items in a given cell, and hence, all empty cells are identified as containing only nondefectives.
    \item In the second stage, for each non-empty cell, a known nonadaptive procedure is used to identify whether the cell has exactly one defective item (and if so, also determine its index) or more than one item.
    \item In the third and fourth stages, the cells having multiple items are merged, and a two-stage group testing procedure is applied (e.g., the one corresponding to Theorem \ref{thm:two-stage} suffices).
\end{itemize}

It is natural to question whether an analogous result holds with two or three stages.  The converse result discussed following Theorem \ref{thm:two-stage} indicates that the answer is negative in the case of two stages and the zero-error criterion.  On the other hand, if one considers the small-error criterion in place of zero-error, a rate of one can be achieved with two stages, as shown in the following result of \cite{Sca18} (see also \cite{damaschke2012randomized} for an earlier result using three stages, \added{and \cite{coja2020optimal} for an analogous result via spatial coupling with polynomial decoding time}). 

\begin{theorem} \label{thm:two-stage-small}
	For any $k = \Theta(n^{\alpha})$ with $\alpha \in (0,1)$, the capacity of two-stage adaptive group testing in the fixed-$T$ setting under the small-error criterion is $C_2 = 1$. 
\end{theorem}

The high-level idea of the proof is straightforward: 
\begin{itemize}
	\item By the approximate recovery result of Theorem \ref{thm:partial_noiseless}, in the first stage, we can find an estimate $\hat{\K}_1$ of cardinality $k$ with at most $\gamma k$ false positives and $\gamma k$ false negatives, where $\gamma > 0$ is an arbitrarily small constant.
    \item In the second stage, we apply any nonadaptive noiseless strategy on the reduced ground set $\{1,\dotsc,n\} \setminus \hat{\K}_1$ to resolve the false negatives.  As long as this strategy achieves a positive rate, the required number of tests will be $O(\gamma k \log n)$, which is negligible since $\gamma$ is arbitrarily small.
    \item Simultaneously in the second stage, we test the items in $\hat{\K}_1$ individually to resolve the false positives.  This only requires $k$ tests.
\end{itemize}
\rev{The results for the noiseless setting surveyed throughout this subsection are summarized in Table \ref{tbl:multi-stage}.}

\begin{table}
\begin{center}
    \begin{tabular}{cccc}
    \hline
    Reference & Rate & \# Stages & Zero-error?  \\
    \hline
    \cite{mezard2011two} & {\small 
    $\begin{cases}
        \frac{1}{\ee \ln 2} & \alpha \le \frac{1}{2} \\
        \ln 2 & \alpha > \frac{1}{2}
    \end{cases}$} & 2 & yes \\
    \cite{damaschke2012randomized} & 1 & 4 & yes  \\
    \cite{damaschke2012randomized} & 1 & 3 & no \\
    \cite{Sca18,coja2020optimal} & 1 & 2 & no \\
    \hline
    \end{tabular}
\end{center}
\caption{\rev{Summary of rates for multi-stage adaptive group testing \added{in the noiseless setting} when $k = \Theta(n^{\alpha})$, depending on the number of stages and whether the error probability is zero (in which case the number of tests may be variable -- the `variable-$T$' setting).  \label{tbl:multi-stage}}}
\end{table}

\subsection{Noisy settings}

The approach described following Theorem \ref{thm:two-stage-small} was adopted in \cite{Sca18} not only for the noiseless setting, but also for noisy settings.  The changes compared to the noiseless setting are outlined as follows. In the first step, one can make use of the approximate recovery results for the noisy setting outlined at the end of Section \ref{sec:partial}.  In the second step, one can use a noisy nonadaptive algorithm that achieves a positive rate, such as the NCOMP algorithm introduced in Section \ref{sec:NCOMP}.  In the third step, testing each item once is no longer sufficient, but testing each item $\Theta(\log_2 k)$ times is enough to combat the noise.

\begin{figure}
\begin{center}
\includegraphics[width=0.6\textwidth]{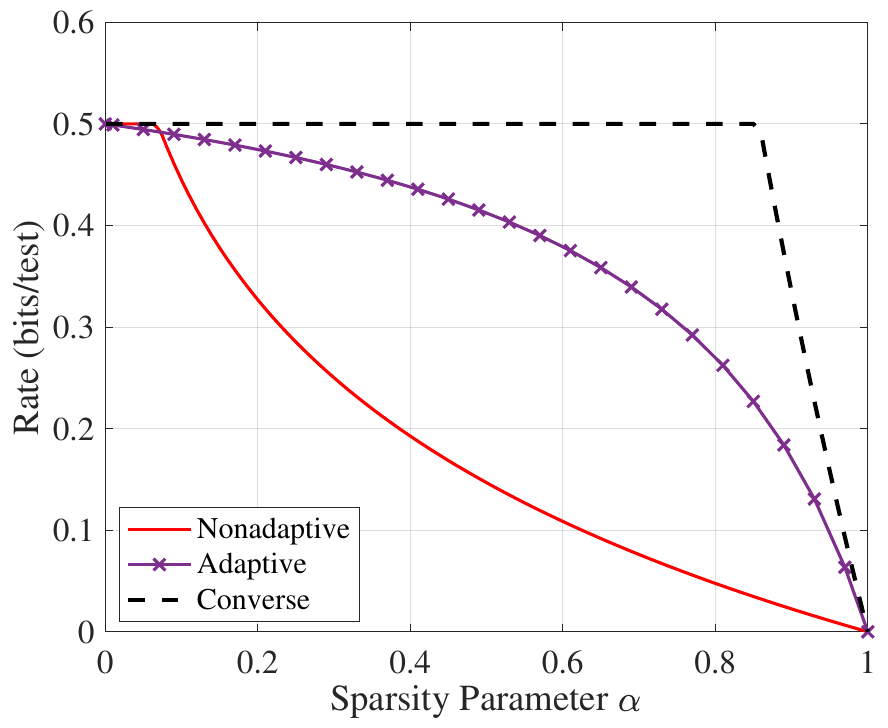} 
\end{center}
\caption{Adaptive and nonadaptive rates for the symmetric noise model (Example \ref{ex:bsc}) with noise level $\rho = 0.11$.  \added{The adaptive rate is obtained by setting $\gamma = 1$ in \cite[Thm.~3]{Sca18}, and a very slight improvement can be obtained by optimizing $\gamma$ therein (see \cite{teo2022noisy} for further discussion).  The nonadaptive rate shown here is from \cite{chen2024exact,coja2024noisy} (as surveyed in Section \ref{sec:high_k_ach}).}\label{fig:noisy_adaptive_rates}} 
\end{figure}

This approach, as well as a slightly refined three-stage version, leads to significant improvements over the best known rates for noisy nonadaptive group testing, as illustrated in Figure \ref{fig:noisy_adaptive_rates}. Another contribution of \cite{Sca18}, also demonstrated in this figure, was to provide an algorithm-independent converse demonstrating that the rate must approach zero as $\alpha \to 1$ (regardless of the number of stages of adaptivity), in stark contrast to the noiseless setting in which a rate of one can be achieved for all $\alpha \in (0,1)$.  While the achievability curve in Figure \ref{fig:noisy_adaptive_rates} does not correspond to a computationally efficient algorithm (since the first adaptive stage in \cite{Sca18} uses an intractable information-theoretic decoder), \added{the same rate was subsequently attained using two efficient approaches:
\begin{itemize}
    \item An efficient variant using 4 stages (or 3 in the noiseless setting) was given in \cite{Sca19}.
    \item A noisy binary search approach was taken in \cite{teo2022noisy}, essentially corresponding to a noisy counterpart of Hwang's generalized binary splitting (see Section \ref{sec:adaptive}).  This approach was also shown to provide reductions in the average number of tests experimentally, but a notable downside is that it is `fully adaptive' rather than having only a few stages.  
\end{itemize}
(These approaches do not permit the slight improvement discussed in the caption of Figure \ref{fig:noisy_adaptive_rates}.)

It remains an interesting question to close the remaining gaps between the achievability and converse bounds in the noisy setting.  While this discussion pertains to the sublinear regime $k = \Theta(n^{\alpha})$, exact thresholds have been established in the \emph{linear regime} $k = \Theta(n)$ \cite{hintze2024noisy}; see Section \ref{sec:linear} for an outline.
}

\section{Universality and counting defectives} \label{sec:universal}

A natural consideration in group-testing is that of \emph{universality} -- that is, how robust the paradigm of group testing is to lack of \emph{a priori} knowledge of the number of defectives $k$ (or suitable upper and/or lower bounds). This question arises naturally when nothing is known in advance regarding the statistical process generating the defectives.

A related issue, as discussed in Section \ref{sec:applications}, is that for certain applications the problem of interest is not to identify defective individuals $\K$, but merely to estimate their total number $k = |\K|$. This focus may arise because identifying the individuals themselves is simply not necessary, is not desirable for reasons of privacy, or is not feasible because of the impracticality of distinguishing between test items (e.g., when studying insect populations). 

The idea of counting the number of defectives using group testing dates back at least as far as  work of Thompson \cite{thompson}, with these original ideas being developed in various works including \cite{sobel3}, \cite{walter}, \cite{chen9} and \cite{swallow}.   
We proceed by discussing the adaptive setting, and then turn to the nonadaptive setting.  In both cases, we only consider noiseless group testing, since it has received by far the most attention in the relevant literature.

\subsection{Adaptive testing}

We first highlight the work of Cheng \cite[Theorem 3.1]{cheng} for adaptive testing, in which the following result was proved.

\begin{theorem}[Exact adaptive defective counting] \label{thm:ctexact} 
For any parameter $c > 1$, there exists an adaptive algorithm that can find the exact number of defective items $k$ with probability at least $1-1/k^{c-1}$, using a number of tests upper bounded by
\begin{equation} \label{eq:ctcheng}
4 k \left( \lceil c \log_2 k + c \rceil + 2 \right).
\end{equation}

\end{theorem}

Cheng's argument is based on a recursive binary splitting argument reminiscent of the binary search technique of Algorithm \ref{alg:bin}. Given a set $A$ that contains at least one defective, we randomly partition it via independent fair coin flips into two subsets $A_1$ and $A_2$ of roughly equal size. Note that this construction is equivalent to taking $p=1/2$ in \eqref{eq:counttoinvert}. Observe that:
\begin{enumerate}
\item \label{it:case1ct} If $A$ contains exactly one defective, one of $A_1$ and $A_2$ will certainly contain no defectives. 
\item \label{it:case2ct} If $A$ contains $d > 1$ defectives, the probability that all of these defectives are placed in the same subset $A_i$ is $1/2^{d-1} \leq 1/2$. 
\end{enumerate}
In the following, we discuss a method for reliably distinguishing these two cases.
Consider testing both subsets $A_1$ and $A_2$. In case \ref{it:case1ct}, one of the two tests will certainly be negative. In case \ref{it:case2ct}, the probability that there is a negative test is at most $1/2$ (since this only occurs if all $d$ defectives lie in the same subset). 

Furthermore, if we repeatedly and independently partition $A$ by using the above coin-tossing procedure $r$ times, the outcomes will be independent between partitions. Hence, our decision rule is simply to ask `did we see $r$ negative tests?'. In case \ref{it:case1ct}, this certainly happens, and in case \ref{it:case2ct}, the probability that this happens is at most $1/2^r$.
In other words, repeatedly randomly partitioning $A$ in this way allows us to efficiently distinguish cases \ref{it:case1ct} and \ref{it:case2ct}.  This allows us to eventually partition the full set of items $\{1,2,\dots, n\}$ into subsets, each of which contains exactly one defective with high probability, so that $k$ is successfully identified.

A standard information-theoretic argument shows that, with no prior information on the number of defectives, to estimate the exact value of $k$ with zero error probability will require at least $\log_2 (n+1)$ tests. In this sense, the question of the optimal order of the number of tests required for exact recovery remains open. Indeed, we note that individual testing of every item will give the exact number of defectives in $n$ tests, which outperforms Cheng's bound \eqref{eq:ctcheng} in the regime where $k = \Theta(n)$.

However, as argued by Falahatgar \etal\ \cite{falahatgar}, the requirement to know the number of defectives \emph{exactly} may be unnecessarily restrictive.  They developed a four-stage adaptive algorithm, for which they proved that that an $O(\log \log k)$ expected number of tests achieves an \emph{approximate recovery} (of $k$) criterion with high probability \cite[Theorem 15]{falahatgar}. The algorithm works by successively refining estimates, with each stage creating a better estimate with a certain probability of error. The first stage finds an estimated number of defectives that (with high probability) lies in the interval  $(k,k^2)$, and later stages tighten this to $(k \epsilon^2, k/\epsilon^2)$, then $(k/4,4k)$, and finally $((1-\eta) k, (1+\eta)k)$ for arbitrarily small $\eta > 0$, using binary search techniques.

Building on the algorithm of \cite{falahatgar}, Bshouty \etal\ \cite{bshouty} improved the performance guarantee by a constant factor, obtaining the following result \cite[Theorem 8]{bshouty}.

\added{
\begin{theorem}[Approximate adaptive defective counting] \label{thm:ctapprox} For any $c > 1$, $\delta \in (0,1)$ and $\eta \in (0,1)$,
there exists a four-stage adaptive algorithm providing an estimate of $k$ that, with probability at least $1-\delta$, lies in the range $\big((1-\eta) k, (1+\eta) k\big)$, using an expected number of tests satisfying
\begin{equation} \label{eq:falahatgar}
 \EE T \le (1- \delta + \delta^c) \log_2 \left( \log_2 k \right)   + O \big( \sqrt{\log \log k} \big) + O \left( \frac{c}{\eta^2} \log \frac{1}{\delta} \right).
\end{equation}
\end{theorem}}

Taking $c$ arbitrarily large gives a leading term that is arbitrarily close to
$(1- \delta) \log_2 \left( \log_2 k \right)$. This result is essentially optimal, since  \cite[Theorem 16]{falahatgar} previously used Fano's inequality to prove an information-theoretic lower bound showing that any such algorithm  requires at least $(1-\delta) \log_2 \left( \log_2 k \right) - 1$ tests.

A comparison between the $\log_2(n+1)$ converse result for exact recovery and the significantly tighter $O( \log \log k)$ approximate recovery bound of Theorem \ref{thm:ctapprox} indicates that the criterion for successful recovery consistently makes a significant difference in this case.
Note that since any subsequent group testing strategy to estimate $\K$ typically requires $\Theta\big(k\log_2 \frac{n}{k} \big)$ tests, it is highly desirable to use $o\big(k\log_2 \frac{n}{k} \big)$ tests in the defective-counting stage, and this is achieved in Theorem \ref{thm:ctapprox} but not Theorem \ref{thm:ctexact} (unless $\log k = o(\log n)$, which is a much stronger requirement than the usual $k = o(n)$).

\subsection{Nonadaptive testing}

\added{Before delving into nonadaptive strategies,} we describe an idea based on random Bernoulli testing that forms part of several such defective-counting papers:
\begin{itemize}
    \item If there are $k$ defectives in some set $A$, and if each item in $A$ lies in a particular test independently with probability $p$, the test will be positive with probability 
    \begin{equation} \label{eq:counttoinvert}
    1 - (1-p)^k.
    \end{equation}
    \item Inverting this relationship, using Bernoulli designs with probability 
    \begin{equation} \label{eq:countinverted}
    p(\ell) := 1 - 2^{-1/\ell},
    \end{equation}
    we would expect half the tests to be positive on average, if the true number of defectives $k$ was equal to $\ell$. Hence, using a Bernoulli design with parameter $p(\ell)$, if empirically we see many more (respectively, many fewer) than half the tests being positive, then this is evidence that $k \gg \ell$ (respectively, $k \ll \ell$). 
\end{itemize}
\added{While the above findings provide a useful starting point,} a notable challenge associated with creating nonadaptive algorithms to count defectives is that (without further information) any value of $k \in \{0,1,\dotsc,n\}$ is possible. Constructions based on \eqref{eq:countinverted} for a particular $p(\ell)$ will struggle to distinguish between putative values $k_1$ and $k_2$ that are both far from $\ell$. To overcome this, Damaschke and Muhammad \cite[Section 4]{damaschke2010competitive} proposed a geometric construction, dividing the range of tests into subintervals of exponentially increasing size and using a series of $p(\ell)$ values tailored to each subinterval.
 
To be more precise, the (nonadaptive) algorithm of \cite{damaschke2010competitive} constructs parallel group tests indexed by integers $t$. Each item lies in the $t$-th test independently with probability 
\begin{equation} \label{eq:dampchoice} 
p_t  := 1 - \left( 1 - \frac{1}{n} \right)^{b^t},
\end{equation}
for some fixed positive $b > 1$. Using \eqref{eq:counttoinvert}, this means that each test is negative with probability $q_t := \left( 1 - 1/n \right)^{b^t k}$. Hence, if $t^*$ is the largest index of a negative pool, we may imagine that $q_{t^*} \simeq 1/2$ and invert this to construct an estimate of $k$. More precisely, \cite{damaschke2010competitive} propose an offset of this, taking
\begin{equation} \label{eq:damkest} \widehat{k} = -\frac{1}{b^{t^*-s} \log_2 \left( 1 - \frac{1}{n} \right)},
\end{equation} for an integer $s$. The following result \cite[Theorem 4.3]{damaschke2010competitive} shows that \eqref{eq:damkest} satisfies certain success criteria.

\begin{theorem}[Approximate nonadaptive defective counting] \label{thm:appdefct}
For a given value\footnote{We assume an upper bound $b \le 2$ to simplify the theorem statement, but this could be replaced by any absolute constant with only minor adjustments.} $b \in (1,2]$ and integer $s$, the estimate $\widehat{k}$ of \eqref{eq:damkest} can be formed using $\log_b n$ tests overall such that:
\begin{enumerate}
\item $\PP( \widehat{k} \leq k) = O \big( \frac{1}{2^{b^s} \log b } \big)$,
\item $\EE( \widehat{k}/k) \leq b^s F(b)$ for a certain explicit function $F$. This function $F$ is monotone increasing in $b$, with $F(b) < 1.466$ for all $b \leq 2$.
\end{enumerate}
\end{theorem}

Observe that the first part bounds the probability that the estimate $\widehat{k}$ underestimates $k$, and the second part shows that $\widehat{k}$ overestimates $k$ by at most an explicit constant factor on average.

\rev{In \cite{damaschke2010nonadapt}, the same authors use an argument based on a related hypothesis testing problem to show that this $\Omega(\log n)$ scaling in the number of tests is essentially optimal. Specifically, for nonadaptive group testing any estimate $\widehat{k}$ with a specified underestimation probability $\PP( \widehat{k} \leq k) \leq \epsilon$ and bounded `competitive ratio' $\EE( \widehat{k}/k) \leq c$ requires a multiple of $\log n$ tests, with a constant factor depending on $\epsilon$ and $c$ \cite[Theorem 1]{damaschke2010nonadapt}.

\added{In subsequent work, Bshouty \cite{bshouty2019lower} showed how to strengthen Theorem \ref{thm:appdefct}, building on the techniques developed in \cite{damaschke2010competitive} and \cite{falahatgar} to provide a \emph{high-probability} accuracy guarantee. In particular, \cite[Theorem 7]{bshouty2019lower} states the following.
\begin{theorem} \label{thm:appdefct2}
Let $\eta > 0$ be a fixed constant, and consider a parameter $\delta > 0$ and number of defectives $k$ that may vary with $n$.  There exists a nonadaptive procedure forming an estimate $\widehat{k}$ using $O \big( \log \frac{1}{\delta} \log n \big)$ tests (with the $\eta$ dependence hidden in $O(\cdot)$ notation), with the property that
\begin{equation} \label{eq:appdefct}
\PP \big( k \leq \widehat{k} \leq (1+\eta)k \big) \geq 1-\delta.
\end{equation}
\end{theorem}
%
Bshouty \cite[Theorem 1]{bshouty2023improved} also provides a counterpart to the above-mentioned result of \cite{damaschke2010nonadapt}: namely, a lower bound on the number of tests required by any (possibly randomized) algorithm to estimate $k$ to within a constant factor.  This lower bound matches the upper bound in Theorem \ref{thm:appdefct2} to within a $\log \log \cdots \log n$ factor for an arbitrarily large number of composed logarithms.}

Returning to the question of universality of group testing (that is, whether we can  recover the defective set $\K$ with no prior knowledge of $k$), we can regard Bshouty's algorithm \cite{bshouty2019lower} as the first stage of a two-stage universal algorithm. In the second stage, we can design the tests using the resulting estimate of $k$, and use the second stage test results to determine $\K$; as a specific example, we could use the COMP algorithm under Bernoulli testing with parameter $p = 1/\widehat{k}$.

\added{Under such a strategy, we could set the `error probability' $\delta$ in the first stage to, say, $\delta = 1/n$, so that Theorem \ref{thm:appdefct2} gives $O( (\log n)^2 )$ scaling in the number of tests.  By comparison, recalling that COMP attains a positive rate when $\widehat{k}$ is within a constant factor of $k$ (see Remark \ref{rem:misspecify}), the number of tests in the second stage is $O(k \log n)$.  Thus, the number of tests in the first stage is asymptotically negligible by comparison provided that $k \gg \log n$.
}


The preceding results (both adaptive and nonadaptive) are summarized in Table \ref{tbl:counting}.

\begin{table}[!ht]
\begin{center}
\begin{tabular}{cccc}
\hline
References & Recovery guarantee & \# Tests & Adaptive? \\
\hline
\cite{cheng} & Exact & $O(k \log k)$ & yes \\
\cite{falahatgar,bshouty} & $(1-\eta)k \le \widehat{k}\le (1 + \eta)k$ & $O(\log \log k)$ & yes \\
\cite{damaschke2012randomized}   & $\widehat{k} \ge k$, $\,\EE(\widehat{k}/k) = O(1)$ & $O(\log n)$ & no \\
\cite{bshouty2019lower}   & $k \le \widehat{k} \le (1+\eta)k$ & $O(\log n)$ & no \\
\hline
\end{tabular}
\end{center}
\caption{\rev{Summary of recovery guarantees for counting defectives, depending on the recovery criteria and availability of adaptive testing.  The results here correspond to the case of a constant nonzero error probability; the precise dependencies on the error probability can be found in the above theorem statements.} \label{tbl:counting}}
\end{table}

}

\subsection{Discussion}

It is worth mentioning a fundamental limitation regarding the quest for universality:  If we require the test design to be completely nonadaptive, then achieving a positive rate when $k = \Theta(n^{\alpha})$ for some $\alpha$ precludes achieving a positive rate for $\alpha' < \alpha$.  This is because the former requirement needs $n = \Omega( n^{\alpha} \log n )$ by the counting bound, and any such scaling on $n$ gives zero rate for $\alpha' < \alpha$.  Hence, having at least some degree of adaptivity is essential to universally attaining a positive rate.

On the other hand, if the number of defectives is known to be upper bounded by some known value $k^*$, then under Bernoulli testing with $p = 1/k^*$, the analysis of COMP (see the Appendix to Chapter \ref{ch:introduction}) leads to vanishing error probability with $T = O(k^* \log n)$ tests, even if the true value of $k$ is much smaller than $k^*$.  More specifically, this can be seen by setting $p = 1/k^*$ in \eqref{eq:COMPprob}, upper bounding $k \le k^*$, and continuing the analysis with $k^*$ in place of $k$.  Therefore, we still have guarantees on the performance when only an upper bound on $k$ is known, but we pay a penalty in the number of tests if that bound is loose.

\added{We note that a more general notion of universality could be taken with respect to the \emph{prior distribution} of defectives, which in general may be non-uniform, may exhibit correlation or clustering structure, and so on (see Section \ref{sec:prior} for an overview).  However, universality with respect to such priors is likely to be significantly more challenging compared to universality with respect to the number of defectives alone, and we are unaware of any existing works attempting this in group testing.
}



Broadly speaking, the above notions of universal group testing can be viewed as a counterpart to information-theoretic problems of universal source coding, where one does not have access to the underlying distribution of the source.  A prominent example of an adaptive universal source coding algorithm is that of Lempel and Ziv, which achieves asymptotically optimal compression for any stationary ergodic source by parsing strings into trees (e.g., see \cite[Chapter 13]{cover}, \cite[Section 6.4]{mackay}). 

There may also be uncertainty regarding the group testing setup in other senses beyond those discussed above.  In particular, in the noisy settings of Chapter \ref{sec:noisy}, we may lack information regarding the parameters of the particular noise model, or even which model applies.  The problem of decoding with no knowledge of the noise model was referred to as `blind group testing' in the work of Huleihel, Elishco, and M\'edard \cite{huleihel}, who proved universal variants of the information-theoretic joint and separate decoding rules (see Sections \ref{sec:noisy} and \ref{sec:separate} respectively).  In the sparse regime with $\alpha \to 0$, they show that their decoders achieve the same asymptotic performance as when the channel is known.  An earlier work of Malyutov and Sadaka \cite{Mal98} showed such a result in the very sparse regime $k = O(1)$.

Again, we can connect this with classical information-theoretic results, namely, for the problem of universal channel coding. Here a transmitter seeks to send a message over a noisy channel, despite not having precise channel statistics other than a guarantee that its capacity exceeds the transmission rate.  The maximum empirical mutual information decoder (e.g., see \cite[p.~100]{csiszarkorner}) is the most well-known decoding method for this scenario, and such a decoder was in fact adopted for group testing in \cite{Mal98}.  The decoder adopted in \cite{huleihel} is slightly different, but still based on empirical probability measures.

\section{Sublinear-time algorithms} \label{sec:sublinear}

The decoding algorithms such as COMP and DD (but not SSS) studied in Chapters \ref{ch:algorithms} and \ref{ch:algorithms_noisy} are efficient, in the sense that they operate in time $O(nT)$.  However, in the case that $n$ is extremely large, or practical considerations require a very fast decoding rule, it is of 
interest to seek algorithms that reduce this runtime further.

To address this question, a recent line of works has considered {\em sublinear-time algorithms} that run in time that is linear or polynomial in $k \log n$, rather than in $n$.  \added{We know from the counting bound that $O(k \log n)$ tests are needed, and if it takes a constant amount of time to read each test result, then the decoding time cannot be reduced below $O(k \log n)$. }

\rev{Early works on sublinear-time decoding for group testing focused on the zero-error recovery criterion \cite{cheraghchi2009noise,indyk2010,ngo2011}, possibly with adversarial noise \cite{cheraghchi2009noise}.  Typical results in these works state that with  $T = O(k^2 \log n)$ tests (which is nearly order-optimal under the zero-error recovery criterion; see Section \ref{sec:zero-error}), one can attain a decoding time of the form $O(T^c)$ for some $c > 1$.}

Our focus in this manuscript is on the small-error recovery criterion, as opposed to zero-error, and we therefore focus on more recent algorithms targeted at this setting.  We immediately see that we cannot directly rely on the main ideas used in Chapters \ref{ch:algorithms} and \ref{ch:algorithms_noisy}; for instance:
\begin{itemize}
    \item Merely traversing the entries of a Bernoulli test matrix requires $\Omega(nT)$ time.  Structured test matrices often permit faster computation, but the only structure inherent in the Bernoulli($p$) case is sparsity, and there are still $\Omega(n \log n)$ entries equal to one in the case that $T = \Omega(k \log n)$ and $p = \Theta\big(\frac{1}{k}\big)$.
    \item Both COMP and DD use the idea of marking items as nondefective if they appear in a negative test.  However, individually marking ${n-k}$ items (or even just a constant fraction thereof) as nondefective already requires linear time.
\end{itemize} 
The first observation suggests either using a highly structured test design, in particular adopting a scheme where the decoder does not need to read the entire test matrix. The second observation suggests that achieving sublinear runtime requires using a decoding algorithm that positively identifies defective items \added{and/or that efficiently marks large numbers of items as being nondefective all at once; doing so one-by-one is no longer feasible.}

We note that the notion of `sublinear time' here applies purely to the decoding algorithm.  Indeed, if we assume that placing an item in a test takes constant time, then {\em encoding} obviously requires at least linear time in total, if each item is tested at least once. Thus, sublinear time algorithms will be of interest in the case that the encoding work naturally parallelizes over the $n$ elements, and/or the decoding time poses more of a bottleneck.

In this section, we discuss two schemes for group testing with sublinear-time decoding time and arbitrarily small error probability:
\begin{description}
  \item[SAFFRON] \cite{lee-pedarsani-ramtin} is a scheme for nonadaptive noiseless testing based on sparse-graph codes. It requires a number of tests and runtime both of order $O(k \log k \log n)$. We also discuss an approximate recovery result, and briefly mention a variant for the noisy setting.
  \item[GROTESQUE] \cite{cai-etal2} is a scheme with three variants: adaptive, two-stage, and nonadaptive. All three variants work for both the noiseless and noisy settings. The adaptive variant performs $O(k \log n)$ tests and requires $O(k \log n)$ runtime, with the former in particular amounting to a positive rate.  In addition, the two-stage variant achieves a positive rate for some scaling regimes of $k$.
\end{description}
At the end of the section, we also discuss some more recent approaches that are known to achieve a positive rate for exact recovery in the nonadaptive setting, unlike SAFFRON and GROTESQUE.

\rev{To the best of our knowledge, GROTESQUE appeared in the literature prior to SAFFRON.  However, we find it natural to first present a simplified form of SAFFRON that is the easiest to analyse, and then move on to the various forms of GROTESQUE.}

\subsection{SAFFRON}

SAFFRON is a scheme for nonadaptive group testing with sublinear decoding time due to Lee, Pedarsani and Ramtin \cite{lee-pedarsani-ramtin}.  \rev{It is based on sparse-graph codes, which were also used earlier in other sparse signal recovery problems (e.g., see \cite{li2015sublinear}).} We present here the simple `singleton-only SAFFRON' version of the scheme. 

The basic idea is as follows: The $T$ tests are split into `bundles' of size $2m$, where $m = \lceil \log_2 n \rceil \sim \log_2 n$. Each item is chosen to either `appear' or `not appear' in each bundle. 
Specifically, items appear in bundles according to an outer Bernoulli design, where each item appears in each bundle independently with probability $p$. 

If item $i$ it does not appear in a given bundle, then it is absent from all $2 m$ tests in that bundle. If item $i$ does appear in a given bundle, it is placed in the tests within that bundle that correspond to the $1$s in the vector $\big(\vec b(i) , \overline{\vec b(i)}\big)\in \{0,1\}^{2m}$, where $\vec b(i) \in \{0,1\}^m$ is the binary expansion of the number $i$, and $\overline{\vec b(i)} = \vec 1 - \vec b(i)$ is $\vec b(i)$ with the $0$s and $1$s reversed. Note that since the vector $\big(\vec b(i) , \overline{\vec b(i)}\big)$ always has weight $m$, any item appearing in a bundle is in exactly $m$ tests within that bundle.  \rev{The idea of encoding binary expansions into tests/measurements was also used earlier in studies of sparse recovery with linear measurements (e.g., see \cite{gilbert2006algorithmic} and the references therein).}

We now describe the decoder. When considering the outputs from a given bundle of $2m$ tests, we first look at the the weight of those outputs -- that is, the number of positive outcomes within the bundle. If the weight is $0$, then no defective items appeared in the bundle; if the weight is exactly $m$, then precisely one defective item appeared in the bundle; if the weight is greater than $m$, then two or more defective items appeared in the bundle. The simplified (singleton-only) SAFFRON decoder considers only those bundles containing precisely one defective. The first $m$ outputs from such a bundle give the binary expansion of the label of the defective item, which is therefore immediately identifiable. Repeating this process for each bundle collects a number of defective items, which form our estimate of the defective set.

The key point here is that the SAFFRON decoder first \emph{detects} a bundle containing exactly one defective item, by calculating the output weight, then affirmatively \emph{identifies} that defective item, using the binary expansion.  Thus, it does not rely on ruling out nondefective items, which takes at least linear time if done one-by-one. Note also that while it would take more than linear time to explicitly read the Bernoulli outer design, it is not necessary to do so, as defective items identify themselves through their binary expansion.  

We then have the following result.

\begin{theorem} \label{th:saffron}
Consider standard nonadaptive group testing with $n$ items, $k$ defectives, and $T$ tests.
Singleton-only SAFFRON succeeds at exact recovery with probability approaching 1 provided that
  \[ T \geq (1+\eta) \,2\ee \,k \ln k \cdot \log_2 n \]
for some $\eta > 0$, and the decoding time is $O(k \log k \, \log n)$.

Furthermore, we have the following approximate recovery result: The output of SAFFRON contains at least $(1-\gamma)k$ defective items with probability approaching $1$ provided that
  \[ T \geq (1 + \eta)\, 2 \ee \,k \ln \bigg(\frac{1}{\gamma}\bigg)\log_2 n \]
for some $\eta > 0$, and the decoding time is $O(k \log(1/\gamma) \log n )$.  

\added{In both of the above cases, with probability $1$, there are no false positives: the output is a subset of the defective set.}
\end{theorem}

\added{This result is slightly different from the main result in \cite{lee-pedarsani-ramtin}, but readily follows from the arguments therein.  We also note a subtle point that the decoding time depends on the model of computation.  The scaling laws are as stated above when it takes a unit amount of time to read each test result, but it was noted in \cite{Price2020} that under a \emph{word-RAM model} of computation, we can interpret the test results as a sequence of length-$O(\log n)$ `words', each of which can be read and processed in $O(1)$ time.  Under this model, the decoding time reduces by a $\log n$ factor.
}

We briefly discuss how the above statements on the number of tests translate into achievable rates when $k = \Theta(n^\alpha)$ with $\alpha \in (0,1)$. In the case of exact recovery, the number of tests is a logarithmic factor higher than the optimal scaling, so the rate is zero. However, the approximate recovery result has a positive rate for fixed $\gamma \in (0,1)$, namely,
  \[ \frac{1}{2\ee \ln \frac{1}{\gamma}} (1-\alpha) \simeq \frac{0.184}{\ln \frac{1}{\gamma}} (1-\alpha). \]
We observe also that this rate tends to zero as $\gamma \to 0$.

\begin{proof}[Proof sketch]
We sketch a proof based on the coupon collector problem.
Write $B$ for the number of bundles, recalling that $T = 2mB \sim 2B \log_2 n$. By picking the Bernoulli parameter as $p = 1/k$, we maximize the average number of bundles containing exactly one defective, and by standard concentration bounds, the actual number is close to the resulting average $\ee^{-1} B$ with high probability.

For exact recovery, we need to `collect' all $k$ defective items. The standard coupon collector problem states that we require $k \ln k$ such bundles to collect the $k$ items (see Remark \ref{rmk:tightness}). Hence, we need $\ee^{-1} B \sim  k \ln k$, and thus $T \sim 2\ee\,k \ln k\,\log_2 n$.

For the approximate recovery criterion, another standard coupon collector result states that that collecting $(1 - \gamma)k$ coupons out of $k$ requires $k \ln (1/\gamma)$ such bundles. The result then follows in the same way.

For each bundle, the outputs are read in time $O(\log n)$, the weight computed in time $O(\log n)$, and the single defective -- if there is one -- identified in time $O(\log n)$. Hence, the running time for the decoder is $O(B \log n) = O(T)$, which is $O(k \log k \log n)$ for exact recovery and $O(k \log(1/\gamma) \log n )$ for approximate recovery.  
\end{proof}

We re-iterate that the above result concerns the simplified `singleton-only' SAFFRON scheme; the full SAFFRON scheme of \cite{lee-pedarsani-ramtin} improves the constant factors in the results of Theorem \ref{th:saffron} as follows:  When a bundle contains two defective items, one of which has been identified elsewhere (e.g., via the singleton approach), the second defective can be then also be identified with high probability. The outer Bernoulli design is also replaced by a design with constant bundles-per-item.

In addition, the authors of \cite{lee-pedarsani-ramtin} give a `robustified SAFFRON' algorithm for noisy group testing. Here, the vectors $\big(\vec b(i), \overline{\vec b(i)}\big)$ are extended with the parity-check bits of a positive-rate code
in order to provide robustness to noise. The resulting scheme is similar to the nonadaptive variant of GROTESQUE described below.

\subsection{GROTESQUE} \label{sec:grotesque}

In this subsection, we give an overview of another sublinear-time algorithm called GROTESQUE (Group Testing, Quick and Efficient) due to Cai, Jahangoshahi, Bakshi and Jaggi \cite{cai-etal2}.  This approach uses expander codes \cite{spielman1996} in its construction, thus highlighting that efficient channel codes (e.g., see \cite{richardson}) can play a role in practical group testing constructions, and complementing the extensive use of information theory for theoretical studies of group testing.

\subsubsection*{Overview of results} \label{sec:sublinear_results}

There are three variations of GROTESQUE with different guarantees on the number of tests and decoding time, corresponding to the fully adaptive, two-stage adaptive, and nonadaptive settings.  We first summarize the respective performance guarantees, and then give an overview of the algorithms themselves.
\begin{theorem}
    There exists an adaptive variant of GROTESQUE using $O(\log n)$ stages of adaptivity that achieves vanishing error probability, performs $O(k \log n)$ tests, and requires $O(k \log n)$ decoding time.
\end{theorem}
Observe that this result gives optimal scaling laws when $k = \Theta(n^{\alpha})$ with $\alpha < 1$.  In particular, the algorithm achieves a positive rate in this regime; however, the rate itself may be low according to the existing proof, which does not optimize the constant factors.
\begin{theorem}
    There exists a nonadaptive variant of GROTESQUE that achieves vanishing error probability with $O(k \log n \, \log k)$ tests and $O(k(\log n + \log^2 k))$ decoding time.  
\end{theorem}
Observe that the number of tests matches that of SAFFRON up to constant factors. In particular, although the rate is zero, the scaling laws are only a log factor away from optimality.  While it may seem unusual for the number of tests to exceed the decoding time (e.g., when $k = O(\log n)$), the idea is that the algorithm can `adaptively' decide which test outcomes to observe, and ultimately leave some tests unobserved.  

\begin{theorem}
    A two-stage adaptive variant of GROTESQUE achieves vanishing error probability with $O(k(\log n + \log^2 k))$ tests and $O(k(\log n + \log^2 k))$ decoding time.
\end{theorem}
This result improves on the number of tests used by the nonadaptive algorithm, and achieves a positive rate whenever $\log^2 k = O(\log n)$. Note, however, that this condition is not satisfied in the regime $k = \Theta(n^\alpha)$ (with $\alpha \in (0,1)$), which has been the focus of most of this monograph.

We briefly mention that all of the above guarantees hold not only in the noiseless setting, but also for the symmetric noise model with a fixed crossover probability in $\big(0,\frac{1}{2}\big)$.

\subsubsection*{Overview of the algorithm variants} 

The basic building block of all three variants of the algorithm are two types of `tests' (in the general sense of the word, rather than the sense of a single group test) that operate on subsets of $\{1,\dotsc,n\}$, described as follows.

\paragraph*{Multiplicity test} A {\em multiplicity test} considers a subset $\mathcal{S}$ of items of size $n' < n$, and only seeks to establish whether the number of defective items in the subset is $0$, $1$, or more than $1$.  To do this, we perform $T_{\mathrm{mul}} = O(\log n)$ tests, where each item in $\mathcal{S}$ is included in each test independently with probability $\frac{1}{2}$.  It is easy to show (see also \eqref{eq:counttoinvert}--\eqref{eq:countinverted} above) that the following holds for each such test:
\begin{itemize}
    \item If $\mathcal{S}$ has no defective items, the output must be $0$ (noiseless case), or have a probability strictly less than $\frac{1}{2}$ of being $1$ (noisy case).
    \item If $\mathcal{S}$ has one defective item, the output is equal to $0$ or $1$ with probability exactly $\frac{1}{2}$ each.
    \item If $\mathcal{S}$ has more than one defective item, then the output equals $1$ with probability at least $\frac{3}{4}$ (noiseless case), or with probability strictly higher than $\frac{1}{2}$ (noisy case).
\end{itemize}
Therefore, by standard concentration bounds, we can correctly categorize $\mathcal{S}$ into these three categories with probability at least $ 1 - O\big(\frac{1}{n^c}\big)$ (for any fixed $c > 0$) using $O(\log n)$ tests.

Notice that the reason that this procedure can be done efficiently is that we pay no attention to which items are included in each test; we merely count the number of positive and negative test outcomes.

\paragraph*{Location test} After a multiplicity test is performed on a subset $\mathcal{S}$, we only apply this step to $\mathcal{S}$ if the set is found to contain exactly one defective item.  If this is indeed the case, we perform a {\em location test} to deduce the index $i \in \{1,\dotsc,n\}$ of that defective item.  

To do this, we perform another set of $T_{\mathrm{loc}} = O(\log n)$ tests on $\mathcal{S}$, but this time we use a structured `test submatrix' of size $T_{\mathrm{loc}} \times |\mathcal{S}|$.  Specifically, we let the columns of this matrix be the codewords of an expander code \cite{spielman1996}; \added{each codeword encodes the associated item index}.  We do not give the details of such a code, but instead highlight its desirable properties:
\begin{itemize}
    \item The decoding time is linear in the block length (i.e., $O(\log n)$);
    \item The error probability decays exponentially in the block length (i.e., $O(n^{-c})$ for some $c > 0$);
    \item The rate is constant (i.e., an item index in $\{1,\dotsc,n\}$ can be reliably identified with a block length $O(\log n)$);
    \item The code is robust to independent random bit flips (i.e., symmetric noise) and/or erasures.
\end{itemize}
These properties suffice to perform a single location test sufficiently reliably using $O(\log n)$ tests and $O(\log n)$ runtime, even in the presence of random noise.

In SAFFRON above, the bundles consisting of binary expansions vectors work a lot like the multiplicity and location step here: A bundle having output weight exactly $m$ certifies that it contains exactly one defective (multiplicity test with $T_\mathrm{mul} = 2m \sim 2 \log_2 n$), and reading off the binary expansion `localizes' the defective item (with no extra steps).

With the preceding building blocks in place, we can now describe the three variations of the algorithm.

\paragraph{Adaptive algorithm} The adaptive algorithm uses $J = O(\log n)$ stages of adaptivity, with all stages except the last using a common procedure.  Specifically, the goal of the first $J-1$ stages is to identify all except at most $\log_2 k$ defective items.  Letting $k_i$ denote the number of unresolved defective items before the $i$-th stage, we randomly partition the $n - k + k_i$ items (excluding resolved defectives) into $2k_i$ groups, and perform a multiplicity test on each such group.  

By standard concentration via McDiarmid's inequality \cite{mcdiarmid}, it can be shown that with high probability, a constant fraction of the groups contain a single defective item.  Assuming the multiplicity tests are successful (which occurs with high probability), all such groups are identified, and the corresponding defective item can then be found via a location test.

When the number of remaining defectives $k_i$ falls sufficiently below $\log_2 n$, the desired concentration behaviour starts to fail to hold.  To address this, in the final stage, we form $O( (\log k)^2 \log \log k )$ groups, each containing $O\big( \frac{n}{\log k} \big)$ unresolved items chosen uniformly at random.  These choices of scaling laws, with suitably-chosen implied constants, ensure that each unresolved defective appears in at least one group by itself with high probability.  As a result, we can identify these remaining items via multiplicity and location tests as above.

By the fact that the number of unresolved defectives decays geometrically in the first $J -1$ stages, it can be shown that these stages collectively only require $O(k \log n)$ tests and runtime.  The final stage requires $O( (\log k)^2 \, \log \log k \, \log n )$ tests and runtime, which is strictly smaller than $O(k \log n)$.

\paragraph{Nonadaptive algorithm} The simplest way to make the algorithm nonadaptive is to note that, since the first stage identifies a random constant fraction of the defective items, repeating that stage independently $O(\log n)$ times is enough to identify all defective items with high probability.  This approach gives the $O(k  \log k \, \log n)$ number of tests stated in Section \ref{sec:sublinear_results}, but requires $O(k  \log k \, \log n)$ runtime instead of the improved $O(k(\log n + \log^2 k))$.  To achieve the latter, one adopts a more sophisticated approach that adaptively chooses which tests to observe; we refer the reader to \cite{cai-etal2} for details.

\paragraph{Two-stage algorithm} Once the previously-mentioned guarantees of the nonadaptive algorithm are in place, analysing the two-stage algorithm is straightforward.  In the first stage, one randomly partitions the $n$ items into $k^3$ bins.  Since there are $k$ defective items, a standard `birthday paradox' argument (see for example \cite[p.~33]{feller}) reveals that with high probability, each such bin contains either $0$ or $1$ defective items.

In the first stage, a `grouped group testing' procedure is applied with $k^3$ `super-items'.  Each super-item corresponds to an entire bin, and testing a super-item amounts to including all the bin's items in the test simultaneously.  Using the above nonadaptive algorithm accordingly with $k^3$ in place of $n$, we see that we can reliably identify the $k$ defective bins using $O(k \log^2 k)$ tests and runtime.  In the second stage, we simply apply the location test separately to each defective bin, thereby identifying the $k$ defective items using $O(k \log n)$ tests and runtime.

\subsubsection{Discussion}

The GROTESQUE algorithm, as described above, assumes exact knowledge of $k$.  However, as highlighted in \cite{cai-etal2}, it can be adapted to the case that $k$ is only known up to a constant factor.  As we saw in Section \ref{sec:universal}, such knowledge can be attained with probability at least $1-\delta$ using one extra stage of adaptivity with only $O\big(\log n \, \log\frac{1}{\delta}\big)$ tests (Theorem \ref{thm:appdefct2}).

\added{As we have hinted earlier, there are significant conceptual similarities between SAFFRON and the nonadaptive version of GROTESQUE, and elements of the two could be combined.  For example, in the noiseless case, one could forgo the expander code in GROTESQUE and use the simpler bit-expansion approach of SAFFRON.}

\added{A related adaptive algorithm called \emph{isolate-and-identify} was also proposed by Wang and Guruswami \cite{wang2024isolate}, using $O(\log k)$ rounds of adaptivity and notably attaining explicit constants in the number of tests (with asymptotic tightness in sufficiently sparse regimes such as $k = (\log n)^{O(1)}$), at the expense of the decoding time having an additional logarithmic factor.}

\added{A subtle point is that the decoding time for expander codes typically refers to the time to recover the codeword, rather than the time to recover the message.  Nevertheless, the decoding times stated above remain justified due to the fact that the mapping from codewords to messages (that is, item indices) can be found via \emph{pre-processing}, which is done completely offline.  Alternatively, if such pre-processing is too expensive (for example, because $n$ is very large), one can perform the mapping in an online manner with only a slightly increased decoding time.  Specifically, even a naive matrix multiplication to map the codeword to the message only requires time quadratic in the block length (rather than linear), thus only incurring an additional $O(\log n)$ factor in the decoding time of GROTESQUE.}


\subsection{Attaining a positive rate with nonadaptive testing} \label{sec:sublinear_pos_rate}

\rev{
A notable limitation of the theoretical guarantees of SAFFRON and the nonadaptive variant of GROTESQUE is that the number of tests is $O(k  \log k \, \log n)$, meaning that the rate is zero unless $k = O(1)$.  Here we briefly highlight two recent works that improved the number of tests at the expense of a higher decoding time, and then another approach that attains $O(k \log n)$ scaling in both.  The results are summarized in Table \ref{tbl:sublinear}. 

\renewcommand{\arraystretch}{1.4}

\begin{table}
\begin{center}
{ \fontsize{9}{11} \selectfont
    \begin{tabular}{cccc}
    \hline
     & \# Tests & Decoding time & Adaptive? \\
    \hline
    {\bf SAFFRON} & $O(k \log k \,\log n)$ & $O(k \log k \,\log n)$ & no \\
    {\bf GROTESQUE} & $O(k \log k \,\log n)$ & $O(k \log n + k\log^2 k)$ & no \\
    {\bf 2-GROTESQUE}  & $O(k \log n + k\log^2 k)$ & $O(k \log n + k\log^2 k)$ & two-stage \\
    {\bf A-GROTESQUE}  & $O(k \log n)$ & $O(k \log n)$ & yes \\
    {\bf Kautz--Singleton} & $O\big(k \log n \, \log \frac{\log n}{\log k} \big)$ & $O\big(k^3 \log n \, \log \frac{\log n}{\log k} \big)$ & no \\
    {\bf BMC} & $O(k \log n)$ & $O(k^2 \log k\, \log n)$ & no \\
    {\bf Binary splitting} & $O(k \log n)$ & $O(k \log n)$ & no \\
    \hline
    \end{tabular}
}
\end{center}
\caption{\rev{Summary of number of tests and decoding times for sublinear-time group testing algorithms in the small-error setting.  `2-GROTESQUE' and `A-GROTESQUE' refer to the two-stage and fully adaptive versions of GROTESQUE, and the other rows are as described above.  \added{Under a word-RAM model of computation, the decoding time of SAFFRON can be improved to $O(k \log k)$ \cite{Price2020}.}} \label{tbl:sublinear}}
\end{table}

\paragraph{Order-optimal tests and higher decoding time} In \cite{inan}, a classical construction of Kautz and Singleton \cite{kautz} was adapted from the zero-error setting to the small-error setting, and was shown to permit exact recovery with $T = O(k \log n \, \log\frac{\log n}{\log k})$ tests and $O(k^3 \log n \, \log\frac{\log n}{\log k})$ decoding time.  The Kautz-Singleton construction is a type of {\em concatenated code}, and will be discussed in more detail in Section \ref{sec:explicit}.  The preceding number of tests amounts to a positive rate whenever $k = \Theta(n^{\alpha})$ for some $\alpha \in (0,1)$, but not in sparser regimes such as $k = O( (\log n)^c )$ (for fixed $c > 0$).

The problem of attaining exact recovery with $T = O(k \log n)$ and sublinear decoding time without further assumptions on $k$ was first solved by Bondorf \etal~\cite{bondorf} via an approach termed {\em bit-mixing coding} (BMC).  This technique tests random `bundles' of items analogously to SAFFRON; the distinction is that instead of seeking to ensure that each defective is the unique one in some bundle corresponding to $O(\log n)$ tests, BMC allows each defective item's index to be encoded in $O( \log n )$ tests with {\em collisions} between the different defective items.  As long as a constant fraction of these tests remains collision-free for each item, the collisions can be controlled using erasure-correcting coding techniques; see \cite{bondorf} for details.  The decoding time of BMC is $O(k^2 \log k \, \log n)$, which improves on that of \cite{inan}, but remains higher than that of SAFFRON and GROTESQUE by a factor of $k$.}


\added{
\paragraph{Order-optimal tests and matching decoding time} The techniques surveyed thus far attain various trade-offs between the number of tests and the decoding time, but none of the \emph{nonadaptive} strategies attain $O(k \log n)$ in both.  This goal was attained in the concurrent works of Price and Scarlett \cite{Price2020} and Cheraghchi and Nakos \cite{Cheraghchi2020}.  The strategies in the two papers are essentially the same, but we follow \cite{Price2020} more closely in our presentation here.  

\begin{figure}[t] 
\begin{center}
\includegraphics[width=0.95\textwidth]{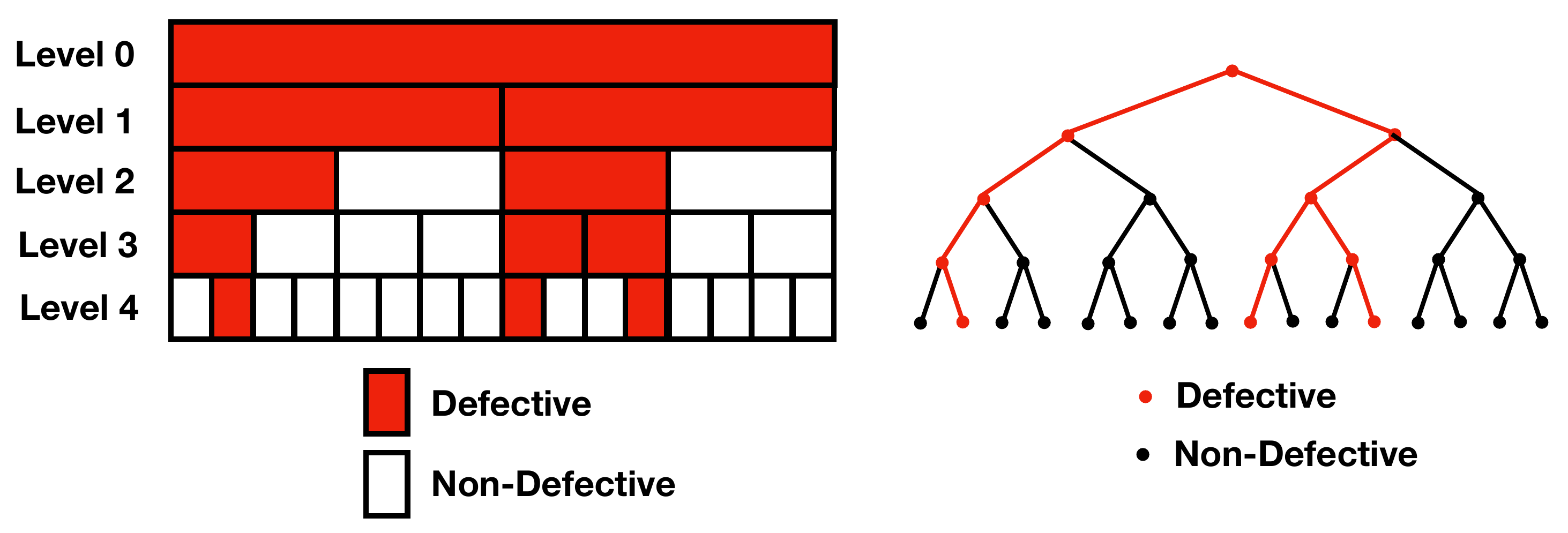}
\end{center}
\caption{\added{Example partitioning of items used for the nonadaptive binary splitting algorithm, with $n=16$ items and $k=3$ defectives.  In the left figure, each level consists of splitting the previous level's groups into two groups of half the size.  The right figure shows and equivalent tree structure, where leaves correspond to individual items and other nodes correspond to group of items, with the root containing all items.  }} \label{fig:DefectiveTree} 
\end{figure}

The strategy can be viewed as roughly emulating Hwang's generalized binary splitting algorithm (described in Section \ref{sec:adaptive}), while performing all the required tests in advance rather than adaptively.  First, we view the items as being recursively partitioned in levels, as illustrated in Figure \ref{fig:DefectiveTree} along with an equivalent binary tree representation.  Level 0 contains a single group with all $n$ items, Level 1 contains two groups of size $n/2$, and so on, up to level $\log_2 n$ containing individual items. (Here we ignore rounding issues for notational simplicity.)  The testing strategy is as follows:
\begin{itemize}
    \item For level $\ell = \log_2 k,\dotsc,\log_2 n - 1$, form $Ck$ tests for a suitable constant $C > 0$, and place each group of that level into one of the tests uniformly random (where `placing a group' means placing all its items).
    \item For level $\ell = \log_2 n$, form $C' \log k$ bundles of $2k$ tests ($2C' k\log k$ tests total), and for each item and each bundle, place the item into a uniformly random test among the $2k$ tests in that bundle.
\end{itemize}
Observe that this gives $O(k\log n)$ tests in total.

Once these tests are performed, the decoding strategy is based on a simple COMP-like approach (see Section \ref{sec:COMP}) that starts from level $\ell = \log_2 k$, keeps only the groups appearing in positive tests, and then recursively works down to the final level, discarding any groups that test negative.  That is, at each level, we keep track of the \emph{possibly defective} groups, and only consider the `children' (from binary splitting) of those groups at the next level.  The analysis of \cite{Price2020} shows that with high probability, only $O\big(k \log \frac{n}{k}\big)$ nodes in the binary tree are `visited' (that is, considered as possible defectives), and by the final level there are only $O(k)$ possibly defective leaf nodes.  The final estimate is then formed by again using COMP decoding (restricted to the possible defectives) on the final level's tests.  Throughout the course of the algorithm, the only test results that need to be checked are the ones that the possibly defective nodes are placed in, thus yielding a decoding time of $O\big(k \log \frac{n}{k}\big) + O(k \log k) = O(k \log n)$.

Formally, a simplified version of the main result of \cite{Price2020} (see also \cite{Cheraghchi2020}) is stated as follows.

\begin{theorem}
    The binary splitting algorithm described above, with suitable choices of the constants $C$ and $C'$, achieves asymptotically vanishing error probability with $O(k \log n)$ tests and $O(k \log n)$ decoding time.
\end{theorem}

The algorithm and its associated guarantee have also been extended in several interesting directions:
\begin{itemize}
    \item Wang, Gabrys, and Guruswami~\cite{Wang2023} developed a variant of the above strategy with a more careful design and analysis that explicitly controls the constant factors in the number of tests, and showed that the constant can be made arbitrarily close to that of COMP (Section \ref{sec:COMP}) while maintaining $O(k \log n)$ decoding time.  In addition, when $k \le O(\sqrt{n})$, they showed that with a modified decoder, the constant can be further improved to be arbitrarily close to that of DD (Section \ref{sec:DD}) with $O(k^2 \log n)$ decoding time.
    \item Price and Scarlett \cite{Price2020} showed that the uniform random placements can be replaced by suitably-chosen hash functions to attain sublinear storage guarantees, at the expense of only a log-factor increase in the decoding time.
    \item Cheraghchi and Nakos \cite{Cheraghchi2020} additionally gave variations for other problems such as compressed sensing and heavy hitters.
    \item Price, Scarlett, and Tan \cite{price2023fast} gave extensions to the noisy setting, as well as a setting with a constrained number of tests per item (see Section \ref{sec:constrained}).  In particular, in the noisy setting with symmetric noise (Example \ref{ex:bsc}), they showed that a suitable variation of the above strategy succeeds with high probability using $O(k \log n)$ tests and $O\big( \big( k \log \frac{n}{k}\big)^{1+\epsilon} \big)$ decoding time, for arbitrarily small $\epsilon > 0$.  The noisy setting was further studied by Li and Mazumdar \cite{LiMazumdar2024}, whose findings include (i) handling addition noise (Example \ref{ex:addition}) while maintaining $O(k \log n)$ decoding time, and (ii)~a refined algorithm that reduces logarithmic terms in the decoding time compared to \cite{price2023fast} under symmetric noise.
\end{itemize}
Regarding the last of these points, while attaining $O(k \log n)$ scaling in both the number of tests and decoding time remains an open problem in the noisy setting, Guruswami and Wang \cite{Guruswami2023} provided a strategy termed \emph{Gacha} based on list decoding and error-correcting codes, which improves the decoding time from $O\big( \big( k \log \frac{n}{k}\big)^{1+\epsilon} \big)$ to $O(k (\log n)^{O(1)})$ in sufficiently sparse scaling regimes, as well as attaining other novel trade-offs between the number of tests, decoding time, and rate of false positives and negatives.
}


\section{The linear regime} \label{sec:linear}

Throughout this survey, we have focused on the sparse regime, where the number of defective items $k$ scales as $k = o(n)$, specifically $k = \Theta(n^\alpha)$ with $\alpha < 1$. However, for many real-world applications, it may be more realistic to assume that each item has a constant probability of being defective as $n \to \infty$, rather than probability tending to $0$. For example, we might assume each soldier has a probability $\beta$ of having syphilis, but we would not expect this probability to decrease as more soldiers join. \added{Along similar lines, disease prevalence (such as for Covid-19) is invariably reported using percentages, and not using `sublinear' notions such as square roots (although sublinear models via Heap's law have been proposed for modelling early stages of pandemics \cite{wang2011evolution}).}  

In this section, we turn our attention to such linear scaling regimes; specifically, we are interested in the asymptotic behaviour of the required number of tests when $k = \Theta(n)$. It will turn out that, in contrast to the sparse regime $k = \Theta(n^\alpha)$ with $\alpha < 1$, the constant term in front of the $n$ is important; thus, we will consider a limiting regime where $k \asym \beta n$, by which we mean that $k/n \to \beta$, for some constant $\beta \in (0,1)$.

The theory of group testing in this regime turns out to be decidedly different to the sparse regime studied throughout the monograph.  We first note that in contrast to \eqref{eq:bincoeffequiv}, the term $\log_2 \binom nk$ from the counting bound (Theorem \ref{thm:bjaconverse}) behaves in this regime as \cite[p.~1187]{cormen}
  \begin{equation} \label{eq:bincoeffequiv2} \log_2 \binom nk \asym nh\left(\frac kn\right) \asym nh(\beta) , \end{equation}
which is linear in $n$. (Here, as before, $h$ is the binary entropy function.) Hence, for algorithms having a nonzero rate, we seek a number of tests scaling as $T = O(n) = O(k)$. In contrast, algorithms and designs requiring $T = \Omega(k \log n)$ tests, as we found before, will have rate $0$.

Moreover, we see here that simply testing each item individually in $T = n$ tests gives a positive rate of 
\begin{equation} \label{eq:indiv}
  \frac{\log_2 \binom nk}{T} = \frac{\log_2 \binom nk}{n} \to h(\beta) .
\end{equation}
In fact, under the combinatorial prior (see Section \ref{sec:whatis}) in which $k$ is known, we only require $T = n-1$ tests, since we will know whether or not the final item is defective by whether we have found $k-1$ or $k$ defectives so far. This still has rate $h(\beta)$, of course. Henceforth, we use the word `optimal' to mean `has an optimal rate' to avoid considering such `second-order' behaviour.

Combining \eqref{eq:indiv} with the counting bound, we see that the capacity $C = C(\beta)$ (or zero-error capacity $C_0$) of group testing in the linear regime is bounded by
  $ h(\beta) \leq C(\beta) \leq 1 $. 

In fact, we will see in this section that for nonadaptive testing, individual testing is optimal, and so we have equality with the lower bound $C(\beta) = C_0(\beta) = h(\beta)$. Even for adaptive testing, individual testing is optimal for large $\beta$, but it can be improved for small $\beta$.

In the rest of this section, we briefly discuss results for group testing with varying degrees of adaptivity.  Recall that under the combinatorial prior we have exactly $k = k(n)$ defectives, with $k/n \to \beta$, while under the i.i.d.~prior each item is independently defective with probability $\beta$.  The results are summarized as follows:
\begin{description}
  \item[Nonadaptive:] For both the zero-error combinatorial model and the small-error i.i.d.~model, individual testing is optimal for all $\beta \in (0,1)$, so the capacity is $C_0(\beta) = C(\beta) = h(\beta)$.
  \item[\added{Two-stage:}] \added{When we define rate with respect to the average number of tests, the rate achieved increases from $\ln 2 \simeq 0.693$ to $1$ as $\beta$ varies from $0$ to $1/2$, and a converse bound shows that this rate is nearly optimal among so-called `conservative' two-stage algorithms \cite{aldridge2020conservative}, which will be defined below.}
  \item[Adaptive:] \mbox{}
  \begin{description}
  \item[Zero-error combinatorial] Individual testing is optimal for $\beta \geq 1 - \log_3 2 \approx 0.369$, giving $C_0(\beta) = h(\beta)$, and this is conjectured to be true for all $\beta \geq 1/3$. For $\beta < 1/3$, there are algorithms giving rates that exceed $h(\beta)$, so individual testing is suboptimal \cite{riccio,hu,aldridge-adaptive}.
  \item[Small-error i.i.d.] Let $\beta^* = (3 - \sqrt 5)/2 \approx 0.382$. Individual testing is optimal for $\beta \geq \beta^*$ giving $C(\beta) = h(\beta)$. For $\beta < \beta^*$, there are algorithms giving rates that exceed $h(\beta)$, so individual testing is suboptimal \cite{FKW,aldridge-adaptive}.
  \end{description}
\end{description}
We also briefly discuss noisy settings in Section \ref{sec:lin_regime_noisy}.

\subsection{Nonadaptive testing} \label{sec:lin_nonada}

We begin with the nonadaptive setting.  The optimality of individual testing for \emph{zero-error} recovery has long been known via the results discussed in Section \ref{sec:zero-error}.  Perhaps more surprisingly, the same is also true under the small-error criterion; this was first shown under the i.i.d.~prior by Aldridge \cite{aldridge-linear}. (This result improved on an earlier converse result of Agarwal, Jaggi and Mazumdar \cite{agarwal-jaggi-mazumdar}.)

\begin{theorem} \label{th:linear}
Consider nonadaptive group testing with an i.i.d.~prior where each of the $n$ items is independently defective with a given probability $\beta \in (0,1)$, independent of $n$. Suppose we use $T < n$ tests. Then there exists a constant $\epsilon = \epsilon(\beta) > 0$, independent of $n$, such that the average error probability is at least $\epsilon$.
\end{theorem}

The key idea of \cite{aldridge-linear} is the following: Suppose that some item $i$ is \emph{masked} (or in the terminology of \cite{aldridge-linear}, \emph{totally disguised}), in that every test containing $i$ also contains a defective item distinct from $i$. Then every test containing $i$ is positive, regardless of whether $i$ is defective or nondefective. Thus, we cannot know whether $i$ is defective or not. \added{(We used this idea earlier in this monograph when analysing SSS in Section \ref{sec:sss} and also in Section \ref{sec:optimal_gt}; but, chronologically, the tools were developed first for the linear regime in \cite{aldridge-linear}.)} In the linear regime, we either guess that $i$ is defective, and are correct with probability $\beta$; guess that $i$ is nondefective, and are correct with probability $1 - \beta$; or take a random choice between the two. In any case, the error probability is bounded below by the constant $\min\{\beta,1 - \beta\} > 0$. Thus, we can attain a converse bound by showing that, again with probability bounded away from $0$, there is such a masked item $i$. The probability that item $i$ is masked is lower bounded as follows:
  \begin{equation} \label{fkg}
    \mathbb P(D_i) \geq \prod_{t \,:\, x_{ti} = 1} \big(1 - (1 - \beta)^{w_t-1}\big)  ,
  \end{equation}
where $w_t$ is the number of items in test $t$ (that is, the weight of test $t$). The bound \eqref{fkg} can easily be shown using the FKG inequality \cite[Lemma 4]{aldridge-linear}, and an elementary but longer proof is given in \cite[Lemma 4]{agarwal-jaggi-mazumdar}.

The proof of Theorem \ref{th:linear} given in \cite{aldridge-linear} first shows that when $T < n$ one can assume, without loss of generality, that there are no tests of weight $0$ or $1$. It then uses \eqref{fkg} to show that, for any design with $w_t \geq 2$ for all $t$, the mean \added{of the log-probability $\log \mathbb P(D_i)$, averaged over all items $i$, is bounded away from $-\infty$.} Hence, \emph{some} item certainly has a probability of being masked that is bounded away from zero, thus proving the theorem.

While Theorem \ref{th:linear} shows that the error probability is bounded away from $0$, the given bound on the error probability can be very small, particularly as $\beta$ decreases \cite{aldridge-linear}.  \added{To address this, a refinement of the result was established by Bay, Price, and Scarlett \cite{bay2022optimal}, stating that if $T < (1-\eta)n$ for arbitrarily small $\eta > 0$, then the error probability in fact approaches one as $n \to \infty$.  Moreover, they showed that this is the case not only in the linear regime, but also in `mildly sublinear' regimes in which $k = \omega\big(\frac{n}{\log n}\big)$.  On the more positive side, when one moves from exact recovery to \emph{approximate recovery} (see Section \ref{sec:partial}), one can significantly outperform individual testing, particularly when the sparsity parameter $\beta$ is small. We omit a detailed discussion here, and refer the interested reader to the work of Tan, Tan, and Scarlett \cite{Tan2022} for the details.}


\subsection{Two-stage testing} \label{sec:conservative}

\added{We now turn to two-stage testing. We consider the average-case number of tests and the notion of rate defined with respect to this average; so we are using $\log_2 \binom{n}{k}/ \EE T$ as the rate in \eqref{eq:rate}, where a fixed number of tests has been replaced by its expected value $\EE T$.}

\subsubsection{Dorfman's strategy}

\added{Dorfman's two-stage algorithm, whose weakness in the sparse regime we previously discussed in Section \ref{sec:adaptive}, becomes competitive again in the linear regime. Recall that Dorfman's algorithm starts by splitting the $n$ items into $n/m$ pools of $m$ items (ignoring rounding issues) for some parameter $m$. Each of these pool is tested. If the test of a pool is negative, we know that the $m$ items are all nondefective; if the test is positive, we then test each of the $m$ items individually to discover their status. This is a two-stage algorithm because we can perform the $n/m$ pooled tests in stage 1 and all the individual tests in stage 2.}

\added{
The number of tests this uses depends on the defectivity prior:
\begin{itemize}
    \item Under the combinatorial prior with $k \sim \beta n$, Dorfman's algorithm takes, in the worst case,
    \begin{equation}
        T = \frac{n}{m} + km = \left(\frac{1}{m} + \beta m\right)n \label{eq:dorf_comb}
    \end{equation}
    tests, since the least favourable outcome is for the $k$ defective items to all be different pools, requiring $km$ stage-2 tests. With $m = 2$, this improves on individual testing for $\beta < \frac14$. 
    \item Under the i.i.d.~prior with parameter $\beta$, Dorfman's algorithm takes on average
    \begin{equation}
        \mathbb ET = \frac{n}{m} \big(1 + (1 - (1-\beta)^m) m\big) = \left(\frac{1}{m} + 1 - (1-\beta)^ m\right)n \label{eq:dorf_iid}
    \end{equation}
    tests: each pool requires $1$ test in the first round, each pool is positive with probability $1 - (1-\beta)^m$, and for each positive outcome there are another $m$ individual tests in the second round. With $m = 3$, this improves on individual testing for $\beta < 1 - 3^{-1/3} \approx 0.307$. (It turns out that $m = 2$ is never optimal here, as it is always beaten by either $m = 3$ or individual testing.)
\end{itemize}
}

\subsubsection{Conservative two-stage testing} 

\added{We now consider a more general class of strategies of which Dorfman's strategy is a special case.  Broadly speaking, we use a first stage that resembles a nonadaptive scheme of the type discussed in Chapter \ref{ch:algorithms}, and use the second stage to resolve any remaining uncertainty.  We restrict attention here to so-called \emph{conservative two-stage testing}, in which the second stage is required to individually test every item that has not been confirmed as nondefective by appearing in a negative test in the first stage. The word `conservative' is used as this is a `safety first' approach that confirms the status of every defective or possibly-defective item with an individual test -- even if an item is logically a `definite defective' (in the terminology of Section \ref{sec:DD}), or even if a more complicated second stage might require fewer tests. (Our use of the term `conservative' follows \cite{aldridge2020conservative,aldridge2022pooled}; other literature has called this `trivial' two-stage testing, including \cite{mezard2011two} studying the sparse regime.)}

\added{Aldridge \cite{aldridge2020conservative} considered conservative two-stage testing in which the first stage follows the trichotomy we set out in Section \ref{sec:basicdefs}: Bernoulli designs (where each item is in each test with probability $p$; the optimal value of $p$ is $1/k =1/\beta n$), constant tests-per-item designs (where each item is in $L$ tests), and doubly regular designs (where each item is in $L$ tests and each test contains $m$ items).}

\added{Dorfman's algorithm corresponds to a doubly regular first stage with $L = 1$. A doubly regular design with $L = 2$ can be pictured by arranging $m^2$ items in an $m \times m$ grid. We perform $2m$ pooled tests: one for each row and one for each column, with the entire row or column being the pool.  Any item that has both its row-test and its column-test positive receives an individual test in the second round. This method was studied in \cite{broder-gupta,aldridge2020conservative}.}

\added{It turns out that the theory in the linear regime has some significant differences to that of the sparse regime. One difference is that the optimal choices of $L$ and $m$ are constant with respect to the problem size, so the fact that they have to be integers needs to be taken into account. Generally there is no closed form expression for the optimal parameters, 
and the rates achieved may not be smooth as a function of the prevalence $\beta$. A second difference is that doubly regular designs become more important in the linear regime. In the sparse regime, the rates of constant tests-per-item and doubly regular designs are identical \cite{Tan2022}, but in the linear regime the doubly regular design outperforms the constant tests-per-item design.}

\added{Aldridge \cite{aldridge2020conservative} gives three theorems, one for each of the three test designs under discussion here.  We focus only on the doubly regular design (first discussed in \cite{broder-gupta}), since it is the one that performs best \cite[Theorem 4]{aldridge2020conservative}.  Specifically, the design of the first stage, with $L$ tests per item and $m$ items per test, is as follows:
\begin{itemize}
    \item Letting $T_1$ be the number of first-stage tests, the test matrix consists of $L$ submatrices of size $\frac{T_1}{L} \times n$ stacked vertically;
    \item Each of the $L$ submatrices is drawn uniformly at random from the set of matrices having column weight exactly 1 and row weight exactly $m$.  (Note that $\frac{T_1}{L} = \frac{n}{m}$ by a double-counting argument.)
\end{itemize}
The result given is for the combinatorial prior with $k \sim \beta n$, but we expect essentially the same behaviour under the i.i.d.~model.  Moreover, while this result considers the average-case number of tests, we expect that a concentration of measure argument could establish matching high-probability behaviour.}

\begin{theorem}
\added{Under a doubly regular design in the first stage with a constant number $L$ of tests per item and a constant number $m$ of items per test, conservative two-stage testing yields the following number of tests on average as $n \to \infty$ when $k \sim \beta n$:
$$\mathbb ET \sim  n\left(\frac{L}{m} + \beta + (1-\beta)\big(1 - (1-\beta)^{m-1}\big)^L \right).$$
In the limit $\beta \to 0$, with optimal choices of $L$ and $m$, this simplifies to
\[ \mathbb ET \sim \frac{\beta\ln \frac{1}{\beta}}{\ln 2}\,n, \]
corresponding to a rate of $\ln 2 \simeq 0.693$.}    
\end{theorem}

\begin{figure} 
    \centering
    \includegraphics[width=0.7\linewidth]{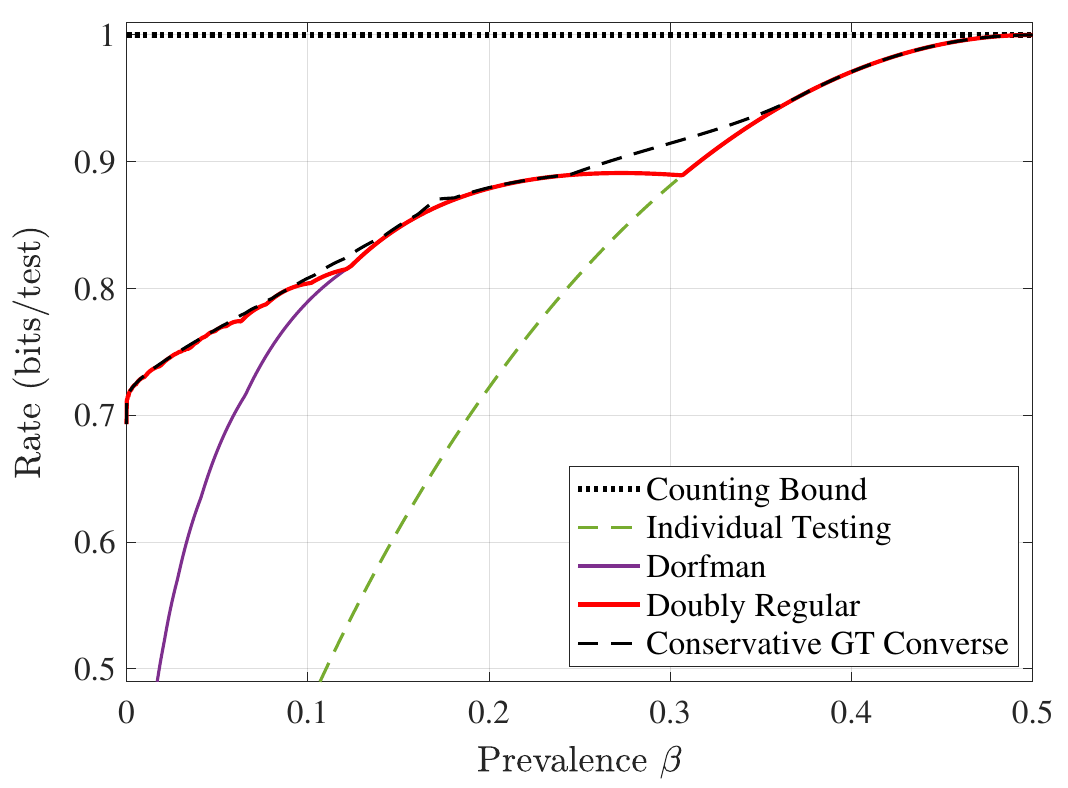}
    \caption{Rates of conservative two-stage schemes for group testing in the linear regime. 
 The converse specific to conservative testing can be found in \cite[Theorem 6]{aldridge2020conservative}, and the result for Dorfman's strategy is given in \eqref{eq:dorf_comb}.}
    \label{fig:conservative}
\end{figure}

\added{Figure \ref{fig:conservative} illustrates several of the results we have discussed. The achievable rates are those of individual testing, Dorfman's algorithm, and conservative two-stage testing with a doubly regular first stage. The upper bounds on rate are the counting bound $R \leq 1$ and an upper bound specific to conservative two-stage testing proved in \cite[Theorem 6]{aldridge2020conservative}. The performance of the doubly regular first stage is very close to optimal (among conservative two-stage methods) in view of this bound.}

\added{Mutesa \emph{et al.}\ \cite{mutesa} studied an adaptive scheme similar to the ones we discussed in this subsection that `usually' only requires two stages, both of which are doubly regular. Their theoretical results were backed up with field trials for Covid testing in Rwanda and South Africa.}

\subsection{Adaptive testing} \label{lin-adap}

We now turn to adaptive testing. Following Aldridge \cite{aldridge-adaptive}, we consider generalized binary splitting algorithms of the same kind as Hwang \cite{hwang}, which we described in Section \ref{sec:adaptive}. Here we consider the following variant. 

\begin{algorithm} The following algorithm finds the defectivity status of every item in a set. The algorithm depends on a parameter $m$; we pick $m = 2^s$ to be a power of $2$ here for convenience.
  \begin{enumerate}
    \item Choose a subset of $m$ items -- say, the first $m$ items.  Test this set.
    \begin{enumerate}
      \item If the test is negative, we find $m = 2^s$ nondefective items using $1$ test.
      \item If the test is positive, perform binary splitting (Algorithm \ref{alg:bin}); we find $1$ defective item and between $0$ and $m-1$ nondefective items using $1+\log_2 m = s+1$ tests.
    \end{enumerate}
  Remove all items whose whose status was discovered in this step.
  \item Repeat Step 1 until no items remain.
  \end{enumerate}
\end{algorithm}

We proceed by discussing the zero-error and small-error settings separately.
  
\subsubsection{Zero-error combinatorial setting} 

As is natural under the zero-error criterion, we take a worst-case analysis and assume that we are always unlucky in step 1(b) and find $0$ nondefective items. Thus, in each stage we find either one defective using $s+1$ tests or $m = 2^s$ nondefectives using one test.  \added{This observation readily yields the following result.

\begin{theorem}
    In the linear regime $k \sim \beta n$, under the combinatorial~prior, the above binary splitting algorithm with parameter $s$ recovers the defective set using at most the following number of tests:
    \begin{equation} \label{comb-lin}
        T \asym (s+1)k + \frac{1}{2^s}(n-k) \asym \left(\frac{1}{2^s} + \left(s+1-\frac{1}{2^s}\right)\beta \right)n.
    \end{equation}
\end{theorem}
}

\begin{figure}[t]
\begin{center}
\includegraphics[width=0.7\textwidth]{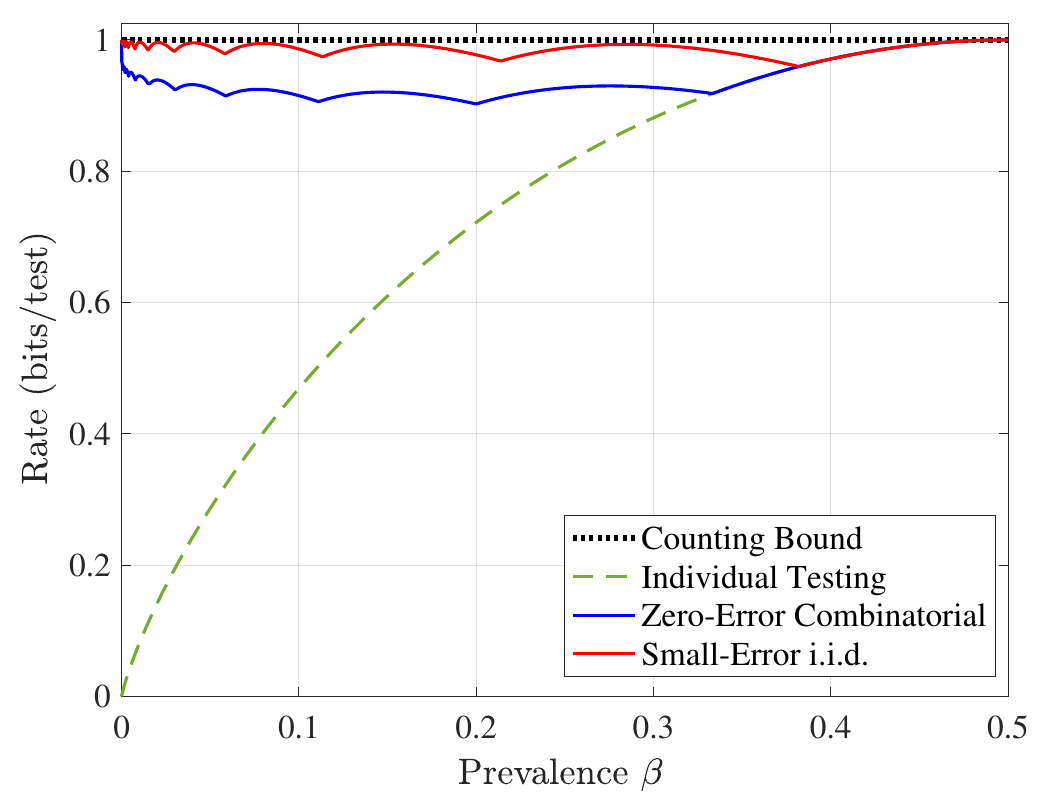}
\end{center}
\caption{\added{Rates of group testing in the linear regime $k \asym \beta n$: an achievable rate of combinatorial zero-error adaptive testing \eqref{comb-lin}; an achievable rate of small-error adaptive testing with an i.i.d.~prior \eqref{prob-lin}; the rate $h(\beta)$ of individual testing, which is the capacity of nonadaptive testing (Theorem \ref{th:linear}); and the counting bound $C \leq 1$.}} \label{fig:linear}
\end{figure}

\rev{The number of tests in \eqref{comb-lin} is linear in $\beta$ for fixed $s$, but only piecewise linear after choosing the optimal value of $s$ for each $\beta$, since $s$ must be an integer.} The resulting rate achieved by this algorithm is shown in Figure \ref{fig:linear}.  \rev{After converting the piecewise linear $T$ into a rate, we now observe `bumps' with endpoints at locations where the optimal value of $s$ in \eqref{comb-lin} changes.}  We obtain a rate of at least $0.9$ for all $\beta \leq 1/2$.

Setting $m = 2^0 = 1$ recovers individual testing, requiring $T = n$ tests. In addition, setting $m=2^1=1$, we see that we outperform individual testing when
  \[ \frac12 + \left(1+1 - \frac12\right)\beta < 1 \]
which is precisely when $\beta < 1/3$. That individual testing is suboptimal for $\beta < 1/3$ was first shown by Hu, Hwang and Wang \cite{hu}, and was also noted in \cite{FKW}. Hu, Hwang and Wang \cite{hu} conjecture that this is the best possible, in the sense that individual testing is optimal for $\beta \geq 1/3$. The best result in this direction is by Riccio and Colbourn \cite{riccio}, who show that individual testing is optimal for $\beta \geq 1 - \log_3 2 \approx 0.369$.

\subsubsection{Small-error i.i.d.~setting} 

By analysing  the average-case behaviour of the above binary splitting algorithm, it can be shown that step 1(b) learns the status of $E$ items on average, where
  \begin{align} 
      E &= \sum_{j=1}^m j\beta (1 - \beta)^{j-1} + m(1 - \beta)^m \notag \\
      &= \frac{1}{\beta} \big(1 + m(1 - \beta)^{m+1} - (m+1)(1 - \beta)^m\big) + m(1 - \beta)^m, \label{eq:bige}
  \end{align}
and that the average number of tests used is
\begin{equation}
    F = 1 +\big(1 - (1 - \beta)^m\big)s. \label{eq:bigf}
\end{equation}
\added{It is then plausible that the average number of tests is $\EE T \sim \frac{F}{E} \cdot n$, and it was proved in \cite{aldridge-adaptive} that this is indeed the case.  Thus far, there is no error probability, but where the small-error criterion comes into play is in establishing concentration around the average with high probability; again see \cite{aldridge-adaptive}.  These results are summarized as follows.

\begin{theorem}
    In the linear regime $k \sim \beta n$, under the i.i.d.~prior, the above binary splitting algorithm recovers the defective set using an average number of tests given by 
    \begin{equation} \label{prob-lin}
        \EE T \asym \frac{F}{E} \,n,
    \end{equation}
    where $E$ and $F$ are defined in \eqref{eq:bige}--\eqref{eq:bigf}.  Moreover, for any $\epsilon > 0$, the number of tests is at most $(1+\epsilon) \EE T$ with probability approaching 1 as $n \to \infty$.
\end{theorem}
}

The corresponding achievable rate is shown in Figure \ref{fig:linear}, and we see that this rate is at least $0.95$ for all $\beta \leq 1/2$. \rev{Again, the `bumps' come from changing integer values of the optimal choice of $s$.}

A similar algorithm is studied in \cite{aldridge-adaptive}, following the work of Zaman and Pippenger \cite{zaman}, where $m$ can be any integer, not just a power of $2$, and an optimal Huffman code to perform the binary splitting. Under such a Huffman code, a defective item will be found in either $\lfloor \log_2 m \rfloor$ or $\lceil \log_2 m \rceil$ tests. After optimizing over all integers $m$, Zaman and Pippenger \cite{zaman} show that this algorithm is optimal among a subset of  adaptive algorithms called \emph{nested} algorithms.

Again, setting $m = 2^0 = 1$ recovers individual testing. Setting $m = 2^1 = 2$ recovers an algorithm of Fischer, Klasner and Wegenera \cite{FKW}, which, as they note, outperforms individual testing when $\beta < \beta^* = (3 - \sqrt 5)/2 \approx 0.382$. Note that $H((1 - \beta)^2) > H(1 - \beta)$ precisely when $\beta < \beta^*$, so intuitively $\beta < \beta^*$ is the regime where a test of two items is `more informative' than a test of one item.  \added{An additional finding in \cite{FKW} is that this threshold $\beta^*$ cannot be improved further, via a converse result showing that individual testing is optimal under the i.i.d.~prior for $\beta \geq \beta^*$.}

\subsection{Noisy settings} \label{sec:lin_regime_noisy}

\added{
We now discuss the noisy setting, focusing mainly on the symmetric noise model in which each test outcome is independently flipped with some probability $\rho \in (0,\frac{1}{2})$.  A notable distinction from the noiseless setting is that testing every item individually once is insufficient, due to the noise.  Nevertheless, we can consider individual testing as a baseline, where we test each item \emph{multiple} times.  A straightforward analysis based on concentration inequalities and the union bound reveals that for this to succeed, we should test each item $\Theta(\log n)$ times, for a total of $\Theta(n \log n)$ tests.  A converse bound due to Scarlett \cite{Sca19} shows that this scaling is optimal among all designs, even when adaptivity is allowed.  While this means that the usual notion of `rate' is zero, it is still interesting to answer the following question: \emph{What is the optimal coefficient to $n \log n$ in the number of tests, up to lower-order asymptotic terms?}

The work of Hintze \etal~\cite{hintze2024noisy} completely resolved this question for the i.i.d.~prior under both nonadaptive and adaptive test designs.  Their results are summarized as follows, with $\beta \in (0,1)$ again denoting the probability of each item being defective:
\begin{itemize}
    \item For nonadaptive designs, the optimal number of tests is
    \begin{equation}
        T \sim \min_{m \in \mathbb{Z}_+} \frac{n \ln n}{ -m \ln\big( 1 - (1-\beta)^{m - 1} (1 - e^{-d(1/2 \mathrel{\|} \rho)}) \big), } \label{eq:noisy_lin_na}
    \end{equation}
    where \[ d(a \mathrel{\|} b) = a \log \frac{a}{b} + (1-a)\log\frac{1-a}{1-b} \] is the binary relative entropy function.
    \item For adaptive designs, the optimal number of tests is
    \begin{equation}
        T \sim \frac{\beta n \ln n}{ d(\rho \mathrel{\|} 1-\rho) },
    \end{equation}
    and moreover, this threshold can be attained using a 3-stage algorithm.
\end{itemize}
An example with $\rho = 0.11$ is shown in Figure \ref{fig:noisy_linear}, where we observe that there is a substantial gap between adaptive and nonadaptive testing.  
We note that \cite{hintze2024noisy} also handles asymmetric binary noise channels, in which $1 \to 0$ flips and $0 \to 1$ flips can have different probabilities.

\begin{figure}[t]
\begin{center}
\includegraphics[width=0.6\textwidth]{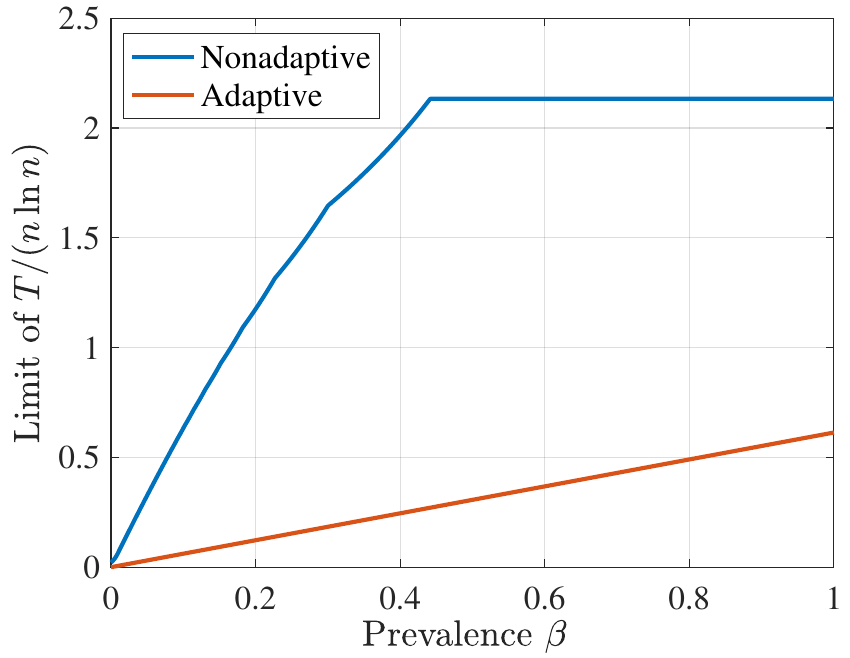}
\end{center}
\caption{\added{Plot of the constant $c$ such that the optimal number of tests is $(c n \log n)(1+o(1))$, for nonadaptive and adaptive testing with noise level $\rho = 0.1$.}} \label{fig:noisy_linear}
\end{figure}

We only provide a few high-level ideas behind the proofs under nonadaptive testing, and refer the interested reader to \cite{hintze2024noisy} for the details.  For the adaptive setting, the achievability part is based on approximate recovery followed by `clean-up' stages, thus bearing conceptual similarity to the techniques outlined in Section \ref{sec:two-stage}.

For the nonadaptive converse, the idea is to consider a `genie-aided' estimator in which each item's defectivity status is estimated via maximum a priori (MAP) decoding \emph{conditioned on the correct status of all other items}.  It is shown that this decoder is at least as good as regular MAP decoding (which is optimal in a non-genie setting), and thus can be analysed for the purpose of a converse.  The decoder is then shown to fail when there are too few tests by upper bounding the number of \emph{non-masked tests} of each item, i.e., tests containing the item but no other defectives.  (For that item, all other tests are in fact ignored by genie-aided MAP.)  The quantity $d(1/2 \mathrel{\|} \rho)$ naturally arises because if item $i$ has $g_i$ non-masked tests, its probability of being misclassified turns out to be roughly $\ee^{-g_i d(1/2 \mathrel{\|} \rho)}$ up to lower-order terms in the exponent.  

For the achievability result, a key idea is to emulate genie-aided decoding using a \emph{pseudo-genie} based on an initial set of $c \log n$ individual tests per item for some constant $c > 0$.  If $c$ is very small then the $c n \log n$ contribution to the number of tests is minimal, yet this still suffices for the vast majority of the pseudo-genie's answers to be correct.  We also note that the parameter $m$ in \eqref{eq:noisy_lin_na} represents a component of the test design in which each item is placed in $m$ tests.

These proof outlines omit a number of other important ideas and subtle issues, for which we again refer the interested reader to \cite{hintze2024noisy}.
}

\section{Group testing with prior statistics} \label{sec:prior}

\added{Throughout the monograph, we have considered prior distributions on the defective set that are relatively `uninformative' other than indicating that defectivity is rare.   In particular, we have considered the combinatorial prior where every defective set of size $k$ is equally likely, and the i.i.d.~prior where every item is defective with the same probability $\qq$.  Under both of these priors, every item has the same probability of being defective, and there is little or no correlation between any two items' defectivity statuses.

Although such properties are natural, they can also be unnecessarily restrictive.  There are several applications in which it is reasonable to expect more prior information to be available, such as some items having a higher chance of defectivity, groups of items exhibiting clustering structure, and so on.  In this section, we outline some studies of such kinds of prior information. }

\subsection{Differing defectivity probabilities}

If certain items are {\em a priori} more likely to be defective than others, then we should ideally try to exploit this additional prior information.   In the following, we describe some test designs and results along these lines.
We focus on the setting described in the following definition, which to the best of our knowledge was first studied in the 1970s in papers such as \cite{garey,hwang2, nebenzahl1973finite}. 

\begin{definition}[Prior defectivity model] \label{def:prior}
Each item $i$ has a (known) prior probability $\qq_i \in [0,1]$ of being defective, and individual items are defective independently of one another. 
\end{definition}

For example, this could model a situation where each individual has a different likelihood of infection according to their prior activity, date of vaccination, and so on.  In the following, we assume that all of the values of $\qq_i$ are known exactly, though we expect the resulting techniques to work well even when they are only known approximately.

Under the model in Definition \ref{def:prior}, the key metrics of success will be expressed in terms of the overall entropy $H_n = \sum_{i=1}^n h(\qq_i)$ (where, as before, $h$ denotes the standard binary entropy) and the average number of defectives $\overline{k}_n = \sum_i \qq_i$. A similar information-theoretic argument to that used to prove the counting bound ({\em cf.}, Theorem \ref{thm:bjaconverse}) shows that at least $H_n$ tests are required to ensure that the success probability tends to one.  We will regard this \emph{entropy bound} as the benchmark for the performance of any algorithm. 

Recall that for any defectivity model, we write $\vc{U} = (U_1, \ldots, U_n)$  for a random vector which encodes the defectivity status of all the items, with $U_i$ being the indicator function of the event that item $i$ is defective, as in Definition \ref{def:udef}. We make the following definition, which generalizes Definition \ref{def:rate} (since in the combinatorial case we have $H(\vc{U}) = \log_2 \binom nk$).

\begin{definition} \label{def:rate2}
Given a random process generating the defectivity vector $\vc{U}$, and $T$ tests, we define the \defn{rate} to be
  \begin{equation} \text{rate} := \frac{H(\vc{U})}T . \end{equation}
\end{definition}

Of course, one may wish to create richer models of defectivity under which $\vc{U}$ is generated. For example, one could imagine individuals represented by the vertices of some graph, and the probability of defectivity of an item being affected by the defectivity status of each individual's graph neighbours, perhaps according to a Markov process.  \added{Some richer models along these lines will be discussed in Section \ref{sec:community}.}

We briefly describe the contributions of two specific papers here. In each case, the key idea is that, by grouping items with a similar value of $\qq_i$ together, we can reduce the problem to one which behaves `locally approximately' like the standard i.i.d.~prior. 
As described in more detail below, splitting strategies are adopted based on binary trees that are balanced (with roughly equal probability on each branch to maximize the information gained from each test). These trees can be understood in analogy with Huffman codes, which are known to achieve optimal lossless data compression (see for example \cite{goldie}).

\subsubsection{Adaptive testing}

Kealy, Johnson and Piechocki \cite{kealy-johnson-piechocki} give a Hwang-type binary splitting algorithm (see Section \ref{sec:adaptive}) in the adaptive case, building on an earlier work of Li \etal\ \cite{li-etal} who developed the {\em Laminar algorithm} for the prior defectivity model (Definition \ref{def:prior}). They discard very low probability items (which are unlikely to be defective anyway, so can be assumed nondefective without wasting tests). The remaining items are grouped together to form a collection of search sets $S_j$ that contain items with $\max_{l,m \in S_j} \qq_l/\qq_m \leq \Gamma$ (for some $\Gamma$), and with total probability $\sum_{i \in S_j} \qq_i \geq 1/2$  wherever possible. Then, one can perform binary splitting over each of these search sets $S_j$ one by one. 

Using this approach, \cite[Theorem 3.9]{kealy-johnson-piechocki} gives a technical condition on $\qq_i$ under which a rate of $1$ (in the sense of Definition \ref{def:rate2}) is achievable in a regime where $\overline{k}_n/H_n \rightarrow 0$. In other words, roughly $H_n$ tests suffice to make the success probability tend to one. This algorithm can be viewed as a non-identical version of Hwang's generalized binary splitting algorithm \cite{hwang}, and this result is a non-identical version of Theorem \ref{thm:adaptcap}.

\added{The techniques outlined above are primarily targeted at the sublinear sparsity regime $\overline{k}_n = o(n)$, but we briefly note that linear sparsity was considered by Attia, Chang, and Tandon~\cite{attia2021hetero}.  Their work adopts the approach of conservative two-stage group testing (see Section \ref{sec:conservative}) while allowing heterogeneity in the defectivity status of the items.}

\subsubsection{Nonadaptive testing}

In the nonadaptive setting, relatively less is known.  As a baseline for the performance, we point out that if the number of defectives behaves as $\overline{k}_n (1+o(1))$ with probability approaching one (as we should usually expect for growing $\overline{k}_n$ due to concentration), then the sufficient number of tests proved for various algorithms in Chapter \ref{ch:algorithms} (e.g., COMP and DD) remain valid.  This is because the analysis of these algorithms was based on fixing a defective set of cardinality $k$ and bounding the probability with respect to the randomness in the test design.  Hence, other than the slight modification of $k$ to $k(1+o(1))$ (which was discussed previously in Remark \ref{rem:misspecify}), the non-uniform prior does not impact the analysis.  It should be noted, however, that using the same number of tests as the uniform setting does not amount to achieving the same rate; the rate can be much smaller for a given number of tests when $H(\vc{U}) \ll k\log_2\frac{n}{k}$.

Nonadaptive test designs  that introduce block structure into the test matrix were explored in \cite{li-etal}.  The performance guarantee given for this approach does not quite amount to a positive rate, as the scaling achieved is \added{$T = O(H(\vc{U}) \log n + \overline{k}_n)$ rather than $T = O(H(\vc{U}))$, but it is only off by a logarithmic factor under the mild assumption $H(\vc{U}) = \Omega(\overline{k}_n)$}.  We refer the interested reader to \cite{li-etal} for further details, and instead focus our attention on providing evidence towards designs that achieve a positive rate, or even a rate of one.

To do so, we consider a simplified setting in which the items are arranged into disjoint groups $G_1,\dotsc,G_m$ whose union equals $ \{1,\dotsc,n\}$.  Suppose that group $j$ contains $n_{j}$ items, each of which is defective with probability $\overline{k}_{j}/n_{j}$, with $\overline{k}_j$ denoting the average number of defectives in the group (in contrast with the above, the dependence on $n$ is left implicit).  Motivated by the idea of using block designs \cite{li-etal}, we can consider a simple approach in which we apply a standard group testing algorithm on each group of items separately.  

Specifically, for group $j$, we fix a number of tests $T_{j}$ and form a $T_{j} \times n_{j}$ i.i.d.~Bernoulli test matrix in which each entry is positive with probability $\nu/\overline{k}_{j}$, for some $\nu > 0$.  We can study each such group using the techniques of the previous chapters, apply a union bound over the $m$ groups to deduce an overall upper bound on the error probability, and note that the total number of tests is $T = \sum_{j=1}^m T_{j}$.

For instance, putting aside computational considerations, suppose that we use the SSS decoding algorithm that achieves a rate of one (in the standard setting) when $k = O(n^{0.409})$ ({\em cf.}, Section \ref{sec:ach_near_const}).  To simplify the analysis, we make the following assumptions:
\begin{itemize} 
    \item Both $n_{j} \to \infty$ and $\overline{k}_{j} \to \infty$ as $n \to \infty$, with $\overline{k}_{j} = O(n_{j}^{0.409})$.  These assumptions readily yield that group $j$ contains $\overline{k}_{j} (1+o(1))$ defectives with probability approaching one.
    \item The number of groups is bounded ($m = O(1)$), so if the error probability for each group vanishes, then so does the overall error probability.
\end{itemize}
By these assumptions and the fact that SSS achieves a rate of one when $k = O(n^{0.409})$ under the near-constant column weight design, one can achieve vanishing error probability with a number of tests satisfying
\[
    T = (1+o(1))\sum_{j=1}^m \overline{k}_{j} \log_2\frac{n_{j}}{\overline{k}_{j}}.
\]
We claim that this in fact corresponds to a rate of one in the non-uniform prior defectivity model.  To see this, note that
\[
    H(\vc{U}) 
        = \sum_{j=1}^m n_{j} \, h\Big( \frac{\overline{k}_{j}}{n_{j}} \Big) 
        = (1+o(1))\sum_{j=1}^m \overline{k}_{j} \log_2\frac{n_{j}}{\overline{k}_{j}},
\]
where we have used $h(\alpha) = (-\alpha \log_2 \alpha)(1+o(1))$ as $\alpha \to 0$. 

It remains an interesting direction for future research to generalize the above approach to general values of $(\qq_1,\dotsc, \qq_n)$ and understand what rates can be achieved, both information-theoretically and with practical decoding techniques.

\added{
\subsection{Community structure} \label{sec:community}

While the setting of differing defectivity probabilities provides a number of interesting insights beyond the standard priors, it is still restricted by the assumption that the status of every item is independent of all others.  In this section, we consider priors that capture \emph{correlations} between items; for example, due to defectives clustering together or otherwise influencing each other.  We first give a brief overview of a number of perspectives that have been explored, and then provide the details on one specific model.

\subsubsection{Overview of the literature}

A line of early works explored group testing with \emph{consecutive defectives} \cite{colbourn1999group,juan2008adaptive}, in which the items are labelled as $\{1,2,\dotsc,n\}$, and the defective set is of the form $\{i^*,i^*+1,\dotsc,i^*+k-1\}$ for some $i^*$.  For example, such a scenario might be encountered in applications where the items are manufactured in sequence and the manufacturing process becomes faulty for a period of time.  See also \cite{bui2022group} for an extension to multiple blocks of defectives.

Following the use of group testing during the Covid-19 pandemic \cite{aldridge2022pooled}, a variety of settings were proposed seeking to capture prior knowledge that may be available in medical applications.  For example, Nikolopoulos \etal~\cite{nikolopoulos2023community} proposed \emph{community-aware group testing} that seeks to exploit knowledge of which individuals share a household, share an office, and so on, and Goenka \etal~\cite{goenka2021contact} and Cao \etal~\cite{cao2023gamp} studied dynamic models based on contact tracing.  Graph-based models have been particularly prevalent \cite{ahn2021adaptive,karimi2022noisy,lau2022model,arasli2023group}, inspired by similar considerations.

Another line of works, more on the theoretical side, has sought to let the prior knowledge be very general by assuming that the defective set lies in some strict subset of the $\binom{n}{k}$ possibilities; see \cite{gonen2022general} for the zero-error criterion and \cite{lau2022model,nikpey2024group} for the small-error criterion.

Since there are too many works to summarize in detail, we will focus our attention on community structure \cite{nikolopoulos2023community}, which we believe to be representative of the general topic and well-aligned with the other parts of this monograph.  Since their setup and terminology are strongly geared towards medical testing, we will momentarily let `individual' be synonymous with `item', and let `infected' be synonymous with `defective'.


\subsubsection{Community-aware group testing}

The setup in \cite{nikolopoulos2023community} is fairly general, allowing communities to overlap with each other, to have different sizes, and to each have a different prevalence of infection.  To simplify the exposition, we focus our attention here on the special case of non-overlapping communities having equal size, equal probability of containing infections, and equal intra-community infection probabilities.

The items are assumed to be arranged in known groups called \emph{communities} (where each `community' might represent a family, a household, an office, or any group with close contact).  The number of communities (or families) is denoted by $F$, 
and each community has $M$ members, so the total number of individuals is $n=FM$.  The infection model is as follows:
\begin{itemize}
    \item Under the combinatorial prior, there are exactly $k_f$ infected communities, each having exactly $k_m$ infected members, and the infected (defective) set is uniformly random over all choices satisfying these conditions. All members of noninfected communities are themselves noninfected.
    \item Under the probabilistic prior, each community is considered to be infected independently with probability $q_f$, and conditioned on the community being infected, each of its members is infected independently with probability $q_m$. All members of noninfected communities are themselves noninfected.
\end{itemize}
These models are related via $k_f \simeq q_f F$ and $k_m \simeq q_m M$, assuming that $F$ and $M$ are large enough to observe concentration around the averages. 

With these models in place, one can readily extend the counting bound in the combinatorial case, and apply an entropy bound in the probabilistic case.  We focus on the former, since it is relatively easier to understand and interpret.

\begin{theorem}
    Under the setup of community-aware group testing with the combinatorial model described above, with parameters $(F,M,k_f,k_m)$, in order to attain asymptotically vanishing error probability as $n \to \infty$, it must be the case that
    $$T \ge (1-\eta)\bigg( \log_2 {\binom{F}{k_f}} + k_f \log_2 {\binom{M}{k_m}} \bigg), $$
    where $\eta > 0$ is arbitrarily small.
\end{theorem}

It is instructive to compare this with the counting bound of $\log_2 {\binom{n}{k}}$ with $k = k_f k_m$, although it should be kept in mind that comparing two lower bounds is not conclusive. (See below for an upper bound.)  To simplify matters, suppose that $k \ll n$ and $k_f \ll F$, so that the comparison is essentially between $k \log_2 \frac{n}{k}$ and $k_f \log_2 \frac{F}{k_f} + k_f \log_2 {\binom{M}{k_m}}$.  We can then consider various cases of interest:
\begin{itemize}
    \item If $k_m \ll M$, then the term $k_f \log_2 {\binom{M}{k_m}}$ roughly simplifies to $k \log_2 \frac{M}{k_m}$, so the main difference is in having $\log_2 \frac{M}{k_m}$ in place of  $\log_2 \frac{n}{k}$.
    \item In the extreme case that $k_m = M$ (entire communities always get infected together), we are only left with $k_f \log_2 \frac{F}{k_f}$, which amounts to a considerable reduction in the leading term from $k = k_f k_m$ to only $k_f$.
    \item In the arguably more realistic case of $k_m = M/2$, the term $k_f \log_2 {\binom{M}{k_m}}$ asymptotically simplifies to $k_f M = 2k_f k_m = 2k$, meaning the main reduction is in the the removal of the logarithmic factor $\log_2 \frac{n}{k}$.
\end{itemize}
Accordingly, we see that the potential reductions are usually in the logarithmic terms, except in extreme scenarios.  However, such reductions can still be quite significant, and this is supported experimentally in \cite{nikolopoulos2023community}.

On the algorithmic side, both adaptive and nonadaptive algorithms were proposed in \cite{nikolopoulos2023community}, but we focus here only on the adaptive case, which is easier to describe and gives a result that can readily be compared with the lower bound.  For clarity, we first focus on a simplified version of their algorithm that is suited to certain specific settings, and then discuss generalizations that can better handle other settings.  The simplified algorithm is described as follows:
\begin{itemize}
    \item[(i)] We first form tests in which entire communities are placed into tests.  Specifically, we consider the $F$ communities as `super-items' and run a regular group testing strategy designed for $F$ `items' and $k_f$ `defectives', where placing a super-item in a test means placing all of its community members.
    \item[(ii)] For each community that is declared infected in step (i), we test each of its members individually.
\end{itemize}
Note that step (ii) relies on the results from step (i), meaning that this approach is inherently adaptive with at least two stages.  In the fully adaptive case, the standard group testing subroutine can be chosen as Hwang's generalized binary splitting (see Section \ref{sec:adaptive}), which incurs $k_f \big( \log_2\frac{F}{k_f} + 1 \big)$ tests in step (i).  Combining this with $k_f M$ individual tests in step (ii), we obtain the following.

\begin{theorem} \label{thm:community_ub}
    Under the setup of community-aware group testing with the combinatorial model described above, with parameters $(F,M,k_f,k_m)$, the adaptive algorithm described above has zero error probability and uses an average number of tests upper bounded as follows:
        $$ \EE T \le k_f \Big( \log_2\frac{F}{k_f} + M + 1 \Big). $$ 
\end{theorem}

We observe that this upper bound asymptotically matches the lower bound when $k_f \ll F$, $k_m = M/2$, and $M \gg 1$, since some simple asymptotic analysis yields 
\[ \log_2 {\binom{F}{k_f}} + k_f \log_2 {\binom{M}{k_m}} = \left( k_f \log_2 \frac{F}{k_f} + k_f M\right)(1+o(1)) . \] 

More generally, the algorithm that we have described is primarily suited to settings where infected communities have an infection rate high enough that individual testing with a community found to be infected is warranted.  If, on the other hand, individual infections are still rare even within an infected community (i.e., $k_m \ll M$), then the $k_f M$ term in the upper bound may be highly suboptimal (just as using $n$ individual tests is highly suboptimal in the standard sparse setting).  
In such cases, a natural improvement is to use group testing in step (ii) as well -- assuming the $k_f$ infected communities are successfully found, one can merge all their members to form $k_f M$ individuals, among which $k_f k_m$ are infected.  We can then apply standard group testing techniques to these individuals, and a suitable generalization of Theorem \ref{thm:community_ub} can readily be obtained.

The situation becomes more complicated when different infected communities have different numbers of infected members.  This can happen due to fluctuations under the probabilistic prior, and more significantly, it arises when we generalize beyond the `symmetric' special case that we are focusing on.  To handle such scenarios, the main algorithm of \cite{nikolopoulos2023community} adopts some additional ideas:
\begin{itemize}
    \item Instead of always testing each entire community together, they propose to randomly take $r$ `representatives' from each community for use in step (i), ignoring any non-representative members initially.
    \item If the analogue of step (i) above declares that some community's set of representatives is infected, then the community is declared `heavily infected', and its members are tested individually in the same way as step (ii) above. 
    \item Even if a given community is not found to be heavily infected, the possibility is still left open that some infections are present in the non-representative members.  Accordingly, in an additional step (iii), further testing is performed on all individuals that were not tested individually.  Since such infections will be rare, a standard group testing algorithm is applied on these, with suitably-chosen parameters.
\end{itemize}

We refer the reader to \cite{nikolopoulos2023community} for a more detailed description, along with significantly more general results than the ones we presented here (including for probabilistic priors, varying infection rates between communities, and overlapping communities), and further comparisons between their upper and lower bounds.

}

\section{Explicit constructions} \label{sec:explicit}

Throughout the monograph, we have focused on randomized test designs, in particularly Bernoulli designs and near-constant column weight designs.  The sublinear-time algorithms described in Section \ref{sec:sublinear} also use randomization in the design stage.  The understanding of explicit deterministic constructions, by contrast, remains relatively limited when it comes to the small-error recovery criterion.  In this section, we give an overview of some progress in this direction.

For the zero-error recovery criterion, several explicit constructions have been proposed.  A prominent early construction, to be described below, was provided by Kautz and Singleton \cite{kautz}, and achieves zero-error reconstruction in the noiseless setting with $T = O\big(k^2 \frac{\log^2 n}{\log^2 k}\big)$ tests.  A related but more recent construction due to Porat and Rothschild \cite{porat} achieves $T = O\big(k^2 \log n\big)$.  Of course, these results are not sufficient for achieving a positive rate when $k = \Theta(n^{\alpha})$ for some $\alpha \in (0,1)$.  The interested reader is referred to \cite{porat} for  more detailed overview of deterministic constructions in the zero-error setting, and to \cite{cheraghchi2010derandomization} for an overview of alternative approaches based on derandomization.

\begin{figure}[t] 
    \begin{center}
        \includegraphics[width=0.95\textwidth]{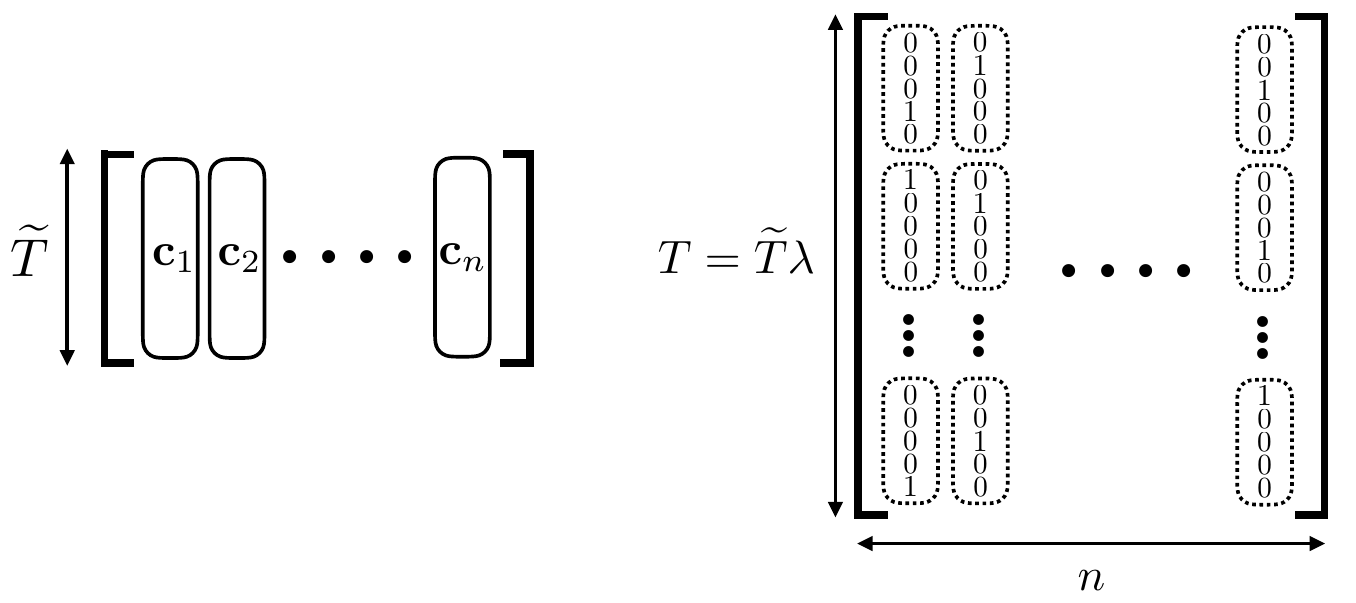}
    \end{center}
    \caption{(Left) The codewords $\{\mathbf{c}_1,\dotsc,\mathbf{c}_n\}$ of a length-$\widetilde{T}$ non-binary code of size $n$ are arranged to form a $\widetilde{T} \times n$ non-binary matrix, with each symbol taking one of $\lambda$ values. (Right) Each non-binary symbol of the matrix on the left is replaced by a length-$\lambda$ binary string with a $1$ in the entry indexing the value (out of $\lambda$ possibilities) of the corresponding non-binary symbol. } \label{fig:explicit} 
\end{figure}

Interestingly, recent developments on achieving a positive rate under the small-error criterion have made use of very similar constructions.  \added{In particular, the construction of Kautz and Singleton \cite{kautz} turns out to be suitable for the small-error criterion when its parameters are suitably modified.}  This construction is based on the idea of  concatenated codes, depicted in Figure \ref{fig:explicit}.  The construction starts with a non-binary channel code ${\cal C} = \{\mathbf{c}_1,\dots,\mathbf{c}_n\}$ containing $n$ codewords of length $\widetilde{T}$, with symbols taking one of $\lambda$ values.  As shown in the left of the figure, these codewords are arranged in columns to form a $\widetilde{T} \times n$ matrix.  To construct the final group testing matrix, each code symbol is replaced by a length-$\lambda$ binary vector with a one in the entry indexing the corresponding non-binary code-symbol and zeros elsewhere.  This produces a constant column weight design, where each of the $n$ items is in exactly $\widetilde{T}$ of the $T = \widetilde{T}\lambda$ tests.
If the original codewords are sufficiently well-separated, then different defective sets should lead to different test outcomes.

For zero-error group testing, the performance of this construction depends crucially on the minimum distance of the code ${\cal C}$.  In contrast, Mazumdar \cite{mazumdar2016} related the (nonzero) error probability of the construction to both the minimum distance and the average distance.  This led to vanishing error probability with a number of tests of the form $T = O\big(k \frac{\log^2 n}{\log k}\big)$ using either of the following two non-binary codes: (i) the above-mentioned construction of Porat and Rothschild \cite{porat}, which achieves the Gilbert-Varshamov bound; (ii) a construction based on algebraic-geometric codes due to Tsfasman, Vl\u{a}dut, and Zink \cite{tsfasman}.  The behaviour $T = O\big(k \frac{\log^2 n}{\log k}\big)$ leads to a positive rate whenever $k = \Theta(n^{\alpha})$ for some $\alpha \in (0,1)$, but the rate vanishes as $\alpha \to 0$.

The original construction of Kautz and Singleton \cite{kautz} used a Reed--Solomon code with $\widetilde{T} = \lambda-1$ and $n = \lambda^k$.  In a recent work, Inan \etal\ \cite{inan} studied a similar construction with Reed-Solomon codes, but proposed different parameters, choosing $\widetilde{T} = \Theta(\log n)$ while still using  $n = \lambda^k$. They performed a direct analysis without relying on distance properties, and showed that one can achieve vanishing error probability with $T = O(k \log n)$ tests as long as $k = \Omega(\log^2 n)$.  As a result, a positive rate is achieved even in the limit $\alpha \to 0$.

Both \cite{mazumdar2016} and \cite{inan} obtained the preceding results using COMP decoding (under a different name).  In addition, in \cite{inan} the same construction was combined with the NCOMP algorithm (see Section \ref{sec:NCOMP}) to obtain the same $T = O(k \log n)$ scaling under the symmetric noise model.  \added{A follow-up work by Inan and Ozgur \cite{Inan2020} additionally showed that the approach of \cite{inan} (with modified parameters) can be combined with bit-mixing coding \cite{bondorf} to attain an explicit design that succeeds with $O(k \log n)$ tests and has sublinear decoding time (\emph{cf.}~Section \ref{sec:sublinear}), namely $O(k^3 \log k + k \log n)$.}

While these results provide important contributions in understanding explicit constructions under the small-error criterion, their focus is on the scaling laws rather than the constant factors or the rates achieved.  It remains an important open challenge to develop achievable rates for explicit constructions that can compete with those of randomized constructions. 

\rev{The preceding results are summarized in Table \ref{tbl:explicit}.}

\begin{table}
\begin{center}
    \begin{tabular}{ccc}
    \hline
    Reference & \# Tests & Zero-error?  \\
    \hline
    Kautz-Singleton \cite{kautz} & $O\big(k^2 \frac{\log^2 n}{\log^2 k} \big)$ & yes  \\
    Porat-Rothschild \cite{porat} & $O(k^2 \log n)$ & yes  \\
    Mazumdar \cite{mazumdar2016} & $O\big(k \frac{\log^2 n}{\log k} \big)$ & no \\
    Inan \etal~\cite{inan} & $O(k \log n)$ & no \\
    \hline
    \end{tabular}
\end{center}
\caption{\rev{Summary of the number of tests required for explicit group testing designs, in the zero-error and small-error settings.  The result in \cite{inan} additionally assumes that $k = \Omega(\log^2 n)$. \label{tbl:explicit}}}
\end{table}

\section{Constrained group testing} \label{sec:constrained}

Thus far, we have assumed that any given group test can contain an arbitrary set of items.  However, in several practical applications, the tests are in fact subject to certain constraints that may make standard designs (e.g., i.i.d.~Bernoulli) infeasible.  In this section, we give an overview of several such constraints and how they are motivated by particular applications of interest.

\subsection{Path constraints on a graph}

The work of Cheraghchi \etal\ \cite{cheraghchi-etal} considers a model in which items may correspond to either nodes or edges in a graph (but not both), and the only tests allowed are those that correspond to a path on that graph.  One of the main motivating applications for this scenario is the problem of network tomography (mentioned previously in Section \ref{sec:applications}), in which the nodes correspond to machines and the edges correspond to connections between machines.  The goal is to identify faults in the network, and this is done by checking whether a packet sent from one machine successfully arrived at a target machine.  There are at least two interesting variants of this setup:
\begin{itemize}
    \item If the goal is to find faulty connections between machines, then one can consider the edges as corresponding to items, and each test as corresponding to sending a packet between two machines along a specific path and checking whether it arrived successfully.  Hence, each test is constrained to be the set of edges along some path in the graph.
    \item If the goal is to find faulty machines, then one can consider the nodes as corresponding to items, and a test again corresponds to sending a packet between two machines along a specific path and checking whether it arrived successfully.  Hence, each test is constrained to be the set of nodes along some path in the graph.
\end{itemize}
The second of these generalizes an earlier model corresponding to the special case of a line graph \cite{cicalese}.

For both of the above models, random testing designs are constructed in \cite{cheraghchi-etal} by performing a {\em random walk} on the graph, thus ensuring that the path constraint is not violated.  It was shown that the $O(k^2 \log n)$ achievability result for zero-error group testing can be generalized to $O(k^2 \tau(n)^2 \log n)$, where $\tau(n) \ge 1$ is a property of the graph known as the mixing time, defined formally in this context in \cite[Definition 6]{cheraghchi-etal}.  For many graphs, $\tau(n)$ is small (e.g., constant or logarithmic in $n$). For example, \cite[Section V]{cheraghchi-etal} discusses the fact that for expander graphs with a constant spectral gap, and for certain Erd\H{o}s-R\'{e}nyi random graphs, $\tau(n) = O(\log n)$ (with high probability).  This means that for graphs of this kind, the additional constraints do not considerably increase the required number of tests in the zero-error setting.

To our knowledge, these constraints have not been studied in the small-error setting, which is the main focus of this monograph.  This poses a potentially interesting direction for future research.

\subsection{Edge constraints on a graph}

A series of works rooted in early studies in theoretical computer science (e.g., see \cite{Aig86,aigner1988searching}) have considered a different type of constraint imposed by a graph $G = (V,E)$.  Here, the edges correspond to items, and $k$ of the $|E|$ items are defective.  However, each test is a group of {\em nodes} rather than edges, and the test outcome is positive if and only if there is at least one defective edge connecting two different nodes in the group. An interesting special case of this problem is obtained when $E$ is the complete graph, and the goal is to identify a subgraph of $k$ edges corresponding to those that are defective. 

As an example application, this might correspond to a scenario where we wish to identify interactions between chemicals, and each test amounts to combining a number of chemicals and observing whether any reaction occurs.

The work of \cite{johann2002group} considers the adaptive setting, and shows that even in the general formulation, it suffices to have $k \log_2 \frac{|E|}{k} + O(k)$ tests, 
which matches  (a slight variation of) the counting bound asymptotically whenever $k = o( |E| )$.  More recently, significant effort has been put into developing algorithms with limited stages of adaptivity.  We refer the reader to \cite{abasi18learning} and the references therein for further details, and highlight a particularly interesting result:   In the special case that $E$ is the complete graph, it is known that when $k$ grows sufficiently slowly compared to $|V|$,\footnote{This is a more stringent requirement than it may seem, since there are $\binom{|V|}{2} \approx \frac{1}{2}|V|^2$ items in total.} any nonadaptive algorithm requires $\Omega(k^2 \log |V|)$ tests even when a small probability of error is tolerated.  This is in stark contrast with standard group testing, where $k^2$ terms only arise under the zero-error criterion (see Section \ref{sec:zero-error}).  On the other hand, it was shown in \cite{li2019learning} that $O(k \log |V|)$ nonadaptive tests indeed suffice in the small-error setting for certain {\em random} (rather than worst-case) graphs with an average of $k$ edges.  \added{See also \cite{austhof2025non} for an extension to an analogous hypergraph learning problem.}

\subsection{Sparse designs} \label{sec:sparse}

Another interesting form of constrained group testing is that in which the number of tests-per-item or items-per-test is limited.  Using the terminology of Gandikota \etal\ \cite{gandikota2016sparse}, we consider the following:
\begin{itemize}
    \item The constraint of {\em $\gamma$-divisibility} requires that each item participates in at most $\gamma$ tests.  If one considers the classical application of testing blood for diseases, this corresponds to the case that each patient's blood sample can only be split into a limited number of smaller subsamples. 
    \item The constraint of {\em $\rho$-sized tests} requires that any given test contains at most $\rho$ items.  This may correspond to equipment limitations that prevent arbitrarily many items from being included in a pool.
\end{itemize}

As noted in \cite{gandikota2016sparse}, the most interesting cases are the regimes $\gamma = o(\log n)$ and $\rho = o\big( \frac{n}{k} \big)$; this is because if one allows $\gamma = O(\log n)$ or $\rho = O\big( \frac{n}{k} \big)$ and the implied constants are not too small, then even standard designs (such as constant column weight designs) can be used that are already near-optimal in the unconstrained sense.

We proceed to briefly outline some of the main existing results.  
Under the $\gamma$-divisibility constraint with $\gamma = o(\log n)$, we have the following:
\begin{itemize}
    \item A simple adaptive algorithm  attains zero error probability using $\gamma k (n/k)^{1/\gamma}$ tests (this was also noted in an early work of Li \cite{li1962}). The algorithm is a `$\gamma$-stage' algorithm in the sense of Section \ref{sec:two-stage}, and is defined recursively. The case $\gamma = 1$ uses individual testing. Then the algorithm for $\gamma$-divisibility is as follows: Partition the $n$ items into $A = k(n/k)^{1/\gamma}$ sets of size $n/A$, and test each set. As usual, all items in negative tests are nondefective. For the remaining items -- of which there are at most $kn/A$ -- continue with the algorithm for $(\gamma - 1)$-divisibility. The first step takes $k(n/k)^{1/\gamma}$ tests, and the worst-case number of tests in each subsequent step is easily checked to also be $k(n/k)^{1/\gamma}$, giving the desired result. The case $\gamma = 2$ gives Dorfman's original adaptive algorithm \cite{dorfman} (see \eqref{eq:dorfmanegg} in Chapter \ref{ch:introduction}).
    \item For nonadaptive testing and the small-error criterion, Gandikota \etal\ \cite{gandikota2016sparse} show that any algorithm requires roughly $\gamma k (n/k)^{1/\gamma}$ tests. In addition, they show that one can attain error probability at most $\epsilon$ using at most $\ee \gamma k (n/\epsilon)^{1/\gamma}$ tests, which behaves similarly to the converse bound but nevertheless leaves a gap in the scaling laws.  The corresponding algorithm and test design are discussed below.
    \item \added{Improved results were obtained by Gebhard \etal\ \cite{Gebhard2022}, giving a lower bound on the number of tests of $\max\big\{\ee^{-1} \gamma k\big( \frac{n}{k} \big)^{1/\gamma}, \gamma k^{1 + 1/\gamma}\big\}$ and an upper bound of $\max\big\{\gamma k\big( \frac{n}{k} \big)^{1/\gamma}, \gamma k^{1 + 1/\gamma}\big\}$, up to lower-order asymptotic terms.  Together, these results establish an exact threshold in sufficiently dense regimes, and a threshold tight to within a factor of $\ee^{-1}$ more generally.}
\end{itemize}
Under the $\rho$-sized tests constraint with $\rho = \Theta\big((n/k)^{\beta} \big)$ for some $\beta \in [0,1)$ and in the sparse regime $k = \Theta(n^\alpha)$, $\alpha < 1$, we have the following results:
\begin{itemize}
    \item Any testing strategy requires at least $n/\rho$ tests, as this many tests is required just to test each item once. Applying the generalized binary splitting algorithm (see Sections \ref{sec:adaptive} and \ref{sec:linear}) starting with sets of size $\rho$ requires
      $\frac{n}{\rho} + k \log_2 \rho + O(k) \asym \frac{n}{\rho}$
    tests, where $n/\rho$ is for testing each set once and $k \log_2 \rho + O(k)$ is for $k$ rounds of binary splitting.
    \item For nonadaptive testing under the small-error criterion, Gandikota \etal\ \cite{gandikota2016sparse} show that any algorithm requires roughly $\frac{1}{1-\beta} \frac{n}{\rho}$ tests. In addition, they show that vanishing error probability can be attained using roughly $\frac{1}{(1-\beta)(1-\alpha)} \, \frac{n}{\rho}$ tests.
    \item \added{Improved results were obtained by Gebhard \etal\ \cite{Gebhard2022} in the special case that $\rho = O(1)$ (so $\beta = 0$), leading to a precise threshold of \[\max\left\{\Big(1 + \Big\lfloor\frac{\alpha}{1-\alpha}\Big\rfloor\Big)\frac{n}{\rho}, \, \frac{2n}{\rho + 1}\right\}\] up to lower-order asymptotic terms.  Analogous improvements were also derived by Tan, Tan, and Scarlett \cite{Tan2022} for the case that $\beta > 0$, but without establishing exact thresholds.}
\end{itemize}

We observe that under both types of constraint, the number of tests required is significantly higher than the usual $O(k \log n)$ scaling.  

The lower bounds of  Gandikota \etal\ \cite{gandikota2016sparse} are based on Fano's inequality, but with more careful entropy bounds than the standard approach of upper bounding the entropy of a test outcome by $1$.  For instance, if each test contains at most  $\rho = o(n/k)$ items, then the probability of the test being positive tends to zero, so the entropy per test is much smaller than one bit.  The upper bounds of \cite{gandikota2016sparse} are based on test designs with exactly $\gamma$ tests per item or exactly $\rho$ items per test, along with the use of the basic COMP algorithm. 

\added{The improved lower bounds in \cite{Gebhard2022,Tan2022} are based on identifying suitable \emph{masking} events, in analogy with the developments outlined in Section \ref{sec:gen_conv}.  The improved upper bounds in \cite{Gebhard2022,Tan2022} are based on improving both the test design and the choice of decoder.  Under the $\gamma$-divisibility constraint, the near-constant tests-per-item design and DD algorithm are the key ingredients, similarly to the unconstrained setting.  Under the $\rho$-sized tests constraint, the design choices are somewhat more delicate: in \cite{Tan2022} a doubly-regular test design is paired with the DD algorithm, whereas in \cite{Gebhard2022} a `nearly regular' test design based on random graph matching is paired with the SCOMP algorithm.}

We briefly mention that a complementary constrained group testing problem has also been considered, in which the number of positive tests (rather than items-per-test or tests-per-item) is constrained \cite{damaschke2016adaptive}.

\section{Computational considerations} \label{sec:computation}

\added{
In Chapter \ref{ch:achievability}, we surveyed a series of works resulting in a precise understanding of what rates are achievable for noiseless nonadaptive group testing: When $k = \Theta(n^{\alpha})$, the best possible rate is $\olR_{\rm NCC} = \min\big\{1,(\ln 2)\frac{1-\alpha}{\alpha}\big\}$.  We saw that this rate can be achieved by various test designs and decoders depending on $\alpha$:
\begin{itemize}
    \item For $\alpha \ge \frac{1}{2}$, it suffices to pair the DD algorithm with the near-constant column weight design;
    \item For $\alpha < \frac{1}{2}$, the near-constant column weight design can still be used, but the decoder adopted in the analysis involves a computationally intractable brute force search;
    \item For all $\alpha \in (0,1)$, the same rate can be achieved with polynomial-time decoding when a spatially coupled test design is used.
\end{itemize}
The findings are similar for the Bernoulli design, albeit with a slightly smaller rate for medium to large values of $\alpha$. 
A similar discussion also holds for approximate recovery (\emph{cf.}, Section \ref{sec:partial}):
\begin{itemize}
    \item The optimal rate is 1 for all $\alpha \in (0,1)$:
    \item Under both the Bernoulli and near-constant column weight designs, this optimal rate can be achieved, but the decoder used is computationally intractable, and the best known practical methods only attain a smaller rate of at most $\ln 2 \approx 0.693$;
    \item Under the spatially coupled test design, the optimal rate can be achieved with polynomial-time decoding.
\end{itemize}
In view of this discussion, the following question naturally arises:
\emph{Is there a computationally efficient decoder that achieves the information-theoretically optimal rates  (for exact recovery with $\alpha < \frac{1}{2}$ or approximate recovery with $\alpha \in (0,1)$) under the Bernoulli design and/or the near-constant column weight design, without incorporating spatial coupling?}  In this section, we survey various works that addressed this question, with an emphasis on certain kinds of \emph{computational hardness} results.  These works focused on approximate recovery (see Section \ref{sec:partial}) with $o(k)$ mistakes, but it is worth noting that any converse bound on the rate for this criterion implies the same converse for exact recovery.

\subsubsection{Evidence from local search algorithms} The works of Iliopoulos and Zadik \cite{iliopoulos2021group} and Lovig and Zadik \cite{lovig2024mcmc} focused on Bernoulli testing under the approximate recovery criterion, and studied the performance of \emph{local search algorithms} that seek to gradually improve the estimate via iterative improvements until some stopping condition is met.  Both works particularly emphasized Monte Carlo methods, which we briefly outlined in Section \ref{sec:monte_carlo}.

While the initial evidence in \cite{iliopoulos2021group} pointed to the possibility of the optimal information-theoretic threshold being attainable via local search algorithms, the follow-up work in \cite{lovig2024mcmc} gave stronger evidence in the opposite direction, suggesting that local search algorithms \emph{cannot} attain a rate exceeding $\ln 2$.  To describe the results of these works in more detail, we first need to outline a property known as the \emph{overlap gap [roperty} (OGP).


In the context of group testing, the Overlap Gap Property (OGP) is related to the following problem: Letting PD be the set of possibly defective items (items not appearing in any negative tests), the goal is to find a size-$k$ set $\Khat \subseteq {\rm PD}$ that minimizes the number of unexplained positive tests (the number of tests that are positive but do not contain any items from $\Khat$).  Informally, the OGP states that the set of near-optimal solutions to this problem forms two distinct clusters; one `close' to the true defective set $\K$ and another `far' from $\K$.  When this is the case, algorithms based on local improvements may get stuck in the `far' cluster and fail to bridge the gap.  Conversely, as discussed in \cite{lovig2024mcmc}, local search algorithms have been observed to succeed in the absence of an OGP in several other contexts (for example, sparse linear regression and planted clique problems).

Letting $\phi(\ell)$ denote the smallest number of unexplained tests over all size-$k$ subsets of PD containing $\ell$ defectives, one of the main results in \cite{iliopoulos2021group} showed that the average-case behaviour of $\phi(\ell)$ is non-decreasing, which suggests the \emph{absence} of the OGP.  However, a refined analysis in \cite{lovig2024mcmc} revealed that the average-case evidence is in fact misleading, and arises due to the average being heavily influenced by rare events.   With a more careful analysis that conditions away such events (namely, events in which certain items appear in `unusually many' tests), it was found that the OGP in fact holds for sufficiently small $\alpha$ when the rate is below a threshold of roughly $0.678$.  As a corollary of this, a certain class of `low-temperature' Markov Chain Monte Carlo methods is shown to fail at such rates; the proof uses connections to a random set cover problem (see our Remark \ref{rem:setcover}).  
%
The rate of $0.678$ is in fact even lower than that of separate decoding of items ($\ln 2 \approx 0.693$), suggesting that the local search algorithms under consideration are strictly suboptimal, even among polynomial-time methods.

It is worth noting, however, that these results are asymptotic in nature, and may not align with empirical findings at small or moderate problem sizes.  In fact, numerical simulations were performed in \cite{iliopoulos2021group} demonstrating that a local search algorithm based on Glauber dynamics can match the performance of the optimal Smallest Satisfying Set (SSS) algorithm at practical problem sizes.  Thus, despite the above theoretical findings, local search methods may still significantly outperform simpler methods such as COMP, DD, and separate decoding of items in practice.


\subsubsection{Evidence from low-degree polynomial algorithms}

The work of Coja-Oghlan \etal~\cite{coja2022statistical} seeks to understand the approximate recovery criterion using an indirect approach that first studies a \emph{detection problem} (see also \cite{truong2020all}): Given a test matrix $\mat X$ and vector of test results $\vec y$, determine whether $\mat X$ and $\vec y$ were produced via the standard noiseless group testing model (under the combinatorial prior on $\K$) or were produced independently of one another (with a suitably defined `null distribution').  It is shown in \cite{coja2022statistical} that this is an easier problem than achieving approximate recovery (with $o(k)$ mistakes) in group testing, meaning that if the detection problem is impossible then so is the approximate recovery problem.

The class of algorithms considered in \cite{coja2022statistical} is that of \emph{low-degree polynomial algorithms}:  Consider the bipartite graph (see Figure \ref{fig:bp_graph} in Section \ref{sec:belprop}) described by $nT$ binary values (1 for an edge, 0 otherwise); apply a degree-$D$ polynomial $f \colon \mathbb{R}^{nT} \to \mathbb{R}$ to that graph (for some $D > 0$); and make the final detection decision based on the resulting output.  Letting $P$ and $Q$ being the two distributions in the detection problem, a natural goal is to have the averages $\mathbb{E}_P[f]$ and $\mathbb{E}_Q[f]$ be separated by at least a constant factor times one standard deviation; since, if this fails, distinguishing the two may be hard or impossible.  This condition is known as \emph{weak separation}.

Informally, a degree-$D$ polynomial algorithm is expected to have runtime exponential in $D$ in the worst case, meaning that if we seek runtime polynomial in $n$, we should target $D = O(\log n)$.  While the class of low-degree polynomial algorithms is not exhaustive, it captures powerful techniques such as spectral methods, as well as other methods providing state-of-the-art guarantees in problems such as planted clique and community detection.  There are also results showing that their failure implies the failure of another prominent class (namely, statistical query algorithms) under suitable assumptions \cite{brennan2021stat}.

With the above discussion in mind, two of the main results in \cite{coja2022statistical} are outlined as follows. As ever, $\alpha$ is the sparsity parameter such that $k = \Theta(n^{\alpha})$.
\begin{itemize}
    \item For the constant column weight design, it is shown that if $T = c k \log_2 \frac{n}{k}$ with $c < \frac{1}{\ln 2}\big( 1 - \frac{\alpha}{2(1-\alpha)} \big)$, then for some $D = n^{\Omega(1)}$, all degree-$D$ polynomial algorithms fail to obtain weak separation.  This suggests a required runtime for weak detection of $\exp(n^{\Omega(1)})$, i.e., higher than polynomial.  As $\alpha \to 0$ the threshold on $c$ approaches $\frac{1}{\ln 2}$, corresponding to a rate of $\ln 2 \approx 0.693$, which matches that of COMP.  This suggests that it is computationally hard to attain a better rate than COMP for small $\alpha$.
    \item Analogous results are given for the Bernoulli design; these are more complicated to state, but they similarly suggest that in low-$\alpha$ regimes there exist rates that are information-theoretically achievable but computationally hard to achieve.  In particular, as $\alpha \to 0$ their results again suggest the hardness of beating the rate $\ln 2 \approx 0.693$, which matches the rate of separate decoding of items. 
\end{itemize}

\subsubsection{Discussion} Overall, while \cite{iliopoulos2021group} and \cite{coja2022statistical} presented opposing evidence, the follow-up work in \cite{iliopoulos2021group} led to a more general agreement of evidence pointing to a computational-statistical gap, at least for small values of $\alpha$.  Both local methods and low-degree polynomial methods are generally very powerful and are known to attain optimal thresholds in other contexts, but it is worth noting that neither of them serve as `exhaustive' models for computation, so it is still conceivable that they could be outperformed by other methods.
}

\section{Other group testing models} \label{sec:other_models}

Throughout the monograph, we have focused primarily on the noiseless model and certain simple random noise models such as symmetric noise, addition noise, and dilution noise.  There are extensive additional models that have been considered previously in the literature, but in most cases, understanding them via the information-theoretic viewpoint remains open.  \added{Here, we provide an overview of two models whose study has been more aligned with the techniques surveyed in this monograph, before providing a partial list of other models.}

\subsection{Quantitative group testing}

Under the {\em quantitative group testing model} (also known as the {\em linear model} or the {\em adder channel model}), each test output provides the \emph{number of defectives} in the test, thus providing significantly more information than the presence or absence of at least one defective.  \added{Quantitative group testing is also equivalent to a \emph{coin-weighing problem} dating back to early work including that of Erd\"os and R\'enyi \cite{erdos-renyi}; see \cite{bshouty4} for a partial historical overview.  Yet another interesting connection is that quantitative group testing can be viewed as a problem of compressive sensing, i.e., sparse recovery from linear measurements \cite{foucart}, but with the notable feature that both the sparse vector and measurement matrix are binary-valued.
}

\added{Compared to standard group testing, the optimal number of (noiseless) tests in quantitative group testing turns out to reduce from $O\big( k \log \frac{n}{k}\big)$ to $O\big( k \frac{\log \frac{n}{k}}{\log k}\big)$ (e.g., see \cite{sebo,bshouty4,wang2016data,Sca17c,gebhard}).  Intuitively, this is because each test outcome is in $\{0,1,\dotsc,k\}$, so can reveal up to $O(\log k)$ bits of information rather than only 1 bit.  In particular, in the sparse regime ($k = \Theta(n^\alpha)$ for $\alpha \in (0,1)$), this means that the optimal number of tests is $O(k)$.  

In the following, until stated otherwise, our discussion is specific to the sparse regime, noiseless testing, and nonadaptive test designs.  
In this setting, in addition to knowing the optimal $O(k)$ scaling, we have a precise understanding of the constant factor: The optimal number of tests is given by $\big( \frac{2k}{\log k}\log\frac{n}{k}\big)(1+o(1))$.  This result comes from combining converse bounds of Djackov \cite{djackov1975} and Lindstr\"om \cite{Lindstrom1975} with an upper bound adapted from El Alaoui \etal~\cite{el2019decoding}; see the work of Feige and Lellouche \cite{feige2020quantitative} for an accessible summary.  The number of tests is notably a factor 2 higher than a lower bound suggested by a simple counting argument, as we will discuss further below.

Even ignoring constant factors, attaining an order-optimal number of tests with \emph{polynomial-time decoding} is not straightforward.  To our knowledge, this goal was first achieved by Hahn-Klimroth and M\"uller \cite{hahn2022near}, using a spatially-coupled test design akin to the one we surveyed in Section \ref{sec:spatial}.  Soleymani and Javidi \cite{soleymani2024non} further gave an algorithm with reduced decoding time, at the expense of the number of tests slightly increasing to $O(k \log \log k)$.

Beyond the sparse regime with noiseless nonadaptive testing, some notable developments are outlined as follows:
\begin{itemize}
    \item In the linear regime $k = \Theta(n)$, the optimal number of tests is $O\big( \frac{k}{\log k} \big)$, which is  sublinear with respect to $k$.  In the case of random Bernoulli testing, corresponding exact constants were established by El Alaoui \etal~\cite{el2019decoding} and Scarlett and Cevher~\cite{Sca17c}.\footnote{\added{See also \cite{lindstrom1964combinatory} for considerably earlier work on the case that there is no pre-specified sparsity level, so the algorithm must work for all $k \le n$.  The works \cite{el2019decoding,Sca17c} are studying a more general `pooled data' problem, of which quantitative group testing is a special case.}}  Despite this sharp information-theoretic understanding, we are unaware of any \emph{efficient} algorithms attaining order-optimal $O\big(\frac{k}{\log k}\big)$ scaling under Bernoulli testing.  In particular, even upon relaxing from exact recovery to approximate recovery, techniques based on approximate message passing use $O(k)$ tests \cite{el2018decoding,tan2023approximate}.  It was further shown in \cite{tan2025quantitative} that when spatial coupling is incorporated, this improves to $o(k)$, albeit with an unspecified precise rate.  On the other hand, \emph{deterministic} nonadaptive designs using $O\big(\frac{k}{\log k}\big)$ tests (when $k = \Theta(n)$) can be found in \cite{lindstrom1965combinatorial,cantor1966determination,wang2016data}, and while computational complexity was not their focus, more recent works \cite{bshouty2012coin,li2023non,tan2025quantitative} noted that they can be decoded efficiently.
    \item Under adaptive testing with $k = o(n)$, the information-theoretic lower bound is halved from $\frac{2k}{\log k}\log\frac{n}{k}$ to $\frac{k}{\log k}\log\frac{n}{k}$ (e.g., see \cite{bshouty4,feige2020quantitative}).  Intuitively, the difference is due to the fact that (i) for adaptive testing, outcomes in $\{0,1,\dotsc,k\}$ could (in principle) reveal up to $\log_2 (k+1) = (\log_2 k)(1+o(1))$ bits per test; (ii) under nonadaptive testing, the number of defectives in a given test concentrates within a window of width at most $O(\sqrt k)$, so the maximum number of bits learned is $\log_2\big( O\big(\sqrt{k}\big) \big)$, which simplifies to $\big(\frac{1}{2} \log_2 k\big)(1+o(1))$.  Bshouty \cite{bshouty4} gave a polynomial-time adaptive algorithm using $\big(\frac{Ck}{\log k}\log\frac{n}{k}\big)(1+o(1))$ tests (when $k = o(n)$), where the constant is proved to be in $[2,4)$ and reported to be $2$.\footnote{\added{More specifically, the constant is $2$ under `favourable rounding' of a suitable submatrix to the next power of two, and \cite[Note of Thm.~4]{bshouty4} states that this rounding issue can be circumvented, but we are unaware of any follow-up work showing this formally.}} It remains open as to whether the constant can be reduced to a smaller value in $[1,2)$, which would imply that adaptivity provably helps.  Numerical evidence supporting this possibility was given by Soleymani and Javidi \cite{soleymani2025learning}.  
    \item Noisy variants of quantitative group testing, under both the exact recovery and approximate recovery criteria, have been studied in \cite{bshouty2012coin,chen2017partial,Sca17c,tan2023approximate,li2023non,hahn2023efficient,tan2025quantitative}, among others.  Without going into detail, we note that at least in the case of exact recovery, the presence of noise typically leads to strictly higher scaling in the required number of tests.
\end{itemize}
}

\subsection{Tropical group testing}

\added{

Another model whose study has aligned with the techniques surveyed this monograph is the \emph{tropical group testing} model of Wang, Gabrys and Vardy \cite{wang2023tropical}.  In this model, items can have different levels of defectivity, and the outcome of a given test is also a (non-binary) defectivity level.  Various results in \cite{wang2023tropical,paligadu2023small}, some of which we outline below, reveal that tropical testing can succeed with fewer tests than standard binary group testing.

The model is as follows: There are $d$ levels of positive defectivity $\{1, 2, \dots, d\}$, from $1$ being the strongest level of defectivity to $d$ being the weakest, with nondefectivity now being denoted by $\infty$ rather than $0$. The output of noiseless tropical testing is the strongest defectivity level -- that is, the minimum associated value -- of any item present in the test.  When $d = 1$, this is equivalent to the standard binary testing model.  The term `tropical' is used due to similar use of the `minimum' operation in the subject of tropical geometry.

An equivalent model also appears in the book of Du and Hwang \cite[Ch.~6]{du-hwang}, along with a simple adaptive algorithm that repeatedly uses binary splitting techniques to identify the current `most defective' item.  However, the only relevant existing work cited in \cite{du-hwang} is that of Hwang and Xu \cite{hwang1987group}, who studied the significantly more specific setting in which one item is `more defective', one item is `less defective', and the remaining $n-2$ items are nondefective.

The tropical model is appropriate for settings where the defectivity of a group is determined by its most defective member -- one may think of the the total capacity of a series of pipes, whose bottleneck is the lowest-capacity pipe.  The tropical model is of particular interest for modelling PCR testing for Covid-19 or other virus-borne illnesses. In PCR testing, the viral concentration in a sample is so weak that it must undergo multiple cycles of amplification until it reaches a detectable concentration. The defectivity level can be then thought of as the number of cycles required to reach a threshold level: the strongest concentrations will require only a small number of cycles, while the weakest concentrations will require many cycles. (A noninfected sample will never reach the threshold no matter how many cycles it goes through, hence denoting it by $\infty$.)  It is often the case that viral concentrations within a collection of individuals differ by so many orders of magnitude that the number of cycles needed by a pooled sample is dictated by the number of cycles required for the most concentrated individual sample, thus leading to the tropical model. Of course, this is not exact; many weak samples could potentially combine to make a stronger pooled sample; see \cite{paligadu2023small} for further discussion of the tropical model as an approximation to PCR testing.

The output of tropical group testing is one of the $d$ positive levels $\{1, 2, \dots, d\}$ or a negative result $\infty$. This is more informative than the simple positive--negative outcome (previously denoted $\{0, 1\}$) of binary group testing, which suggests that the tropical model might require fewer tests than the binary model. On the other hand, in tropical group testing, existing works have aimed to not only recover which items are defective, but also recover the defectivity level of each such item; this will require more information than learning the defective set alone, suggesting that more tests could be needed.

Compared to binary group testing, or even quantitative group testing, the study of tropical group testing is in its relatively early stages.  Under the Bernoulli test design, Paligadu, Johnson, and Aldridge \cite{paligadu2023small} give several counterparts to the algorithms and results of \cite{aldridge-baldassini-johnson} that we surveyed in Chapter \ref{ch:algorithms}. We briefly outline some of the results of \cite{paligadu2023small}; throughout, we write $k_1,\dotsc,k_d$ for the number of items having each positive defectivity level, $k = \sum_{i=1}^d k_i$ for the total number of defectives, and $n$ for the total number of items.

First, a `tropical COMP' algorithm is defined as follows: Declare each item's defectivity level to be the weakest defectivity level (that is, the highest value) of any test it appears in.  This precisely recovers the usual COMP algorithm when the number of levels is $d = 1$.  Under the Bernoulli design, tropical COMP succeeds with the same asymptotic number of tests as the binary case, $(\ee k \ln n)(1+o(1))$. 

Second, \cite{paligadu2023small} defines and analyses a `tropical DD' algorithm. For tropical DD, the number of tests is somewhat more complicated, as it is written as the maximum of a number of terms (one per defectivity level). The number of tests can be significantly smaller than COMP, and is even seen to be asymptotically optimal in certain parameter regimes via an analogous argument to that of SSS (Section \ref{sec:sss}).  Tropical DD is seen to outperform both tropical COMP and classical DD in numerical experiments, and these experiments also suggest further improvements via a `tropical SCOMP' algorithm.

Finally, a simple counting argument leads to an analogue of the counting bound with $\log_2 {\binom{n}{k}}$ replaced by $\log_{d+1}{\binom{n}{\mathbf{k}}}$, where ${\binom{n}{\mathbf{k}}} = \frac{n!}{k_1!\dotsc k_d!  (n - k)!}$ is the multinomial coefficient.  When $k = \Theta(n^{\alpha})$ for $\alpha \in (0,1)$, this scales as $\Theta\big( \frac{k \log n}{\log (d+1)} \big)$, as opposed to $\Theta(k \log n)$ for the binary model.  The denominator of $\log(d+1)$ has a similar interpretation to that in the quantitative group testing model outlined above.

These results leave several interesting questions open for future work.  For example, further analogous results to those of the binary model could be sought in the form of information-theoretic achievability results, improvements via near-constant tests-per-item designs, improved converse bounds for arbitrary nonadaptive designs, and so on.}

\subsection{Other models}

\added{Beyond the quantitative and tropical models, we provide a partial list of other models that have been considered in the literature.}
\begin{itemize}
    \item The {\em semi-quantitative model} \cite{emad2014} lies in between the two extremes of the standard model and the quantitative model.  The model has a number of thresholds, and we get to observe the largest of those thresholds that the number of defective items in the test $|\{i \in \K \,:\, x_i = 1\}|$ is greater than or equal to. 
    See also \cite{emad2016code,li2023finding,cheraghchi2021semi} for follow-up work on this model.
    \item Various forms of {\em threshold group testing} output a $0$ if there are too few defectives, output a $1$ if there are sufficiently many defectives, and exhibit either random \cite{laarhoven-1} or adversarial \cite{damaschke2006threshold,cheraghchi2013improved} behaviour in between the two corresponding thresholds.  We presented a simple randomized version of this model in Example \ref{ex:threshold}.
    \item Other models have been considered with more than two types of items, with a prominent example being {\em group testing with inhibitors} introduced by Farach \etal\, \cite{farach} and discussed in \cite[Chapter 8]{du-hwang2}.  In the simplest case, each item is either defective, an inhibitor, or neither of the two, and the test is positive if and only if it contains at least one defective but no inhibitors.  In other words, inhibitors may `hide' the fact that the test contains one or more defective items. Some results concerning this model are provided, for example, in \cite{deBonis1998improved,ganesan2015}.
    \item Beyond the tropical group testing model outlined above, other models motivated by PCR and medical testing can be found in \cite{ghosh2021compressed,gabrys2020ac,aldridge2022pooled} and the references therein.
    \item \added{Various frameworks have been studied for understanding highly general models in a unified manner.  In particular, in \emph{complex group testing} \cite{torney1999sets,chen2008upper,chin2013non}, as well as an equivalent problem based on hypergraph learning \cite{angluin2008learning,abasi2018non}, the problem is described by a \emph{general set} of subsets of items whose simultaneous inclusion in a test leads to a positive outcome.  Another example is \emph{generalized group testing} \cite{cheng2022generalized}, in which the probability of a positive test is an arbitrary increasing function of the number of defectives in the test (see also \cite{morjaria2025density} for a related density-dependent model).}
\end{itemize}

While some of the models listed above can be studied under the information-theoretic framework considered in Chapter \ref{ch:achievability}, characterizing the number of tests still requires the non-trivial step of bounding mutual information terms, for example, as in Theorem \ref{thm:general_noise}.  This has been done for the quantitative group testing model in \cite{Sca17c}, and in certain random models for threshold group testing in \cite{laarhoven-1}.  

Despite this, several upper and lower bounds on the number of tests required in the above models have indeed been developed, often using rather different approaches compared to the standard setting.  \rev{Hence, it is of significant interest to further study what the information theory perspective can provide for these models, potentially building on the concepts and techniques surveyed in this monograph.}

\chapter{Conclusion and Open Problems} \label{ch:conclusion}

In this monograph, we have surveyed recent theoretical and algorithmic developments in group testing, with an emphasis on achievable rates \added{characterizing the asymptotic number of bits learned per test by a successful group testing strategy.  We have particularly focused on nonadaptive testing in the small-error regime with a number of defectives scaling as $k = \Theta(n^{\alpha})$.}

In the noiseless setting (Chapter \ref{ch:algorithms}), we presented the achievable rates of COMP and DD, considering both the Bernoulli design and the near-constant tests-per-item design.  We also observed that the SCOMP and linear programming (LP) algorithms to perform better experimentally while achieving rates at least as high as DD (which in turn exceed those of COMP).  In the noisy setting (Chapter \ref{ch:algorithms_noisy}), we presented noisy variants of COMP, DD, and LP, as well as two additional algorithms: separate decoding of items, which is convenient to analyse theoretically; and belief propagation, which performs very well experimentally but currently lacks a theoretical analysis.  

\added{In Chapter \ref{ch:achievability}, we surveyed the information-theoretic limits of group testing, again considering the Bernoulli design and the near-constant tests-per-item design.  In the noiseless setting, we concluded that the information-theoretic threshold for the near-constant tests-per-item design is in fact optimal among \emph{all} nonadaptive designs, and moreover, that this threshold can be attained in a computationally efficient manner using a spatially coupled design.  We also surveyed analogous developments in the noisy setting.}


In Chapter \ref{ch:other-topics}, we surveyed a wide range of important variants of the standard group testing problem, including approximate recovery, multi-stage adaptive algorithms, counting defectives, sublinear-time decoding, linear sparsity, non-uniform prior statistics, explicit constructions, constrained test designs, computational considerations, and other group testing models.

\medskip

\added{

We conclude the monograph by discussing some open (and formerly open) problems.  We first restate the problems posed in the first edition and indicate to what extent they have been resolved.  We then present a few further open problems that are new to this edition.
}

\section*{Open problems from the first edition}

In the following, we detail the open problems from the first edition, as well as indicating the extent to which they have been solved.

\begin{openprob} \label{openprob1} {\bf (Solved)}
\emph{What is the capacity of nonadaptive group testing in the sparse regime, where $k = \Theta(n^{\alpha})$ with $\alpha \in (0,1)$?}

\added{As outlined above and detailed in Chapter \ref{ch:achievability}, this problem is now settled due to the work of Coja-Oghlan \etal~\cite{coja2020optimal}: The optimal rate (the capacity) is that attained by the near-constant column weight design, $\olR_{\rm NCC} =  \min\big\{1,(\ln 2)\frac{1-\alpha}{\alpha}\big\}$.  Analogous results can also be found in \cite{bay2022optimal} for denser regimes, including ``mildly sublinear'' regimes such as $k = \Theta\big( \frac{n}{\log n} \big)$.}
\end{openprob}


\begin{openprob} {\bf (Remains open)}
\emph{What more can be said about finite-size group testing problems using information-theoretic methods?}

Results regarding the rate of group testing indicate how the number $T = T(n)$ of tests required behaves as $n \to \infty$. However, in practice, we might be interested in a fairly modest number of items, perhaps of the order $100$ to $10,000$.  In such cases, results concerning the asymptotic rate may be of limited value, particularly if $\log_2\binom nk/T$ only converges slowly to the maximum achievable rate.  A natural question is then what we say about the number of tests required with such `finite $n$'.
Information-theoretic approaches to `finite blocklength' results in channel coding were pioneered by Polyanskiy, Poor and Verd\'u \cite{ppv2010}, and similar ideas may be useful for better understanding group testing.

One existing work in this direction is \cite{johnsonconverse}, which built on the ideas of \cite{ppv2010} to develop converse results that generalize Theorem \ref{thm:bjaconverse}.  However,
in stark contrast with channel coding, counterparts for the {\em achievability} part appear to be unexplored.  \added{Some of the analyses surveyed in this monograph, such as those of COMP (Section \ref{sec:COMP}) and two-stage conservative testing (Section \ref{sec:conservative}), are sufficiently simple that analogous non-asymptotic bounds could be inferred without too much difficulty.  However, other analyses, notably including information-theoretic achievability (Section \ref{sec:pf_ach}), are heavily reliant on asymptotics, often including terms that decay to zero very slowly.}
\end{openprob}

\begin{openprob} {\bf (Solved)}
\emph{Find a practical decoding algorithm (and a nonadaptive test design) that achieves a rate higher than $\ln 2 \approx 0.693$, or prove that no such algorithm exists.}

\added{We stated this problem based on the observation that prominent practical algorithms such as COMP and DD achieve rates of at most $\ln 2$ for all $\theta \in (0,1)$.

If we take `practical' to mean `polynomial-time', then the above-mentioned results of \cite{coja2020optimal} provide much more than what this open problem asked, giving a polynomial-time method based on spatial coupling that attains the \emph{optimal rate} for all $\alpha \in (0,1)$.  However, this spatial coupling technique has not been tested experimentally to the best of our knowledge, so we pose Open Problem \ref{op:spatial_practical} below addressing its practical performance.}

\end{openprob}
       

\begin{openprob} {\bf (Solved)}
\emph{Establish improved achievable rates and converse results for noisy nonadaptive group testing.}

\added{This problem was addressed in detail concurrently by Chen and Scarlett \cite{chen2024exact} and Coja-Oghlan \etal~\cite{coja2024noisy}, as we outlined in Section \ref{sec:high_k_ach}.  An important remaining open problem regarding converse bounds for general designs is given as Open Problem \ref{op:noisy_conv} below.  We also present Open Problem \ref{op:noisy_ada} regarding \emph{adaptive} testing in the sublinear sparsity regime (see Section \ref{sec:lin_regime_noisy} for the linear regime).}

\end{openprob}




\begin{openprob} This problem has two parts:

    \emph{(a) {\bf (Solved)} What are the fundamental limits of noiseless group testing in the linear regime with exact recovery and a constant error probability in $(0,1)$?}
    
    \emph{(b) {\bf (Remains open)} What are the fundamental limits of noiseless group testing in the linear regime with approximate recovery?}
    
    \added{In Section \ref{sec:linear}, we discussed the linear regime $k \sim \beta n$, and showed that in the nonadaptive setting with exact recovery and the small-error criterion (i.e., asymptotically vanishing error probability), it is optimal to test every item individually.  However, these results do not rule out the possibility of using fewer tests when the allowed error probability is a fixed constant, e.g., $0.1$ or $0.5$.  As discussed in Section \ref{sec:lin_nonada}, it is now known from \cite{bay2022optimal} that allowing some error probability does not help much.  Specifically, for arbitrarily small $\eta > 0$, at least $n(1-\eta)$ tests are needed to avoid having the error probability approach one as $n \to \infty$.  
    
    On the other hand, Tan, Tan, and Scarlett~\cite{Tan2022} showed that significantly fewer tests suffice for high-probability \emph{approximate recovery}, particularly when $\beta$ is small in the relation $k \sim \beta n$.  However, there still remains significant room for further results, including both converse results and improved achievability results.  Hence, we label this problem as still being largely open.}
    
    
    
\end{openprob}

\begin{openprob} {\bf (Solved)}
\emph{Find a nonadaptive group testing algorithm that succeeds with $O(k \log n)$ tests and has $O(k \log n)$ decoding time.}

\added{This goal was concurrently achieved by Price and Scarlett \cite{price2023fast} and Cheraghchi and Nakos \cite{Cheraghchi2020}, as we summarized in Section \ref{sec:sublinear_pos_rate}.}


\end{openprob}

\begin{openprob} This problem has two parts:

\emph{
(a) {\bf (Partially solved)} Establish precise achievable rates for group testing algorithms with sublinear decoding time.}

\emph{(b) {\bf (Remains open)} Establish precise achievable rates for group testing algorithms with an explicit construction of the test matrix.}

\added{Bounds on the number of tests for group testing with sublinear-time decoding (Section \ref{sec:sublinear}) and/or explicit test designs (Section \ref{sec:explicit}) are usually given using $O(\cdot)$ notation, and thus do not provide explicit achievable rates.  Naturally, understanding the constant factors in detail would significantly sharpen our understanding of these settings. 

This goal was addressed for sublinear-time decoding by Wang and Guruswami \cite{Wang2023}, who obtained constants arbitrarily close to those of the COMP and DD algorithms, and left open the problem of obtaining further improvements.  When it comes to explicit designs, there has been little or no progress in this direction, to the best of our knowledge.  In both settings, further progress may be based on trying to further refine the analysis of existing methods, or it may be that that new algorithmic ideas are needed in order to obtain the best rates possible.}
\end{openprob}

\begin{openprob} {\bf (Remains open)} 
\emph{Prove the Hu--Hwang--Wang conjecture, that individual testing is optimal for adaptive combinatorial zero-error group testing when $k \geq n/3$.}

Adaptive testing in the linear regime where $k \asym \beta n$ was discussed in Section \ref{lin-adap}. In particular, for zero-error testing (with the combinatorial prior), we saw that one could improve on testing each item individually when $\beta < \frac13$. Hu, Hwang and Wang \cite{hu} conjecture that this result is tight, in the sense that one requires $T \geq n-1$ when $\beta \geq \frac13$. (Recall that, since $k$ is known, we need not test the final item, hence we need only $n-1$ tests for `individual testing'.) The best known result is that individual testing is optimal for $\beta \geq 1 - \log_3 2 \approx 0.369$ \cite{riccio}.
\end{openprob}


\begin{openprob} {\bf (Remains open)}
\emph{What does the information theory perspective have to offer other non-standard group testing models, and more general structured signal recovery problems?}

This question was intentionally open-ended, and we emphasize that information theory has indeed already played a major role in extensive problems concerning structured signal recovery and high-dimensional statistics.  \added{Moreover, since the first edition, there has been substantial progress in various group testing models including quantitative group testing \cite{hahn2022near, hahn2023efficient, tan2023approximate, soleymani2024non} and tropical group testing \cite{wang2023tropical,paligadu2023small}, among others.  Nevertheless, we expect that there is still substantial room for further work, so we mark this problem as generally still remaining open.}


\end{openprob}

\section*{Newly added open problems}

\added{
We conclude this chapter by presenting three open problems that are new to the second edition.

\begin{openprob} \label{op:spatial_practical}
    \emph{What is the experimental performance of the asymptotically optimal spatially coupled group testing design, and to what extent can this design be beneficial in applications?}

    The spatially coupled design outlined in Section \ref{sec:spatial} provided a major theoretical breakthrough in attaining the optimal rates of noiseless group testing while permitting polynomial-time decoding. (See also Section \ref{sec:high_k_ach} regarding the noisy setting.)  However, to our knowledge, such designs have not been tested experimentally at practical values of the number of items $n$.  We expect that some practical adjustments might be required, e.g., avoiding fixed parameter choices such as $\ell = \sqrt{\log n}$ and $s = \log \log n$ and instead treating these as tunable parameters.  Nevertheless, it would be interest to determine whether the design can be preferable to the regular near-constant column weight design (or other designs that perform well in practice) after such adjustments for practical problem sizes.
\end{openprob}
}

\added{
\begin{openprob} \label{op:noisy_conv}
\emph{For noisy group testing, is the rate achieved by the near-constant column weight design equal to the optimal rate among all nonadaptive designs?} 

In the noiseless setting, the corresponding question is known to have an affirmative answer (see Section \ref{sec:optimal_gt}).  Thus, it is fairly natural to expect the same in the noisy setting.  On the other hand, the noisy setting has a significantly more complicated achievable rate expression with distinct error events arising, and new phenomena are observed such as the choice $\nu = \ln 2$ (in the design parameter $L = \nu T/k$) no longer necessarily being optimal.  Thus, it remains entirely conceivable that other designs can do better.

Both a positive and negative answer to the asymptotic optimality of the near-constant column design would have significant implications.  In particular, a positive answer would imply that the spatially coupled design from \cite{coja2024noisy} is also asymptotically optimal, meaning that we have a polynomial-time strategy that attains the optimal rate.  On the other hand, a negative answer would imply that noiseless and noisy group testing have deeper fundamental differences than the current results suggest.
\end{openprob}
}

\added{
\begin{openprob} \label{op:noisy_ada}
    \emph{What are the optimal rates of noisy adaptive group testing in the sublinear sparsity regime?} 

    This question is related to two directions in noisy settings that have already been solved --- exact rates are known for the two most widely-used \emph{nonadaptive} test designs in the sublinear sparsity regime (see Section \ref{sec:high_k_ach}), and optimal thresholds for adaptive testing are known in the \emph{linear sparsity regime} (see Section \ref{sec:lin_regime_noisy}).  However, for adaptive testing in the sublinear regime, non-minor gaps remain between the best known upper and lower bounds on the optimal rate (see Section \ref{sec:two-stage}).  It is unclear to what extent the techniques from the linear regime are applicable; for instance, the algorithm in \cite{hintze2024noisy} assigns each item some individual tests, but in the sublinear regime we do not have enough tests to do so.
\end{openprob}
}

\addcontentsline{toc}{chapter}{References}
\bibliographystyle{habbrv}
\bibliography{bibliography/bibliography}

\begin{thebibliography}{100}
\expandafter\ifx\csname url\endcsname\relax
  \def\url#1{\texttt{#1}}\fi
\expandafter\ifx\csname doi\endcsname\relax
  \def\doi#1{\burlalt{\textsc{doi}:
  \texttt{\detokenize{#1}}}{http://dx.doi.org/#1}}\fi
\expandafter\ifx\csname urlprefix\endcsname\relax\def\urlprefix{\textsc{url:
  }}\fi
\expandafter\ifx\csname href\endcsname\relax
  \def\href#1#2{#2}\fi
\expandafter\ifx\csname burlalt\endcsname\relax
  \def\burlalt#1#2{\href{#2}{#1}}\fi

\bibitem{abasi18learning}
H.~Abasi and N.~H. Bshouty.
\newblock On learning graphs with edge-detecting queries.
\newblock In {\em International Conference on Algorithmic Learning Theory
  (ALT)}, pages 3--30, 2019.
\newblock \urlprefix\url{https://proceedings.mlr.press/v98/abasi19a}.

\bibitem{abasi2018non}
H.~Abasi, N.~H. Bshouty, and H.~Mazzawi.
\newblock Non-adaptive learning of a hidden hypergraph.
\newblock {\em Theoretical Computer Science}, 716:15--27, 2018.
\newblock \doi{10.1016/j.tcs.2017.11.019}.

\bibitem{agarwal-jaggi-mazumdar}
A.~Agarwal, S.~Jaggi, and A.~Mazumdar.
\newblock Novel impossibility results for group-testing.
\newblock In {\em IEEE International Symposium on Information Theory (ISIT)},
  pages 2579--2583, 2018.
\newblock \doi{10.1109/ISIT.2018.8437471}.

\bibitem{agarwal2020group}
A.~Agarwal, O.~Milenkovic, S.~Pattabiraman, and J.~Ribeiro.
\newblock Group testing with runlength constraints for topological molecular
  storage.
\newblock In {\em IEEE International Symposium on Information Theory (ISIT)},
  pages 132--137. IEEE, 2020.
\newblock \doi{10.1109/ISIT44484.2020.9174502}.

\bibitem{ahn2021adaptive}
S.~Ahn, W.-N. Chen, and A.~Özgür.
\newblock Adaptive group testing on networks with community structure: The
  stochastic block model.
\newblock {\em IEEE Transactions on Information Theory}, 69(7):4758--4776,
  2023.
\newblock \doi{10.1109/TIT.2023.3247520}.

\bibitem{Aig86}
M.~Aigner.
\newblock Search problems on graphs.
\newblock {\em Discrete Applied Mathematics}, 14(3):215--230, 1986.
\newblock \doi{10.1016/0166-218X(86)90026-0}.

\bibitem{aigner1988searching}
M.~Aigner and E.~Triesch.
\newblock Searching for an edge in a graph.
\newblock {\em Journal of Graph Theory}, 12(1):45--57, 1988.
\newblock \doi{10.1002/jgt.3190120106}.

\bibitem{aksoylar}
C.~Aksoylar, G.~K. Atia, and V.~Saligrama.
\newblock Sparse signal processing with linear and nonlinear observations: {A}
  unified {S}hannon-theoretic approach.
\newblock {\em IEEE Transactions on Information Theory}, 63(2):749--776, 2017.
\newblock \doi{10.1109/TIT.2016.2605122}.

\bibitem{aldridge-thesis}
M.~Aldridge.
\newblock {\em Interference mitigation in large random wireless networks}.
\newblock PhD thesis, University of Bristol, 2011, arXiv:
  \burlalt{\texttt{1109.1255}}{http://arxiv.org/abs/1109.1255}.

\bibitem{aldridge}
M.~Aldridge.
\newblock The capacity of {B}ernoulli nonadaptive group testing.
\newblock {\em IEEE Transactions on Information Theory}, 63(11):7142--4148,
  2017.
\newblock \doi{10.1109/TIT.2017.2748564}.

\bibitem{aldridge-2}
M.~Aldridge.
\newblock On the optimality of some group testing algorithms.
\newblock In {\em IEEE International Symposium on Information Theory (ISIT)},
  pages 3085--3089, 2017.
\newblock \doi{10.1109/ISIT.2017.8007097}.

\bibitem{aldridge-linear}
M.~{Aldridge}.
\newblock Individual testing is optimal for nonadaptive group testing in the
  linear regime.
\newblock {\em IEEE Transactions on Information Theory}, 65(4):2058--2061,
  2019.
\newblock \doi{10.1109/TIT.2018.2873136}.

\bibitem{aldridge-adaptive}
M.~Aldridge.
\newblock Rates of adaptive group testing in the linear regime.
\newblock In {\em IEEE International Symposium on Information Theory (ISIT)},
  pages 236--240, 2019.
\newblock \doi{10.1109/ISIT.2019.8849712}.

\bibitem{aldridge2020conservative}
M.~Aldridge.
\newblock Conservative two-stage group testing in the linear regime, 2020,
  arXiv: \burlalt{\texttt{2005.06617}}{http://arxiv.org/abs/2005.06617}.

\bibitem{aldridge-baldassini-gunderson}
M.~Aldridge, L.~Baldassini, and K.~Gunderson.
\newblock Almost separable matrices.
\newblock {\em Journal of Combinatorial Optimization}, 33(1):215--236, 2017.
\newblock \doi{10.1007/s10878-015-9951-1}.

\bibitem{aldridge-baldassini-johnson}
M.~Aldridge, L.~Baldassini, and O.~T. Johnson.
\newblock Group testing algorithms: {B}ounds and simulations.
\newblock {\em IEEE Transactions on Information Theory}, 60(6):3671--3687,
  2014.
\newblock \doi{10.1109/TIT.2014.2314472}.

\bibitem{aldridge2022pooled}
M.~Aldridge and D.~Ellis.
\newblock Pooled testing and its applications in the {COVID-19} pandemic.
\newblock In M.~d.~C. Boado-Penas, J.~Eisenberg, and
  {\c{S}}.~{\c{S}}ahin‬‬‬, editors, {\em Pandemics: Insurance and Social
  Protection}, pages 217--249. Springer, 2022.
\newblock \doi{10.1007/978-3-030-78334-1_11}.
\newblock Extended version, arXiv:
  \href{https://arxiv.org/abs/2105.08845}{2105.08845}.

\bibitem{allemann}
A.~Allemann.
\newblock An efficient algorithm for combinatorial group testing.
\newblock In H.~Aydinian, F.~Cicalese, and C.~Deppe, editors, {\em Information
  Theory, Combinatorics, and Search Theory: {I}n Memory of Rudolf Ahlswede},
  pages 569--596. Springer, 2013.
\newblock \doi{10.1007/978-3-642-36899-8_29}.

\bibitem{ambainis}
A.~Ambainis, A.~Belovs, O.~Regev, and R.~De~Wolf.
\newblock Efficient quantum algorithms for (gapped) group testing and junta
  testing.
\newblock In {\em ACM-SIAM Symposium On Discrete Algorithms (SODA)}, pages
  903--922, 2016.
\newblock \doi{10.1137/1.9781611974331.ch65}.

\bibitem{angluin2008learning}
D.~Angluin and J.~Chen.
\newblock Learning a hidden graph using {$O(\log n)$} queries per edge.
\newblock {\em Journal of Computer and System Sciences}, 74(4):546--556, 2008.
\newblock \doi{10.1016/j.jcss.2007.06.006}.

\bibitem{arasli2023group}
B.~Arasli and S.~Ulukus.
\newblock Group testing with a graph infection spread model.
\newblock {\em Information}, 14(1):48, 2023.
\newblock \doi{10.3390/info14010048}.

\bibitem{ash}
R.~B. Ash.
\newblock {\em Information Theory}.
\newblock Dover Publications Inc., New York, 1990.

\bibitem{atia2}
G.~K. Atia, S.~Aeron, E.~Ermis, and V.~Saligrama.
\newblock On throughput maximization and interference avoidance in cognitive
  radios.
\newblock In {\em IEEE Consumer Communications and Networking Conference
  (CCNC)}, pages 963--967, 2008.
\newblock \doi{10.1109/ccnc08.2007.222}.

\bibitem{atia-saligrama}
G.~K. Atia and V.~Saligrama.
\newblock Boolean compressed sensing and noisy group testing.
\newblock {\em IEEE Transactions on Information Theory}, 58(3):1880--1901,
  2012.
\newblock \doi{10.1109/TIT.2011.2178156}.
\newblock See also \cite{atia-corr}.

\bibitem{atia-corr}
G.~K. {Atia}, V.~{Saligrama}, and C.~{Aksoylar}.
\newblock Correction to `{B}oolean compressed sensing and noisy group testing'.
\newblock {\em IEEE Transactions on Information Theory}, 61(3):1507--1507,
  2015.
\newblock \doi{10.1109/TIT.2015.2392116}.

\bibitem{atici}
A.~At{\i}c{\i} and R.~A. Servedio.
\newblock Quantum algorithms for learning and testing juntas.
\newblock {\em Quantum Information Processing}, 6(5):323--348, 2007.
\newblock \doi{10.1007/s11128-007-0061-6}.

\bibitem{attia2021hetero}
M.~A. Attia, W.-T. Chang, and R.~Tandon.
\newblock Heterogeneity aware two-stage group testing.
\newblock {\em IEEE Transactions on Signal Processing}, 69:3977--3990, 2021.
\newblock \doi{10.1109/TSP.2021.3093785}.

\bibitem{austhof2025non}
B.~Austhof, L.~Reyzin, and E.~Tani.
\newblock Non-adaptive learning of random hypergraphs with queries, 2025,
  arXiv: \burlalt{\texttt{2501.12771}}{http://arxiv.org/abs/2501.12771}.

\bibitem{baldassini-johnson-aldridge}
L.~Baldassini, O.~T. Johnson, and M.~Aldridge.
\newblock The capacity of adaptive group testing.
\newblock In {\em IEEE International Symposium on Information Theory (ISIT)},
  pages 2676--2680, 2013.
\newblock \doi{10.1109/ISIT.2013.6620712}.

\bibitem{balding}
D.~J. Balding, W.~J. Bruno, D.~C. Torney, and E.~Knill.
\newblock A comparative survey of non-adaptive pooling designs.
\newblock In T.~Speed and M.~S. Waterman, editors, {\em Genetic Mapping and DNA
  Sequencing}, pages 133--154. Springer, 1996.
\newblock \doi{10.1007/978-1-4612-0751-1_8}.

\bibitem{bay2022optimal}
W.~H. Bay, E.~Price, and J.~Scarlett.
\newblock Optimal non-adaptive probabilistic group testing in general sparsity
  regimes.
\newblock {\em Information and Inference: A Journal of the IMA},
  11(3):1037--1053, 2022.
\newblock \doi{10.1093/imaiai/iaab020}.

\bibitem{berger2002}
T.~Berger and V.~I. Levenshtein.
\newblock Asymptotic efficiency of two-stage disjunctive testing.
\newblock {\em IEEE Transactions on Information Theory}, 48(7):1741--1749,
  2002.
\newblock \doi{10.1109/TIT.2002.1013122}.

\bibitem{berger2}
T.~Berger, N.~Mehravari, D.~Towsley, and J.~Wolf.
\newblock Random multiple-access communication and group testing.
\newblock {\em IEEE Transactions on Communications}, 32(7):769--779, 1984.
\newblock \doi{10.1109/TCOM.1984.1096146}.

\bibitem{black2024clustering}
H.~Black, E.~Lee, A.~Mazumdar, and B.~Saha.
\newblock Clustering with non-adaptive subset queries.
\newblock In {\em Advances in Neural Information Processing Systems},
  volume~37, pages 56446--56478, 2024.
\newblock \urlprefix\url{https://dl.acm.org/doi/10.5555/3737916.3739714}.

\bibitem{blais}
E.~Blais.
\newblock Testing juntas: {A} brief survey.
\newblock In O.~Goldreich, editor, {\em Property Testing: {C}urrent research
  and surveys}, pages 32--40. Springer, 2010.
\newblock \doi{10.1007/978-3-642-16367-8_4}.

\bibitem{bloom}
B.~H. Bloom.
\newblock Space/time trade-offs in hash coding with allowable errors.
\newblock {\em Communications of the ACM}, 13(7):422--426, 1970.
\newblock \doi{10.1145/362686.362692}.

\bibitem{bondorf}
S.~Bondorf, B.~Chen, J.~Scarlett, H.~Yu, and Y.~Zhao.
\newblock Sublinear-time non-adaptive group testing with {$O(k \log n)$} tests
  via bit-mixing coding, 2020.
\newblock \doi{10.1109/TIT.2020.3046113}.

\bibitem{brennan2021stat}
M.~S. Brennan, G.~Bresler, S.~Hopkins, J.~Li, and T.~Schramm.
\newblock Statistical query algorithms and low degree tests are almost
  equivalent.
\newblock In {\em Conference on Learning Theory}, pages 774--774, 2021.
\newblock \urlprefix\url{https://proceedings.mlr.press/v134/brennan21a}.

\bibitem{broder-gupta}
A.~Z. Broder and R.~Kumar.
\newblock A note on double pooling tests, 2020, arXiv:
  \burlalt{\texttt{2004.01684}}{http://arxiv.org/abs/2004.01684}.

\bibitem{bshouty4}
N.~H. Bshouty.
\newblock Optimal algorithms for the coin weighing problem with a spring scale.
\newblock In {\em Conference on Learning Theory}, 2009.
\newblock
  \urlprefix\url{https://www.learningtheory.org/colt2009/papers/004.pdf}.

\bibitem{bshouty2012coin}
N.~H. Bshouty.
\newblock On the coin weighing problem with the presence of noise.
\newblock In {\em International Workshop on Approximation Algorithms for
  Combinatorial Optimization}, pages 471--482. Springer, 2012.
\newblock \doi{10.1007/978-3-642-32512-0_40}.

\bibitem{bshouty2019lower}
N.~H. Bshouty.
\newblock Lower bound for non-adaptive estimation of the number of defective
  items.
\newblock In {\em International Symposium on Algorithms and Computation
  (ISAAC)}. Schloss-Dagstuhl-Leibniz Zentrum f{\"u}r Informatik, 2019.
\newblock \doi{10.4230/LIPIcs.ISAAC.2019.2}.

\bibitem{bshouty2023improved}
N.~H. Bshouty.
\newblock Improved lower bound for estimating the number of defective items.
\newblock In {\em International Conference on Combinatorial Optimization and
  Applications}, pages 303--315. Springer, 2023.
\newblock \doi{10.1007/978-3-031-49611-0_22}.

\bibitem{bshouty}
N.~H. Bshouty, V.~E. Bshouty-Hurani, G.~Haddad, T.~Hashem, F.~Khoury, and
  O.~Sharafy.
\newblock Adaptive group testing algorithms to estimate the number of
  defectives.
\newblock In {\em International Conference on Algorithmic Learning Theory
  (ALT)}, pages 93--110, 2018.
\newblock \urlprefix\url{http://proceedings.mlr.press/v83/bshouty18a.html}.

\bibitem{bshouty2}
N.~H. Bshouty and A.~Costa.
\newblock Exact learning of juntas from membership queries.
\newblock {\em Theoretical Computer Science}, 742:82--97, 2018.
\newblock \doi{10.1016/j.tcs.2017.12.032}.
\newblock Algorithmic Learning Theory.

\bibitem{nyt}
Q.~Bui, S.~Kliff, and M.~Sanger-Katz.
\newblock How to test more people for coronavirus without actually needing more
  tests.
\newblock {\em The New York Times}, 2020.
\newblock
  \urlprefix\url{https://www.nytimes.com/interactive/2020/07/27/upshot/coronavirus-pooled-testing.html}.

\bibitem{bui2022group}
T.~V. Bui, Y.~M. Chee, J.~Scarlett, and V.~K. Vu.
\newblock Group testing with blocks of positives.
\newblock In {\em IEEE International Symposium on Information Theory (ISIT)},
  pages 1082--1087. IEEE, 2022.
\newblock \doi{10.1109/ISIT50566.2022.9834611}.

\bibitem{busschbach}
P.~Busschbach.
\newblock Constructive methods to solve problems of $s$-surjectivity, conflict
  resolution, coding in defective memories, 1984.
\newblock Rapport Interne ENST 84 D005.

\bibitem{cai-etal2}
S.~Cai, M.~Jahangoshahi, M.~Bakshi, and S.~Jaggi.
\newblock Efficient algorithms for noisy group testing.
\newblock {\em IEEE Transactions on Information Theory}, 63(4):2113--2136,
  2017.
\newblock \doi{10.1109/TIT.2017.2659619}.

\bibitem{cantor1966determination}
D.~G. Cantor and W.~Mills.
\newblock Determination of a subset from certain combinatorial properties.
\newblock {\em Canadian Journal of Mathematics}, 18:42--48, 1966.
\newblock \doi{10.4153/CJM-1966-007-2}.

\bibitem{cao2023gamp}
S.-J. Cao, R.~Goenka, C.-W. Wong, A.~Rajwade, and D.~Baron.
\newblock Group testing with side information via generalized approximate
  message passing.
\newblock {\em IEEE Transactions on Signal Processing}, 71:2366--2375, 2023.
\newblock \doi{10.1109/TSP.2023.3287671}.

\bibitem{chan-etal-1}
C.~L. Chan, P.~H. Che, S.~Jaggi, and V.~Saligrama.
\newblock Non-adaptive probabilistic group testing with noisy measurements:
  {N}ear-optimal bounds with efficient algorithms.
\newblock In {\em Annual Allerton Conference on Communication, Control, and
  Computing}, pages 1832--1839, 2011.
\newblock \doi{10.1109/ALLERTON.2011.6120391}.

\bibitem{chan-etal-3}
C.~L. Chan, S.~Jaggi, V.~Saligrama, and S.~Agnihotri.
\newblock Non-adaptive group testing: {E}xplicit bounds and novel algorithms.
\newblock {\em IEEE Transactions on Information Theory}, 60(5):3019--3035,
  2014.
\newblock \doi{10.1109/TIT.2014.2310477}.

\bibitem{chen9}
C.~L. Chen and W.~H. Swallow.
\newblock Using group testing to estimate a proportion, and to test the
  binomial model.
\newblock {\em Biometrics}, 46(4):1035--1046, 1990.
\newblock \doi{10.2307/2532446}.

\bibitem{chen2008upper}
H.-B. Chen, H.-L. Fu, and F.~K. Hwang.
\newblock An upper bound of the number of tests in pooling designs for the
  error-tolerant complex model.
\newblock {\em Optimization Letters}, 2:425--431, 2008.
\newblock \doi{10.1007/s11590-007-0070-5}.

\bibitem{chen-hwang}
H.-B. Chen and F.~K. Hwang.
\newblock Exploring the missing link among $d$-separable, $\bar{d}$-separable
  and $d$-disjunct matrices.
\newblock {\em Discrete Applied Mathematics}, 155(5):662--664, 2007.
\newblock \doi{10.1016/j.dam.2006.10.009}.

\bibitem{chen10}
H.-B. Chen and F.~K. Hwang.
\newblock A survey on nonadaptive group testing algorithms through the angle of
  decoding.
\newblock {\em Journal of Combinatorial Optimization}, 15(1):49--59, 2008.
\newblock \doi{10.1007/s10878-007-9083-3}.

\bibitem{chen2024exact}
J.~Chen and J.~Scarlett.
\newblock Exact thresholds for noisy non-adaptive group testing.
\newblock In {\em ACM-SIAM Symposium on Discrete Algorithms (SODA)}, 2025.
\newblock \doi{10.1137/1.9781611978322.159}.

\bibitem{chen2017partial}
W.-N. Chen and I.-H. Wang.
\newblock Partial data extraction via noisy histogram queries: Information
  theoretic bounds.
\newblock In {\em IEEE International Symposium on Information Theory (ISIT)},
  pages 2488--2492, 2017.
\newblock \doi{10.1109/ISIT.2017.8006977}.

\bibitem{cheng2022generalized}
X.~Cheng, S.~Jaggi, and Q.~Zhou.
\newblock Generalized group testing.
\newblock In {\em International Conference on Artificial Intelligence and
  Statistics (AISTATS)}, pages 10777--10835, 2022.
\newblock \urlprefix\url{https://proceedings.mlr.press/v151/cheng22a.html}.

\bibitem{cheng}
Y.~Cheng.
\newblock An efficient randomized group testing procedure to determine the
  number of defectives.
\newblock {\em Operations Research Letters}, 39(5):352--354, 2011.
\newblock \doi{10.1016/j.orl.2011.07.001}.

\bibitem{cheraghchi2009noise}
M.~Cheraghchi.
\newblock Noise-resilient group testing: {L}imitations and constructions.
\newblock In {\em International Symposium on Fundamentals of Computation
  Theory}, pages 62--73, 2009.
\newblock \doi{10.1007/978-3-642-03409-1_7}.

\bibitem{cheraghchi2010derandomization}
M.~Cheraghchi.
\newblock Derandomization and group testing.
\newblock In {\em Allerton Conference on Communication, Control, and
  Computing}, pages 991--997, 2010.
\newblock \doi{10.1109/ALLERTON.2010.5707017}.

\bibitem{cheraghchi2013improved}
M.~Cheraghchi.
\newblock Improved constructions for non-adaptive threshold group testing.
\newblock {\em Algorithmica}, 67(3):384--417, 2013.
\newblock \doi{10.1007/s00453-013-9754-7}.

\bibitem{cheraghchi2021semi}
M.~Cheraghchi, R.~Gabrys, and O.~Milenkovic.
\newblock Semiquantitative group testing in at most two rounds.
\newblock In {\em IEEE International Symposium on Information Theory (ISIT)},
  pages 1973--1978, 2021.
\newblock \doi{10.1109/ISIT45174.2021.9518270}.

\bibitem{cheraghchi-etal}
M.~Cheraghchi, A.~Karbasi, S.~Mohajer, and V.~Saligrama.
\newblock Graph-constrained group testing.
\newblock {\em IEEE Transactions on Information Theory}, 58(1):248--262, 2012.
\newblock \doi{10.1109/TIT.2011.2169535}.

\bibitem{Cheraghchi2020}
M.~Cheraghchi and V.~Nakos.
\newblock Combinatorial group testing and sparse recovery schemes with
  near-optimal decoding time.
\newblock In {\em IEEE Annual Symposium on Foundations of Computer Science
  (FOCS)}, pages 1203--1213, 2020.
\newblock \doi{10.1109/FOCS46700.2020.00115}.

\bibitem{chin2013non}
F.~Y. Chin, H.~C. Leung, and S.-M. Yiu.
\newblock Non-adaptive complex group testing with multiple positive sets.
\newblock {\em Theoretical Computer Science}, 505:11--18, 2013.
\newblock \doi{10.1016/j.tcs.2013.04.011}.

\bibitem{chvatal}
V.~Chvatal.
\newblock A greedy heuristic for the set-covering problem.
\newblock {\em Mathematics of Operations Research}, 4(3):233--235, 1979.
\newblock \doi{10.1287/moor.4.3.233}.

\bibitem{ciampiconi2020maxsat}
L.~Ciampiconi, B.~Ghosh, J.~Scarlett, and K.~S. Meel.
\newblock A {MaxSAT}-based framework for group testing.
\newblock In {\em AAAI Conference on Artificial Intelligence}, pages
  10144--10152, 2020.
\newblock \doi{10.1609/aaai.v34i06.6574}.

\bibitem{cicalese}
F.~Cicalese, P.~Damaschke, and U.~Vaccaro.
\newblock Optimal group testing algorithms with interval queries and their
  application to splice site detection.
\newblock {\em International Journal of Bioinformatics Research and
  Applications}, 1(4):363--388, 2005.
\newblock \doi{10.1504/IJBRA.2005.008441}.

\bibitem{clifford}
R.~Clifford, K.~Efremenko, E.~Porat, and A.~Rothschild.
\newblock Pattern matching with don't cares and few errors.
\newblock {\em Journal of Computer and System Sciences}, 76(2):115--124, 2010.
\newblock \doi{10.1016/j.jcss.2009.06.002}.

\bibitem{cohen2020efficient}
A.~Cohen, A.~Cohen, and O.~Gurewitz.
\newblock Efficient data collection over multiple access wireless sensors
  network.
\newblock {\em IEEE/ACM Transactions on Networking}, 28(2):491--504, 2020.
\newblock \doi{10.1109/TNET.2020.2964764}.

\bibitem{cohen2021serial}
A.~Cohen, N.~Shlezinger, S.~Salamatian, Y.~C. Eldar, and M.~M{\'e}dard.
\newblock Serial quantization for sparse time sequences.
\newblock {\em IEEE Transactions on Signal Processing}, 69:3299--3314, 2021.
\newblock \doi{10.1109/TSP.2021.3083985}.

\bibitem{coja}
A.~Coja-Oghlan, O.~Gebhard, M.~Hahn-Klimroth, and P.~Loick.
\newblock Information-theoretic and algorithmic thresholds for group testing.
\newblock {\em IEEE Transactions on Information Theory}, 66(12):7911--7928,
  2020.
\newblock \urlprefix\url{10.1109/TIT.2020.3023377}.

\bibitem{coja2020optimal}
A.~Coja-Oghlan, O.~Gebhard, M.~Hahn-Klimroth, and P.~Loick.
\newblock Optimal group testing.
\newblock In {\em Conference on Learning Theory}, pages 1374--1388, 2020.
\newblock
  \urlprefix\url{https://proceedings.mlr.press/v125/coja-oghlan20a.html}.

\bibitem{coja2022statistical}
A.~Coja-Oghlan, O.~Gebhard, M.~Hahn-Klimroth, A.~S. Wein, and I.~Zadik.
\newblock Statistical and computational phase transitions in group testing.
\newblock In {\em Conference on Learning Theory}, pages 4764--4781, 2022.
\newblock \urlprefix\url{https://proceedings.mlr.press/v178/coja-oghlan22a}.

\bibitem{coja2024noisy}
A.~Coja-Oghlan, M.~Hahn-Klimroth, L.~Hintze, D.~Kaaser, L.~Krieg, M.~Rolvien,
  and O.~Scheftelowitsch.
\newblock Noisy group testing via spatial coupling, 2025.
\newblock \doi{10.1017/S0963548324000336}.

\bibitem{coja2022efficient}
A.~Coja-Oghlan, M.~Hahn-Klimroth, P.~Loick, and M.~Penschuck.
\newblock Efficient and accurate group testing via belief propagation: An
  empirical study.
\newblock In {\em International Symposium on Experimental Algorithms}.
  Schloss-Dagstuhl-Leibniz Zentrum f{\"u}r Informatik, 2022.
\newblock \doi{10.4230/LIPIcs.SEA.2022.8}.

\bibitem{colbourn1999group}
C.~J. Colbourn.
\newblock Group testing for consecutive positives.
\newblock {\em Annals of Combinatorics}, 3(1):37--41, 1999.
\newblock \doi{10.1007/BF01609873}.

\bibitem{cormen}
T.~H. Cormen, C.~E. Leiserson, R.~L. Rivest, and C.~Stein.
\newblock {\em Introduction to Algorithms}.
\newblock The MIT Press, 3rd edition edition, 2009.
\newblock \urlprefix\url{https://dl.acm.org/doi/book/10.5555/1614191}.

\bibitem{cormode}
G.~Cormode and S.~Muthukrishnan.
\newblock What's hot and what's not: {T}racking most frequent items
  dynamically.
\newblock {\em ACM Transactions on Database Systems (TODS)}, 30(1):249--278,
  2005.
\newblock \doi{10.1145/1061318.1061325}.

\bibitem{cover}
T.~M. Cover and J.~A. Thomas.
\newblock {\em Elements of Information Theory}.
\newblock Wiley-Interscience, 2nd edition edition, 2006.
\newblock \doi{10.1002/047174882X}.

\bibitem{csiszarkorner}
I.~Csisz{\'a}r and J.~K{\"o}rner.
\newblock {\em Information Theory: {C}oding Theorems for Discrete Memoryless
  Systems}.
\newblock Cambridge University Press, 2011.
\newblock \doi{10.1017/CBO9780511921889}.
\newblock 2nd edition.

\bibitem{curnow}
R.~N. Curnow and A.~P. Morris.
\newblock Pooling {DNA} in the identification of parents.
\newblock {\em Heredity}, 80(1):101--109, 1998.
\newblock \doi{10.1038/sj.hdy.6882420}.

\bibitem{damaschke2006threshold}
P.~Damaschke.
\newblock Threshold group testing.
\newblock In {\em General Theory of Information Transfer and Combinatorics},
  pages 707--718. Springer, 2006.
\newblock \doi{10.1007/11889342_45}.

\bibitem{damaschke2016adaptive}
P.~Damaschke.
\newblock Adaptive group testing with a constrained number of positive
  responses improved.
\newblock {\em Discrete Applied Mathematics}, 205:208--212, 2016.
\newblock \doi{10.1016/j.dam.2016.01.010}.

\bibitem{damaschke2010nonadapt}
P.~Damaschke and A.~S. Muhammad.
\newblock Bounds for nonadaptive group tests to estimate the amount of
  defectives.
\newblock In {\em Combinatorial Optimization and Applications (COCOA)}, pages
  117--130, 2010.
\newblock \doi{10.1007/978-3-642-17461-2_10}.

\bibitem{damaschke2010competitive}
P.~Damaschke and A.~S. Muhammad.
\newblock Competitive group testing and learning hidden vertex covers with
  minimum adaptivity.
\newblock {\em Discrete Mathematics, Algorithms and Applications},
  2(03):291--311, 2010.
\newblock \doi{10.1142/S179383091000067X}.

\bibitem{damaschke2012randomized}
P.~Damaschke and A.~S. Muhammad.
\newblock Randomized group testing both query-optimal and minimal adaptive.
\newblock In {\em International Conference on Current Trends in Theory and
  Practice of Computer Science}, pages 214--225, 2012.
\newblock \doi{10.1007/978-3-642-27660-6_18}.

\bibitem{deBonis2005optimal}
A.~De~Bonis, L.~Gasieniec, and U.~Vaccaro.
\newblock Optimal two-stage algorithms for group testing problems.
\newblock {\em SIAM Journal on Computing}, 34(5):1253--1270, 2005.
\newblock \doi{10.1137/S0097539703428002}.

\bibitem{deBonis1998improved}
A.~De~Bonis and U.~Vaccaro.
\newblock Improved algorithms for group testing with inhibitors.
\newblock {\em Information Processing Letters}, 67(2):57--64, 1998.
\newblock \doi{10.1016/S0020-0190(98)00088-X}.

\bibitem{debonis}
A.~De~Bonis and U.~Vaccaro.
\newblock Constructions of generalized superimposed codes with applications to
  group testing and conflict resolution in multiple access channels.
\newblock {\em Theoretical Computer Science}, 306(1-3):223--243, 2003.
\newblock \doi{10.1016/S0304-3975(03)00281-0}.

\bibitem{debonis2}
A.~De~Bonis and U.~Vaccaro.
\newblock {$\epsilon$}-almost selectors and their applications to
  multiple-access communication.
\newblock {\em IEEE Transactions on Information Theory}, 63(11):7304--7319,
  2017.
\newblock \doi{10.1109/TIT.2017.2750178}.

\bibitem{djackov1975}
A.~Djackov.
\newblock On a search model of false coins.
\newblock In I.~Csiszár and P.~Elias, editors, {\em Topics in Information
  Theory}, volume~16 of {\em Colloquia Mathematica Societatis J{\'a}nos
  Bolyai}, pages 163--170. North-Holland, 1975.

\bibitem{dorfman}
R.~Dorfman.
\newblock The detection of defective members of large populations.
\newblock {\em The Annals of Mathematical Statistics}, 14(4):436--440, 1943.
\newblock \doi{10.1214/aoms/1177731363}.

\bibitem{du-hwang}
D.-Z. Du and F.~Hwang.
\newblock {\em Combinatorial Group Testing and Its Applications}.
\newblock World Scientific, 2nd edition edition, 1999.
\newblock \doi{10.1142/4252}.

\bibitem{du-hwang2}
D.-Z. Du and F.~K. Hwang.
\newblock {\em Pooling Designs and Nonadaptive Group Testing: {I}mportant Tools
  for {DNA} Sequencing}.
\newblock World Scientific, 2006.
\newblock \doi{10.1142/6122}.

\bibitem{dyachkov_lectures}
A.~G. D'yachkov.
\newblock Lectures on designing screening experiments, 2004, arXiv:
  \burlalt{\texttt{1401.7505}}{http://arxiv.org/abs/1401.7505}.
\newblock Lecture Note Series 10, POSTECH.

\bibitem{dyachkov-rykov}
A.~G. D'yachkov and V.~V. Rykov.
\newblock Bounds on the length of disjunctive codes.
\newblock {\em Problemy Peredachi Informatsii}, 18(3):7--13, 1982.
\newblock
  \urlprefix\url{https://www.researchgate.net/publication/268498029_Bounds_on_the_length_of_disjunctive_codes}.
\newblock Translation: \emph{Problems of Information Transmission}. 18(3):
  166--171.

\bibitem{dyachkov1983survey}
A.~G. D'yachkov and V.~V. Rykov.
\newblock A survey of superimposed code theory.
\newblock {\em Problems of Control and Information Theory}, 12(4):1--13, 1983.
\newblock
  \urlprefix\url{https://www.researchgate.net/publication/235008674_Survey_of_Superimposed_Code_Theory}.

\bibitem{el2018decoding}
A.~El~Alaoui, A.~Ramdas, F.~Krzakala, L.~Zdeborov{\'a}, and M.~I. Jordan.
\newblock Decoding from pooled data: Phase transitions of message passing.
\newblock {\em IEEE Transactions on Information Theory}, 65(1):572--585, 2018.
\newblock \doi{10.1109/TIT.2018.2855698}.

\bibitem{el2019decoding}
A.~El~Alaoui, A.~Ramdas, F.~Krzakala, L.~Zdeborov{\'a}, and M.~I. Jordan.
\newblock Decoding from pooled data: Sharp information-theoretic bounds.
\newblock {\em SIAM Journal on Mathematics of Data Science}, 1(1):161--188,
  2019.
\newblock \doi{10.1137/18M1183339}.

\bibitem{emad2014poisson}
A.~Emad and O.~Milenkovic.
\newblock Poisson group testing: {A} probabilistic model for nonadaptive
  streaming {B}oolean compressed sensing.
\newblock In {\em IEEE International Conference on Acoustics, Speech and Signal
  Processing (ICASSP), 2014}, pages 3335--3339, 2014.
\newblock \doi{10.1109/ICASSP.2014.6854218}.

\bibitem{emad2014}
A.~Emad and O.~Milenkovic.
\newblock Semiquantitative group testing.
\newblock {\em IEEE Transactions on Information Theory}, 60(8):4614--4636,
  2014.
\newblock \doi{10.1109/TIT.2014.2327630}.

\bibitem{emad2016code}
A.~Emad and O.~Milenkovic.
\newblock Code construction and decoding algorithms for semi-quantitative group
  testing with nonuniform thresholds.
\newblock {\em IEEE Transactions on Information Theory}, 62(4):1674--1687,
  2016.
\newblock \doi{10.1109/TIT.2016.2524002}.

\bibitem{emad2015semiquantitative}
A.~Emad, K.~R. Varshney, and D.~M. Malioutov.
\newblock A semiquantitative group testing approach for learning interpretable
  clinical prediction rules.
\newblock In {\em Signal Processing with Adaptive Sparse Structured
  Representations (SPARS)}, 2015.
\newblock
  \urlprefix\url{https://krvarshney.github.io/pubs/EmadVM_spars2015.pdf}.

\bibitem{engels2021practical}
J.~Engels, B.~Coleman, and A.~Shrivastava.
\newblock Practical near neighbor search via group testing.
\newblock {\em Advances in Neural Information Processing Systems},
  34:9950--9962, 2021.
\newblock \urlprefix\url{https://dl.acm.org/doi/10.5555/3540261.3541022}.

\bibitem{erdos-renyi2}
P.~Erd\H{o}s and A.~R\'{e}nyi.
\newblock On a classical problem of probability theory.
\newblock {\em A Magyar Tudom\'{a}nyos Akad\'{e}mia Matematikai Kutat\'{o}
  Int\'{e}zet\'{e}nek K\"{o}zlem\'{e}nyei}, 6:215--220, 1961.
\newblock \urlprefix\url{http://www.renyi.hu/~p_erdos/1961-09.pdf}.

\bibitem{erdos-renyi}
P.~Erd{\H{o}}s and A.~R{\'e}nyi.
\newblock On two problems of information theory.
\newblock {\em A Magyar Tudom\'{a}nyos Akad\'{e}mia Matematikai Kutat\'{o}
  Int\'{e}zet\'{e}nek K\"{o}zlem\'{e}nyei}, 8:229--243, 1963.
\newblock \urlprefix\url{https://www.renyi.hu/~p_erdos/1963-12.pdf}.

\bibitem{erlich2}
Y.~Erlich, A.~Gilbert, H.~Ngo, A.~Rudra, N.~Thierry-Mieg, M.~Wootters,
  D.~Zielinski, and O.~Zuk.
\newblock Biological screens from linear codes: {T}heory and tools.
\newblock {\em bioRxiv}, 2015.
\newblock \doi{10.1101/035352}.

\bibitem{erlich}
Y.~Erlich, A.~Gordon, M.~Brand, G.~Hannon, and P.~Mitra.
\newblock Compressed genotyping.
\newblock {\em IEEE Transactions on Information Theory}, 56(2):706--723, 2010.
\newblock \doi{10.1109/TIT.2009.2037043}.

\bibitem{falahatgar}
M.~Falahatgar, A.~Jafarpour, A.~Orlitsky, V.~Pichapati, and A.~T. Suresh.
\newblock Estimating the number of defectives with group testing.
\newblock In {\em IEEE International Symposium on Information Theory (ISIT)},
  pages 1376--1380, 2016.
\newblock \doi{10.1109/ISIT.2016.7541524}.

\bibitem{farach}
M.~{Farach}, S.~{Kannan}, E.~{Knill}, and S.~{Muthukrishnan}.
\newblock Group testing problems with sequences in experimental molecular
  biology.
\newblock In {\em Proceedings of Compression and Complexity of Sequences},
  pages 357--367, 1997.
\newblock \doi{10.1109/SEQUEN.1997.666930}.

\bibitem{feige2020quantitative}
U.~Feige and A.~Lellouche.
\newblock Quantitative group testing and the rank of random matrices, 2020,
  arXiv: \burlalt{\texttt{2006.09074}}{http://arxiv.org/abs/2006.09074}.

\bibitem{Fei54}
A.~Feinstein.
\newblock A new basic theorem of information theory.
\newblock {\em Transactions of the IRE Professional Group on Information
  Theory}, 4(4):2--22, 1954.
\newblock \doi{10.1109/TIT.1954.1057459}.

\bibitem{feller}
W.~Feller.
\newblock {\em An Introduction to Probability Theory and Its Applications},
  volume~I.
\newblock John Wiley \& Sons, 3rd edition, 1968.

\bibitem{finucan}
H.~M. Finucan.
\newblock The blood testing problem.
\newblock {\em Journal of the Royal Statistical Society Series C (Applied
  Statistics)}, pages 43--50, 1964.
\newblock \doi{10.2307/2985222}.

\bibitem{FKW}
P.~Fischer, N.~Klasner, and I.~Wegenera.
\newblock On the cut-off point for combinatorial group testing.
\newblock {\em Discrete Applied Mathematics}, 91(1):83--92, 1999.
\newblock \doi{10.1016/S0166-218X(98)00119-X}.

\bibitem{foucart}
S.~Foucart and H.~Rauhut.
\newblock {\em A Mathematical Introduction to Compressive Sensing}.
\newblock Birkh{\"a}user, 2013.
\newblock \doi{10.1007/978-0-8176-4948-7}.

\bibitem{friedlina}
V.~L. Freidlina.
\newblock On a design problem for screening experiments.
\newblock {\em Teoriya Veroyatnostei i ee Primeneniya}, 20(1):100--114, 1975.
\newblock \doi{10.1137/1120008}.
\newblock Translation: \emph{Theory of Probability \& Its Applications}. 20(1):
  102--115.

\bibitem{furedi}
Z.~F{\"u}redi.
\newblock On $r$-cover-free families.
\newblock {\em Journal of Combinatorial Theory, Series A}, 73(1):172--173,
  1996.
\newblock \doi{10.1006/jcta.1996.0012}.

\bibitem{furon}
T.~Furon.
\newblock The illusion of group testing.
\newblock Technical Report RR-9164, Inria Rennes Bretagne Atlantique, 2018.
\newblock \urlprefix\url{https://hal.inria.fr/hal-01744252}.

\bibitem{furon2012mcmc}
T.~{Furon}, A.~{Guyader}, and F.~{C\'{e}rou}.
\newblock Decoding fingerprints using the {M}arkov {C}hain {M}onte {C}arlo
  method.
\newblock In {\em IEEE International Workshop on Information Forensics and
  Security (WIFS)}, pages 187--192, 2012.
\newblock \doi{10.1109/WIFS.2012.6412647}.

\bibitem{gabrys2020ac}
R.~Gabrys, S.~Pattabiraman, V.~Rana, J.~Ribeiro, M.~Cheraghchi, V.~Guruswami,
  and O.~Milenkovic.
\newblock {AC--DC}: Amplification curve diagnostics for {Covid-19} group
  testing, 2020, arXiv:
  \burlalt{\texttt{2011.05223}}{http://arxiv.org/abs/2011.05223}.

\bibitem{gandikota2016sparse}
V.~Gandikota, E.~Grigorescu, S.~Jaggi, and S.~Zhou.
\newblock Nearly optimal sparse group testing.
\newblock In {\em Annual Allerton Conference on Communication, Control, and
  Computing}, pages 401--408, 2016.
\newblock \doi{10.1109/ALLERTON.2016.7852259}.

\bibitem{ganesan}
A.~Ganesan, S.~Jaggi, and V.~Saligrama.
\newblock Learning immune-defectives graph through group tests.
\newblock In {\em IEEE International Symposium on Information Theory (ISIT)},
  pages 66--70, 2015.
\newblock \doi{10.1109/ISIT.2015.7282418}.

\bibitem{ganesan2015}
A.~Ganesan, S.~Jaggi, and V.~Saligrama.
\newblock Non-adaptive group testing with inhibitors.
\newblock In {\em IEEE Information Theory Workshop (ITW)}, pages 1--5, 2015.
\newblock \doi{10.1109/ITW.2015.7133108}.

\bibitem{garey}
M.~R. Garey and F.~K. Hwang.
\newblock Isolating a single defective using group testing.
\newblock {\em Journal of the American Statistical Association},
  69(345):151--153, 1974.
\newblock \doi{10.2307/2285514}.

\bibitem{gastwirth}
J.~L. Gastwirth and P.~A. Hammick.
\newblock Estimation of the prevalence of a rare disease, preserving the
  anonymity of the subjects by group testing: {A}pplication to estimating the
  prevalence of {AIDS} antibodies in blood donors.
\newblock {\em Journal of Statistical Planning and Inference}, 22(1):15--27,
  1989.
\newblock \doi{10.1016/0378-3758(89)90061-X}.

\bibitem{gebhard}
O.~Gebhard, M.~Hahn-Klimroth, D.~Kaaser, and P.~Loick.
\newblock On the parallel reconstruction from pooled data.
\newblock In {\em 2022 IEEE International Parallel and Distributed Processing
  Symposium (IPDPS)}, pages 425--435, 2022.
\newblock \doi{10.1109/IPDPS53621.2022.00048}.

\bibitem{Gebhard2022}
O.~Gebhard, M.~Hahn-Klimroth, O.~Parczyk, M.~Penschuck, M.~Rolvien,
  J.~Scarlett, and N.~Tan.
\newblock Near-optimal sparsity-constrained group testing: Improved bounds and
  algorithms.
\newblock {\em IEEE Transactions on Information Theory}, 68(5):3253--3280,
  2022.
\newblock \doi{10.1109/TIT.2022.3141244}.

\bibitem{gebhard2021improved}
O.~Gebhard, O.~Johnson, P.~Loick, and M.~Rolvien.
\newblock Improved bounds for noisy group testing with constant tests per item.
\newblock {\em IEEE Transactions on Information Theory}, 68(4):2604--2621,
  2021.
\newblock \doi{10.1109/TIT.2021.3138489}.

\bibitem{ghosh2021compressed}
S.~Ghosh, R.~Agarwal, M.~A. Rehan, S.~Pathak, P.~Agarwal, Y.~Gupta, S.~Consul,
  N.~Gupta, R.~Goenka, A.~Rajwade, and M.~Gopalkrishnan.
\newblock A compressed sensing approach to pooled {RT}-{PCR} testing for
  {COVID}-19 detection.
\newblock {\em IEEE Open Journal of Signal Processing}, 2:248--264, 2021.
\newblock \doi{10.1109/OJSP.2021.3075913}.

\bibitem{ghosh2023efficient}
S.~Ghosh, S.~Saxena, and A.~Rajwade.
\newblock Efficient neural network based classification and outlier detection
  for image moderation using compressed sensing and group testing, 2023, arXiv:
  \burlalt{\texttt{2305.07639}}{http://arxiv.org/abs/2305.07639}.

\bibitem{gilbert2012}
A.~C. Gilbert, B.~Hemenway, A.~Rudra, M.~J. Strauss, and M.~Wootters.
\newblock Recovering simple signals.
\newblock In {\em Information Theory and Applications Workshop (ITA)}, pages
  382--391, 2012.
\newblock \doi{10.1109/ITA.2012.6181772}.

\bibitem{gilbert}
A.~C. Gilbert, M.~A. Iwen, and M.~J. Strauss.
\newblock Group testing and sparse signal recovery.
\newblock In {\em Asilomar Conference on Signals, Systems and Computers}, pages
  1059--1063, 2008.
\newblock \doi{10.1109/ACSSC.2008.5074574}.

\bibitem{gilbert2006algorithmic}
A.~C. Gilbert, M.~J. Strauss, J.~A. Tropp, and R.~Vershynin.
\newblock Algorithmic linear dimension reduction in the $\ell_1$ norm for
  sparse vectors.
\newblock In {\em Allerton Conference on Communication, Control, and
  Computing}, 2006.
\newblock
  \urlprefix\url{https://web.eecs.umich.edu/~martinjs/papers/GSTV06-allerton.pdf}.

\bibitem{gilbert2007one}
A.~C. Gilbert, M.~J. Strauss, J.~A. Tropp, and R.~Vershynin.
\newblock One sketch for all: Fast algorithms for compressed sensing.
\newblock In {\em ACM Symposium on Theory of Computing (STOC)}, pages 237--246,
  2007.
\newblock \doi{10.1145/1250790.1250824}.

\bibitem{gille}
C.~Gille, K.~Grade, and C.~Coutelle.
\newblock A pooling strategy for heterozygote screening of the {$\Delta$F508}
  cystic fibrosis mutation.
\newblock {\em Human Genetics}, 86(3):289--291, 1991.
\newblock \doi{10.1007/BF00202411}.

\bibitem{goenka2021contact}
R.~Goenka, S.-J. Cao, C.-W. Wong, A.~Rajwade, and D.~Baron.
\newblock Contact tracing enhances the efficiency of {COVID}-19 group testing.
\newblock In {\em IEEE International Conference on Acoustics, Speech and Signal
  Processing (ICASSP)}, pages 8168--8172, 2021.
\newblock \urlprefix\url{10.1109/ICASSP39728.2021.9414034}.

\bibitem{goldie}
C.~Goldie and R.~Pinch.
\newblock {\em Communication Theory}.
\newblock Cambridge University Press, 1991.
\newblock \doi{10.1017/CBO9781139172448}.

\bibitem{gonen2022general}
M.~Gonen, M.~Langberg, and A.~Sprintson.
\newblock Group testing on general set-systems.
\newblock In {\em IEEE International Symposium on Information Theory (ISIT)},
  pages 874--879, 2022.
\newblock \doi{10.1109/ISIT50566.2022.9834789}.

\bibitem{goodrich2}
M.~T. Goodrich, M.~J. Atallah, and R.~Tamassia.
\newblock Indexing information for data forensics.
\newblock In {\em Applied Cryptography and Network Security}, pages 206--221,
  2005.
\newblock \doi{10.1007/11496137_15}.

\bibitem{goodrich}
M.~T. Goodrich and D.~S. Hirschberg.
\newblock Improved adaptive group testing algorithms with applications to
  multiple access channels and dead sensor diagnosis.
\newblock {\em Journal of Combinatorial Optimization}, 15(1):95--121, 2008.
\newblock \doi{10.1007/s10878-007-9087-z}.

\bibitem{Guruswami2023}
V.~Guruswami and H.-P. Wang.
\newblock Nonadaptive noise-resilient group testing with order-optimal tests
  and fast-and-reliable decoding, 2024, arXiv:
  \burlalt{\texttt{2311.08283}}{http://arxiv.org/abs/2311.08283}.

\bibitem{hahn2023efficient}
M.~Hahn-Klimroth, D.~Kaaser, and M.~Rau.
\newblock Efficient approximate recovery from pooled data using doubly regular
  pooling schemes, 2023, arXiv:
  \burlalt{\texttt{2303.00043}}{http://arxiv.org/abs/2303.00043}.

\bibitem{hahn2022near}
M.~Hahn-Klimroth and N.~M{\"u}ller.
\newblock Near optimal efficient decoding from pooled data.
\newblock In {\em Conference on Learning Theory}, pages 3395--3409, 2022.
\newblock
  \urlprefix\url{https://proceedings.mlr.press/v178/hahn-klimroth22a.html}.

\bibitem{Han03}
T.~S. Han.
\newblock {\em Information-Spectrum Methods in Information Theory}.
\newblock Springer--Verlag, 2003.
\newblock \doi{10.1007/978-3-662-12066-8}.

\bibitem{harvey2007}
N.~J. Harvey, M.~Patrascu, Y.~Wen, S.~Yekhanin, and V.~W. Chan.
\newblock Non-adaptive fault diagnosis for all-optical networks via
  combinatorial group testing on graphs.
\newblock In {\em IEEE International Conference on Computer Communications
  (INFOCOM)}, pages 697--705. IEEE, 2007.
\newblock \doi{10.1109/INFCOM.2007.87}.

\bibitem{hayes}
J.~F. Hayes.
\newblock An adaptive technique for local distribution.
\newblock {\em IEEE Transactions on Communications}, 26(8):1178--1186, 1978.
\newblock \doi{10.1109/TCOM.1978.1094204}.

\bibitem{hintze2024noisy}
L.~Hintze, L.~Krieg, O.~Scheftelowitsch, and H.~Zhu.
\newblock Noisy group testing in the linear regime: Exact thresholds and
  efficient algorithms.
\newblock In {\em Conference on Learning Theory}, pages 2805--2821, 2025.
\newblock \urlprefix\url{https://proceedings.mlr.press/v291/hintze25a.html}.

\bibitem{hong}
E.~S. Hong and R.~E. Ladner.
\newblock Group testing for image compression.
\newblock {\em IEEE Transactions on Image Processing}, 11(8):901--911, 2002.
\newblock \doi{10.1109/TIP.2002.801124}.

\bibitem{hong2}
Y.-W. Hong and A.~Scaglione.
\newblock Group testing for sensor networks: {T}he value of asking the right
  questions.
\newblock In {\em Asilomar Conference on Signals, Systems and Computers},
  volume~2, pages 1297--1301, 2004.
\newblock \doi{10.1109/ACSSC.2004.1399362}.

\bibitem{hu}
M.~C. Hu, F.~K. Hwang, and J.~K. Wang.
\newblock A boundary problem for group testing.
\newblock {\em SIAM Journal on Algebraic Discrete Methods}, 2(2):81--87, 1981.
\newblock \doi{10.1137/0602011}.

\bibitem{huleihel}
W.~{Huleihel}, O.~{Elishco}, and M.~{M\'edard}.
\newblock Blind group testing.
\newblock {\em IEEE Transactions on Information Theory}, 65(8):5050--5063,
  2019.
\newblock \doi{10.1109/TIT.2019.2906607}.

\bibitem{hwang1987group}
F.~Hwang and Y.~Xu.
\newblock Group testing to identify one defective and one mediocre item.
\newblock {\em Journal of Statistical Planning and Inference}, 17:367--373,
  1987.
\newblock \doi{0378-3758(87)90127-3}.

\bibitem{hwang}
F.~K. Hwang.
\newblock A method for detecting all defective members in a population by group
  testing.
\newblock {\em Journal of the American Statistical Association},
  67(339):605--608, 1972.
\newblock \doi{10.1080/01621459.1972.10481257}.

\bibitem{hwang2}
F.~K. Hwang.
\newblock A generalized binomial group testing problem.
\newblock {\em Journal of the American Statistical Association},
  70(352):923--926, 1975.
\newblock \doi{10.1080/01621459.1975.10480324}.

\bibitem{iliopoulos2021group}
F.~Iliopoulos and I.~Zadik.
\newblock Group testing and local search: Is there a computational-statistical
  gap?
\newblock In {\em Conference on Learning Theory}, pages 2499--2551. PMLR, 2021.
\newblock \urlprefix\url{https://proceedings.mlr.press/v134/iliopoulos21a}.

\bibitem{inan2019group}
H.~A. Inan, S.~Ahn, P.~Kairouz, and A.~Ozgur.
\newblock A group testing approach to random access for short-packet
  communication.
\newblock In {\em IEEE International Symposium on Information Theory (ISIT)},
  pages 96--100, 2019.
\newblock \doi{10.1109/ISIT.2019.8849823}.

\bibitem{inan2018energy}
H.~A. Inan, P.~Kairouz, and A.~Ozgur.
\newblock Energy-limited massive random access via noisy group testing.
\newblock In {\em IEEE International Symposium on Information Theory (ISIT)},
  pages 1101--1105, 2018.
\newblock \doi{10.1109/ISIT.2018.8437557}.

\bibitem{inan}
H.~A. Inan, P.~Kairouz, M.~Wootters, and A.~Özgür.
\newblock On the optimality of the {K}autz-{S}ingleton construction in
  probabilistic group testing.
\newblock {\em IEEE Transactions on Information Theory}, 65(9):5592--5603,
  2019.
\newblock \doi{10.1109/TIT.2019.2902397}.

\bibitem{Inan2020}
H.~A. Inan and A.~Ozgur.
\newblock Strongly explicit and efficiently decodable probabilistic group
  testing.
\newblock In {\em IEEE International Symposium on Information Theory (ISIT)},
  pages 525--530, 2020.
\newblock \doi{10.1109/ISIT44484.2020.9174462}.

\bibitem{indyk}
P.~Indyk.
\newblock Deterministic superimposed coding with applications to pattern
  matching.
\newblock In {\em 38th Annual Symposium on Foundations of Computer Science
  (FOCS)}, pages 127--136, 1997.
\newblock \doi{10.1109/SFCS.1997.646101}.

\bibitem{indyk2010}
P.~Indyk, H.~Q. Ngo, and A.~Rudra.
\newblock Efficiently decodable non-adaptive group testing.
\newblock In {\em ACM-SIAM Symposium on Discrete Algorithms (SODA)}, pages
  1126--1142, 2010.
\newblock \doi{10.1137/1.9781611973075.91}.

\bibitem{jain2023probabilistic}
S.~Jain, M.~Cardone, and S.~Mohajer.
\newblock Probabilistic group testing in distributed computing with attacked
  workers.
\newblock In {\em IEEE International Symposium on Information Theory (ISIT)},
  pages 1615--1620, 2023.
\newblock \doi{10.1109/ISIT54713.2023.10206705}.

\bibitem{johann2002group}
P.~Johann.
\newblock A group testing problem for graphs with several defective edges.
\newblock {\em Discrete Applied Mathematics}, 117(1-3):99--108, 2002.
\newblock \doi{10.1016/S0166-218X(01)00181-0}.

\bibitem{johnsonconverse}
O.~T. Johnson.
\newblock Strong converses for group testing from finite blocklength results.
\newblock {\em IEEE Transactions on Information Theory}, 63(9):5923--5933,
  2017.
\newblock \doi{10.1109/TIT.2017.2697358}.

\bibitem{johnson-aldridge-scarlett}
O.~T. Johnson, M.~Aldridge, and J.~Scarlett.
\newblock Performance of group testing algorithms with near-constant
  tests-per-item.
\newblock {\em IEEE Transactions on Information Theory}, 65(2):707--723, 2019.
\newblock \doi{10.1109/TIT.2018.2861772}.

\bibitem{juan2008adaptive}
J.~S.-T. Juan and G.~J. Chang.
\newblock Adaptive group testing for consecutive positives.
\newblock {\em Discrete Mathematics}, 308(7):1124--1129, 2008.
\newblock \doi{10.1016/j.disc.2007.04.002}.

\bibitem{kahng2006new}
A.~B. Kahng and S.~Reda.
\newblock New and improved bist diagnosis methods from combinatorial group
  testing theory.
\newblock {\em IEEE Transactions on Computer-Aided Design of Integrated
  Circuits and Systems}, 25(3):533--543, 2006.
\newblock \doi{10.1109/TCAD.2005.854635}.

\bibitem{kainkaryam}
R.~M. Kainkaryam and P.~J. Woolf.
\newblock Pooling in high-throughput drug screening.
\newblock {\em Current Opinion on Drug Discovery and Development}, 12(3):339,
  2009.
\newblock \urlprefix\url{https://www.ncbi.nlm.nih.gov/pubmed/19396735}.

\bibitem{karimi2022noisy}
E.~Karimi, A.~Heidarzadeh, K.~R. Narayanan, and A.~Sprintson.
\newblock Noisy group testing with side information.
\newblock In {\em Asilomar Conference on Signals, Systems, and Computers},
  pages 867--871. IEEE, 2022.
\newblock \urlprefix\url{10.1109/IEEECONF56349.2022.10052078}.

\bibitem{Karp1972}
R.~M. Karp.
\newblock {\em Reducibility among combinatorial problems}, pages 85--103.
\newblock Springer, 1972.
\newblock \doi{10.1007/978-1-4684-2001-2_9}.

\bibitem{Kaspi2015}
Y.~Kaspi, O.~Shayevitz, and T.~Javidi.
\newblock Searching for multiple targets with measurement dependent noise.
\newblock In {\em IEEE International Symposium on Information Theory (ISIT)},
  pages 969--973, 2015.
\newblock \doi{10.1109/ISIT.2015.7282599}.

\bibitem{katholi}
C.~R. Katholi, L.~To\'{e}, A.~Merriweather, and T.~R. Unnasch.
\newblock Determining the prevalence of {\em {o}nchocerca volvulus} infection
  in vector populations by polymerase chain reaction screening of pools of
  black flies.
\newblock {\em Journal of Infectious Diseases}, 172(5):1414--1417, 1995.
\newblock \doi{10.1093/infdis/172.5.1414}.

\bibitem{katona1973combinatorial}
G.~O.~H. Katona.
\newblock Combinatorial search problems.
\newblock In J.~N. Srivastava, editor, {\em A Survey of Combinatorial Theory},
  pages 285--308. North-Holland, 1973.
\newblock \doi{10.1016/B978-0-7204-2262-7.50028-4}.

\bibitem{kautz}
W.~Kautz and R.~Singleton.
\newblock Nonrandom binary superimposed codes.
\newblock {\em IEEE Transactions on Information Theory}, 10(4):363--377, 1964.
\newblock \doi{10.1109/TIT.1964.1053689}.

\bibitem{kealy-johnson-piechocki}
T.~Kealy, O.~Johnson, and R.~Piechocki.
\newblock The capacity of non-identical adaptive group testing.
\newblock In {\em Annual Allerton Conference on Communication, Control, and
  Computing}, pages 101--108, 2014.
\newblock \doi{10.1109/ALLERTON.2014.7028442}.

\bibitem{khan}
S.~A. Khan, P.~Chowdhury, P.~Choudhury, and P.~Dutta.
\newblock Detection of west nile virus in six mosquito species in synchrony
  with seroconversion among sentinel chickens in {I}ndia.
\newblock {\em Parasites and Vectors}, 10(1):13, 2017.
\newblock \doi{10.1186/s13071-016-1948-9}.

\bibitem{khattab}
S.~Khattab, S.~Gobriel, R.~Melhem, and D.~Mosse.
\newblock Live baiting for service-level {DoS} attackers.
\newblock In {\em IEEE International Conference on Computer Communications
  (INFOCOM)}, pages 171--175, 2008.
\newblock \doi{10.1109/INFOCOM.2008.43}.

\bibitem{knill1996mcmc}
E.~Knill, A.~Schliep, and D.~Torney.
\newblock Interpretation of pooling experiments using the {M}arkov chain
  {M}onte {C}arlo method.
\newblock {\em Journal of Computational Biology}, 3:395--406, 1996.
\newblock \doi{10.1089/cmb.1996.3.395}.

\bibitem{komlos}
J.~Komlos and A.~Greenberg.
\newblock An asymptotically fast nonadaptive algorithm for conflict resolution
  in multiple-access channels.
\newblock {\em IEEE Transactions on Information Theory}, 31(2):302--306, 1985.
\newblock \doi{10.1109/TIT.1985.1057020}.

\bibitem{laarhoven-1}
T.~Laarhoven.
\newblock Asymptotics of fingerprinting and group testing: {T}ight bounds from
  channel capacities.
\newblock {\em IEEE Transactions on Information Forensics and Security},
  10(9):1967--1980, 2015.
\newblock \doi{10.1109/TIFS.2015.2440190}.

\bibitem{lau2022model}
I.~Lau, J.~Scarlett, and Y.~Sun.
\newblock Model-based and graph-based priors for group testing.
\newblock {\em IEEE Transactions on Signal Processing}, 70:6035--6050, 2022.
\newblock \doi{10.1109/TSP.2022.3229942}.

\bibitem{lee-pedarsani-ramtin}
K.~Lee, K.~Chandrasekher, R.~Pedarsani, and K.~Ramchandran.
\newblock {SAFFRON}: {A} fast, efficient, and robust framework for group
  testing based on sparse-graph codes.
\newblock {\em IEEE Transactions on Signal Processing}, 67(17):4649--4664,
  2019.
\newblock \doi{10.1109/TSP.2019.2929938}.

\bibitem{li1962}
C.~H. Li.
\newblock A sequential method for screening experimental variables.
\newblock {\em Journal of the American Statistical Association},
  57(298):455--477, 1962.
\newblock \doi{10.1080/01621459.1962.10480672}.

\bibitem{li-etal}
T.~Li, C.~L. Chan, W.~Huang, T.~Kaced, and S.~Jaggi.
\newblock Group testing with prior statistics.
\newblock In {\em IEEE International Symposium on Information Theory (ISIT)},
  pages 2346--2350, 2014.
\newblock \doi{10.1109/ISIT.2014.6875253}.

\bibitem{LiMazumdar2024}
X.~Li and A.~Mazumdar.
\newblock Noisy nonadaptive group testing with binary splitting: New test
  design and improvement on {P}rice-{S}carlett-{T}an's scheme.
\newblock In {\em IEEE International Symposium on Information Theory (ISIT)},
  2025.
\newblock \doi{10.1109/ISIT63088.2025.11195538}.

\bibitem{li2015sublinear}
X.~{Li}, S.~{Pawar}, and K.~{Ramchandran}.
\newblock Sub-linear time compressed sensing using sparse-graph codes.
\newblock In {\em IEEE International Symposium on Information Theory (ISIT)},
  pages 1645--1649, 2015.
\newblock \doi{10.1109/ISIT.2015.7282735}.

\bibitem{li2023finding}
Y.-H. Li, R.~Gabrys, J.~Sima, I.~Shomorony, and O.~Milenkovic.
\newblock Finding a burst of positives via nonadaptive semiquantitative group
  testing.
\newblock In {\em IEEE International Symposium on Information Theory (ISIT)},
  pages 1848--1853, 2023.
\newblock \doi{10.1109/ISIT54713.2023.10206886}.

\bibitem{li2023non}
Y.-H. Li and I.-H. Wang.
\newblock Non-adaptive combinatorial quantitative group testing with
  adversarially perturbed measurements, 2022, arXiv:
  \burlalt{\texttt{2101.12653}}{http://arxiv.org/abs/2101.12653}.

\bibitem{li2019learning}
Z.~Li, M.~Fresacher, and J.~Scarlett.
\newblock Learning {E}rd{\H{o}}s--{R}{\'{e}}nyi random graphs via edge
  detecting queries.
\newblock {\em Advances in Neural Information Processing Systems}, 32, 2019.
\newblock \urlprefix\url{https://dl.acm.org/doi/10.5555/3454287.3454324}.

\bibitem{liang2021neural}
W.~Liang and J.~Zou.
\newblock Neural group testing to accelerate deep learning.
\newblock In {\em IEEE International Symposium on Information Theory (ISIT)},
  pages 958--963, 2021.
\newblock \doi{10.1109/ISIT45174.2021.9518038}.

\bibitem{lindstrom1964combinatory}
B.~Lindstr{\"o}m.
\newblock On a combinatory detection problem {I}.
\newblock {\em A Magyar Tudom{\'a}nyos Akad{\'e}mia Matematikai Kutat{\'o}
  Int{\'e}zet{\'e}nek K{\"o}zlem{\'e}nyei}, 9(1--2):195--207, 1964.
\newblock \urlprefix\url{https://real-j.mtak.hu/515/}.

\bibitem{lindstrom1965combinatorial}
B.~Lindstr{\"o}m.
\newblock On a combinatorial problem in number theory.
\newblock {\em Canadian Mathematical Bulletin}, 8(4):477--490, 1965.
\newblock \doi{10.4153/CMB-1965-034-2}.

\bibitem{Lindstrom1975}
B.~Lindstr{\"o}m.
\newblock Determining subsets by unramified experiments.
\newblock In J.~N. Srivastava, editor, {\em A Survey of Statistical Designs and
  Linear Models}, pages 407--418. North-Holland, 1975.

\bibitem{lo}
C.~Lo, M.~Liu, J.~P. Lynch, and A.~C. Gilbert.
\newblock Efficient sensor fault detection using combinatorial group testing.
\newblock In {\em IEEE International Conference on Distributed Computing in
  Sensor Systems (DCOSS)}, pages 199--206, 2013.
\newblock \doi{10.1109/DCOSS.2013.57}.

\bibitem{lovig2024mcmc}
M.~Lovig and I.~Zadik.
\newblock On the {MCMC} performance in {B}ernoulli group testing and the random
  max set-cover problem, 2024, arXiv:
  \burlalt{\texttt{2410.09231}}{http://arxiv.org/abs/2410.09231}.

\bibitem{luo}
J.~Luo and D.~Guo.
\newblock Neighbor discovery in wireless ad hoc networks based on group
  testing.
\newblock In {\em Annual Allerton Conference on Communication, Control, and
  Computing}, pages 791--797, 2008.
\newblock \doi{10.1109/ALLERTON.2008.4797638}.

\bibitem{ma}
L.~Ma, T.~He, A.~Swami, D.~Towsley, K.~K. Leung, and J.~Lowe.
\newblock Node failure localization via network tomography.
\newblock In {\em Proceedings of the 2014 Conference on Internet Measurement
  Conference (IMC)}, pages 195--208, 2014.
\newblock \doi{10.1145/2663716.2663723}.

\bibitem{mackay}
D.~J.~C. MacKay.
\newblock {\em Information Theory, Inference and Learning Algorithms}.
\newblock Cambridge University Press, 2003.
\newblock \doi{10.2277/0521642981}.

\bibitem{macula1998two}
A.~J. Macula.
\newblock Probabilistic nonadaptive and two-stage group testing with relatively
  small pools and {DNA} library screening.
\newblock {\em Journal of Combinatorial Optimization}, 2(4):385--397, 1998.
\newblock \doi{10.1023/A:1009732820981}.

\bibitem{macula}
A.~J. Macula.
\newblock Probabilistic nonadaptive group testing in the presence of errors and
  {DNA} library screening.
\newblock {\em Annals of Combinatorics}, 3(1):61--69, 1999.
\newblock \doi{10.1007/BF01609876}.

\bibitem{macula2004}
A.~J. Macula and L.~J. Popyack.
\newblock A group testing method for finding patterns in data.
\newblock {\em Discrete Applied Mathematics}, 144(1):149--157, 2004.
\newblock \doi{10.1016/j.dam.2003.07.009}.

\bibitem{madej}
T.~Madej.
\newblock An application of group testing to the file comparison problem.
\newblock In {\em International Conference on Distributed Computing Systems},
  pages 237--243, 1989.
\newblock \doi{10.1109/ICDCS.1989.37952}.

\bibitem{malioutov-malyutov}
D.~M. Malioutov and M.~Malyutov.
\newblock Boolean compressed sensing: {LP} relaxation for group testing.
\newblock In {\em IEEE International Conference on Acoustics, Speech and Signal
  Processing (ICASSP)}, pages 3305--3308, 2012.
\newblock \doi{10.1109/ICASSP.2012.6288622}.

\bibitem{malioutov}
D.~M. Malioutov and K.~R. Varshney.
\newblock Exact rule learning via {B}oolean compressed sensing.
\newblock In {\em International Conference on Machine Learning}, pages
  765--773, 2013.
\newblock \urlprefix\url{http://proceedings.mlr.press/v28/malioutov13.html}.

\bibitem{Malioutov2017}
D.~M. Malioutov, K.~R. Varshney, A.~Emad, and S.~Dash.
\newblock Learning interpretable classification rules with {B}oolean compressed
  sensing.
\newblock In T.~Cerquitelli, D.~Quercia, and F.~Pasquale, editors, {\em
  Transparent Data Mining for Big and Small Data}, pages 95--121. Springer,
  2017.
\newblock \doi{10.1007/978-3-319-54024-5_5}.

\bibitem{Mal98}
M.~Malyutov and H.~Sadaka.
\newblock Maximization of {ESI}. {J}aynes principle in testing significant
  inputs of linear model.
\newblock {\em Random Operators and Stochastic Equations}, 6(4):311--330, 1998.
\newblock \doi{10.1515/rose.1998.6.4.311}.

\bibitem{malyutov-1}
M.~B. Malyutov.
\newblock The separating property of random matrices.
\newblock {\em Matematicheskie Zametki}, 23(1):155--167, 1978.
\newblock \doi{10.1007/BF01104893}.
\newblock Translation: \emph{Mathematical Notes of the Academy of Sciences of
  the USSR}. 23(1):84--91.

\bibitem{malyutov}
M.~B. Malyutov.
\newblock Search for sparse active inputs: {A} review.
\newblock In H.~Aydinian, F.~Cicalese, and C.~Deppe, editors, {\em Information
  Theory, Combinatorics, and Search Theory: In Memory of Rudolf Ahlswede},
  pages 609--647. Springer, 2013.
\newblock \doi{10.1007/978-3-642-36899-8_31}.

\bibitem{Mal80}
M.~B. Malyutov and P.~S. Mateev.
\newblock Planning of screening experiments for a nonsymmetric response
  function.
\newblock {\em Matematicheskie Zametki}, 27:109--127, 1980.
\newblock \doi{10.1007/BF01149816}.
\newblock Translation: \emph{Mathematical Notes of the Academy of Sciences of
  the USSR}. 27(1):57--68.

\bibitem{mazumdar2016}
A.~Mazumdar.
\newblock Nonadaptive group testing with random set of defectives.
\newblock {\em IEEE Transactions on Information Theory}, 62(12):7522--7531,
  2016.
\newblock \doi{10.1109/TIT.2016.2613870}.

\bibitem{mcdiarmid}
C.~McDiarmid.
\newblock On the method of bounded differences.
\newblock In J.~Siemons, editor, {\em Surveys in Combinatorics 1989: {I}nvited
  Papers at the Twelfth British Combinatorial Conference}, pages 148--188.
  Cambridge University Press, 1989.
\newblock \doi{10.1017/CBO9781107359949.008}.

\bibitem{mcmorrow}
D.~McMorrow and J.~Scarlett.
\newblock Optimal non-adaptive group testing with one-sided error guarantees,
  2025, arXiv: \burlalt{\texttt{2506.10374}}{http://arxiv.org/abs/2506.10374}.

\bibitem{mezard}
M.~M{\'e}zard, M.~Tarzia, and C.~Toninelli.
\newblock Group testing with random pools: {P}hase transitions and optimal
  strategy.
\newblock {\em Journal of Statistical Physics}, 131(5):783--801, 2008.
\newblock \doi{10.1007/s10955-008-9528-9}.

\bibitem{mezard2011two}
M.~M{\'e}zard and C.~Toninelli.
\newblock Group testing with random pools: {O}ptimal two-stage algorithms.
\newblock {\em IEEE Transactions on Information Theory}, 57(3):1736--1745,
  2011.
\newblock \doi{10.1109/TIT.2010.2103752}.

\bibitem{morjaria2025density}
R.~Morjaria, S.~Bulusu, V.~Gandikota, and S.~Jaggi.
\newblock Density-dependent group testing.
\newblock In {\em International Conference on Artificial Intelligence and
  Statistics (AISTATS)}, pages 3628--3636, 2025.
\newblock \urlprefix\url{https://proceedings.mlr.press/v258/morjaria25a.html}.

\bibitem{mossel}
E.~Mossel, R.~O'Donnell, and R.~A. Servedio.
\newblock Learning functions of $k$ relevant variables.
\newblock {\em Journal of Computer and System Sciences}, 69(3):421--434, 2004.
\newblock \doi{10.1016/j.jcss.2004.04.002}.

\bibitem{mourad}
R.~Mourad, Z.~Dawy, and F.~Morcos.
\newblock Designing pooling systems for noisy high-throughput protein-protein
  interaction experiments using {B}oolean compressed sensing.
\newblock {\em IEEE/ACM Transactions on Computational Biology and
  Bioinformatics}, 10(6):1478--1490, 2013.
\newblock \doi{10.1109/TCBB.2013.129}.

\bibitem{mutesa}
L.~Mutesa, P.~Ndishimye, Y.~Butera, J.~Souopgui, A.~Uwineza, R.~Rutayisire,
  E.~L. Ndoricimpaye, E.~Musoni, N.~Rujeni, T.~Nyatanyi, E.~Ntagwabira,
  M.~Semakula, C.~Musanabaganwa, D.~Nyamwasa, M.~Ndashimye, E.~Ujeneza, I.~E.
  Mwikarago, C.~M. Muvunyi, J.~B. Mazarati, S.~Nsanzimana, N.~Turok, and
  W.~Ndifon.
\newblock A pooled testing strategy for identifying {SARS}-{CoV}-2 at low
  prevalence.
\newblock {\em Nature}, 589:276–280, 2020.
\newblock \doi{10.1038/s41586-020-2885-5}.

\bibitem{nebenzahl1973finite}
E.~Nebenzahl and M.~Sobel.
\newblock Finite and infinite models for generalized group-testing with unequal
  probabilities of success for each item.
\newblock In {\em Discriminant Analysis and Applications}, pages 239--289.
  Elsevier, 1973.
\newblock \doi{10.1016/B978-0-12-154050-0.50020-4}.

\bibitem{ngo2011}
H.~Q. Ngo, E.~Porat, and A.~Rudra.
\newblock Efficiently decodable error-correcting list disjunct matrices and
  applications.
\newblock In {\em International Colloquium on Automata, Languages and
  Programming (ICALP)}, pages 557--568, 2011.
\newblock \doi{10.1007/978-3-642-22006-7_47}.

\bibitem{nikolopoulos2023community}
P.~Nikolopoulos, S.~R. Srinivasavaradhan, T.~Guo, C.~Fragouli, and S.~N.
  Diggavi.
\newblock Community-aware group testing.
\newblock {\em IEEE Transactions on Information Theory}, 69(7):4361--4383,
  2023.
\newblock \doi{10.1109/TIT.2023.3250119}.

\bibitem{nikpey2024group}
H.~Nikpey, S.~Sarkar, and S.~S. Bidokhti.
\newblock Group testing with general correlation using hypergraphs.
\newblock In {\em IEEE International Symposium on Information Theory (ISIT)},
  pages 3225--3230, 2024.
\newblock \doi{10.1109/ISIT57864.2024.10619244}.

\bibitem{niles2023all}
J.~Niles-Weed and I.~Zadik.
\newblock It was ``all'' for ``nothing'': Sharp phase transitions for noiseless
  discrete channels.
\newblock {\em IEEE Transactions on Information Theory}, 69(8):5188--5202,
  2023.
\newblock \doi{10.1109/TIT.2022.3225802}.

\bibitem{paligadu2023small}
V.~Paligadu, O.~Johnson, and M.~Aldridge.
\newblock Small error algorithms for tropical group testing, 2024.
\newblock \doi{10.1109/TIT.2024.3445271}.

\bibitem{ppv2010}
Y.~Polyanskiy, H.~V. Poor, and S.~Verd\'{u}.
\newblock Channel coding rate in the finite blocklength regime.
\newblock {\em IEEE Transactions on Information Theory}, 56(5):2307--2359,
  2010.
\newblock \doi{10.1109/TIT.2010.2043769}.

\bibitem{porat}
E.~Porat and A.~Rothschild.
\newblock Explicit nonadaptive combinatorial group testing schemes.
\newblock {\em IEEE Transactions on Information Theory}, 57(12):7982--7989,
  2011.
\newblock \doi{10.1109/TIT.2011.2163296}.

\bibitem{Price2020}
E.~Price and J.~Scarlett.
\newblock A fast binary splitting approach to non-adaptive group testing.
\newblock In {\em International Conference on Randomization and Computation
  (RANDOM)}, 2020.
\newblock \doi{10.4230/LIPIcs.APPROX/RANDOM.2020.13}.

\bibitem{price2023fast}
E.~Price, J.~Scarlett, and N.~Tan.
\newblock Fast splitting algorithms for sparsity-constrained and noisy group
  testing.
\newblock {\em Information and Inference: A Journal of the IMA},
  12(2):1141--1171, 2023.
\newblock \doi{10.1093/imaiai/iaac031}.

\bibitem{riccio}
L.~Riccio and C.~J. Colbourn.
\newblock Sharper bounds in adaptive group testing.
\newblock {\em Taiwanese Journal of Mathematics}, 4(4):669--673, 2000.
\newblock \doi{10.11650/twjm/1500407300}.

\bibitem{richardson}
T.~Richardson and R.~Urbanke.
\newblock {\em Modern Coding Theory}.
\newblock Cambridge University Press, 2008.
\newblock \doi{10.1017/CBO9780511791338}.

\bibitem{ruszinko}
M.~Ruszink{\'o}.
\newblock On the upper bound of the size of the $r$-cover-free families.
\newblock {\em Journal of Combinatorial Theory, Series~A}, 66(2):302--310,
  1994.
\newblock \doi{10.1016/0097-3165(94)90067-1}.

\bibitem{Sak2025}
M.~H. Sak, R.~Y. Liu, E.~E. Kwan, and E.~N. Jacobsen.
\newblock Accelerating the discovery of multicatalytic cooperativity.
\newblock {\em Nature}, 2025.
\newblock \doi{10.1038/s41586-025-09813-2}.

\bibitem{Sca19}
J.~Scarlett.
\newblock An efficient algorithm for capacity-approaching noisy adaptive group
  testing.
\newblock In {\em IEEE International Symposium on Information Theory (ISIT)},
  2019.
\newblock \doi{10.1109/ISIT.2019.8849310}.

\bibitem{Sca18}
J.~Scarlett.
\newblock Noisy adaptive group testing: Bounds and algorithms.
\newblock {\em IEEE Transactions on Information Theory}, 65(6):3646--3661,
  2019.
\newblock \doi{10.1109/TIT.2018.2883604}.

\bibitem{scarlett-cevher-3}
J.~Scarlett and V.~Cevher.
\newblock Converse bounds for noisy group testing with arbitrary measurement
  matrices.
\newblock In {\em IEEE International Symposium on Information Theory (ISIT)},
  pages 2868--2872, 2016.
\newblock \doi{10.1109/ISIT.2016.7541823}.

\bibitem{scarlett-cevher-2}
J.~Scarlett and V.~Cevher.
\newblock Phase transitions in group testing.
\newblock In {\em ACM-SIAM Symposium on Discrete Algorithms (SODA)}, pages
  40--53, 2016.
\newblock \doi{10.1137/1.9781611974331.ch4}.

\bibitem{scarlett-cevher-4}
J.~Scarlett and V.~Cevher.
\newblock How little does non-exact recovery help in group testing?
\newblock In {\em IEEE International Conference on Acoustics, Speech and Signal
  Processing (ICASSP)}, pages 6090--6094, 2017.
\newblock \doi{10.1109/ICASSP.2017.7953326}.

\bibitem{scarlett-cevher-1}
J.~Scarlett and V.~Cevher.
\newblock Limits on support recovery with probabilistic models: {A}n
  information-theoretic framework.
\newblock {\em IEEE Transactions on Information Theory}, 63(1):593--620, 2017.
\newblock \doi{10.1109/TIT.2016.2606605}.

\bibitem{Sca17c}
J.~Scarlett and V.~Cevher.
\newblock Phase transitions in the pooled data problem.
\newblock In {\em Advances in Neural Information Processing Systems (NeurIPS)},
  pages 377--385, 2017.
\newblock \urlprefix\url{https://dl.acm.org/doi/10.5555/3294771.3294807}.

\bibitem{Sca17b}
J.~Scarlett and V.~Cevher.
\newblock Near-optimal noisy group testing via separate decoding of items.
\newblock {\em IEEE Journal of Selelected Topics in Signal Processing},
  2(4):625--638, 2018.
\newblock \doi{10.1109/JSTSP.2018.2844818}.

\bibitem{scarlett-johnson}
J.~Scarlett and O.~Johnson.
\newblock Noisy non-adaptive group testing: A (near-)definite defectives
  approach.
\newblock {\em IEEE Transactions on Information Theory}, 66(6):3775--3797,
  2020.
\newblock \doi{10.1109/TIT.2020.2970184}.

\bibitem{schliep2003dna}
A.~{Schliep}, D.~C. {Torney}, and S.~{Rahmann}.
\newblock Group testing with {DNA} chips: Generating designs and decoding
  experiments.
\newblock In {\em IEEE Bioinformatics Conference}, pages 84--91, 2003.
\newblock \doi{10.1109/CSB.2003.1227307}.

\bibitem{sebo}
A.~Seb{\H o}.
\newblock On two random search problems.
\newblock {\em Journal of Statistical Planning and Inference}, 11(1):23--31,
  1985.
\newblock \doi{10.1016/0378-3758(85)90022-9}.

\bibitem{sejdinovic-johnson}
D.~Sejdinovic and O.~T. Johnson.
\newblock Note on noisy group testing: {A}symptotic bounds and belief
  propagation reconstruction.
\newblock In {\em Annual Allerton Conference on Communication, Control, and
  Computing}, pages 998--1003, 2010.
\newblock \doi{10.1109/ALLERTON.2010.5707018}.

\bibitem{sham}
P.~Sham, J.~S. Bader, I.~Craig, M.~O'Donovan, and M.~Owen.
\newblock {DNA} pooling: {A} tool for large-scale association studies.
\newblock {\em Nature Reviews Genetics}, 3(11):862--871, 2002.
\newblock \doi{10.1038/nrg930}.

\bibitem{shangguan}
C.~Shangguan and G.~Ge.
\newblock New bounds on the number of tests for disjunct matrices.
\newblock {\em IEEE Transactions on Information Theory}, 62(12):7518--7521,
  2016.
\newblock \doi{10.1109/TIT.2016.2614726}.

\bibitem{shannon-zero}
C.~{Shannon}.
\newblock The zero error capacity of a noisy channel.
\newblock {\em IRE Transactions on Information Theory}, 2(3):8--19, 1956.
\newblock \doi{10.1109/TIT.1956.1056798}.

\bibitem{Sha57}
C.~E. Shannon.
\newblock Certain results in coding theory for noisy channels.
\newblock {\em Information and Control}, 1(1):6--25, 1957.
\newblock \doi{10.1016/S0019-9958(57)90039-6}.

\bibitem{sharma2}
A.~Sharma and C.~R. Murthy.
\newblock Group testing-based spectrum hole search for cognitive radios.
\newblock {\em IEEE Transactions on Vehicular Technology}, 63(8):3794--3805,
  2014.
\newblock \doi{10.1109/TVT.2014.2305978}.

\bibitem{shental}
N.~Shental, A.~Amir, and O.~Zuk.
\newblock Identification of rare alleles and their carriers using compressed
  se(que)nsing.
\newblock {\em Nucleic Acids Research}, 38(19):e179, 2010.
\newblock \doi{10.1093/nar/gkq675}.

\bibitem{shi}
M.~Shi, T.~Furon, and H.~J{\'e}gou.
\newblock A group testing framework for similarity search in high-dimensional
  spaces.
\newblock In {\em ACM International Conference on Multimedia}, pages 407--416,
  2014.
\newblock \doi{10.1145/2647868.2654895}.

\bibitem{sobel3}
M.~Sobel and R.~Elashoff.
\newblock Group testing with a new goal, estimation.
\newblock {\em Biometrika}, pages 181--193, 1975.
\newblock \doi{10.2307/2334502}.

\bibitem{sobel}
M.~Sobel and P.~A. Groll.
\newblock Group testing to eliminate efficiently all defectives in a binomial
  sample.
\newblock {\em Bell Labs Technical Journal}, 38(5):1179--1252, 1959.
\newblock \doi{10.1002/j.1538-7305.1959.tb03914.x}.

\bibitem{sobel2}
M.~Sobel and P.~A. Groll.
\newblock Binomial group-testing with an unknown proportion of defectives.
\newblock {\em Technometrics}, 8(4):631--656, 1966.
\newblock \doi{10.1080/00401706.1966.10490408}.

\bibitem{soleymani2024non}
M.~Soleymani and T.~Javidi.
\newblock A non-adaptive algorithm for the quantitative group testing problem.
\newblock In {\em Conference on Learning Theory}, pages 4574--4592, 2024.
\newblock \urlprefix\url{https://proceedings.mlr.press/v247/soleymani24a.html}.

\bibitem{soleymani2025learning}
M.~Soleymani and T.~Javidi.
\newblock Learning to ask: Decision transformers for adaptive quantitative
  group testing, 2025, arXiv:
  \burlalt{\texttt{2509.01723}}{http://arxiv.org/abs/2509.01723}.

\bibitem{spielman1996}
D.~A. Spielman.
\newblock Linear-time encodable and decodable error-correcting codes.
\newblock {\em IEEE Transactions on Information Theory}, 42(6):1723--1731,
  1996.
\newblock \doi{10.1109/18.556668}.

\bibitem{stan}
M.~R. Stan, P.~D. Franzon, S.~C. Goldstein, J.~C. Lach, and M.~M. Ziegler.
\newblock Molecular electronics: {F}rom devices and interconnect to circuits
  and architecture.
\newblock {\em Proceedings of the IEEE}, 91(11):1940--1957, 2003.
\newblock \doi{10.1109/JPROC.2003.818327}.

\bibitem{sterrett}
A.~Sterrett.
\newblock On the detection of defective members of large populations.
\newblock {\em The Annals of Mathematical Statistics}, 28(4):1033--1036, 1957.
\newblock \doi{10.1214/aoms/1177706807}.

\bibitem{swallow}
W.~H. Swallow.
\newblock Group testing for estimating infection rates and probabilities of
  disease transmission.
\newblock {\em Phytopathology}, 75(8):882--889, 1985.
\newblock \doi{10.1094/Phyto-75-882}.

\bibitem{tan2025quantitative}
N.~Tan, P.~P. Cobo, and R.~Venkataramanan.
\newblock Quantitative group testing and pooled data in the linear regime with
  sublinear tests.
\newblock {\em IEEE Transactions on Information Theory}, 2025.
\newblock \doi{10.1109/TIT.2025.3611276}.

\bibitem{tan2023approximate}
N.~Tan, P.~Pascual~Cobo, J.~Scarlett, and R.~Venkataramanan.
\newblock Approximate message passing with rigorous guarantees for pooled data
  and quantitative group testing.
\newblock {\em SIAM Journal on Mathematics of Data Science}, 6(4):1027--1054,
  2024.
\newblock \doi{10.1137/23M1604928}.

\bibitem{Tan2022}
N.~Tan, W.~Tan, and J.~Scarlett.
\newblock Performance bounds for group testing with doubly-regular designs.
\newblock {\em IEEE Transactions on Information Theory}, 69(2):1224--1243,
  2022.
\newblock \doi{10.1109/TIT.2022.3210593}.

\bibitem{teo2022noisy}
B.~Teo and J.~Scarlett.
\newblock Noisy adaptive group testing via noisy binary search.
\newblock {\em IEEE Transactions on Information Theory}, 68(5):3340--3353,
  2022.
\newblock \doi{10.1109/TIT.2022.3140604}.

\bibitem{thompson}
K.~H. Thompson.
\newblock Estimation of the proportion of vectors in a natural population of
  insects.
\newblock {\em Biometrics}, 18(4):568--578, 1962.
\newblock \doi{10.2307/2527902}.

\bibitem{tilghman}
M.~Tilghman, D.~Tsai, T.~P. Buene, M.~Tomas, S.~Amade, D.~Gehlbach, S.~Chang,
  C.~Ignacio, G.~Caballero, S.~Espitia, S.~May, E.~V. Noormahomed, and
  D.~Smith.
\newblock Pooled nucleic acid testing to detect antiretroviral treatment
  failure in {HIV}-infected patients in {M}ozambique.
\newblock {\em Journal of Acquired Immune Deficiency Syndromes}, 70(3):256,
  2015.
\newblock \doi{10.1097/QAI.0000000000000724}.

\bibitem{torney1999sets}
D.~C. Torney.
\newblock Sets pooling designs.
\newblock {\em Annals of Combinatorics}, 3(1):95--101, 1999.
\newblock \doi{doi.org/10.1007/BF01609879}.

\bibitem{truong2020all}
L.~V. Truong, M.~Aldridge, and J.~Scarlett.
\newblock On the all-or-nothing behavior of {B}ernoulli group testing.
\newblock {\em IEEE Journal on Selected Areas in Information Theory},
  1(3):669--680, 2020.
\newblock \doi{10.1109/JSAIT.2020.3039790}.

\bibitem{tsfasman}
M.~A. Tsfasman, S.~G. Vl{\u{a}}dut, and T.~Zink.
\newblock Modular curves, {S}himura curves, and {G}oppa codes, better than
  {V}arshamov-{G}ilbert bound.
\newblock {\em Mathematische Nachrichten}, 109(1):21--28, 1982.
\newblock \doi{10.1002/mana.19821090103}.

\bibitem{tu}
X.~M. Tu, E.~Litvak, and M.~Pagano.
\newblock On the informativeness and accuracy of pooled testing in estimating
  prevalence of a rare disease: {A}pplication to {HIV} screening.
\newblock {\em Biometrika}, pages 287--297, 1995.
\newblock \doi{10.2307/2337408}.

\bibitem{ubaru2020multilabel}
S.~Ubaru, S.~Dash, A.~Mazumdar, and O.~Gunluk.
\newblock Multilabel classification by hierarchical partitioning and
  data-dependent grouping.
\newblock {\em Advances in Neural Information Processing Systems},
  33:22542--22553, 2020.
\newblock \urlprefix\url{https://dl.acm.org/doi/10.5555/3495724.3497614}.

\bibitem{ubaru2017multilabel}
S.~Ubaru and A.~Mazumdar.
\newblock Multilabel classification with group testing and codes.
\newblock In {\em International Conference on Machine Learning}, volume~70,
  pages 3492--3501, 2017.
\newblock \urlprefix\url{https://proceedings.mlr.press/v70/ubaru17a.html}.

\bibitem{varanasi}
M.~K. Varanasi.
\newblock Group detection for synchronous {G}aussian code-division
  multiple-access channels.
\newblock {\em IEEE Transactions on Information Theory}, 41(4):1083--1096,
  1995.
\newblock \doi{10.1109/18.391251}.

\bibitem{vazirani}
V.~V. Vazirani.
\newblock {\em Approximation Algorithms}.
\newblock Springer, 2001.
\newblock \doi{10.1007/978-3-662-04565-7}.

\bibitem{wadayama17}
T.~Wadayama.
\newblock Nonadaptive group testing based on sparse pooling graphs.
\newblock {\em IEEE Transactions on Information Theory}, 63(3):1525--1534,
  2017.
\newblock \doi{10.1109/TIT.2016.2621112}.
\newblock See also \cite{wadayama2}.

\bibitem{wadayama2}
T.~Wadayama.
\newblock Comments on `nonadaptive group testing based on sparse pooling
  graphs'.
\newblock {\em IEEE Transactions on Information Theory}, 64(6):4686--4686,
  2018.
\newblock \doi{10.1109/TIT.2018.2827463}.

\bibitem{walter}
S.~D. Walter, S.~W. Hildreth, and B.~J. Beaty.
\newblock Estimation of infection rates in populations of organisms using pools
  of variable size.
\newblock {\em American Journal of Epidemiology}, 112(1):124--128, 1980.
\newblock \doi{10.1093/oxfordjournals.aje.a112961}.

\bibitem{wang2018}
C.~Wang, Q.~Zhao, and C.-N. Chuah.
\newblock Optimal nested test plan for combinatorial quantitative group
  testing.
\newblock {\em IEEE Transactions on Signal Processing}, 66(4):992--1006, 2018.
\newblock \doi{10.1109/TSP.2017.2780053}.

\bibitem{Wang2023}
H.-P. Wang, R.~Gabrys, and V.~Guruswami.
\newblock Quickly-decodable group testing with fewer tests:
  {P}rice–{S}carlett’s nonadaptive splitting with explicit scalars.
\newblock In {\em IEEE International Symposium on Information Theory (ISIT)},
  pages 1609--1614, 2023.
\newblock \doi{10.1109/ISIT54713.2023.10206843}.

\bibitem{wang2023tropical}
H.-P. Wang, R.~Gabrys, and A.~Vardy.
\newblock Tropical group testing.
\newblock {\em IEEE Transactions on Information Theory}, 69(9):6098--6120,
  2023.
\newblock \urlprefix\url{10.1109/TIT.2023.3282847}.

\bibitem{wang2024isolate}
H.-P. Wang and V.~Guruswami.
\newblock Isolate and then identify: Rethinking adaptive group testing.
\newblock In {\em IEEE International Symposium on Information Theory (ISIT)},
  pages 3231--3236, 2024.
\newblock \doi{10.1109/ISIT57864.2024.10619098}.

\bibitem{wang2016data}
I.-H. Wang, S.-L. Huang, K.-Y. Lee, and K.-C. Chen.
\newblock Data extraction via histogram and arithmetic mean queries:
  Fundamental limits and algorithms.
\newblock In {\em IEEE International Symposium on Information Theory (ISIT)},
  pages 1386--1390, 2016.
\newblock \doi{10.1109/ISIT.2016.7541526}.

\bibitem{wang11}
J.~Wang, E.~Lo, and M.~L. Yiu.
\newblock Identifying the most connected vertices in hidden bipartite graphs
  using group testing.
\newblock {\em IEEE Transactions on Knowledge and Data Engineering},
  25(10):2245--2256, 2013.
\newblock \doi{10.1109/TKDE.2012.178}.

\bibitem{wang2011evolution}
L.~Wang, X.~Li, Y.-Q. Zhang, Y.~Zhang, and K.~Zhang.
\newblock Evolution of scaling emergence in large-scale spatial epidemic
  spreading.
\newblock {\em {PLoS ONE}}, 6(7):e21197, 2011.
\newblock \doi{10.1371/journal.pone.0021197}.

\bibitem{wolf}
J.~K. Wolf.
\newblock Born again group testing: {M}ultiaccess communications.
\newblock {\em IEEE Transactions on Information Theory}, 31(2):185--191, 1985.
\newblock \doi{10.1109/TIT.1985.1057026}.

\bibitem{wu5}
S.~Wu, S.~Wei, Y.~Wang, R.~Vaidyanathan, and J.~Yuan.
\newblock Achievable partition information rate over noisy multi-access
  {B}oolean channel.
\newblock In {\em IEEE International Symposium on Information Theory (ISIT)},
  pages 1206--1210, 2014.
\newblock \doi{10.1109/ISIT.2014.6875024}.

\bibitem{wu4}
S.~Wu, S.~Wei, Y.~Wang, R.~Vaidyanathan, and J.~Yuan.
\newblock Partition information and its transmission over {B}oolean
  multi-access channels.
\newblock {\em IEEE Transactions on Information Theory}, 61(2):1010--1027,
  2015.
\newblock \doi{10.1109/TIT.2014.2375211}.

\bibitem{xhemrishi2023fedgt}
M.~Xhemrishi, J.~Östman, A.~Wachter-Zeh, and A.~Graell~i Amat.
\newblock {FedGT}: Identification of malicious clients in federated learning
  with secure aggregation.
\newblock {\em IEEE Transactions on Information Forensics and Security},
  20:2577--2592, 2025.
\newblock \doi{10.1109/TIFS.2025.3539964}.

\bibitem{xu}
W.~Xu, M.~Wang, E.~Mallada, and A.~Tang.
\newblock Recent results on sparse recovery over graphs.
\newblock In {\em Asilomar Conference on Signals, Systems and Computers}, pages
  413--417, 2011.
\newblock \doi{10.1109/ACSSC.2011.6190031}.

\bibitem{xuan}
Y.~Xuan, I.~Shin, M.~T. Thai, and T.~Znati.
\newblock Detecting application denial-of-service attacks: {A}
  group-testing-based approach.
\newblock {\em IEEE Transactions on Parallel and Distributed Systems},
  21(8):1203--1216, 2010.
\newblock \doi{10.1109/TPDS.2009.147}.

\bibitem{zaman}
N.~Zaman and N.~Pippenger.
\newblock Asymptotic analysis of optimal nested group-testing procedures.
\newblock {\em Probability in the Engineering and Informational Sciences},
  30(4):547--552, 2016.
\newblock \doi{10.1017/S0269964816000267}.

\bibitem{zhanghuang}
W.~{Zhang} and L.~{Huang}.
\newblock On or many-access channels.
\newblock In {\em IEEE International Symposium on Information Theory (ISIT)},
  pages 2638--2642, 2017.
\newblock \doi{10.1109/ISIT.2017.8007007}.

\bibitem{zhou2025twenty}
L.~Zhou and A.~O. Hero.
\newblock Twenty questions with random error.
\newblock {\em Foundations and Trends in Communications and Information
  Theory}, 22(4):394--604, 2025.
\newblock \doi{10.1561/0100000144}.

\end{thebibliography}

\end{document}